\documentclass{elsarticle}

\usepackage{epsfig}
\usepackage{amssymb}
\usepackage{graphicx}
\usepackage{nccpic}
\usepackage{amsmath}
\usepackage{pdfpages}

\newcommand{\bfl}{\begin{flushleft}}
\newcommand{\efl}{\end{flushleft}}
\newcommand{\beq}{\begin{equation}}
\newcommand{\eeq}{\end{equation}}

\newcommand{\be}{\begin{equation}}
\newcommand{\ee}{\end{equation}}

\newcommand{\bea}{\begin{eqnarray}}
\newcommand{\eea}{\end{eqnarray}}

\newcommand{\nnl}{\nonumber\\}
 
 \def\gsim{\mathrel{\rlap{\lower0.2em\hbox{$\sim$}}\raise0.2em\hbox{$>$}}}
 \def\gtrless{\mathrel{\rlap{\lower0.0em\hbox{$>$}}\raise0.41em\hbox{$<$}}}
 \def\lessgtr{\mathrel{\rlap{\lower0.0em\hbox{$<$}}\raise0.41em\hbox{$>$}}}

\journal{Progress in Particle and Nuclear Physics}

\begin{document}

\begin{frontmatter}

\title{Effective QCD and transport description of dilepton and photon
production in heavy-ion collisions and elementary processes}

\author[a]{O.~Linnyk}
\author[b]{E.~L.~Bratkovskaya}
\author[a]{W.~Cassing}

% Address

\address[a]{Institut f\"ur Theoretische Physik, %
  Justus Liebig Universit\"at Giessen, %
  %\\
  %Heinrich--Buff--Ring 16, %
  35392 Giessen, %
  Germany}

\address[b]{Institut f\"ur Theoretische Physik, %
 Goethe Universit{\" a}t Frankfurt am Main, %
 %\\
 %Max-von-Laue-Str. 1, %
 60438 Frankfurt am Main, %
 Germany }

%%%abstract
\begin{abstract}
In this review we address the dynamics of relativistic heavy-ion
reactions and in particular the information obtained from
electromagnetic probes that stem from the partonic and hadronic
phases. The {out-of-equilibrium} description of strongly interacting
relativistic fields is based on the theory of Kadanoff and Baym.
For the modeling of the partonic phase  we
introduce an effective dynamical quasiparticle model (DQPM) for QCD
{\em in equilibrium}. In the DQPM, the widths and masses of the
dynamical quasiparticles are controlled by transport coefficients
that can be compared to the corresponding quantities from lattice
QCD. The resulting off-shell transport approach is denoted by
Parton-Hadron-String Dynamics (PHSD) which includes covariant
dynamical transition rates for hadronization and keeps track of the
hadronic interactions in the final phase. It is shown that the PHSD
captures the bulk dynamics of heavy-ion collisions from lower SPS to
LHC energies and thus provides a solid basis for the evaluation of
the electromagnetic emissivity on the basis of the same dynamical
parton propagators that are employed for the dynamical evolution of
the partonic system. The production of {\em direct} photons in
elementary processes and heavy-ion reactions is discussed and the present
status of the photon $v_2$ ``puzzle" -- a large elliptic flow $v_2$
of the {\em direct} photons experimentally observed in  heavy-ion
collisions - is addressed for nucleus-nucleus reactions at RHIC and
LHC energies. The role of hadronic and partonic sources for the
photon spectra and the flow coefficients $v_2$ and $v_3$ are
considered as well as the possibility to subtract the QGP signal
from the experimental observables. Furthermore, the production of
$e^+e^-$ or $\mu^+ \mu^-$ pairs in elementary processes and A+A reactions is
addressed. The calculations within the PHSD from SIS to LHC energies
show an increase of the low mass dilepton yield essentially due to
the in-medium modification of the $\rho$-meson and at the lowest
energy also due to a multiple regeneration of $\Delta$-resonances.
Furthermore, pronounced traces of the partonic degrees-of-freedom
are found in the intermediate dilepton mass regime (1.2 GeV $< M < $
3 GeV) at relativistic energies, which will also shed light on the
nature of the very early degrees-of-freedom in nucleus-nucleus
collisions.
\end{abstract}


\end{frontmatter}
\tableofcontents
%\listoffigures
%\listoftables

%-------------------------------------------------------------------
%

% part of the review in PPNP 2015
%%%%%%%

\section{Introduction}
Present experiments at the Relativistic Heavy-Ion Collider (RHIC) or
the Large Hadron Collider (LHC) have reached for short time scales
the conditions met in the first micro-seconds in the evolution of
the universe after the 'Big Bang'. The 'Big Bang' scenario implies
that on these time scales the entire state has emerged from a
partonic system of quarks, antiquarks and gluons -- a quark-gluon
plasma (QGP) -- to color neutral hadronic matter consisting of
interacting hadronic states (and resonances) in which the partonic
degrees-of-freedom are confined. The nature of confinement and the
dynamics of this phase transition  is still an outstanding question
of today's physics. Early concepts of the QGP were guided by the
idea of a weakly interacting system of massless partons which might
be described by perturbative QCD (pQCD). However, experimental
observations at RHIC and LHC indicated that the new medium created
in ultra-relativistic heavy-ion collisions is interacting more
strongly than hadronic matter. It is presently widely accepted that
this medium is an almost perfect liquid of partons as suggested
experimentally from the strong radial expansion and the scaling of
the elliptic flow $v_2(p_T)$ of mesons and baryons with the number
of constituent quarks and antiquarks. While the last years have been
devoted to explore the collective and transport properties of this
partonic medium, the present focus lies on the electromagnetic
emissivity of the new type of matter, i.e. its emission of {\em
direct} photons or dilepton pairs. Since the system is initially far
from equilibrium and no clear evidence has been achieved so far that
an early equilibration at times of the order of 0.5 - 1.0 fm/c is
achieved, microscopic studies based on non-equilibrium dynamics are
mandatory.

Non-equilibrium many-body theory or transport theory has become a
major topic of research in nuclear physics, in cosmological particle
physics as well as condensed matter physics. The multidisciplinary
aspect arises due to a common interest to understand the various
relaxation phenomena of quantum dissipative systems. Important
questions in nuclear and particle physics at the highest energy
densities are: i) how do nonequilibrium systems in extreme
environments  evolve, ii) how do they eventually thermalize, iii)
how phase transitions occur in real time with possibly
nonequilibrium remnants? The dynamics of heavy-ion collisions at
various bombarding energies provide the laboratory of choice for
research on nonequilibrium quantum many-body physics and
relativistic quantum-field theories, since the initial state of a
collision resembles an extreme non-equilibrium configuration while the
final state might even exhibit a certain degree of thermalization.

Especially the powerful method of the `Schwinger-Keldysh'
\cite{Schwinger:1960qe,BAKSHI:1963BN,Keldysh:1964ud,Keldysh:1964udi,Cr68} or `closed
time path' (CTP) real-time Greens functions -- being the essential
degrees-of-freedom -- has been shown to provide an appropriate basis
for  the formulation of the complex problems in the various  areas
of nonequilibrium quantum many-body physics. Within this framework
one can derive  valid approximations - depending, of course, on the
problem under consideration - by preserving  overall consistency
relations. Originally, the resulting causal Dyson-Schwinger equation
of motion for the one-particle Greens functions (or two-point
functions), i.e. the Kadanoff-Baym (KB) equations \cite{KB}, have
served as the underlying scheme for deriving various transport
phenomena and generalized transport equations. For review articles
on the Kadanoff-Baym equations in the various areas of
nonequilibrium quantum physics we refer the reader to Refs.
\cite{Bo,DuBois,Danielewicz:1982kk,Chou:1984es,RAMMER:1986ZZ,Calzetta:1986cq,Haug}.

On the other hand, kinetic transport theory is a convenient tool to
study many-body nonequilibrium systems, non-relativistic or
relativistic. Kinetic equations, which do play the central role in
more or less all practical  simulations, can be derived from KB
equations within suitable approximations. Hence, a major impetus in
the past has been to derive semi-classical Boltzmann-like transport
equations within the standard quasi-particle approximation.
Additionally, off-shell extensions by means of a gradient expansion
in the space-time inhomogeneities - as already introduced by
Kadanoff and Baym \cite{KB} - have been formulated for various
directions in physics: from a relativistic electron-photon plasma
\cite{BB72} to the transport of nucleons at intermediate heavy-ion
reactions \cite{botmal}, for transport of particles in scalar
$\Phi^4$-theory \cite{Calzetta:1986cq,danmrow} to the transport of
partons in high-energy heavy-ion reactions
\cite{Makhlin:1994ew,Makhlin:1998zi,Geiger:1995ak,Geiger:1996ym,BD98,BI99}.
We recall that on the formal level of the KB-equations the various
forms assumed for the self-energy have to fulfill consistency
relations in order to preserve symmetries of the fundamental
Lagrangian \cite{KB,knoll1,knoll2}. This allows also for a unified
treatment of stable and unstable (resonance) particles. In this
review we will briefly sketch the derivation of the KB equations and
of the  off-shell transport equations in first-order gradient
expansion in Section 2.

The perspectives to solve QCD in Minkowski space for out of
equilibrium configurations and non-vanishing quark densities will be
very low also in the next years such that effective approaches are
necessary to model the dominant properties of QCD in equilibrium,
i.e. the thermodynamic quantities as well as transport coefficients.
To this aim the dynamical quasiparticle model (DQPM) has been
introduced which is based on partonic propagators with sizeable
imaginary parts (or broad spectral functions). We will briefly
recall the basic definitions in the DQPM and its results for the QCD
equation of state  in Section 3.

By merging off-shell transport theory (Section 2) and the DQPM
(Section 3) we obtain the Parton-Hadron-String-Dynamics (PHSD)
transport approach that incorporates additionally a dynamical
transition from partonic to hadronic degrees-of-freedom  without
violating the second law of thermodynamics (Section 4). This
approach will provide the background dynamics for the study of the
electromagnetic emissivity from relativistic heavy-ion collisions.
Before proceeding to the actual results on the photon and dilepton
production, the PHSD calculations are first confronted with
differential hadron spectra and single-particle collective flow
coefficients $v_2$ and $v_3$ of charged hadrons produced in
heavy-ion collisions from the lower Super-Proton-Synchrotron (SPS)
to LHC energies.

Section 5 is devoted to the implementation of the photon and
dilepton production channels in a non-equilibrium off-shell transport
approach and the computation of the individual production cross sections.
The radiation from the interactions of the broad
quasiparticles as effective degrees-of-freedom in the QGP is
calculated, thus employing the same effective propagators in the
evaluation of the electromagnetic radiation as used for the time
evolution of the partonic system. Additionally, electromagnetic radiation by
hadron decays and  their mutual interactions is implemented for an
extended number of hadronic species. In particular, the photon
production in the bremsstrahlung processes such as
$\pi+\pi\to\pi+\pi+\gamma$ is calculated beyond the soft-photon
approximation (SPA) using the  full cross sections within a
one-boson-exchange (OBE) model.

In Section 6, we present the results on the production of real
photons in  heavy-ion reactions where we will focus on the
transverse momentum spectra (at midrapidity) and the collective flow
of the {\em direct} photons in order to shed some light on the
present ``photon $v_2$ puzzle". Results from PHSD for the triangular
flow $v_3$ at RHIC and LHC energies will be presented, too. In
Section 7, we continue with the production of dilepton pairs from
elementary and heavy-ion collisions from SIS up to LHC energies.
Special attention here is paid to the low-mass enhancement of the
dilepton pairs (0.2 GeV$ < M < 0.7$~GeV) and to the intermediate
mass range (1.2 GeV $< M <$ 3 GeV)  in order to explore the possible
contribution from a QGP phase and/or in-medium modifications of
vector mesons in the hadronic phase. A summary in Section 8
completes this review.

\section{Relativistic dynamics of many-body systems and off-shell transport}

\begin{figure}[h]
\begin{center}
{\includegraphics[width=8cm]{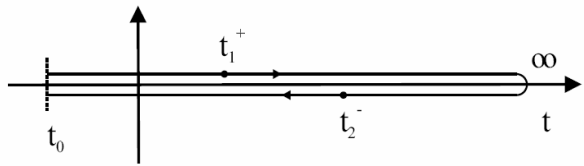} }
\end{center}
\caption{The Keldysh-contour for the time integration in the
Heisenberg picture.} \label{abb:kontur}
\end{figure}

 Relativistic formulations of the many-body problem are
essentially described within covariant field theory. Since the
fields themselves are distributions in space-time $x= (t,{\bf x})$
one  uses the Heisenberg picture for convenience. In the Heisenberg
picture the time evolution of the system is described by
time-dependent operators that are evolved  with the help of the
time-evolution operator $\hat U(t,t')$ which follows
 \begin{equation}
 \label{SEI1}
i\frac{\partial\hat U(t,t_0)}{\partial t}=\hat H(t)\hat{U}(t,t_0)  ,
\end{equation} with ${\hat H}(t)$ denoting the Hamilton operator of the system at time $t$.
 Eq. (\ref{SEI1}) is formally solved by
\be \hat U(t,t_0)=T\left(\exp\left[-i \int_{t_0}^{t} \mathrm{d}z
~\hat H(z)\right] \right) = \sum_{n=0}^\infty \frac{T[-i
\int_{t_0}^{t} \mathrm{d}z ~\hat H(z)]^n}{n!}   , \ee
where $T$ denotes the time-ordering operator, which is also denoted
as Dyson series. Let's assume that the initial state is given by
some density matrix $\hat\rho$, which may be a pure or mixed state,
then the time evolution of any operator ${\hat O}$ in the Heisenberg
picture from time $t_0$ to $t$ is given by \be O(t)=\langle \hat
O_H(t)\rangle=\mathrm{Tr}\left(\hat\rho\,\hat
O_H(t)\right)=\mathrm{Tr}\left(\hat \rho\,\hat U(t_0,t)\hat O\,\hat
U(t,t_0)\right) =\mathrm{Tr}\left(\hat \rho\,\hat
U^\dagger(t,t_0)\hat O\,\hat U(t,t_0)\right) . \label{gl:dichte} \ee
Eq. (\ref{gl:dichte}) implies that first the system is evolved from $t_0$ to $t$ and
then backward from $t$ to  $t_0$. This may be expressed as a time
integral along the (Keldysh-)contour shown in Fig. \ref{abb:kontur}.

\subsection{Two-point functions}

 Now Green functions on the Keldysh-contour may have time arguments
on the same branch of the contour or on opposite branches. This
gives four possibilities for the Green functions defined -- in case
of a field theory with only scalar fields $\phi(x)$ (for sake of illustration) --  by
\begin{eqnarray}
iG^c(x,y)&=iG^{++}(x,y)=\left\langle \right.\hat T^c(\phi(x)\phi(y))\left.\right\rangle \label{gl:green1}\\
iG^< (x,y)&=iG^{+-}(x,y)=\left\langle \phi(y)\phi(x)\right\rangle \\
iG^>(x,y)&=iG^{-+}(x,y)=\left\langle \phi(x)\phi(y)\right\rangle \\
iG^a(x,y)&=iG^{--}(x,y)=\left\langle \right.\hat
T^a(\phi(x)\phi(y))\left.\right\rangle  ,\label{green2}
\end{eqnarray} which are not independent! Here $x=(x^0, {\bf x})$ and $y=(y^0, {\bf y})$. Time-ordering has to be fulfilled if both time arguments are on the same axis. The causal time-ordering operator $T^c$ places fields at later times to the left while the anticausal operator  $T^a$ places fields at later times to the right.
The Green functions  $G^>$ and $G^< $  are denoted as{ \bf Wightman
functions} and will play the essential role in the dynamical
description of the system. One may also write the Green function on
the Keldysh-contour in terms of a 2x2 matrix
\begin{eqnarray}
G(x,y)=\bordermatrix{  & + & - \cr  + & G^c(x,y) & G^{<}(x,y) \cr -
& G^{>}(x,y) & G^{a}(x,y)}  .
\end{eqnarray}
Note that the Green functions defined in Eqs. (\ref{gl:green1}) to
(\ref{green2}) are two-point functions, i.e. they correspond to a
single-particle degree-of-freedom!

 The further derivation starts with the Dyson equation for $G(x,y)$,
\begin{equation} \label{DS0}
G(x,y)=G_0(x,y)+[G_0 \Sigma G](x,y) ,
\end{equation} with $G_0(x,y)$ denoting the bare Green function.
The selfenergy $\Sigma(x,y)$ has the meaning of a one-body
mean-field potential and in lowest order for fermions is given by
the Hartree-Fock potential ($\times 2M$) since in the relativistic
case $\Sigma$ has the dimension $[$energy$]^2$.

 The relation to the one-body density matrix $\rho$ - as
employed in density-matrix theory  \cite{Wang} - is given by \be
\rho({\bf x}, {\bf x}';t) = i G^<( {\bf x}, {\bf x}';t,t) , \ee since
the time diagonal Green function can be identified with an integral
over the energy variable $\omega$ using
\be  G^<( {\bf x}, {\bf x}';\omega,t) = \int_{-\infty}^\infty d (\tau-\tau') \ \exp(i \omega (\tau-\tau'))
\ G^<( {\bf x}, {\bf x}';\tau,\tau') \ee
 (for $t=(\tau+\tau')/2$), i.e.
\be G^<( {\bf x}, {\bf x}';t) = \int_{-\infty}^\infty \frac{d \omega}{2 \pi}
\ G^<( {\bf x}, {\bf x}';\omega,t) . \label{DS0a} \ee Two-point
functions  $F$ on the closed-time-path (CTP) generally can be
expressed by retarded and advanced components as
\begin{equation}  \label{gl:retadv}
F^{{R}}(x,y)=F^c(x,y)-F^<(x,y)=F^>(x,y)-F^a(x,y)  , \end{equation} $$
F^{{A}}(x,y)=F^c(x,y)-F^>(x,y)=F^<(x,y)-F^a(x,y)$$
 giving in particular the relation \be
 F^{{R}}(x,y)-F^{{A}}(x,y) = F^>(x,y)- F^<(x,y) .
 \ee
Note that the advanced and retarded components of the Green
functions only contain spectral and no statistical information, \be
G^{{R/A}}(x,y) = G_0(x,y) \ \delta(t_1-t_2) \pm \Theta(\pm(t_1-t_2))\
[G^>(x,y)-G^<(x,y)]. \ee

\subsection{The Dyson-Schwinger equation}

 The Dyson-Schwinger equation (\ref{DS0}) on the closed-time path reads in matrix form:
\begin{equation}
%\begin{split}\left( \begin{array}{rr} 0 & 1 \\ 1 & 0 \end{array} \right)
\left( \begin{array}{rr} G^c(x,y) & G^<(x,y)\\ G^>(x,y) & G^a(x,y)
\end{array} \right) = \left(  \begin{array}{rr} G^c_0(x,y) &
G^<_0(x,y) \\ G^>_0(x,y) & G^a_0(x,y) \end{array} \right) +
\nonumber \ee \be \left(  \begin{array}{rr} G^c_0(x,x') &
G^<_0(x,x') \\ G^>_0(x,x') & G^a_0(x,x') \end{array}  \right) \odot
\left(  \begin{array}{rr} \Sigma^c(x',y') & -\Sigma^<(x',y')\\
-\Sigma^>(x',y') & \Sigma^a(x',y')  \end{array} \right) \ \odot
\left(  \begin{array}{rr} G^c(y',y) & G^<(y',y)\\ G^>(y',y) &
G^a(y',y)  \end{array} \right) ,~~~~~ \label{eq:dyson} \ee where the
symbol $\odot$ stands for an intermediate integration over
space-time on the CTP, i.e. $x'$ or $y'$. The selfenergy $\Sigma$ on
the CPT is defined along Eq. (\ref{gl:retadv}) and incorporates
interactions of higher order. In lowest order $\Sigma/2M$ is given
by the Hartree or Hartree-Fock mean-field in the non-relativistic
limit (in case of fermions) but it follows a nonperturbative
expansion \cite{Bo}.

\subsection{Kadanoff-Baym equations}

 To derive the {\bf Kadanoff-Baym equations} one multiplies Eq. (\ref{eq:dyson})
 with the inverse free Green function (operator) $G_{0x}^{-1}= -(\partial_\mu^x \partial^\mu_x + m^2)$ from the left.
This gives four equations which can be cast into the form:
\be
-(\partial_\mu^x\partial^\mu_x+m^2)G^{{R/A}}(x,y)=\delta(x-y)+
\Sigma^{R/A}(x,x')\odot G^{R/A}(x',y) ,\label{gl:kaba1} \ee \be
-(\partial_\mu^x\partial^\mu_x+m^2)G^{<} (x,y)=\Sigma^{R}(x,x')
\odot G^{<}(x',y)+\Sigma^{<}(x,x') \odot G^{A}(x',y)  ,
\label{gl:kaba2} \ee \be -(\partial_\mu^x\partial^\mu_x+m^2)G^{>}
(x,y) =\Sigma^{R}(x,x')\odot G^{>}(x',y)+\Sigma^{>}(x,x') \odot
G^{A}(x',y)  . \label{gl:kaba2b} \ee The propagation of the Green
functions in the variable $y$ is defined by the adjoint equations:
\be -(\partial_\mu^y\partial^\mu_y+m^2)G^{{R/A}}(x,y)=\delta(x-y)+
G^{R/A}(x,x') \odot\Sigma^{R/A}(x',y) ,\label{gl:kaba3} \ee \be
-(\partial_\mu^y\partial^\mu_y+m^2)G^<(x,y)=G^{R}(x,x')\odot\Sigma^<(x',y)+
G^<(x,x')\odot \Sigma^{A}(x',y)  , \ee \be
-(\partial_\mu^y\partial^\mu_y+m^2)G^>(x,y)=G^{R}(x,x')\odot\Sigma^>(x',y)
+G^>(x,x')\odot \Sigma^{A}(x',y)  . \label{gl:kaba4} \ee Note again
that the evolution of the retarded/advanced Green functions only
depends on retarded/advanced quantities.

\subsection*{Definition of selfenergies}

For the solution of the KB equations the computation/fixing of the (two-point) selfenergies $\Sigma$ is mandatory.
In the context of field theory the latter is extracted from the effective action
\begin{equation}
\Gamma[G]=\Gamma^0[G_0] +\frac{i}2[\ln(1- G_0 \Sigma) + G
\Sigma]+\Phi[G]  \label{gl:effektive Wirkung}
\end{equation}
assuming a vanishing vacuum expectation value  $\langle 0|\phi(x)|0\rangle$.
 Here $\Gamma^0[G_0]$ only depends on the free Green function $G_0$ and
can be considered as constant in the following. Note that all
internal and external integrations in (\ref{gl:effektive Wirkung})
have to be performed over the CTP. In $\Phi[G]$ all closed
two-particle irreducible (2PI) diagrams are included in lowest
(nontrivial) order. We recall that 2PI diagrams are those that cannot be separated
in two disjunct diagrams by cutting two propagator lines; formally
this implies that after second order differentiation with respect to
$G$ no separate diagrams survive.

 For the derivation of selfenergies one now considers
the variation of the action $\Gamma[G]$ with respect to $G$
requiring $\delta \Gamma= 0$,
\begin{equation}
\delta \Gamma=0= \frac{i}{2}\Sigma\,\delta G
-\frac{i}{2}\frac{G_0}{1-G_0\,\Sigma}\delta\Sigma
+\frac{i}{2}G\,\delta\Sigma +\delta\Phi =\frac{i}{2}\Sigma\,\delta G
-\frac{i}{2} \underbrace{\frac{1}{G_0^{-1}-\Sigma}}_{=G}
\delta\Sigma +\frac{i}{2}G\,\delta\Sigma +\delta\Phi
=\frac{i}{2}\Sigma\,\delta G +\delta \Phi . \ee
\begin{equation}
\Rightarrow \Sigma=2i\frac{\delta \Phi}{\delta G}  .
~~~~~~~~~~~~~~~~~~~~~~~~~~~\label{gl:selbstenergie}
\end{equation}
%Note that $iG^<$ plays the role of the one-body density matrix in non-relativistic formulations.
The selfenergies thus are obtained by opening of a
propagator-line in the irreducible diagrams $\Phi$. Note that this
definition of the selfenergy preserves all conservation laws of
the theory (as well as causality) and does not introduce additional conserved
currents. In principle the $\Phi$-functional includes irreducible diagrams up to infinite order,
but here we will consider only the contributions up to second order in the coupling (2PI).
 For our present purpose this approximation
is sufficient since we include the leading mean-field effects as
well as the leading order scattering processes that pave the way to
thermalization.

\subsection{Spectral function}

 The spectral function of the fields $\phi$ is of particular interest since it follows from the
field commutator at unequal times and reflects the quantization of the theory. For scalar, symmetric fields
$\phi$ it is given by
\begin{equation}
A(x,y)=\left<\right.[\phi(x),\phi(y)]_-\left.\right>=i[G^>(x,y)-G^<(x,y)]
=i[G^R(x,y)-G^A(x,y)] .\label{gl:spektral0}
\end{equation}
For homogenous systems in space we have  in momentum-time representation
\begin{equation} \label{specfu}
A({\bf p},t_1,t_2)=i[G^>({\bf p},t_1,t_2)-G^<({\bf p}, t_1,t_2)]=
i\left[-[G^<({\bf p},t_1,t_2)]^*-G^<({\bf p}, t_1,t_2)\right] . \ee
The quantity (\ref{specfu}) is displayed in Fig. \ref{spc} (l.h.s.)
as a function of $\Delta t=t_1-t_2$ and ${t} = (t_1+t_2)/2$ for a
low lying momentum mode in case of the $\phi^4$-theory  for strong
coupling $\lambda$ as evaluted numerically in Ref.
\cite{Juchem:2003bi}. We observe a damped oscillation in $\Delta t$
(for $\Delta t \geq 0$) in all cases with characteristic time scale
$1/\gamma$ which practically does not depend on the average time ${
t}$. This pattern is very similar for all momentum modes (cf. Ref.
\cite{Juchem:2003bi}).

The spectral function in energy-momentum representation is obtained
by Fourier transformation with respect to the time difference
$\Delta t = (t_1-t_2)$ for each average time $t$:
\begin{equation} \label{FTOM}
 A({\bf p},p_0,t)=\int_{-\infty}^\infty {d} \Delta t
~\exp(i\Delta t\,p_0) A({\bf p},t_1=t+\Delta t/2,t_2=t-\Delta t/2) .
\end{equation}
Since the spectral function essentially shows a damped oscillation in $t_1-t_2$ (cf. Fig. \ref{spc}, l.h.s.)) this implies  that the Fourier transform (\ref{FTOM}) is of
relativistic Breit-Wigner shape with a width $\gamma$ that describes the decay in the relative time $\Delta t$  (r.h.s. of Fig. \ref{spc}). The spectral shape can be well approximated by \be \label{bw}
A (p_0, {\bf p})  \ = \frac{\gamma}{2\tilde{E}}
\biggl(\frac{1}{(p_0 - \tilde{E})^2 + \gamma^2} -
\frac{1}{(p_0 + \tilde{E})^2 + \gamma^2} \biggr)
 = \frac{2 p_0 \gamma}{(p_0^2
- {\bf p}^2 - M^2)^2 + 4 \gamma^2 p_0^2} \end{equation}
with ${\tilde E}^2 = {\bf p}^2 + M^2- \gamma^2$ where $M$ denotes the mass of the degrees-of-freedom. We will come back to this functional form in Section 3.

\begin{figure}
\includegraphics[width=0.49\textwidth]{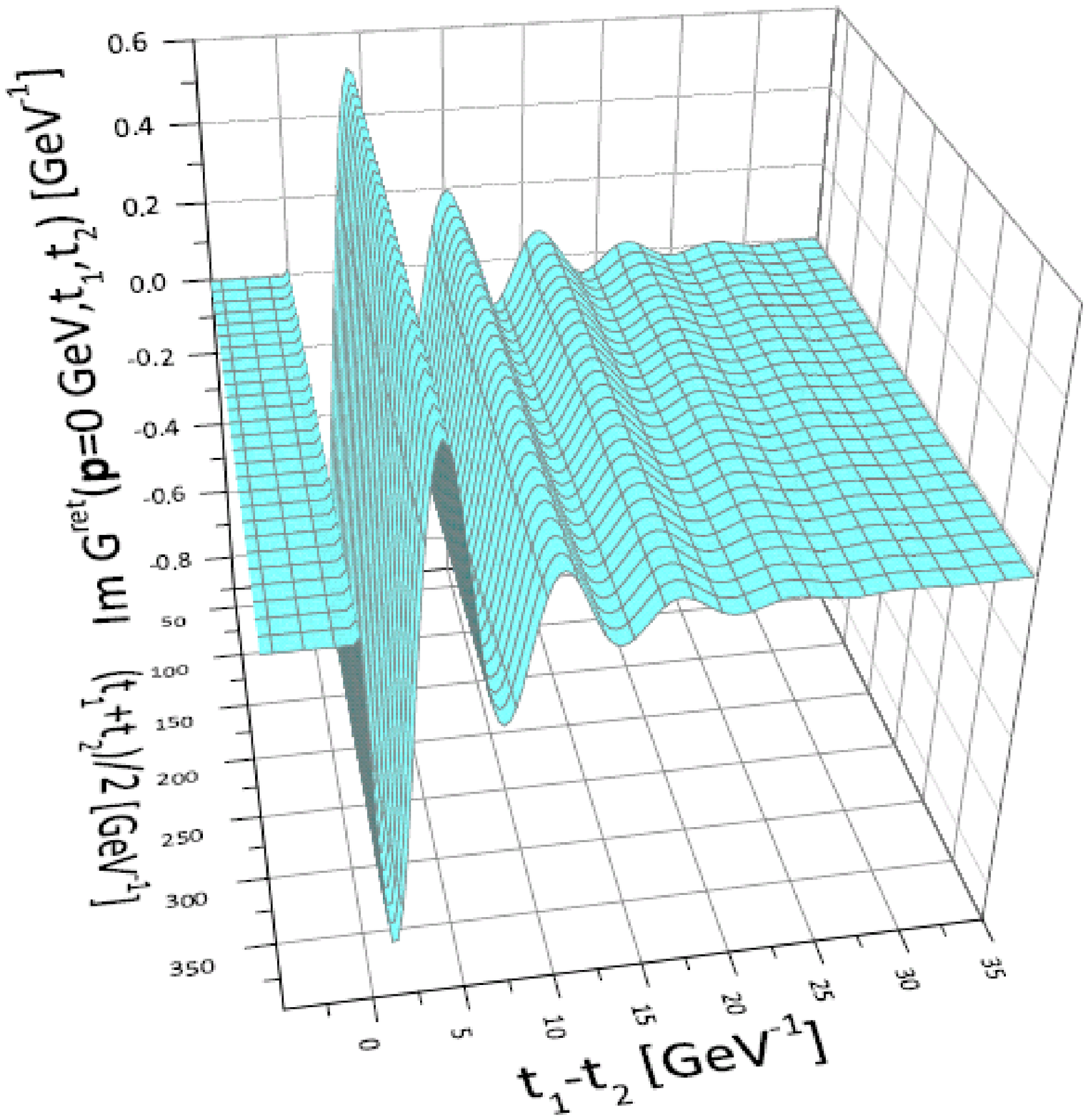} \hspace{-1cm}
\includegraphics[width=0.49\textwidth]{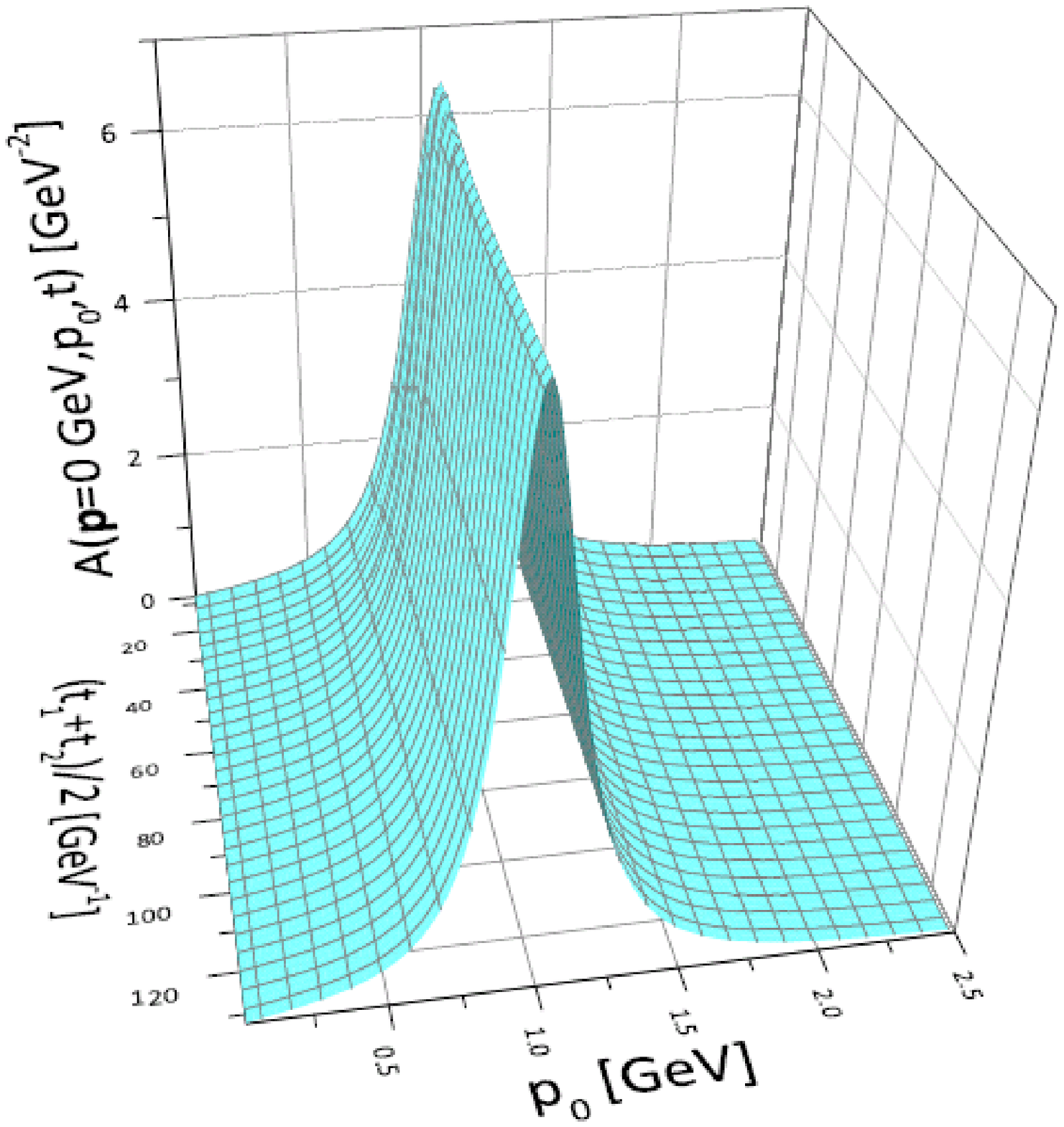}
   \caption{(l.h.s.) The imaginary part of the retarded Green function as a function
   of $t_1-t_2$ and the average time $(t_1+t_2)/2$ for $\phi^4$ -theory in
   strong coupling as emerging from the Kadanoff-Baym approach (cf. Ref. \protect\cite{Juchem:2003bi}.
   (r.h.s.) The Fourier transform (\ref{FTOM}) in energy $p_0$ for the momentum mode ${\bf p}$ = 0 is displayed on the l.h.s.
   in case of $\phi^4$-theory for strong coupling.  }
   \label{spc}
\end{figure}

\subsection{The equilibrium distribution}

 Now we introduce the energy and momentum-dependent distribution
function $ N({\bf p},p_0,\bar{t})$ at any system time $\bar{t}$ in case of scalar bosons by the definition
\bea i \, G^{<}({\bf p},p_0,\bar{t}) & = & A({\bf p},p_0,\bar{t}) \;
\; N({\bf p},p_0,\bar{t}) \, , \nnl[0.1cm]
i \, G^{>}({\bf p},p_0,\bar{t}) & = & A({\bf p},p_0,\bar{t}) \; [
\: N({\bf p},p_0,\bar{t}) + 1 \: ] \, , \label{distdef} \eea
since $ G^{<}({\bf p},p_0,\bar{t})$ and  $G^{>}({\bf p},p_0,\bar{t})$ are known
from the integration of the Kadanoff-Baym equations as well as $A({\bf p},p_0,\bar{t})$.
In equilibrium (at temperature $T$) the Green functions obey the
Kubo-Martin-Schwinger relation (KMS) for all momenta ${\bf p}$,
\bea G^{>}_{eq}({\bf p},p_0) \; = \; e^{p_0/T} \;
G^{<}_{eq}({\bf p},p_0) \qquad \qquad \forall \;\; {\bf p} \, .
\label{kms} \eea
If there exists a conserved quantum number in the theory we have,
furthermore, a contribution of the corresponding chemical
potential in the exponential function which leads to a shift of
arguments: $p_0/T \rightarrow (p_0 - \mu)/T$. In case of $\phi^4$-theory,
however, there is no conserved quantum number and thus the
equilibrium state has  $\mu = 0$.

 From the KMS condition of the Green functions (\ref{kms})  the equilibrium form of the distribution function
(\ref{distdef}) at temperature $T$ is obtained as
\bea N_{eq}({\bf p},p_0) \; = \; N_{eq}(p_0) \; = \;
\frac{1}{e^{p_0/T} - 1} \; = \; N_{bose}(p_0/T) \, ,
\label{distequi} \eea from
$$
\frac{G^<}{G^>} = e^{-p_0/T} = \frac{N_{eq}}{N_{eq}+1},
$$ which is the well-known Bose distribution.
As is obvious from Eq. (\ref{distequi}) the equilibrium distribution
can only be a function of energy $p_0$ and not of the momentum
variable ${\bf p}$ in addition \cite{Juchem:2003bi}.

\subsection{Derivation of the off-shell relativistic transport theory}

Formal derivations of off-shell transport equations have been
presented more than 50 years ago by Kadanoff and Baym \cite{KB} but
actual solutions have barely been addressed
\cite{Morawetz:1998em,Kohler:2001zv}. This Subsection is devoted to
a brief derivation of generalized transport equations in first order
gradient expansion including a generalized test-particle ansatz for
the solution of the off-shell transport equation following Ref.
\cite{Juchem:2004cs}.

%\section{\label{sec:general_transport_derivation} Derivation
%of off-shell transport theory}
 The derivation of generalized transport equations starts
by rewriting the Kadanoff-Baym equation for the Wightman functions
in coordinate space ($ x_1\!=\!(t_1,{\bf x}_1), x_2\!=\!(t_2,{\bf
x}_2) $) (\ref{gl:kaba2}) as
\bea   \label{eq:kbcs1} [ \, \partial^{\mu}_{x_1}
\partial_{\mu}^{x_1} + m^2 + \Sigma^{\delta}(x_1) \, ] \;\: i
G^{\gtrless}(x_1,x_2) \; = \; i\,I_1^{\gtrless}(x_1,x_2) , \eea
where the collision terms on the r.h.s. of Eq. (\ref{eq:kbcs1}) are given in
$D = d+1$ space-time dimensions by convolution integrals over
coordinate-space selfenergies and Green functions:
\bea \label{eq:i1cs} I_1^{\gtrless}(x_1,x_2) = - \!\!
\int_{t_0}^{t_1} \!\!\! d^{D}\!z \; \left[ \Sigma^{>}(x_1,z) -
\Sigma^{<}(x_1,z)  \right] G^{\gtrless}(z,x_2)  +  \!\!
\int_{t_0}^{t_2} \!\!\! d^{D}\!z \Sigma^{\gtrless}(x_1,z)  \left[
 G^{>}(z,x_2) - G^{<}(z,x_2) \right]. \eea
In the general case of an arbitrary (scalar) quantum field theory
$\Sigma^{\delta}$ is the local (non-dissipative tadpole) part of the path
self-energy while $\Sigma^{\gtrless}$ resemble the non-local
collisional self-energy contributions. In the representation
(\ref{eq:i1cs}) the integration boundaries are exclusively given
for the time coordinates, while the integration over the spatial
coordinates extends over the whole spatial volume from $- \infty$
to $+ \infty$ in $d$ dimensions.

Since transport theories   are formulated in phase-space one changes
to the Wigner representation via Fourier transformation with respect
to the rapidly varying ('intrinsic') relative coordinate $\Delta x =
x_1 - x_2$ and treats the system evolution in terms of the
('macroscopic') mean space-time coordinate $x = (x_1 + x_2)/2$ and
the four-momentum $p = (p_0,{\bf p})$. The functions in Wigner space
are obtained as
\bea \label{eq:wignertrafo} \bar{F}(p,x) \; = \;
\int_{-\infty}^{\infty} \!\!\! d^{D}\!\Delta x \;\; \;
e^{+i\:\Delta x_{\mu}\:p^{\mu}} \;\; F(x_1=x+\Delta
x/2,\,x_2=x-\Delta x/2) \, . \eea
For the formulation of transport theory in the Wigner representation
we have to focus not only on the transformation properties of
ordinary two-point functions as given in Eq. (\ref{eq:wignertrafo}),
but also of convolution integrals as appearing in  Eq.
(\ref{eq:i1cs}). A convolution integral in $D$ dimensions (for
arbitrary functions $F, G$),
\bea \label{eq:conv} H(x_1,x_2) \; = \; \int_{-\infty}^{\infty}
 d^{D}\!z \  F(x_1,z) \  G(z,x_2) \eea
transforms as
\bea \bar{H}(p,x) & = & \int_{-\infty}^{\infty} \!\!\! d^{D}\!\Delta
x \;\; \; e^{+i\:\Delta x_{\mu}\:p^{\mu}} \;\; H(x_1,x_2)
 =  \int_{-\infty}^{\infty} \!\!\! d^{D}\!\Delta x \;\; \;
e^{+i\:\Delta x_{\mu}\:p^{\mu}} \;\; \int_{-\infty}^{\infty} \!\!\!
d^{D}\!z \;\; F(x_1,z) \;\; G(z,x_2) \nnl & = & \left. \;
e^{+i\,\frac{1}{2} \, (\partial_{p^{\phantom{\prime}}}^{\mu} \!\!
\cdot \:
 \partial^{x^{\prime}}_{\mu}
 \: - \;
 \partial_{x^{\phantom{\prime}}}^{\mu} \!\! \cdot \:
 \partial^{p^{\prime}}_{\mu} ) } \;\:
\left[ \; \bar{F}(p,x) \;\; \bar{G}(p^{\prime},x^{\prime}) \right]
\right|_{x^{\prime} = x,\: p^{\prime} = p} . \eea
In accordance with the standard assumption of transport theory we
assume that all functions only smoothly evolve in the mean
space-time coordinates and thus restrict to first order
derivatives. All terms proportional to second or higher order
derivatives in the mean space-time coordinates (also mixed ones)
will be dropped. Thus the Wigner transformed convolution integrals
(\ref{eq:conv}) are given in {\it first order gradient
approximation} by,
\bea \label{eq:firstordergrad} \bar{H}(p,x) \; = \; \bar{F}(p,x)
\;\; \bar{G}(p,x) \; + \; i \, \frac{1}{2} \: \{ \, \bar{F}(p,x)
\, , \, \bar{G}(p,x) \, \} \; + \; {\cal O}(\partial^2_x) \, ,
\eea
using the relativistic generalization of the Poisson bracket
\bea
\label{eq:def_poisson}
\{ \, \bar{F}(p,x) \, , \, \bar{G}(p,x) \, \}
\; : = \;
\partial^{p}_{\mu} \, \bar{F}(p,x) \cdot
\partial_{x}^{\mu} \, \bar{G}(p,x) \; - \;
\partial_{x}^{\mu} \, \bar{F}(p,x) \cdot
\partial^{p}_{\mu} \, \bar{G}(p,x) \; .
\eea
In order to obtain the dynamics for the spectral functions within
the approximate (first order gradient) scheme we start with the
Dyson-Schwinger equations for the retarded and advanced Green
functions in coordinate space (\ref{gl:kaba1}). -- Note that the
convolution integrals in (\ref{gl:kaba1}) extend over the whole
space and time range in contrast to the equations of motion for the
Wightman functions given in Eqs. (\ref{gl:kaba2}) and
(\ref{gl:kaba2b})! -- The further procedure consists in the
following steps:
\\
i) First we transform the above equations into the Wigner
representation and apply the first order gradient approximation. In
this limit the convolution integrals yield the product terms and the
general Poisson bracket of the selfenergies and the Green functions
$\{\, \Sigma^{R/A}, G^{R/A} \,\}$. We, further on, represent both
equations in terms of real quantities by the decomposition of the
retarded and advanced Green functions and selfenergies as
\bea
\begin{array}{ccccccc}
\bar{G}^{R/A} &\!\!=\!\!& \Re\,\bar{G}^{R} \,\pm\,
i\,\Im\,\bar{G}^{R} &\!\!=\!\!& \Re\,\bar{G}^{R} \,\mp\, i\,\bar{A}
/ 2\; , \phantom{aaaaaa} \bar{A} &\!\!=\!\!& \mp \, 2 \,
\Im\,\bar{G}^{R/A} \, ,
\\[0.1cm]
\bar{\Sigma}^{R/A} &\!\!=\!\!& \Re\,\bar{\Sigma}^{R} \,\pm\,
i\,\Im\,\bar{\Sigma}^{R} &\!\!=\!\!& \Re\,\bar{\Sigma}^{R} \,\mp\,
i\,\bar{\Gamma} / 2 \; , \phantom{aaaaaa} \bar{\Gamma} &\!\!=\!\!&
\mp \, 2\, \Im\,\bar{\Sigma}^{R/A} \, .
\end{array}
\phantom{aa} \eea
We find that in Wigner space the real parts of the retarded and
advanced Green functions and selfenergies are equal, while the
imaginary parts have opposite sign and are proportional to the
spectral function $\bar{A}$ and the width $\bar{\Gamma}$,
respectively. The next step consists in \\  ii) the separation of
the real part and the imaginary part of the two equations for the
retarded and advanced Green functions, that have to be fulfilled
independently. Thus we obtain four real-valued equations for the
self-consistent retarded and advanced Green functions. In the last
step \\  iii) we get simple relations by linear combination of these
equations, i.e. by adding/subtrac\-ting the relevant equations.

 This finally leads to two algebraic relations for the
spectral function $\bar{A}$ and the real part of the retarded Green
function $Re\,\bar{G}^{R}$ in terms of the width $\bar{\Gamma}$ and
the real part of the retarded self-energy $Re\,\bar{\Sigma}^{R}$ as
\cite{Juchem:2004cs}:
\bea \label{eq:specrel1} [ \, p_0^2 - {\bf p}^{\,2} - m^2 -
\bar{\Sigma}^{\delta} + \Re\,\bar{\Sigma}^{R} \, ] \;
\Re\,\bar{G}^{R} & = & 1 \: + \: \frac{1}{4} \: \bar{\Gamma} \;
\bar{A} \, ,
\\
\label{eq:specrel2} [ \, p_0^2 - {\bf p}^{\,2} - m^2 -
\bar{\Sigma}^{\delta} + \Re\,\bar{\Sigma}^{R} \, ] \; \bar{A} & = &
\bar{\Gamma} \; \Re\,\bar{G}^{R} \, . \eea
Note that all terms with first order gradients have disappeared in
Eqs. (\ref{eq:specrel1}) and (\ref{eq:specrel2}). A first consequence
of (\ref{eq:specrel2}) is a direct relation between the real and
the imaginary parts of the retarded/advanced Green function, which
reads (for $\bar{\Gamma} \neq 0$):
\bea \label{ins1}
 \Re\,\bar{G}^{R} \; = \; \frac{p_0^2 -
{\bf p}^{\,2} - m^2 - \bar{\Sigma}^{\delta} - \Re\,\bar{\Sigma}^{R}
}{\bar{\Gamma}} \; \bar{A} \; . \eea
Inserting Eq. (\ref{ins1}) in Eq. (\ref{eq:specrel1})  we end up with the
following result for the spectral function and the real part of
the retarded Green function
\bea \label{eq:specorder0} \bar{A} \; = \; \frac{\bar{\Gamma}}{[ \,
p_0^2 - {\bf p}^{\,2} - m^2 - \bar{\Sigma}^{\delta} -
\Re\,\bar{\Sigma}^{R} \, ]^2 + \bar{\Gamma}^2/4} & = &
\frac{\bar{\Gamma}}{\bar{M}^2 + \bar{\Gamma}^2/4} \, , \phantom{aaa}
\\
\label{eq:regretorder0} \Re\,\bar{G}^{R} \; = \; \frac{[ \, p_0^2 -
{\bf p}^{\,2} - m^2 - \bar{\Sigma}^{\delta} - \Re\,\bar{\Sigma}^{R}
\, ]} {[ \, p_0^2 - {\bf p}^{\,2} - m^2 - \bar{\Sigma}^{\delta} -
\Re\,\bar{\Sigma}^{R} \, ]^2 + \bar{\Gamma}^2/4} & = &
\frac{\bar{M}}{\bar{M}^2 + \bar{\Gamma}^2/4} \, , \eea
where we have introduced the mass-function $\bar{M}(p,x)$ in
Wigner space:
\bea \label{eq:massfunction} \bar{M}(p,x) & = & p_0^2 - {\bf
p}^{\,2} - m^2 - \bar{\Sigma}^{\delta}(x) -
\Re\,\bar{\Sigma}^{R}(p,x) \; . \eea
The  spectral function (\ref{eq:specorder0}) shows a typical
Breit-Wigner shape with energy- and momentum-dependent self-energy
terms. Although the above equations are purely algebraic solutions
and contain no derivative terms, they are valid up to the first
order in the gradients!

 In addition, subtraction of the real parts and adding up the
imaginary parts lead to the time evolution equations
\bea \label{eq:specorder1} p^{\mu} \, \partial_{\mu}^x \, \bar{A} &
= & \frac{1}{2} \, \{ \, \bar{\Sigma}^{\delta} +
\Re\,\bar{\Sigma}^{R} \, ,
   \, \bar{A} \, \}
\: + \: \frac{1}{2} \, \{ \, \bar{\Gamma} \, , \, \Re\,\bar{G}^{R}
\, \} \, ,
\\
p^{\mu} \, \partial_{\mu}^x \, \Re\,\bar{G}^{R} & = &
\label{eq:regretorder1} \frac{1}{2} \, \{ \, \bar{\Sigma}^{\delta} +
\Re\,\bar{\Sigma}^{R} \, ,
   \, \Re\,\bar{G}^{R} \, \}
\: - \: \frac{1}{8} \, \{ \, \bar{\Gamma} \, , \, \bar{A} \, \} \, .
\eea
The Poisson bracket containing the mass-function $\bar{M}$ leads
to the well-known drift operator
$p^{\mu}\,\partial^{x}_{\mu}\,\bar{F}$ (for an arbitrary function
$\bar{F}$), i.e.
\bea \label{eq:mass_poisson} \{\, \bar{M} \, , \, \bar{F} \,\} & = &
\{\, p_0^2 - {\bf p}^{\,2} - m^2 - \bar{\Sigma}^{\delta} -
\Re\,\bar{\Sigma}^{R} \, , \, \bar{F} \,\}  =  2 \, p^{\mu} \,
\partial_{\mu}^x \: \bar{F} \: - \: \{\,\bar{\Sigma}^{\delta} +
\Re\,\bar{\Sigma}^{R} \, , \, \bar{F} \,\} \; , \eea
such that the first order equations (\ref{eq:specorder1}) and
(\ref{eq:regretorder1}) can be written in a more comprehensive
form as
\bea \label{eq:specorder1final} \{ \, \bar{M} \, , \, \bar{A} \, \}
& = & \{ \, \bar{\Gamma} \, , \, \Re\,\bar{G}^{R} \, \} \, ,
\\
\{ \, \bar{M} \, , \, \Re\,\bar{G}^{R} \, \} & = &
\label{eq:regretorder1final} - \, \frac{1}{4} \, \{ \, \bar{\Gamma}
\, , \, \bar{A} \, \} \, . \eea
When inserting (\ref{eq:specorder0}) and (\ref{eq:regretorder0})
we find that these first order time evolution equations are {\em
solved} by the algebraic expressions. In this case the following
relations hold \cite{Juchem:2004cs}:
\bea \{\, \bar{M} \, , \, \bar{A} \,\} \; = \; \{\, \bar{\Gamma} \,
, \, \Re\,\bar{G}^{R} \,\} & = & \{\, \bar{M} \, , \, \bar{\Gamma}
\,\} \;\; \frac{\bar{M}^2 - \bar{\Gamma}^2/4} {[\, \bar{M}^2 +
\bar{\Gamma}^2/4 \,]^2} \, ,
\\
\{\, \bar{M} \, , \, \Re\,\bar{G}^{R} \,\} \; = \; - \, \frac{1}{4}
\, \{\, \bar{\Gamma} \, , \, \bar{A} \,\} & = & \{\, \bar{M} \, , \,
\bar{\Gamma} \,\} \;\; \frac{ \bar{M} \, \bar{\Gamma} / 2 } {[\,
\bar{M}^2 + \bar{\Gamma}^2/4 \,]^2} \, . \eea
Thus we have derived the proper structure of the spectral function
(\ref{eq:specorder0}) within the first-order gradient (or
semiclassical) approximation. Together with the explicit form for
the real part of the retarded Green function
(\ref{eq:regretorder0}) we now have fixed the dynamics of the
spectral properties, which is consistent up to first order in the
gradients.

%%%\subsection{ Kadanoff-Baym transport}
 As a next step we rewrite the memory terms in the collision
integrals  (\ref{eq:i1cs}) such that the time integrations extend from $- \infty$ to
$+ \infty$. In this respect we consider the initial time $t_0 = -
\infty$ whereas the upper time boundaries $t_1, t_2$ are taken into
account by $\Theta$-functions, i.e.
\bea \label{eq:i1csnew} I_1^{\gtrless}(x_1,x_2) \: = \: & - & \!\!
\int_{-\infty}^{\infty} \!\!\! d^{D}x^{\prime} \;\;\;
\Theta(t_1-t^{\prime}) \: \left[ \, \Sigma^{>}(x_1,x^{\prime}) -
\Sigma^{<}(x_1,x^{\prime}) \, \right] \;\;
G^{\gtrless}(x^{\prime},x_2) \nnl & + & \!\! \int_{-\infty}^{\infty}
\!\!\! d^{D}x^{\prime} \;\;\; \Sigma^{\gtrless}(x_1,x^{\prime}) \;
\; \Theta(t_2-t^{\prime}) \: \left[ \, G^{>}(x^{\prime},x_2) -
G^{<}(x^{\prime},x_2) \, \right] \nnl \: = \: & - & \!\!
\int_{-\infty}^{\infty} \!\!\! d^{D}x^{\prime} \;\;\;\;
\Sigma^{R}(x_1,x^{\prime}) \; G^{\gtrless}(x^{\prime},x_2) \: + \:
\Sigma^{\gtrless}(x_1,x^{\prime}) \; G^{A}(x^{\prime},x_2) \; .
\phantom{aaaa} \eea
We now perform the analogous steps as invoked before for the
retarded and advanced Dyson-Schwinger equations. We start with a
first order gradient expansion of the Wigner transformed
Kadanoff-Baym equation using (\ref{eq:i1csnew}) for the memory
integrals. Again we separate the real and the imaginary parts in the
resulting equation, which have to be satisfied independently. At the
end of this procedure we obtain a generalized transport equation:
\bea \label{eq:general_transport} \underbrace{ \phantom{\frac{1}{1}}
\!\!\! 2\,p^{\mu}\:\partial^{x}_{\!\mu} \: i\bar{G}^{\gtrless} \, -
\, \{ \, \bar{\Sigma}^{\delta} \!+\! \Re\,\bar{\Sigma}^{R} ,
   \, i \bar{G}^{\gtrless} \, \} }
\, - \, \{ \, i\bar{\Sigma}^{\gtrless} \, , \, \Re\,\bar{G}^{R} \,
\} & = & i\bar{\Sigma}^{<} \; i\bar{G}^{>} \, - \, i\bar{\Sigma}^{>}
\; i\bar{G}^{<} \nnl[0.cm] \{ \, \bar{M} \, , \, i
\bar{G}^{\gtrless} \, \}  \;\; \qquad \qquad \, - \, \{ \,
i\bar{\Sigma}^{\gtrless} \, , \, \Re\,\bar{G}^{R} \, \} & = &
i\bar{\Sigma}^{<} \; i\bar{G}^{>} \, - \, i\bar{\Sigma}^{>} \;
i\bar{G}^{<} \phantom{aaaaaaa} \eea
as well as a generalized mass-shell equation
\bea \label{eq:general_mass} \underbrace{ \phantom{\frac{1}{1}}
\!\!\!\! [ \, p^2 - m^2 - \bar{\Sigma}^{\delta} -
\Re\,\bar{\Sigma}^{R} \, ]}_{\bar{M}} \;\, i \bar{G}^{\gtrless} \: =
\: i\bar{\Sigma}^{\gtrless} \; \Re\,\bar{G}^{R} \, + \, \frac{1}{4}
\, \{ \, i\bar{\Sigma}^{>} , \, i\bar{G}^{<} \, \} \, - \,
\frac{1}{4} \, \{ \, i\bar{\Sigma}^{<} , \, i\bar{G}^{>} \, \}
\phantom{aaa} \eea
with the mass-function $\bar{M}$ specified in Eq.
(\ref{eq:massfunction}). Since the Green function
$G^{\gtrless}(x_1,x_2)$ consists of an antisymmetric real part and a
symmetric imaginary part with respect to the relative coordinate
$x_1-x_2$, the Wigner transform of this function is purely
imaginary. It is thus convenient to represent the Wightman functions
in Wigner space  by the real-valued quantities $i
\bar{G}^{\gtrless}(p,x)$. Since the collisional selfenergies obey
the same symmetry relations in coordinate space and in phase-space,
they will be kept also as $i \bar{\Sigma}^{\gtrless}(p,x)$ further
on.

 In the transport equation (\ref{eq:general_transport}) one
recognizes on the l.h.s. the drift term
$p^{\mu}\:\partial^{x}_{\!\mu} \: i\bar{G}^{\gtrless}$, as well as
the Vlasov term with the local self-energy $\bar{\Sigma}^{\delta}$
and the real part of the retarded self-energy
$Re\,\bar{\Sigma}^{R}$. On the other hand the r.h.s. represents the
collision term with its typical `gain and loss' structure. The loss
term $i\bar{\Sigma}^{>} \; i\bar{G}^{<}$ (proportional to the Green
function itself) describes the scattering out of a respective
phase-space cell, whereas the gain term $i\bar{\Sigma}^{<} \;
i\bar{G}^{>}$ takes into account scatterings into the actual cell.
The last term on the l.h.s. $\{\, i\bar{\Sigma}^{\gtrless} ,
\Re\,\bar{G}^{R} \, \}$ is very {\em peculiar} since it does not
contain directly the distribution function $i\bar{G}^{<}$. This
second Poisson bracket vanishes in the quasiparticle approximation
and thus does not appear in the on-shell Boltzmann limit. As
demonstrated in detail in Refs. \cite{KB,Juchem:2003bi}
 the second Poisson bracket $\{ \,
i\bar{\Sigma}^{\gtrless} , \Re\,\bar{G}^{R} \, \}$ governs the
evolution of the off-shell dynamics for nonequilibrium systems.

 Although the generalized transport equation
(\ref{eq:general_transport}) and the generalized mass-shell equation
(\ref{eq:general_mass}) have been derived from the same
Kadanoff-Baym equation in a first order gradient expansion, both
equations are not exactly equivalent \cite{botmal,Juchem:2003bi}.
Instead, they deviate from each other by contributions of second
gradient order, which are hidden in the term $\{\,
i\bar{\Sigma}^{\gtrless} , \Re\,\bar{G}^{R} \, \}$. A consistency,
however, can be achieved by rewriting the self-energy
$\bar{\Sigma}^{<}$ by $\bar{G}^< \cdot \bar{\Gamma} / \bar{A}$ in
the Poisson bracket term $\{ \bar{\Sigma}^{<} , \Re\,\bar{G}^{R}
\}$. The generalized transport equation (\ref{eq:general_transport})
then can be written in short-hand notation
\bea \label{eq:general_transport_bm} \frac{1}{2} \, \bar{A} \;
\bar{\Gamma} \; \left[ \, \{ \, \bar{M} \, , \, \, i\bar{G}^{<} \,
\} \: - \: \frac{1}{\bar{\Gamma}} \; \{ \, \bar{\Gamma} \, , \,
\bar{M} \cdot i\bar{G}^{<} \, \} \, \right] \; = \;
i\bar{\Sigma}^{<} \; i\bar{G}^{>} \, - \, i\bar{\Sigma}^{>} \;
i\bar{G}^{<} \;\;\;\;\;\; \eea
with the mass-function $\bar{M}$ (\ref{eq:massfunction}). The
transport equation (\ref{eq:general_transport_bm}) within the
Botermans-Malfliet (BM) form resolves the discrepancy between the
generalized mass-shell equation (\ref{eq:general_mass}) and the
generalized transport equation in its original Kadanoff-Baym form
(\ref{eq:general_transport}).

\subsection{Test-particle representation and numerical solution}

 The generalized transport equation (\ref{eq:general_transport_bm})
allows to extend the traditional on-shell transport approaches for
which efficient numerical recipes have been set up. In order to
obtain a practical solution to the transport equation
(\ref{eq:general_transport_bm}) we use a test-particle ansatz for
the Green function $G^{<}$, more specifically for the real and
positive semi-definite quantity (using ${\bar G}=G, {\bar
\Sigma}=\Sigma, {\bar \Gamma}=\Gamma$),
\be F({x,p}) \; =  i \, G^{<}(x,p) \; \sim \; \sum_{i=1}^{N} \;
\delta^{(3)} ({{\bf x}} \, - \, {{\bf X}}_i (t)) \; \; \delta^{(3)}
({{\bf p}} \, - \, {{\bf P}}_i (t)) \; \; \delta(p_0 - \,
\epsilon_i(t)) \: , \label{test-particle} \ee
where the sum over $i$ describes the sum over all (properly normalized) testparticles.
In the most general case (where the self energies depend on
four-momentum $P$, time $t$ and the spatial coordinates ${\bf X}$)
the equations of motion for the test-particles $i$ read
\hspace{-0.9cm} \bea \label{eomr} \frac{d {\vec X}_i}{dt} \! & = &
\!  \frac{1}{1 - C_{(i)}} \, \frac{1}{2 \epsilon_i} \: \left[ \, 2
\, {\vec P}_i \, + \, {\vec \nabla}_{P_i} \, \Re \Sigma^{R}_{(i)} \,
+ \, \frac{ \epsilon_i^2 - {\vec P}_i^2 - M_0^2 - \Re
\Sigma^{R}_{(i)}}{\Gamma_{(i)}} \: {\vec \nabla}_{P_i} \,
\Gamma_{(i)} \: \right],
\\[0.1cm]
\label{eomp} \frac{d {\vec P}_i}{d t} \! & = & \! -
\frac{1}{1-C_{(i)}} \, \frac{1}{2 \epsilon_{i}} \: \left[ {\vec
\nabla}_{X_i} \, \Re \Sigma^{R}_i \: + \: \frac{\epsilon_i^2 - {\vec
P}_i^2 - M_0^{2} - \Re \Sigma^{R}_{(i)}}{\Gamma_{(i)}} \: {\vec
\nabla}_{X_i} \, \Gamma_{(i)} \: \right],
\\[0.1cm]
\label{eome} \frac{d \epsilon_i}{d t} \!  & = & \! \phantom{- }
\frac{1}{1 - C_{(i)}} \, \frac{1}{2 \epsilon_i} \: \left[
\frac{\partial \Re \Sigma^{R}_{(i)}}{\partial t} \: + \:
\frac{\epsilon_i^2 - {\vec P}_i^2 - M_0^{2} - Re
\Sigma^{R}_{(i)}}{\Gamma_{(i)}} \: \frac{\partial
\Gamma_{(i)}}{\partial t} \right], \eea
where the notation $F_{(i)}$ implies that the function is taken at
the coordinates of the test-particle, i.e. $F_{(i)} \equiv F(t,{\bf
X}_{i}(t),{\bf P}_{i}(t),\epsilon_{i}(t))$.

 In Eqs. (\ref{eomr}-\ref{eome}), a common multiplication factor
$(1-C_{(i)})^{-1}$ appears, which contains the energy derivatives
of the retarded self energy
\bea \label{correc} C_{(i)} \: = \: \frac{1}{2 \epsilon_i} \left[
\frac{\partial}{\partial \epsilon_i} \, \Re \Sigma^{R}_{(i)} \: + \:
\frac{\epsilon_i^2 - {\vec P}_i^2 - M_0^2 - Re
\Sigma^{R}_{(i)}}{\Gamma_{(i)}} \: \frac{\partial }{\partial
\epsilon_i} \, \Gamma_{(i)} \right] . \eea
It yields a shift of the system time $t$ to the 'eigentime' of
particle $i$ defined by $\tilde{t}_{i} = t /(1-C_{(i)})$. As the
reader immediately verifies, the derivatives with respect to the
'eigentime', i.e. $d {\bf X}_i / d \tilde{t}_i$, $d {\bf P}_i / d
\tilde{t}_i$ and $d \epsilon_i / d \tilde{t}_i$ then emerge without
this renormalization factor for each test-particle $i$ when
neglecting higher order time derivatives in line with the
semiclassical approximation scheme. Note that the test-particle
equations of motion (presented above) should not be applied for
arbitrary selfenergies $\Sigma^{R}$ and width $\Gamma$ since the
theory must obey micro-causality. This leads to severe constraints
for the selfenergies \cite{knoll1,knoll2,Rauber:2014mca}.

 Some limiting cases should be mentioned explicitly: In
case of a momentum-independent 'width' $\Gamma(x)$ we take $M^{2} =
P^2 - Re \Sigma^{R}$ as an independent variable instead of $P_0$,
which then fixes the energy (for given ${\bf P}$ and $M^{2}$) to
\bea p_{0}^{2} \; = \; {\bf p}^{2} \: + \: M^{2} \: + \: \Re \Sigma
(x,{\bf p},M^2)^{R} \, . \label{energyfix} \eea
Eq. (\ref{eome}) then turns to ($\Delta M_i^2 = M_i^2 - M_0^2$)
\bea \label{eomm} \frac{d \Delta M_i^2}{dt} \; = \; \frac{\Delta
M_i^2}{\Gamma_{(i)}} \; \frac{d \Gamma_{(i)}}{dt} \hspace{1cm}
\leftrightarrow \hspace{1cm} \frac{d}{dt} \ln \left( \frac{\Delta
M_i^2}{\Gamma_{(i)}} \right) = 0 \eea
for the time evolution of the test-particle $i$ in the invariant
mass squared. In case of $\Gamma = const.$ the familiar equations of
motion for test-particles in on-shell transport approaches are
regained. We mention in passing that in the Parton-Hadron-String
Dynamics (PHSD) transport approach \cite{CasBrat,BrCa11} the width
of partonic degrees-of-freedom (so far) is taken as momentum independent such
that the simple limit (\ref{eomm}) applies (see below).

% part of the review in PPNP 2015

%\section{Thermodynamics and transport properties of QCD}
\section{Dynamical quasiparticle model for hot QCD}

Early concepts of the Quark-Gluon-Plasma (QGP) were guided by the
idea of a weakly interacting system of massless partons which might
be described by perturbative QCD (pQCD). However, experimental
observations at RHIC indicated that the new medium created in
ultrarelativistic Au+Au collisions is interacting more strongly than
hadronic matter. It is presently widely accepted that this medium is
an almost perfect liquid of partons  as extracted experimentally
from the strong radial expansion and the scaling of the elliptic
flow $v_2(p_T)$ of mesons and baryons with the number of constituent
quarks and antiquarks. At vanishing quark chemical potential $\mu_q$
the QCD problem can be addressed at zero and finite temperature by
lattice QCD calculations on a 3+1 dimensional torus with a suitable
discretization of the QCD action on the euclidian lattice. These
calculations so far have provided valuable information on the QCD
equation of state, chiral symmetry restoration and various
correlators that can be attributed/related to transport
coefficients. Due to the Fermion 'sign'-problem lQCD calculations at
finite $\mu_q$ are presently not robust and one has to rely on
nonperturbative - but effective - models to obtain information in
the ($T$, $\mu_q$) plane or for systems out-off equilibrium.

\subsection{Quasiparticle properties}

As demonstrated above a consistent dynamical approach for the
description of strongly interacting systems - also out-off
equilibrium - can be formulated on the basis of Kadanoff-Baym (KB)
equations  or off-shell transport equations in phase-space
representation (cf. Section 2), respectively. In the KB theory the
field quanta are described in terms of dressed propagators with
complex selfenergies \cite{Cassing:2008nn}. Whereas the real part of
the selfenergies can be related to mean-field potentials (of Lorentz
scalar, vector or tensor type), the imaginary parts  provide
information about the lifetime and/or reaction rates of time-like
'particles'. Once the proper (complex) selfenergies of the
degrees-of-freedom are known, the time evolution of the system is
fully governed  by off-shell transport equations (cf. Section 2).
The determination/extraction of complex selfenergies for the
partonic degrees-of-freedom can be performed within the Dynamical
QuasiParticle Model (DQPM) by fitting lattice QCD calculations in
thermal equilibrium. The DQPM postulates retarded propagators of the
quark and gluon degrees-of-freedom in the form
\begin{equation}
\label{propdqpm} G^{R} (\omega, {\bf p}) = \frac{1}{\omega^2 - {\bf
p}^2 - M^2 + 2 i \gamma \omega}
\end{equation}
using $\omega=p_0$.
In the scope of the DQPM the running coupling (squared) is
approximated by
\begin{equation}
g^2(T/T_{c})=\frac{48\pi^2}{(11N_{c}-2N_{f})\ln[\lambda^2(T/T_{c}-T_{s}/T_{c})^2]},
\label{running}
\end{equation}
where the parameters $\lambda \approx 2.42$ and $T_{s}/T_{c} \approx
0.56$ have to be extracted from a fit to the lattice data in Fig. 3
(r.h.s.) (see below). In Eq.
(\ref{running}), $N_{c}=3$ stands for the number of colors, $T_c$ is
the critical temperature ($\approx 158$ MeV), while $N_{f} (=3)$
denotes the number of flavors.  The parameter $T_s$ is essentially
important for the infrared enhancement of the coupling at low
temperature $T$. It has been demonstrated in
Ref.~\cite{Peshier:2005pp} that this functional form for the strong
coupling $\alpha_s = g^2/(4\pi)$ is in accordance with the lQCD
calculations of the Bielefeld group for the long range part of the
$q - \bar{q}$ potential. Furthermore, it matches the
hard-thermal-loop (HTL) limit for high temperatures $T$.

 The dynamical quasiparticle mass (for gluons and quarks)
is assumed to be given by the HTL  thermal mass in the asymptotic
high-momentum regime, i.e. for gluons
\begin{equation} \label{Mg9}
M^2_{g}(T)=\frac{g^2}{6}\left(\left(N_{c}+\frac{1}{2}N_{f}\right)T^2
+\frac{N_c}{2}\sum_{q}\frac{\mu^{2}_{q}}{\pi^2}\right)\ ,
\end{equation}
and for quarks (antiquarks)
\begin{equation} \label{Mq9}
M^2_{q(\bar
q)}(T)=\frac{N^{2}_{c}-1}{8N_{c}}g^2\left(T^2+\frac{\mu^{2}_{q}}{\pi^2}\right)\
,
\end{equation}
but with the coupling  given in Eq. (\ref{running}).  The dynamical masses (\ref{Mq9})
in the QGP are large compared to the bare masses of the light
($u,d$) quarks and adopted in the form (\ref{Mq9}) for the ($u,d$) quarks.
The strange quark has a larger bare mass which also enters to some
extent the dynamical mass $M_s(T)$. This essentially suppresses the channel
$g \rightarrow s + {\bar s}$ relative to the channel $g \rightarrow u + {\bar u}$
or $d + {\bar d}$ and controls the strangeness ratio in the QGP. Empirically we have used
$M_s(T)= M_u(T)+ \Delta M = M_d(T)+ \Delta M$ where $\Delta M$ =
35 MeV, which has been fixed once in comparison to experimental
data for the $K^+/\pi^+$ ratio in central Au+Au collisions at
$\sqrt{s_{NN}}$ = 17.3 GeV.  Furthermore, the effective
quarks, antiquarks and gluons in the DQPM have finite widths, which
for $\mu_{q}=0$ are adopted in the form
\begin{eqnarray}\label{widthg}
\gamma_{g}(T)&=&\frac{1}{3}N_{c}\frac{g^2T}{8\pi}\ln\left(\frac{2c}{g^2}+1\right)
, \label{widthq} \ \ \  \gamma_{q(\bar
q)}(T)=\frac{1}{3}\frac{N^{2}_{c}-1}{2N_{c}}\frac{g^2T}{8\pi}
\ln\left(\frac{2c}{g^2}+1\right),
\end{eqnarray}
where $c=14.4$  is related to a magnetic cut-off, which is one of
the parameters of the DQPM. Furthermore, we assume that the width
of the strange quark is the same as that for the light ($u,d$)
quarks.

 The physical processes contributing to the width
$\gamma_g$ are both $gg \leftrightarrow gg$, $gq \leftrightarrow gq$
scattering as well as splitting and fusion reactions $gg
\leftrightarrow g$, $gg \leftrightarrow ggg$, $ggg \leftrightarrow
gggg$ or $g \leftrightarrow q \bar{q}$ etc. On the fermion side
elastic fermion-fermion scattering $pp \leftrightarrow pp$, where
$p$ stands for a quark $q$ or antiquark $\bar{q}$, fermion-gluon
scattering $pg \leftrightarrow pg$, gluon bremsstrahlung $pp
\leftrightarrow pp+g$ or quark-antiquark fusion $q \bar{q}
\leftrightarrow g$ etc. emerge. Note, however, that the explicit
form of (\ref{widthg}) is derived for hard two-body scatterings
only.

\subsection{Spectral functions}

In the DQPM the parton spectral functions are no
longer $\delta$-functions in the invariant mass squared but taken
as (cf. Eq. (\ref{bw}) in Section 2)
\begin{equation}
\label{20} \rho_{j}(\omega,{\bf p}\!)=\!\frac{\gamma_{j}}{2 E_j}\!
\left(\frac{1}{(\omega-E_j)^2+\gamma^{2}_{j}}
-\frac{1}{(\omega+E_j)^2+\gamma^{2}_{j}}\right)
\end{equation}
\\[0.1cm]
separately for quarks, antiquarks and gluons ($j = q,\bar q,g$).
Here $E_{j}^2({\bf p}^2)={\bf p}^2+M_{j}^{2}-\gamma_{j}^{2}$, where the parameters $\gamma_{j},
M_{j}$ from the DQPM have been described above. The spectral
function (\ref{20}) is antisymmetric in $\omega$ and normalized as
\begin{equation}
\int\limits_{-\infty}^{\infty}\frac{d\omega}{2\pi}\
2 \omega \ \rho_{j}(\omega,{\bf p})=
2 \int\limits_{0}^{\infty}\frac{d\omega}{2\pi}\ 2\omega
\rho_{j}(\omega,{\bf p})=1\
\end{equation} as mandatory for quantum field theory. \\

 The actual gluon mass $M_g$ and width $\gamma_g$ --
employed as input in the further calculations -- as well as the
quark mass $M_q$ and width $\gamma_q$ are depicted in Fig.
\ref{fig1} (l.h.s.) as a function of $T/T_c$.  Note that for $\mu_q$
= 0 the DQPM gives
\begin{equation} \label{qma}
M_q = \frac{2}{3} M_g, \hspace{1cm} \gamma_q = \frac{4}{9} \gamma_g
\ . \end{equation}

\begin{figure}
{\includegraphics*[width=0.455\textwidth]{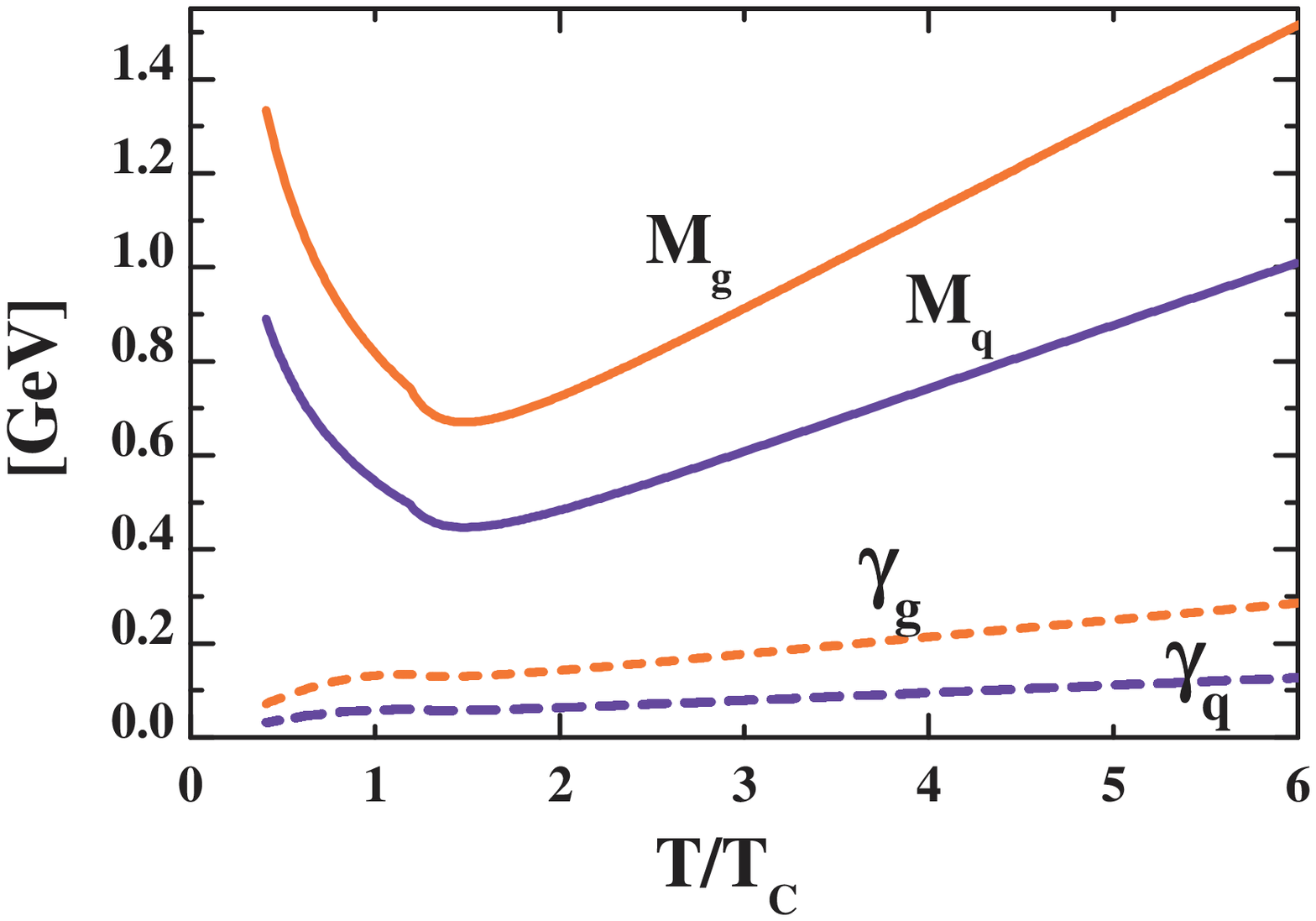}} \hspace{1cm}
\includegraphics*[width=0.42\textwidth]{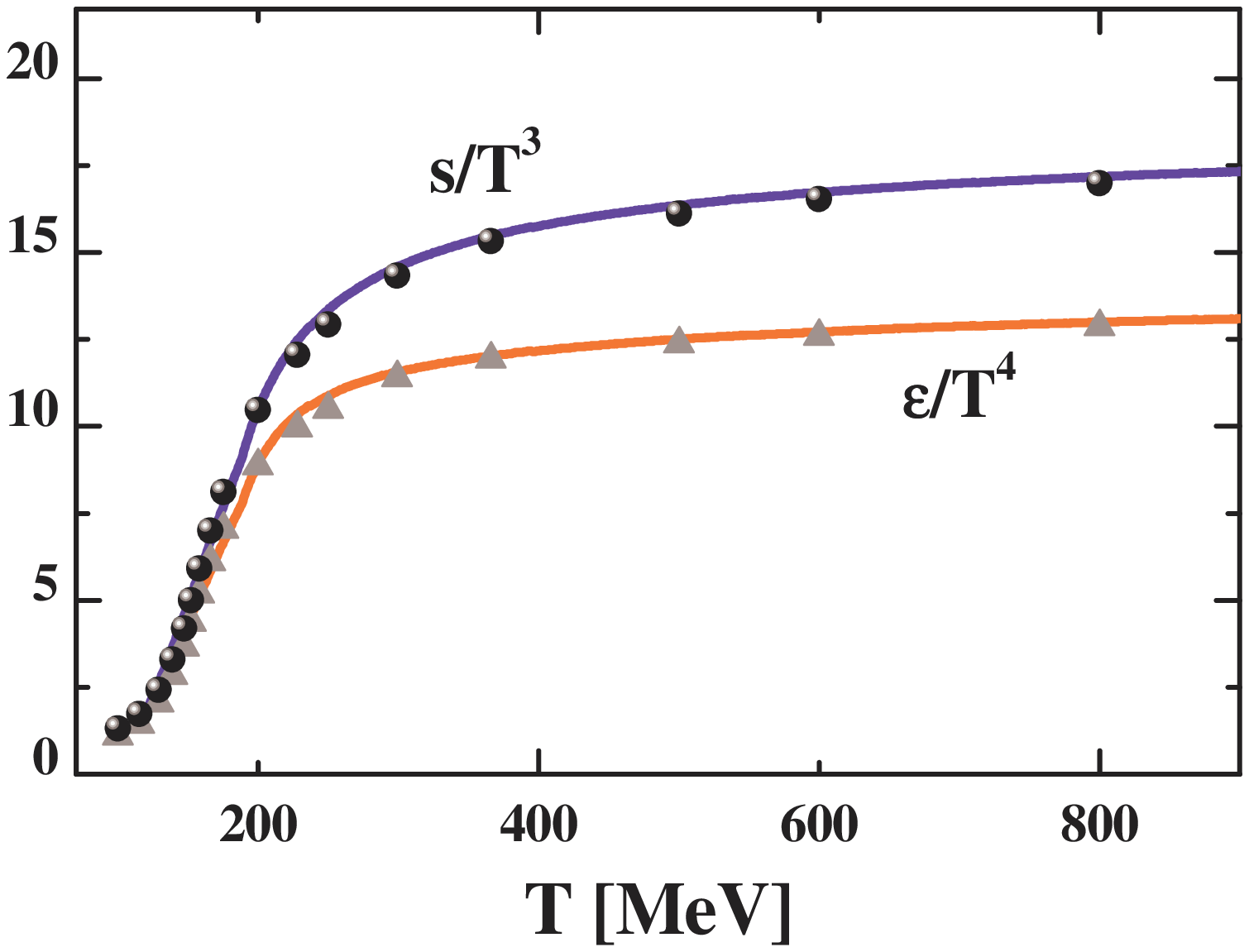}
\caption{{\it (l.h.s.)} The effective gluon mass $M_g$ and width
$\gamma_g$ as function of the scaled temperature $T/T_c$
(upper red lines). The lower blue lines show the corresponding quantities for
quarks. {\it (r.h.s.)} The scaled entropy density $s(T)/T^3$ (upper blue
line) and scaled energy density $\epsilon(T)/T^4$ (lower red line) from
the DQPM in comparison to the lQCD results from the BMW group (full
dots and triangles) \protect\cite{Aoki:2009sc}. The figures are taken
from Ref. \protect\cite{BrCa11}.}
\label{fig1}  \label{fig4}
\end{figure}

 These variations of the masses with the temperature $T$ -- that appear drastic in
Fig. \ref{fig1} (l.h.s.) -- become, however,
rather smooth if viewed as a function of the scalar parton density
$\rho_s$ defined (in thermal equilibrium) by
\bea \label{rhos} \rho_s({T}/{T_c}) & = & d_g \int_0^\infty
\frac{d\omega}{2 \pi} \int \frac{d^3 p}{(2 \pi)^3}\ 2  \sqrt{p^2} \
\rho_g(\omega, {\bf p}) \ n_B(\omega/T) \ \Theta(p^2) \nnl && + d_q
\int_0^\infty  \frac{d\omega}{2 \pi} \int \frac{d^3 p}{(2 \pi)^3} \
2  \sqrt{p^2} \  \rho_q(\omega, {\bf p}) \ n_F((\omega-\mu_q)/T) \
\Theta(p^2) \nnl && + d_{\bar q} \int_0^\infty  \frac{d\omega}{2
\pi} \int \frac{d^3 p}{(2 \pi)^3} \ 2  \sqrt{p^2} \  \rho_{\bar
q}(\omega, {\bf p}) \ n_F((\omega+\mu_q)/T) \ \Theta( p^2) \ , \eea
where $n_B$ and $n_F$ denote the Bose and Fermi functions,
respectively, while $\mu_q$ stands for the quark chemical potential.
The number of transverse gluonic degrees-of-freedom is $d_g=16$
while the fermion degrees-of-freedom amount to $d_q=d_{\bar q}=2 N_c
N_f=18$ in case of three flavors ($N_f$=3). The function $\Theta(
p^2)$ (with $p^2 = \omega^2 - {\bf p}^2$) projects on {\bf
time-like} four-momenta since only this fraction of the
four-momentum distribution can be propagated within the light cone.

\subsection{Thermodynamics of QCD}

 With the quasiparticle properties (or propagators) chosen as described above, one can
evaluate the entropy density $s(T)$, the pressure $P(T)$ and energy
density $\epsilon(T)$ in a straight forward manner by starting with
the entropy density in the quasiparticle limit from Baym
\cite{Baym}, \bea  \label{sdqp} \hspace{0.5cm}
  s^{dqp}
  &=&
  - d_g \!\int\!\!\frac{d \omega}{2 \pi} \frac{d^3p}{(2 \pi)^3}
  \frac{\partial n_B}{\partial T}
   \left( \Im\ln(-\Delta^{-1}) + \Im\Pi\,\Re\Delta \right) \nnl &&
   - d_q \!\int\!\!\frac{d \omega}{2 \pi} \frac{d^3p}{(2 \pi)^3}
  \frac{\partial n_F((\omega-\mu_q)/T)}{\partial T}
   \left( \Im\ln(-S_q^{-1}) + \Im\Sigma_q\,\Re S_q \right)
   \!, \nnl &&
   - d_{\bar q} \!\int\!\!\frac{d \omega}{2 \pi} \frac{d^3p}{(2 \pi)^3}
  \frac{\partial n_F((\omega+\mu_q)/T)}{\partial T}
   \left( \Im\ln(-S_{\bar q}^{-1}) + \Im\Sigma_{\bar q}\,\Re S_{\bar q} \right)
   \!, \eea
 where $n_B(\omega/T) = (\exp(\omega/T)-1)^{-1}$ and
$n_F((\omega-\mu_q)/T) = (\exp((\omega-\mu_q)/T)+1)^{-1}$ denote the
Bose and Fermi distribution functions, respectively, while $\Delta
=(P^2-\Pi)^{-1}$, $S_q = (P^2-\Sigma_q)^{-1}$ and $S_{\bar q} =
(P^2-\Sigma_{\bar q})^{-1}$ stand for the full (scalar)
quasiparticle propagators of gluons $g$, quarks $q$ and antiquarks
${\bar q}$.  In Eq. (\ref{sdqp}) $\Pi$ and $\Sigma = \Sigma_q
\approx \Sigma_{\bar q}$ denote the (retarded) quasiparticle
selfenergies. In principle, $\Pi$ as well as $\Delta$ are Lorentz
tensors and should be evaluated in a nonperturbative framework. The
DQPM treats these degrees-of-freedom as independent scalar fields
with scalar selfenergies  which are assumed to be identical for
quarks and antiquarks. Note that one has to treat quarks and
antiquarks separately in Eq. (\ref{sdqp}) as their abundance differs
at finite quark chemical potential $\mu_q$.

 Since the nonperturbative evaluation of the propagators
and selfenergies in QCD is a formidable task [and addressed in
Dyson-Schwinger (DS) Bethe-Salpeter (BS) approaches]  an alternative
and practical procedure is to use physically motivated {\em
Ans\"atze} with Lorentzian spectral functions for quarks\footnote{In
the following the abbreviation is used that 'quarks' denote quarks
and antiquarks if not specified explicitly.} and gluons as in
(\ref{20}). With this choice the complex selfenergies  $\Pi =
M_g^2-2i \omega \gamma_g$ and $\Sigma_{q} ({\bf q}) = M_{q} ({\bf
q})^2 - 2 i \gamma_{q} ({\bf q})$ are fully defined via (\ref{Mg9}),
(\ref{Mq9}), (\ref{widthg}). Note that the retarded propagator
(\ref{propdqpm}), \beq G_R^{-1} = \omega^2 - {\bf p}^2 - M^2 + 2i
\gamma \omega, \eeq corresponds to the propagator of a damped
harmonic oscillator (with an additional ${\bf p}^2$)  and preserves
microcausality also for $\gamma > M$ \cite{Rauber:2014mca}, i.e. in
case of overdamped motion. Although the 'Ansatz' for the parton
propagators is not QCD we will demonstrate that a variety of QCD
observables on the lattice are compatible with this choice.

 Since within the DQPM the real and imaginary parts of the propagators
$\Delta$ and $S$ now are fixed analytically the entropy density
(\ref{sdqp}) can be evaluated numerically. As we deal with a
grandcanonical ensemble the Maxwell relations give
 (at $\mu_q$ = 0),
\begin{equation}
\label{pressure} s =\frac{\partial P}{\partial T} \ ,
\end{equation} such that the pressure can be obtained by integration
of the entropy density $s$ over $T$, where one tacitly identifies
the 'full' entropy density $s$ with the quasiparticle entropy
density $s^{dqp}$ (\ref{sdqp}). The starting point for the
integration in $T$ is chosen between 100 MeV $< T < $ 120 MeV where
the entropy density is approximated by that of a noninteracting
pion, $\eta$ and kaon gas.

 The energy density $\epsilon$ then follows from the
thermodynamical relation \begin{equation}
\label{eps} \epsilon = T s -P \end{equation} (for $\mu_q$ = 0) and
thus is also fixed by the entropy $s(T)$ as well as the
interaction measure
\begin{equation} \label{wint} W(T): = \epsilon(T) - 3P(T) = Ts - 4
P \end{equation} that vanishes for massless and noninteracting
degrees-of-freedom.

A direct comparison of the resulting entropy
density $s(T)$ (\ref{sdqp}) - using (66) to (70) - and energy
density $\epsilon(T)$ (\ref{eps}) from the DQPM with lQCD results from the BMW
group \cite{Aoki:2009sc}  is presented in Fig. \ref{fig4} (r.h.s.). Both results have
been divided by $T^3$ and $T^4$, respectively, to demonstrate the
scaling with temperature. The agreement is sufficiently good. A satisfactory agreement also
holds for the dimensionless 'interaction measure', i.e. $(\epsilon -
3 P)/T^4$ (cf. Fig. \ref{figpot}, l.h.s.).

\subsection{Partonic mean-field potentials from the DQPM}

\begin{figure}
\centering \includegraphics*[width=75mm]{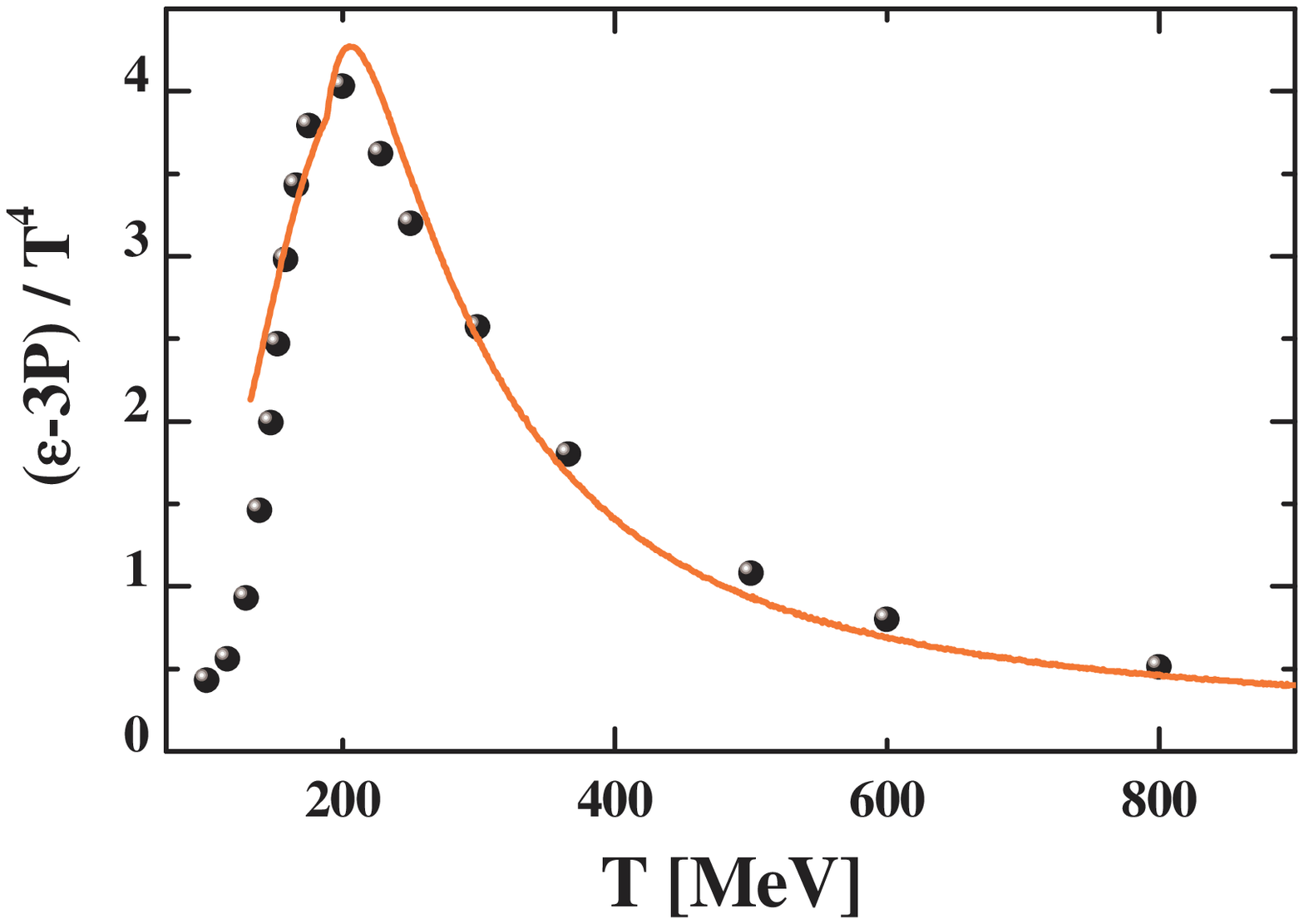} \hspace{0.5cm}
\includegraphics*[width=75mm]{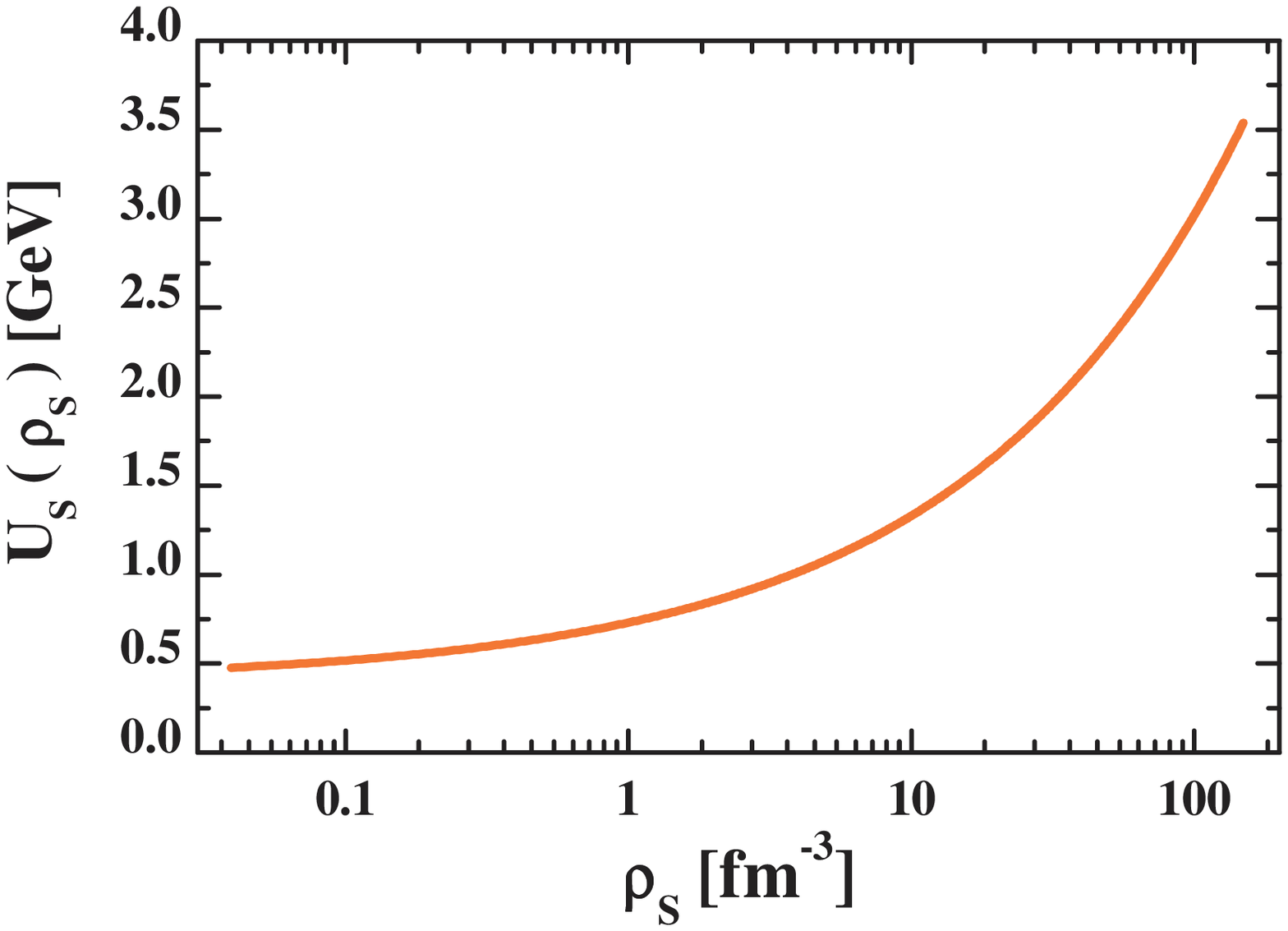} \caption{(l.h.s.) The
interaction measure $(\epsilon - 3 P)/T^4$ from the DQPM in
comparison to the lQCD results from \protect\cite{Aoki:2009sc}.
(r.h.s.)  The scalar mean-field (\ref{uss}) for quarks and
antiquarks from the DQPM as a function of the scalar parton density
$\rho_s$ for $\mu_q=0$.  The figures are taken from Ref.
\protect\cite{BrCa11}.} \label{figpot}
\end{figure}

The DQPM uniquely defines a potential energy density,
\begin{equation} \label{Vp}
V_p(T,\mu_q) = T^{00}_{g-}(T,\mu_q) + T^{00}_{q-}(T,\mu_q)
+ T^{00}_{{\bar q}-}(T,\mu_q) , \end{equation}
where the different contributions $T^{00}_{j-}$ correspond to the space-like part
of the energy-momentum tensor component $T^{00}_{j}$ of parton $j
= g, q, \bar{q}$ \cite{Cassing:2008nn}. It is found that this quantity is practically
independent on the quark chemical potential (for moderate $\mu_q$)
when displayed as a function of the scalar density $\rho_s$ instead of $T$ and $\mu_q$
separately. Note that the field quanta involved in (\ref{Vp}) are
virtual and thus correspond to partons exchanged in interaction
diagrams.

 A scalar mean-field $U_s(\rho_s)$ for quarks and
antiquarks is defined by the derivative \cite{Cassing:2008nn},
\begin{equation} \label{uss}
U_s(\rho_s) = \frac{d V_p(\rho_s)}{d \rho_s} ,
\end{equation}
which is evaluated numerically within the DQPM. The result  is
displayed in Fig. \ref{figpot} (r.h.s.) as a function of the parton
scalar density $\rho_s$ (\ref{rhos}) and shows that the scalar
mean-field is in the order of a few GeV for $\rho_s > 10$ fm$^{-3}$.
This mean-field (\ref{uss}) is employed in the  PHSD transport
calculations and determines the force on a quasiparticle $j$, i.e. $
\sim M_j/E_j \nabla U_s(x) = M_j/E_j \ d U_s/d \rho_ s \ \nabla
\rho_s(x)$ where the scalar density $\rho_s(x)$ is determined
numerically on a space-time grid (see below).

\subsection{DQPM at finite quark chemical potential}

Since the coupling (squared) in the DQPM is a function of $T/T_c$ and in
the Hard-Thermal-Loop approximation depends on
 \be T^* (T,\mu_q)= \sqrt{T^2 + \mu_q^2/\pi^2} ,
 \ee
a straight forward extension of the DQPM to finite $\mu_q$ is to
consider the coupling as a function of $T^*/T_c(\mu_q)$ with a
$\mu_q$-dependent critical temperature $T_c(\mu_q)$,
\begin{equation}
\label{equ:Sec2.8} \frac{T_c(\mu_q)}{T_c(\mu_q=0)} =
\sqrt{1 -  \alpha \ \mu_q^2} \approx 1 -  \alpha/2 \ \mu_q^2 + \cdots
\end{equation}
with $\alpha \approx$ 8.79 GeV$^{-2}$.  The expression of
$T_c(\mu_q)$ in Eq. (\ref{equ:Sec2.8}) is obtained by requiring a
constant energy density $\epsilon$ for the system at $T=T_c(\mu_q)$
where $\epsilon$ at $T_c(\mu_q = 0) \approx 0.158$ GeV is fixed by a
lattice QCD calculation at $\mu_q = 0$ \cite{Aoki:2009sc}. The
coefficient in front of the $\mu_q^2$-dependent part can be compared
to recent lQCD calculations at finite   (but small) $\mu_B$ which
gives~\cite{Bonati:2014rfa}:
 \begin{equation} \label{eli}
 \frac{T_c(\mu_B)}{T_c} = 1 - \kappa \left( \frac{\mu_B}{T_c} \right) ^2 + \cdots
 \end{equation} with $\kappa$ = 0.013(2). Rewriting Eq. (\ref{equ:Sec2.8})
 in the form (\ref{eli}) and  using $\mu_B \approx 3 \mu_q$ we get $\kappa_{DQPM} \approx 0.0122$ which
 compares very well with the lQCD result.  Consequently one has to expect an approximate scaling of the DQPM results
 if the partonic width is assumed to have the form,
\begin{eqnarray}
\label{equ:Sec2.9} & & {}  \gamma_g (T,\mu_q) = \frac{1}{3} N_c
\frac{g^2 (T^*/T_c(\mu_q))}{8 \pi} \, T \ \ln \left(\frac{2 c}{g^2
(T^*/T_c(\mu_q))} + 1 \right),
\nonumber\\
& & \gamma_q (T,\mu_q) = \frac{1}{3} \frac{N_c^2 - 1}{2 N_c}
\frac{g^2 (T^*/T_c(\mu_q))}{8 \pi} \, T \ \ln \left(\frac{2 c}{g^2
(T^*/T_c(\mu_q))} + 1 \right) .
\end{eqnarray}
% where $g^2(T/T_c)$ has been replaced by $g^2(T^*/T_c(\mu_q))$.
This choice leads to an approximate independence of the potential
energies per degree-of-freedom as a function of (moderate) $\mu_q$.
Nevertheless, the conjecture (\ref{equ:Sec2.9}) should be explicitly
controlled by future lQCD studies for $N_f$=3 at finite quark
chemical potential. Unfortunately, this task is presently out of
reach and one has to live with the uncertainty in (\ref{equ:Sec2.9})
which is assumed in the following investigations.

We point out, furthermore,  that in general the quasiparticle masses
$M_j$ as well as the widths $\gamma_j$ might depend also on the
momentum ${\bf q}$ relative to the medium at rest and approach the
perturbative values at high $q^2$. So far, the momentum dependence
of the complex selfenergy cannot reliably be extracted from the lQCD
results in thermodynamic equilibrium which are essentially sensitive
to momenta in the order of a few times the temperature. This is
presently an open issue and will have to be re-addressed in future.

On the basis of the DQPM for the partonic phase a relativistic
off-shell transport approach has been developed in the past decade
that gives approximately the same dynamics as the DQPM for partonic
systems in equilibrium but also contains interacting hadrons and a
dynamical transition between hadronic and partonic
degrees-of-freedom. This approach that can also be employed for
systems out of equilibrium -- such as heavy-ion collisions -- is
denoted by Parton-Hadron-String Dynamics (PHSD). The detailed set up
of PHSD as well as its comparison to heavy-ion data from low SPS to
LHC energies is described in the next Section.

% part of the review in PPNP 2015

\section{The PHSD approach}

The Parton-Hadron-String-Dynamics  approach is a microscopic
covariant transport model that incorporates effective partonic as
well as hadronic degrees-of-freedom and involves a dynamical
description of the hadronization process from partonic to hadronic
matter. Whereas the hadronic part is essentially equivalent to the
conventional HSD approach \cite{Ehehalt,Cass99} the partonic
dynamics is based on the Dynamical Quasiparticle Model
\cite{Peshier:2005pp,Andre05,Cassing06,Cassing:2007nb} which
describes QCD properties in terms of single-particle Green's
functions.  With the (essentially three) DQPM parameters for the
temperature-dependent effective coupling (\ref{running}) fixed by
lattice QCD results -- as described in Section 3 -- the approach is
fully defined in the partonic phase.

One might ask whether the quasiparticle properties -- fixed in
thermal equilibrium -- should be appropriate also for the
nonequilibrium configurations. This question is nontrivial and can
only be answered by detailed investigations e.g. on the basis of
Kadanoff-Baym equations. We recall that such studies have been
summarized in Ref.~\cite{Cassing:2008nn} for strongly interacting
scalar fields that initially are far off-equilibrium and simulate
momentum distributions of colliding systems at high relative
momentum. The results for the effective parameters $M$ and $\gamma$,
which correspond to the time-dependent pole mass and width of the
propagator, indicate that the quasiparticle properties - except for
the very early off-equilibrium configuration - are close to the
equilibrium mass and width even though the phase-space distribution
of the particles is far from equilibrium (cf. Figs. 8 to 10 in Ref.
\cite{Cassing:2008nn}). Accordingly, we will adopt the equilibrium
quasiparticle properties also for phase-space configurations out of
equilibrium as appearing in relativistic heavy-ion collisions. The
reader has to keep in mind that this approximation is well
motivated, however, not fully equivalent to the exact solution.

On the hadronic side PHSD includes explicitly the  baryon octet and
decouplet, the $0^-$- and $1^-$-meson nonets as well as selected
higher resonances as in HSD~\cite{Ehehalt,Cass99}. Hadrons of higher
masses ($>$ 1.5 GeV in case of baryons and $>$ 1.3 GeV in case of
mesons) are treated as 'strings' (color-dipoles) that  decay to the
known (low-mass) hadrons according to the JETSET algorithm
\cite{JETSET}. We discard an explicit recapitulation of the string
formation and decay and refer the reader to the original work
\cite{JETSET}.

\subsection{Hadronization}
\label{sect:hadroniz}
Whereas the dynamics of partonic as well as hadronic systems is fixed by the DQPM or HSD, respectively,
the change in the degrees-of-freedom has to be specified in line with the lattice QCD equation of state.
The hadronization, i.e. the transition from partonic to hadronic
degrees-of-freedom, has been introduced in Refs.
\cite{CasBrat,Cass08} and is repeated here for
completeness. The hadronization is implemented in PHSD by local
covariant transition rates e.g. for $q+\bar{q}$ fusion to a mesonic state $m$ of four-momentum
$p= (\omega, {\bf p})$ at space-time point $x=(t,{\bf x})$:
\begin{eqnarray}
&&\phantom{a}\hspace*{-5mm} \frac{d N_m(x,p)}{d^4x d^4p}= Tr_q
Tr_{\bar q} \
  \delta^4(p-p_q-p_{\bar q}) \
  \delta^4\left(\frac{x_q+x_{\bar q}}{2}-x\right) %\nonumber\\ && \times
  \omega_q \ \rho_{q}(p_q)
   \  \omega_{\bar q} \ \rho_{{\bar q}}(p_{\bar q})
\nonumber \\
&& \times |v_{q\bar{q}}|^2 \ W_m(x_q-x_{\bar q},(p_q-p_{\bar q})/2)
\, \, N_q(x_q, p_q) \
  N_{\bar q}(x_{\bar q},p_{\bar q}) \ \delta({\rm flavor},\, {\rm color}).
\label{trans}
\end{eqnarray}
In Eq. (\ref{trans}) we have introduced the shorthand notation,
\begin{equation}
Tr_j = \sum_j \int d^4x_j \int \frac{d^4p_j}{(2\pi)^4} \ ,
\end{equation}
where $\sum_j$ denotes a summation over discrete quantum numbers
(spin, flavor, color); $N_j(x,p)$ is the phase-space density of
parton $j$ at space-time position $x$ and four-momentum $p$.  In Eq.
(\ref{trans}) $\delta({\rm flavor},\, {\rm color})$ stands
symbolically for the conservation of flavor quantum numbers as well
as color neutrality of the formed hadronic state $m$ which can be
viewed as a color-dipole or 'pre-hadron'.  Furthermore, $v_{q{\bar
q}}(\rho_p)$ is the effective quark-antiquark interaction  from the
DQPM  (displayed in Fig. 10 of Ref. \cite{Cassing:2007nb}) as a
function of the local parton ($q + \bar{q} +g$) density $\rho_p$ (or
energy density). Furthermore, $W_m(x,p)$ is the dimensionless
phase-space distribution of the formed 'pre-hadron', i.e.
\begin{equation} \label{Dover:1991zn} W_m(\xi,p_\xi) =
\exp\left( \frac{\xi^2}{2 b^2} \right)\ \exp\left( 2 b^2 (p_\xi^2-
(M_q-M_{\bar q})^2/4) \right)
\end{equation} with $\xi = x_1-x_2 = x_q - x_{\bar q}$ and $p_\xi = (p_1-p_2)/2
= (p_q - p_{\bar q})/2$ (which had been previously introduced in
Eq. (2.14) of Ref. \cite{Dover:1991zn}). The width parameter $b$ is
fixed by $\sqrt{\langle r^2 \rangle} = b$ = 0.66 fm (in the rest
frame) which corresponds to an average rms radius of mesons. We note
that the expression (\ref{Dover:1991zn}) corresponds to the limit of
independent harmonic oscillator states and that the final
hadron-formation rates are approximately independent of the
parameter $b$ within reasonable variations. By construction the
quantity (\ref{Dover:1991zn}) is Lorentz invariant; in the limit of
instantaneous 'hadron formation', i.e. $\xi^0=0$, it provides a
Gaussian dropping in the relative distance squared $({\bf r}_1 -
{\bf r}_2)^2$. The four-momentum dependence reads explicitly (except
for a factor $1/2$)
\begin{equation} (E_1 - E_2)^2 - ({\bf p}_1 - {\bf p}_2)^2 -
(M_1-M_2)^2 \leq 0
\end{equation} and leads to a negative argument of the second
exponential in Ed. (\ref{Dover:1991zn}) favoring the fusion of
partons with low relative momenta $p_q - p_{\bar q}= p_1-p_2$.

Some comments on the hadronization scheme are in order: The
probability for a quark to hadronize is essentially proportional to
the timestep $dt$ in the calculation, the number of possible
hadronization partners in the volume $dV \sim$ 5 fm$^3$ and the
transition matrix element squared (apart from the gaussian overlap
function). For temperatures above $T_c$ the probability is rather
small ($\ll$ 1) but for temperatures close to $T_c$ and below $T_c$
the matrix element  becomes very large since it essentially scales
with the effective coupling squared (66) which is strongly enhanced
in the infrared. For a finite timestep $dt$ -- as used in the
calculations -- the probability becomes larger than 1 which implies
that the quark has to hadronize with some of the potential
antiquarks in the actual timestep if the temperature or energy
density becomes too low. Furthermore, the gluons practically freeze
out close to $T_c$ since the mass difference between quarks and
gluons increases drastically with decreasing temperature and the
reaction channel $g \leftrightarrow q + {\bar q}$ is close to
equilibrium. This implies that all partons hadronize. Due to
numerics some 'leftover' partons may occur at the end of the
calculations which are 'forced' to hadronize by increasing the
volume $dV$ until they have found a suitable partner. In practice
the 'forced' hadronization was only used for LHC energies where the
computational time was stopped at $\sim$ 1000 fm/c when partons with
rapidities close to projectile or target rapidity did not yet
hadronize due to time dilatation ($\gamma_{cm} \approx $ 1400).

Related transition rates  (\ref{trans}) are
defined for the fusion of three off-shell quarks ($q_1+q_2+q_3
\leftrightarrow B$) to a color neutral baryonic ($B$ or $\bar{B}$)
resonances of finite width (or strings) fulfilling energy and
momentum conservation as well as flavor current conservation (cf.
Section 2.3 in Ref. \cite{CasBrat}). In contrast to the familiar
coalescence models  this hadronization scheme solves the problem of
simultaneously fulfilling all conservation laws and the constraint
of entropy production. For further details we refer the reader to
Refs. \cite{CasBrat,Cass08}.

\subsection{Initial conditions} \label{sect:init}

The initial conditions for the parton/hadron dynamical system have
to be specified additionally.  In order to describe relativistic
heavy-ion reactions we start with two nuclei in their
'semi-classical' groundstate, boosted towards each other with a
velocity $\beta$ (in $z$-direction), fixed by the bombarding energy.
The initial phase-space distributions of the projectile and target
nuclei are determined in the local Thomas-Fermi limit as in the HSD
transport approach~\cite{Ehehalt,Cass99} or the UrQMD model
\cite{Bass:1998ca,Bleicher:1999xi}. We recall that at relativistic
energies the initial interactions of two nucleons are well described
by the excitation of two color-neutral strings which decay in time
to the known hadrons (mesons, baryons, antibaryons) \cite{JETSET}.
Initial hard processes - i.e. the short-range high-momentum transfer
reactions that can be well described by perturbative QCD - are
treated in PHSD (as in HSD) via PYTHIA 5.7 \cite{PYTHIA}. The novel
element in PHSD (relative to HSD) is the 'string melting concept' as
also used in the AMPT model \cite{Lin:2004en} in a similar context.
However, in PHSD the strings (or possibly formed hadrons) are only
allowed to 'melt' if the local energy density $\epsilon(x)$ (in the
local rest frame) is above  the transition energy density
$\epsilon_c$ which in the present DQPM version is $\epsilon_c \approx 0.5
$ GeV/fm$^3$. The mesonic strings then decay to quark-antiquark
pairs according to an intrinsic quark momentum distribution,
\begin{equation} \label{mom0}
F({\bf q}) \sim \exp(- 2 b^2 {\bf q}^2) \ ,
\end{equation}
in the meson rest-frame (cf. Eq. (\ref{trans}) for the inverse
process). The parton final four-momenta are selected randomly
according to the momentum distribution (\ref{mom0}) (with $b$= 0.66
fm), and the parton-energy distribution is fixed by the DQPM at
given energy density $\epsilon(\rho_s)$ in the local cell with
scalar parton density $\rho_s$. The flavor content of the $q\bar{q}$
pair is fully determined by the flavor content of the initial
string. By construction the 'string melting' to massive partons
conserves energy and momentum as well as the flavor content. In
contrast to Ref. \cite{Lin:2004en} the partons are of finite mass --
in line with their local spectral function -- and obtain a random
color $c= (1,2,3)$ or $(r,b,g)$ in addition. Of course, the color
appointment is color neutral, i.e. when selecting a color $c$ for
the quark randomly the color for the antiquark is fixed by $-c$. The
baryonic strings melt analogously into a quark and a diquark while
the diquark, furthermore, decays to two quarks. Dressed gluons are
generated by the fusion of nearest neighbor $q+ {\bar q}$ pairs ($q+
{\bar q} \rightarrow g$) that are flavor neutral until the ratio of
gluons to quarks reaches the value $N_g/(N_q + N_{\bar q})$ given by
the DQPM for the energy density of the local cell. This
'recombination' is performed for all cells in space during the
passage time of the target and projectile (before the calculation
continues with the next timestep) and conserves the four-momentum as
well as the flavor currents. We note, however, that the initial
phase in PHSD is dominated by quark and antiquark
degrees-of-freedom.

 Apart from proton-proton, proton-nucleus or
nucleus-nucleus collisions the PHSD approach can also be employed to
study the properties of the interacting hadron/parton system in a
finite box with periodic boundary conditions (cf. Section 4.4). To this aim the system
is initialized by a homogeneous distribution of test-particles in a
finite box with a momentum distribution close to a thermal one. Note
that in PHSD the system cannot directly be initialized by a
temperature and chemical potential since these 'Lagrange parameters'
can only be determined when the system has reached a thermal and
chemical equilibrium, i.e. when all forward and backward reaction
rates have become equal.

\subsection{System evolution}

\begin{figure} \centering
 \includegraphics*[width=80mm]{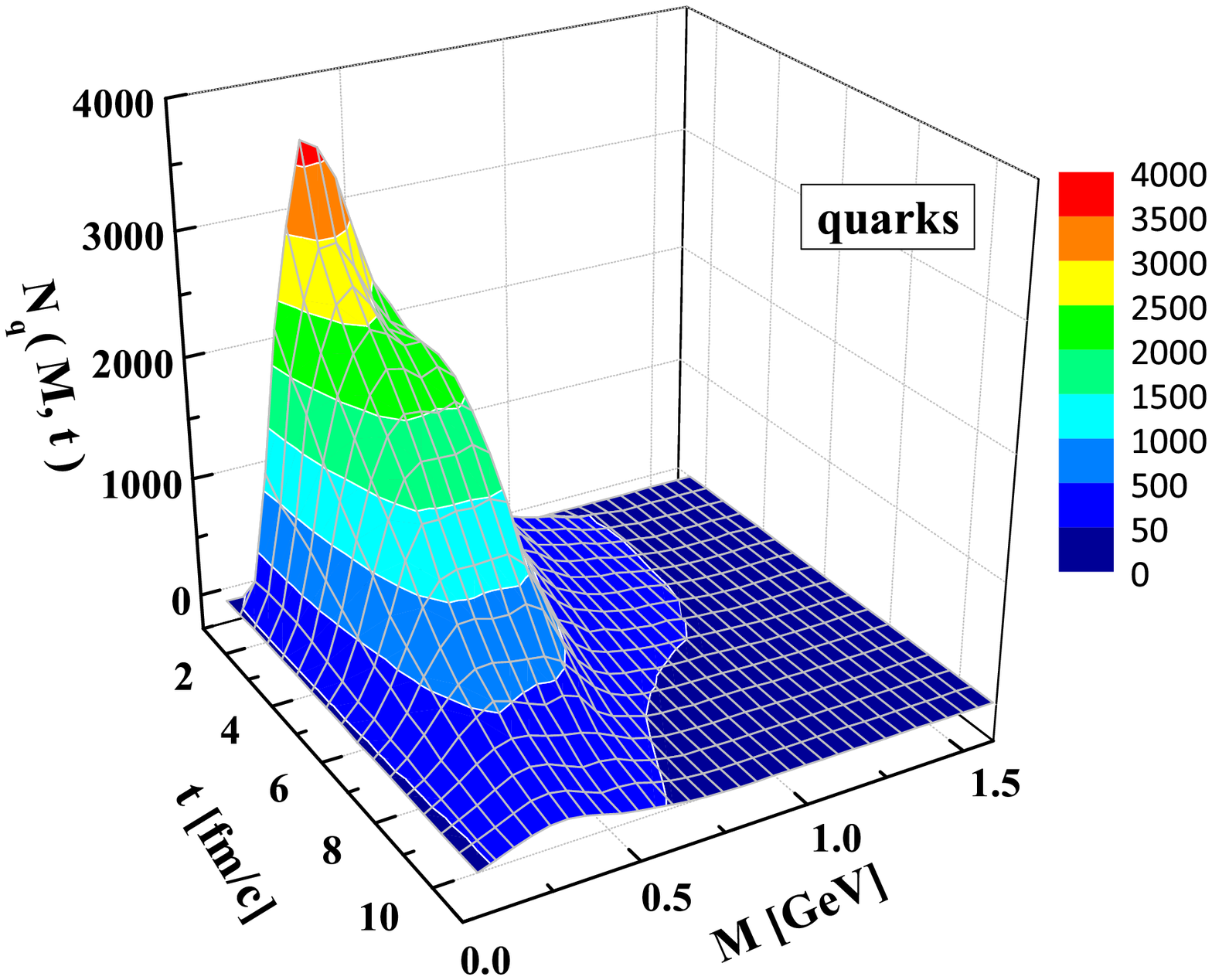} \includegraphics*[width=80mm]{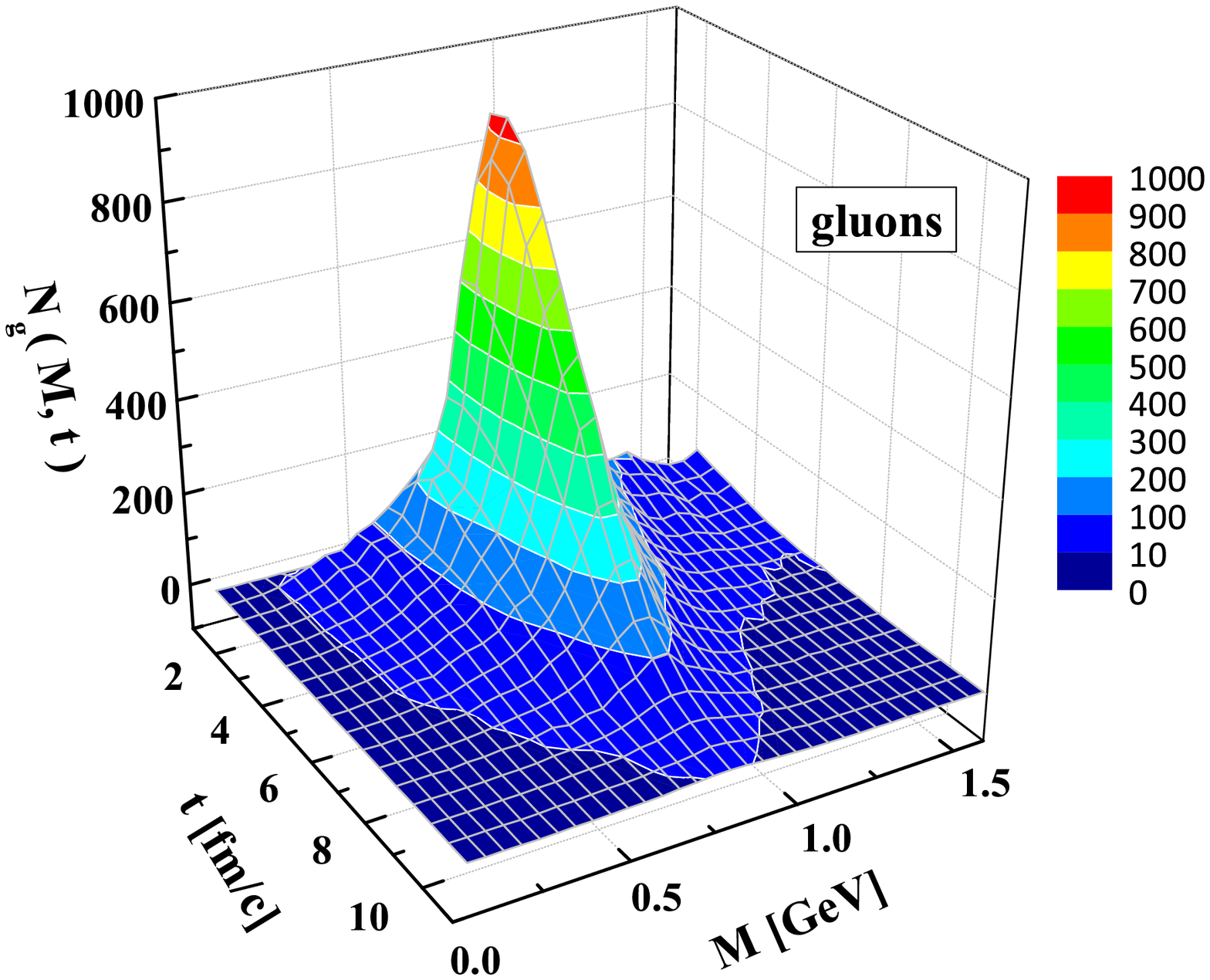}
 \caption{The time-dependent mass distributions for quarks (+ antiquarks) (l.h.s.)
and gluons (r.h.s.) for a central Au+Au collision at $\sqrt{s}$ =
200 GeV and $b$=1 fm at midrapidity ($|y| \leq 1$).  The figures are
taken from Ref. \protect\cite{BrCa11}.} \label{speq}
\end{figure}

The dynamical evolution of the system is entirely described by the
transport dynamics in PHSD incorporating the off-shell propagation
of the partonic quasiparticles according to Refs.
\cite{Juchem:2003bi,Cassing:2008nn} (Section 2.7) as well as the
transition to resonant hadronic states (or 'strings') via  Eq.
(\ref{trans}). The time integration for the test-particle-equations
of motion (cf. Eqs. (\ref{eomr}), (\ref{eomp}), (\ref{eome})) is
performed in the same way as in case of hadronic off-shell
transport, where (in view of the presently momentum-independent
width $\gamma$) the simple relation (\ref{eomm}) is employed. For
the collisions of partons two variants are at our disposal: i)
geometrical collision criteria as used in standard hadronic
transport, ii) the in-cell method developed in Ref.~\cite{Lang}. The
latter can easily be extended to describe $2 \leftrightarrow 3$ or
$1 \leftrightarrow 3$ processes etc. in a covariant way
\cite{Cassing:2001ds} and is the better choice at high particle
densities (cf. Refs.~\cite{Xu:2008av,Xu:2005tm,Xu:2004mz}). The
hadronization is performed by integrating the rate equations (e.g.
(\ref{trans})) in space and time which are discretized on a
four-dimensional grid by $\Delta t$ and $\Delta V(t) = \Delta x(t)
\Delta y(t) \Delta z(t) $. In beam direction we use an initial grid
size $\Delta z = 1/\gamma_{cm}$ fm with $\gamma_{cm}$ denoting the
Lorentz-$\gamma$ factor in the nucleon-nucleon center-of-mass system
while in the transverse direction we use $\Delta x = \Delta y$ = 1
fm. The grid size is increased dynamically during the transport
calculation such that all particles are included on the actual grid.
This practically implies that the grid boundary in beam direction
approximately moves with the velocity of light. In each time step
$\Delta t$ and cell $\Delta V$ the integrals in (\ref{trans}) and
the respective integrals for baryon (antibaryon) formation
 are evaluated by a sum over all (time-like)
test-particles using (e.g. for the quark density)
\begin{equation} \label{rho_DV}
\phantom{a}\hspace*{-10mm} \frac{1}{\Delta V} \int_{\Delta V} d^3x
\int_{-\infty}^\infty \frac{d \omega_q}{2 \pi} \ 2 \omega_q
\int_{-\infty}^\infty \frac{d^3  p_q}{(2\pi)^3} \
\rho_q(\omega_q,p_q)\ {\tilde N}_q(x,p_q) = \frac{1}{\Delta V}
\sum_{J_q \ {\rm in} \  \Delta V}  1 \ =  \ \rho_q(\Delta V) \ ,
\end{equation}
where the sum over $J_q$ implies a sum over all test-particles of
type $q$ (here quarks) in the local volume $\Delta V$ in each
parallel run. In Eq. (\ref{rho_DV}) ${\tilde N}$ denotes the
occupation number in phase space which in thermal equilibrium is
given by Bose- or Fermi-functions, respectively. In case of other
operators like the scalar density, energy density etc. the number 1
in Eq. (\ref{rho_DV}) has to be replaced by $\sqrt{P^2_J}/\omega_J$,
$\omega_J$ etc.  In order to obtain lower numerical fluctuations the
integrals are averaged over the  parallel runs (typically  50 at
RHIC energies). For each individual test-particle (i.e. $x_q$ and
$p_q$ fixed) the additional integrations in Eq. (\ref{trans})  give
a probability for a hadronization process to happen; the actual
event then is selected by a Monte Carlo algorithm. Energy-momentum
conservation fixes the four-momentum $p$ of the hadron produced and
its space-time position $x$ is determined by (\ref{trans}). The
final state is either a hadron with flavor content fixed by the
fusing quarks (and/or antiquarks) or by a string of invariant mass
$\sqrt{s}$ (with the same flavor), if $\sqrt{s}$ is above 1.3 GeV
for mesonic or above 1.5 GeV for baryonic quark content.

On the partonic side the following elastic and inelastic
interactions are included in PHSD $qq \leftrightarrow qq$, $\bar{q}
\bar{q} \leftrightarrow \bar{q}\bar{q}$, $gg \leftrightarrow gg$,
$gg \leftrightarrow g$, $q\bar{q} \leftrightarrow g$, $q g \leftrightarrow q g$,
$g \bar{q} \leftrightarrow g \bar{q}$  exploiting
'detailed-balance' with interaction rates again from the DQPM
\cite{CasBrat,Cassing:2007nb}. Partonic reactions such as $g+q \leftrightarrow q$
or $g+g \leftrightarrow q+ {\bar q}$  have been discarded
in the actual calculations due to their low rates since
the large mass of the gluon leads leads to a strong
mismatch in the energy thresholds between the initial and final channels.  On the other hand the evaluation of
photon and dilepton production is calculated perturbatively and channels
like $g+q \rightarrow  q+\gamma$ are included. In this case the probability for
photon (dilepton) production from each
channel is added up and integrated over space and time (Sections 5 - 7).

Numerical tests of the parton
dynamics with respect to conservation laws, interaction rates in and
out-off equilibrium in a finite box with periodic boundary conditions
have been presented in Ref. \cite{Ozvenchuk:2012fn}. In fact, in
Ref. \cite{Ozvenchuk:2012fn} it was shown that the PHSD
calculations 'in the box' give practically the same results in
equilibrium as the DQPM. We note in passing that the total energy
is conserved in the box calculations up to about 3 digits while in
the heavy-ion collisions addressed here in the following the violation
of energy conservation is typically less than 1 \% \cite{CasBrat}.

For illustration of the parton dynamics we display the time
evolution of the quark and gluon distributions in mass for a central
Au + Au collision at $\sqrt{s_{NN}}$ = 200 GeV in Fig. \ref{speq}
which shows the number of 'particles' as a function of invariant
mass $M$ and time $t$ at midrapidity ($|y| \leq 1$). Note that by
integration over $M$ one obtains the number of quarks (+ antiquarks)
$N_q(t)$ and gluons $N_g(t)$ in the rapidity interval $|y| \leq 1$
while dividing by $N_q(t)$ and $N_g(t)$, respectively, an estimate
for the particle spectral functions is obtained. Note that the mass
distributions displayed here are the product of the spectral
functions and the occupation numbers in a restricted phase space.
Due to a moderate variation of the partons pole mass and width with
the scalar density $\rho_s$  the shapes of the partonic mass
distributions do not change very much in time. The average quark
mass is about 0.5 GeV while the average gluon mass is only slightly
less than 1 GeV. Note, however, that the width of the mass function
- which reflects the actual interaction rate per parton -
 remains significant for all times up to hadronization.

\subsection{Transport coefficients of the QGP}

 The evaluation of transport coefficients can be performed
in different ways and is usually performed by evaluating the
temporal decay of correlators in the Kubo formalism \cite{Kubo,Kuboi}.
However, the results do not differ very much from those in the
relaxation time approximation (RTA) which is easier to work out. We
will thus focus on the latter approximation in this review for
brevity.

\subsubsection*{Shear and bulk viscosities in the relaxation time approximation}

The starting hypothesis of the relaxation time approximation  is
that the collision integral can be approximated (linearized) by
\begin{equation}
C[f]=-\frac{f-f^{eq}}{\tau}=: -\gamma (f-f^{eq}),
\end{equation}
where $\tau$ is the relaxation time and $f^{eq}$ the equilibrium
distribution. In this approach it has been shown that the shear and
bulk viscosities (without mean-field or potential effects) can be
written as (e.g. in Ref.~\cite{Chakraborty:2010fr}):
\begin{equation}
\eta=\frac{1}{15T}\sum\limits_{a}\int\frac{d^3p}{(2\pi)^3}\frac{|{\bf
p}|^4}{E_a^2}\tau_a(E_a)f^{eq}_a(E_a/T),
\end{equation}
\begin{equation}
\zeta=\frac{1}{9T}\sum\limits_{a}
\int\frac{d^3p}{(2\pi)^3} \ \frac{\tau_a(E_a)}{E_a^2}\bigl[(1-3v_s^2)E_a^2-M_a^2\bigr]f^{eq}_a(E_a/T),
\end{equation}
where the sum is over particles of different type $a$ (in our case,
$a=q,\bar q,g$). In the PHSD transport approach the relaxation time
can be expressed in the following way:
\begin{equation}
\tau_a=\gamma^{-1}_a,
\end{equation}
where $\gamma_a$ is the width of particles of type $a=q,\bar q,g$,
 defined by Eq. (\ref{widthg}).  In our numerical
simulation the volume $V$ averaged shear and bulk viscosities assume the
following expressions:
\begin{equation}
\eta=\frac{1}{15TV}\sum\limits_{i=1}^{N}\frac{|{\bf
p}_i|^4}{E_i^2}\gamma^{-1}_i, \mbox{\hspace{0.5 cm}}
%\end{equation}
%
%\begin{equation}
\zeta=\frac{1}{9TV}\sum\limits_{i=1}^{N}
\frac{\gamma^{-1}_i}{E_i^2}\bigl[(1-3v_s^2)E_i^2-M_i^2\bigr]^2,
\end{equation}
where the speed of sound $v_s=v_s(T)$ is taken from the DQPM using
\begin{equation} \label{vss} v_s^2 = \frac{\partial P}{\partial \epsilon}. \end{equation}

\begin{figure} \centering
\includegraphics[width=0.55\textwidth]{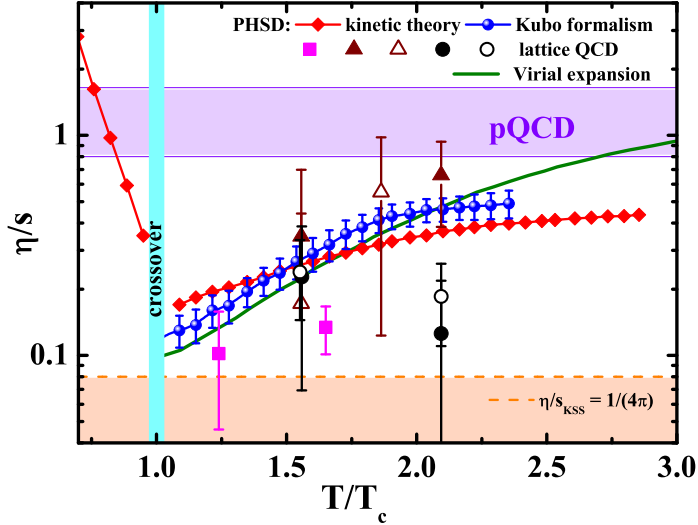}
\caption{  The shear viscosity to entropy density ratio $\eta/s$ as
a function of temperature of the system obtained by the PHSD
simulations using different methods: the relaxation time
approximation (red line$+$diamonds) and the Kubo formalism (blue
line$+$dots). The others symbols denote lattice QCD data for pure
$SU_c(3)$ gauge theory from different sources.
 The orange
dashed line demonstrates the Kovtun-Son-Starinets bound
$(\eta/s)_{KSS}=1/(4\pi).$ For comparison, the results in the virial
expansion approach (solid green line) \cite{Mattiello} are shown as
a function of temperature.  The figure is taken from Ref.
\protect\cite{Ozvenchuk:2012kh}.}\label{shear}
\end{figure}

In Fig.~\ref{shear} we present the shear viscosity to entropy
density ratio $\eta/s$ as a function of the temperature of the
system extracted from the PHSD simulations in the box employing
different methods: the relaxation time approximation (red
line$+$diamonds) and the Kubo formalism (blue line$+$dots). For
comparison, the results from the virial expansion approach (green
line) \cite{Mattiello}   and lattice QCD data for pure $SU_c(3)$
gauge theory are shown as a function of temperature, too.

In the absence of the chemical potential there should be no
consideration of vector or tensor fields, only scalar fields. This
affects the bulk viscosity, but not the shear viscosity. The
expression for the bulk viscosity with potential effects is~\cite{Chakraborty:2010fr}
\begin{eqnarray}
\zeta
&=&\frac{1}{T}\sum\limits_{a}\int\frac{d^3p}{(2\pi)^3}\frac{\tau_a(E_a)}{E_a^2}f^{eq}_a(E_a/T) %\\ &\times &
\Bigl[\Bigl(\frac{1}{3}-v_s^2\Bigr)|{\bf
p}|^2-v_s^2\Bigl(M_a^2-T^2\frac{d(M_a^2)}{d(T^2)}\Bigr)\Bigr]^2.
\end{eqnarray}

\noindent
 In the numerical simulation the volume averaged bulk viscosity with
mean-field effects is calculated as
\begin{equation}
\zeta=\frac{1}{TV}\sum\limits_{i=1}^{N}
\frac{\gamma^{-1}_i}{E_i^2}\Bigl[\Bigl(\frac{1}{3}-v_s^2\Bigr)|{\bf
p}|^2-v_s^2\Bigl(M_i^2-T^2\frac{d(M_i^2)}{d(T^2)}\Bigr)\Bigr]^2.
\end{equation}
Using the DQPM expressions for masses of quarks and gluons
(\ref{Mg9}) and (\ref{Mq9}), we can calculate the derivative
$d(M^2)/d(T^2)$ as well as $v_s^2$ according to Eq. (\ref{vss}). For
the actual results we refer the reader to Fig.~\ref{bulk} (l.h.s.),
where the bulk viscosity to entropy density ratio $\zeta/s$ is
presented as a function of temperature of the system extracted from
the PHSD simulations in the box using the relaxation time
approximation including mean-field effects (red line$+$diamonds) and
without potential effects (blue line$+$open triangles). The r.h.s.
of Fig.~\ref{bulk} shows the bulk to shear viscosity ratio as a
function of temperature. Let us stress that contrary to $\eta/s$,
the ratio $\zeta/s$ has a maximum close to $T_c$.

\begin{figure}
\includegraphics[width=0.49\textwidth]{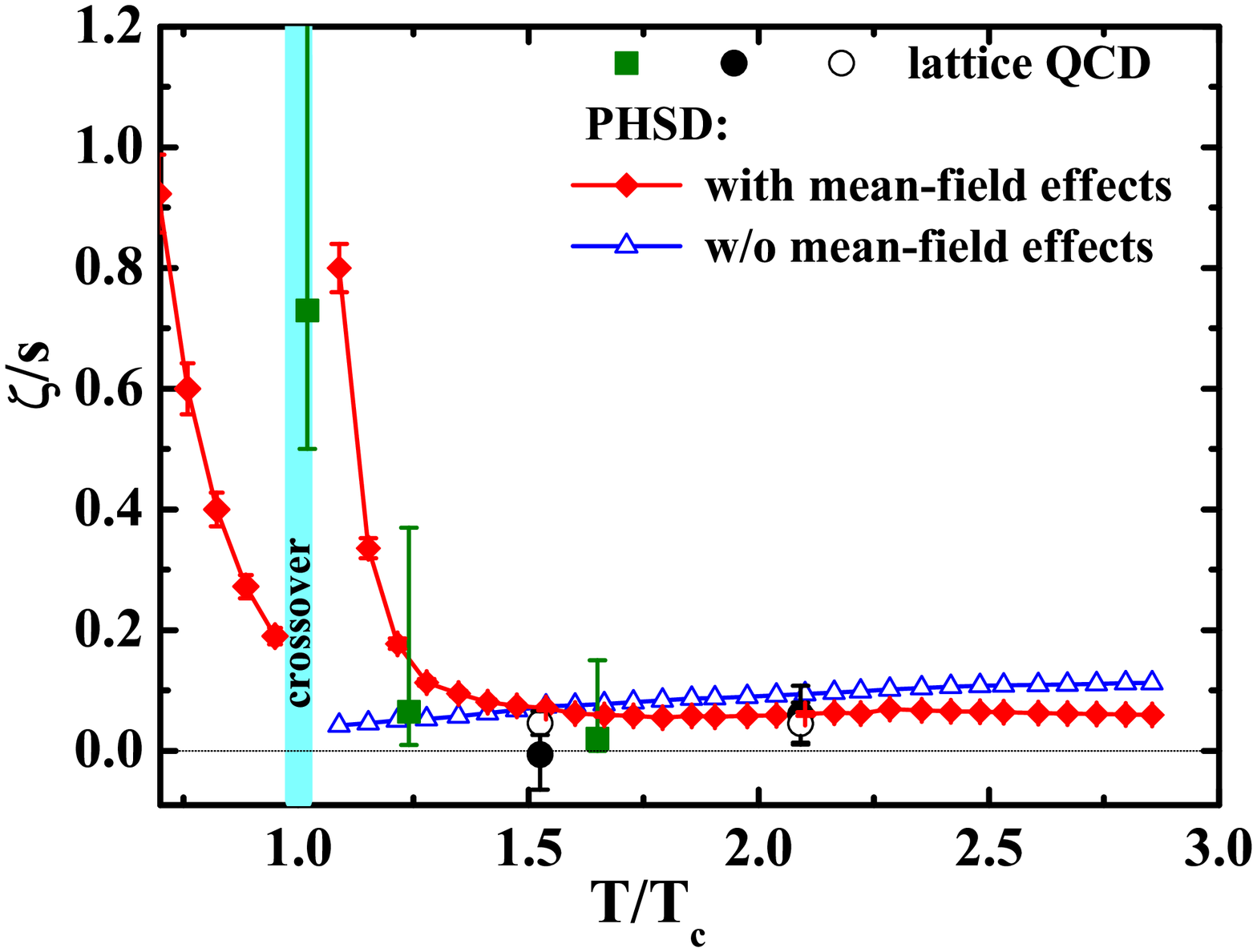} \includegraphics[width=0.49\textwidth]{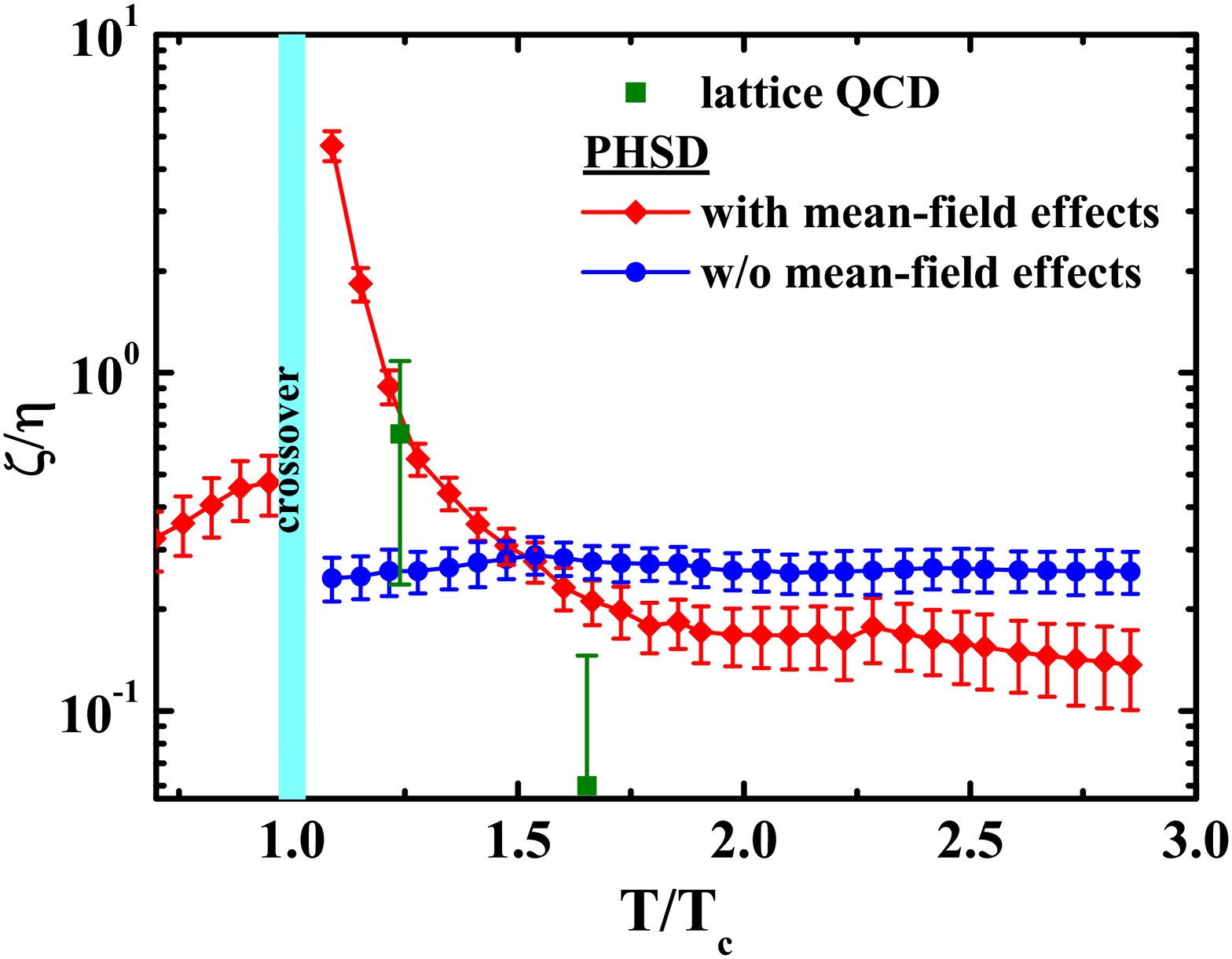}
\caption{(l.h.s.) The bulk viscosity to entropy density ratio
$\zeta/s$ as a function of temperature of the system extracted from
the PHSD simulations in the box using the relaxation time
approximation including mean-field effects (red line$+$diamonds) and
without potential effects (blue line$+$open triangles). The other
symbols show the available lattice QCD data from different sources
(r.h.s.). The bulk to shear viscosity ratio as a function of
temperature as obtained by the PHSD simulations in the box employing
the relaxation time approximation including mean-field effects (red
line$+$diamonds) and without potential effects (blue
line$+$circles). Figures taken from
Ref.~\protect\cite{Ozvenchuk:2012kh}. }\label{bulk}
\end{figure}

\subsubsection*{Electric conductivity}

\begin{figure}[t] \centering
     \includegraphics[width=0.55\textwidth]{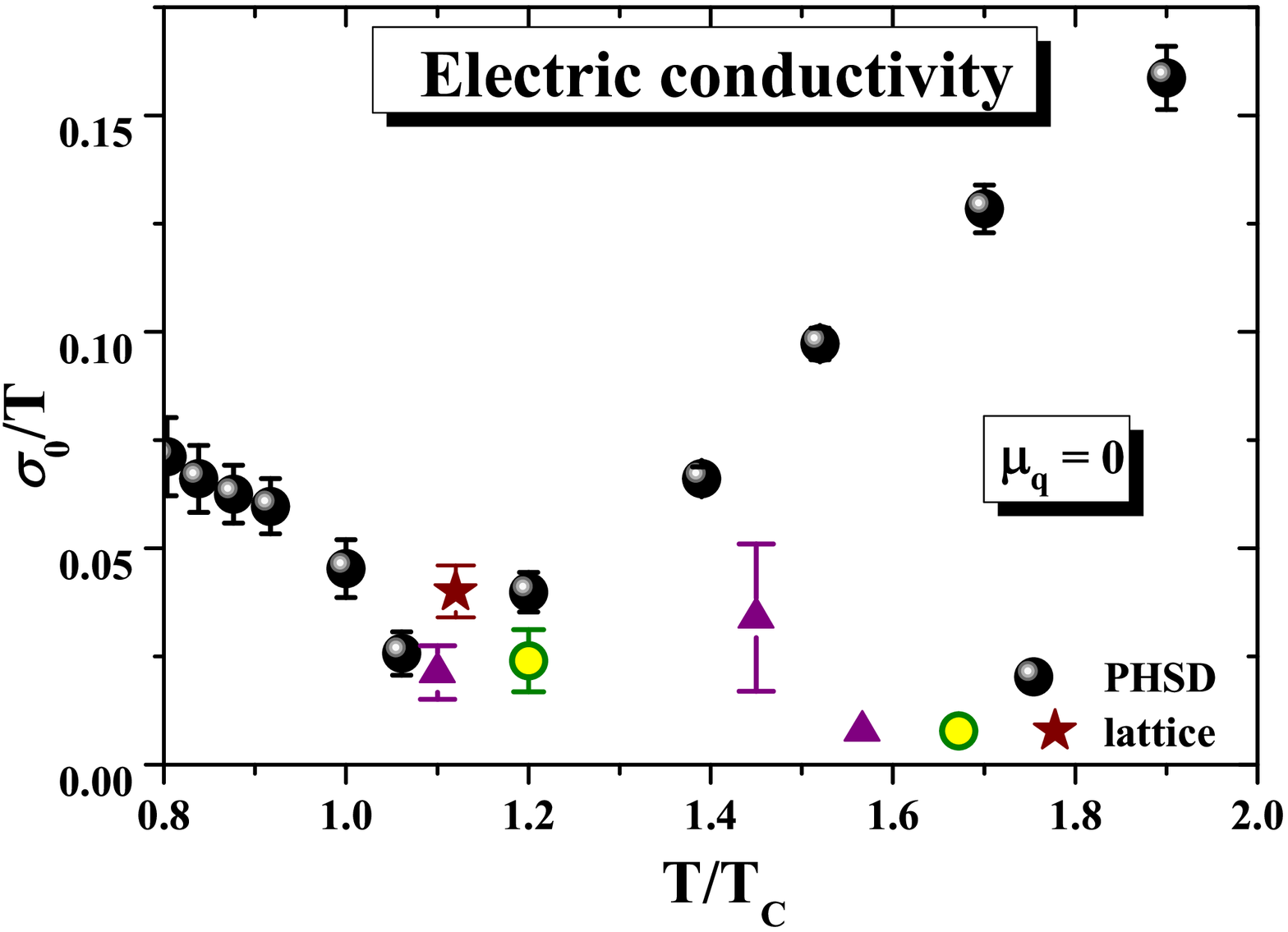}
\caption{ The dimensionless ratio of electric conductivity over
temperature $\sigma_0/T$ (\ref{e4}) as a function of the scaled
temperature $T/T_c$ for $\mu_q=0$ in comparison to recent lattice
QCD results. The figure is taken from Ref.
\protect\cite{Cassing:2013iz}.} \label{figg1}
\end{figure}

Whereas shear and bulk viscosities of hot QCD matter at finite
temperature $T$ presently are roughly known, the electric
conductivity $\sigma_0(T,\mu_q)$ is a further macroscopic quantity
of interest since it controls the electromagnetic emissivity of the
plasma.  First results from lattice calculations on the
electromagnetic correlator have provided results that varied by more
than an order of magnitude. Furthermore, the conductivity dependence
on the temperature $T$ (for $T\!\!>\!\!T_c$) is widely unknown, too,
as well as its dependence on $\mu_q$. The electric conductivity
$\sigma_0$ is also important for the creation of electromagnetic
fields in ultra-relativistic nucleus-nucleus collisions from
partonic degrees-of-freedom, since $\sigma_0$ specifies the
imaginary part of the electromagnetic (retarded) propagator and
leads to an exponential decay of the propagator in
time $\sim \! \exp(-\sigma_0 (t-t')/({\hbar} c))$.\\

In order to include the effects from an external electric field
${\bf E}$ or magnetic field ${\bf B}$ on the charged degrees-of-freedom, the propagation of each
charged test-particle $j$ in the PHSD is performed with the additional Lorentz
force in the equation of motion:
\begin{equation} \label{e1}
\frac{d}{dt} {\bf p}^j = q_j e ({\bf E} + \frac{{\bf p}^j}{E^j}
\times {\bf B}),
\end{equation}
where $q_j$ denotes the fractional charge of the test-particle ($\pm
1/3, \pm 2/3$) and $E^j$ its energy. We recall that the external
electric field will lead to an acceleration of positively and
negatively charged particles in opposite directions while the
particle scatterings/interactions will damp this acceleration and
eventually lead to an equilibrium current if the external field is
of moderate strength.
The  electric current density $j_z(t)$ (for an external electric
field in $z$-direction) is calculated by
\begin{equation} \label{e2} j_z(t) = \frac{1}{VN} \sum_{k=1}^N \sum_{j=1}^{N_k(t)} \ e q_j
\frac{p_z^j(t)}{E_j(t)} . \end{equation}   The summation in (\ref{e2}) is
carried out over $N$ ensemble members $k=1 \dots N$ while $N_k(t)$
denotes the time-dependent number of 'physical' ($u,d,s$) quarks and
antiquarks that varies  with time $t$ due to the processes $q +
\bar{q} \leftrightarrow g \leftrightarrow q'+{\bar q}'$ in a single
member of the ensemble (run). The number of runs $N$ is typically
taken as a few hundred which gives a current $j_z(t)$ practically
independent on the number of ensemble members $N$. We recall that
(without external fields) each run of the ensemble is a
micro-canonical simulation of the dynamics as inherent in the PHSD
transport approach which strictly conserves the total four-momentum
as well as all discrete conservation laws (e.g. net fermion number
for each flavor etc.). A note of caution has to be given, since due
to an external field we deal with an open system with increasing
energy density (temperature) in time. Therefore we employ
sufficiently small external fields $eE_z$, such that the energy
increase during the computation time (in each run) stays below 2\%
and the increase in temperature below 1 MeV. For the details we refer the
reader to Refs. \cite{Cassing:2013iz,Steinert:2013fza}.

We find that for constant electric fields up to $e E_z$ = 50 MeV/fm
a stable electric current $j_{eq}$ emerges that is $\sim E_z$.
Accordingly,  we obtain the conductivity
$\sigma_0(T,\mu_q)$ from the ratio of the stationary current density
$j_{eq}$ and the electric field strength as
\begin{equation} \label{e4}
\frac{\sigma_0(T,\mu_q)}{T} = \frac{j_{eq}(T,\mu_q)}{E_z T} \ .
\end{equation}
The results for the dimensionless ratio (\ref{e4}) at $\mu_q=0$
 are displayed in Fig. \ref{figg1} by the full dots as a
function of the scaled temperature $T/T_c$ in comparison to
recent lattice QCD results  and suggest a minimum in the ratio
$\sigma_0(T,\mu_q=0)/T$ close to the critical temperature $T_c$
followed by an approximate linear rise up to 2 $T_c$. The
recent lQCD results are roughly compatible with the PHSD
predictions.

Within PHSD (or the DQPM) also the dependence of the electrical conductivity on the quark chemical
potential can be evaluated \cite{Steinert:2013fza}.
The numerical result could be fitted by a quadratic correction
%(solid line in Fig.~\ref{pic:condTvsmu})
\begin{equation} \label{parabol}
\frac{\sigma_0(T,\mu_q)}{T} =  \frac{\sigma_0(T,\mu_q=0)}{T} \left(1
+ a(T) \mu_q^2 \right)
\end{equation} with $a(T) \approx$ 11.6 \  GeV$^{-2}$ for $T=0.2$ GeV.
This result comes about as follows: We recall that the electric
conductivity of gases, liquids and solid states is described in the
relaxation time approach by the Drude formula,
\begin{equation} \label{eq7} \sigma_0 = \frac{e^2 n_e \tau}{m_e^*} ,
\end{equation}
where $n_e$ denotes the density of non-localized charges, $\tau$ is
the relaxation time of the charge carriers in the medium and $m_e^*$
their effective mass. This expression can be directly computed for
partonic degrees-of-freedom within the DQPM, which matches
 the quasiparticles properties to lattice QCD results in
equilibrium. In the DQPM, the relaxation time for quarks/antiquarks
is given by $\tau = 1/\gamma_q(T,\mu_q)$, where $\gamma_q(T,\mu_q)$
is the width of the quasiparticle spectral function (\ref{widthg}).
Furthermore, the spectral distribution for the mass
of the quasiparticle has a finite pole mass $M_q(T,\mu_q)$ that is
also fixed in the DQPM (\ref{Mq9}) as well as the density of ($u,
\bar{u}, d, \bar{d}, s, \bar{s}$) quarks/antiquarks as a function of
temperature $T$ and chemical potential $\mu_q$. The latter is given
by an expression similar to the scalar density $\rho_s$ in
(\ref{rhos}) but $\sqrt{p^2}$ replaced by $\omega$. Thus, we obtain
for the dimensionless ratio (\ref{e4}) the expression
\begin{equation} \label{e8}
\frac{\sigma_0 (T,\mu_q)}{T} \approx \frac{2}{9} \frac{e^2
n_{q+{\bar q}}(T,\mu_q)}{M_q(T,\mu_q) \gamma_q(T,\mu_q) T} ,
\end{equation}
where $n_{q+{\bar q}}(T,\mu_q)$ denotes the total density of quarks
and antiquarks and the pre-factor $2/9$ reflects the flavor averaged
fractional quark charge squared $(\sum_f q_f^2)/3$. As found in Ref. \cite{Steinert:2013fza}
the DQPM results match well with the explicit PHSD calculations in the box also for
finite $\mu_q$ since  PHSD in equilibrium is a suitable transport
realization of the DQPM.
 In the DQPM we have $\gamma_q(T,\mu_q) \approx \gamma_q(T,\mu_q =0)$
and $M_q(T,\mu_q) \approx M_q(T,\mu_q=0)$ for $\mu_q \leq $ 100 MeV,
however, \begin{equation} \label{D5} n_{q+{\bar q}}(T,\mu_q) \approx
n_{q+{\bar q}}(T,\mu_q=0) \left( 1 + a(T) \mu_q^2 \right)
\end{equation} with the same coefficient $a(T)$ as in  Eq. (\ref{parabol}).

\begin{figure}
\centering {\psfig{figure=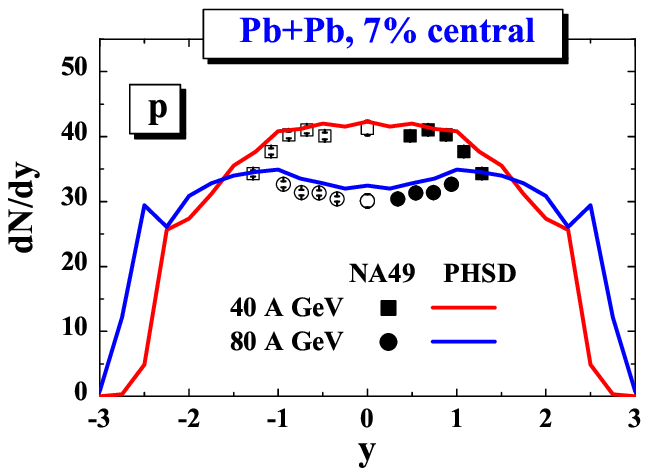,width=8.cm}}
{\psfig{figure=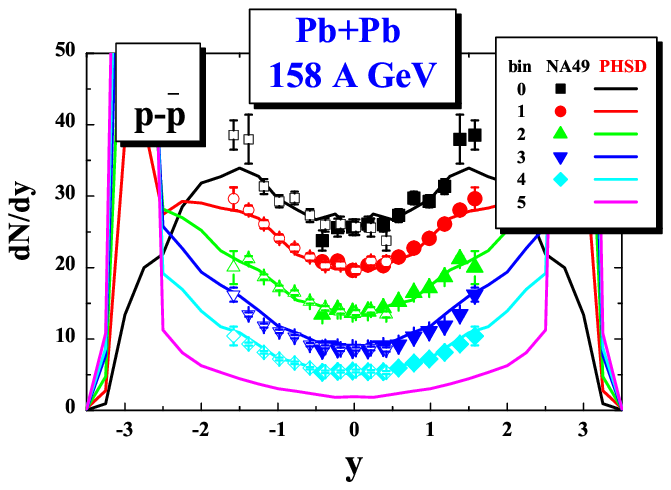,width=8.cm}} \caption{The proton rapidity
distributions for central (7\%) Pb+Pb collisions at 40 and 80
(l.h.s.) in comparison to the data from Ref. \protect\cite{NA49yp}.
The r.h.s. of the figure presents the net-proton rapidity
distribution at 158 A$\cdot$GeV for different centrality bins (bin 0
--  0-5\%; bin 1  -- 5-12\%; bin 2 -- 12.5-23.5\%; bin 3 --
23.5-33.5\%; bin 4 --  33.5-43.5\% and bin 5 -- 43.5-78.5\% central
events) from PHSD (solid  lines) in comparison to  the experimental
data from the NA49 Collaboration \protect\cite{NA49yp2}. The figures
are taken from Ref. \protect\cite{CasBrat}.} \label{fig15g}
\end{figure}

 The temperature dependence of the expansion coefficient $a(T)$ is
found to be $\sim 1/T^2$ such that  the ratio $\sigma_0/T$ can be
approximated by
\begin{equation} \label{expand}
\frac{\sigma_0(T,\mu_q)}{T} \approx  \frac{\sigma_0(T,\mu_q=0)}{T}
\left(1 + c_{\sigma_0} \frac{\mu_q^2}{T^2} \right) .
\end{equation}
 A fit to the coefficient $c_{\sigma_0}$ in the temperature
range 170 MeV$ \leq T \leq $ 250 MeV gives $c_{\sigma_0} \approx
0.46$. This completes our study on the stationary electric
conductivity $\sigma_0$ which can be well understood in its
variation with $T$ and $\mu_q$ within the DQPM or PHSD, respectively.
We note that the conductivity $\sigma_0$ controls the electromagnetic
emissivity of systems in thermal equilibrium at low photon momentum (see Section 5.3).

\subsection{Application to Au+Au or Pb+Pb collisions}

In this Subsection we employ the PHSD approach  to nucleus-nucleus
collisions from $\sqrt{s_{NN}}$ = 5.5 GeV to 2.76 TeV. Note that at
RHIC or more specifically LHC energies other initial conditions
(e.g. a color-glass condensate (CGC) \cite{Larry,Larryi}) might be
necessary. In the present work we discard such alternative initial
conditions and explore to what extent the present initial conditions
(described in Section~\ref{sect:init}) are compatible with
differential measurements by the various collaborations at the SPS,
RHIC or LHC. A more detailed comparison to results from CGC initial
conditions in Pb-Pb collisions at $\sqrt{s_{NN}}$ = 2.76 TeV may be
found in Ref. \cite{Konchakovski:2014fya}.

\subsubsection*{Particle spectra in comparison to experiment}

\begin{figure}
\hspace{0.9cm} {\psfig{figure=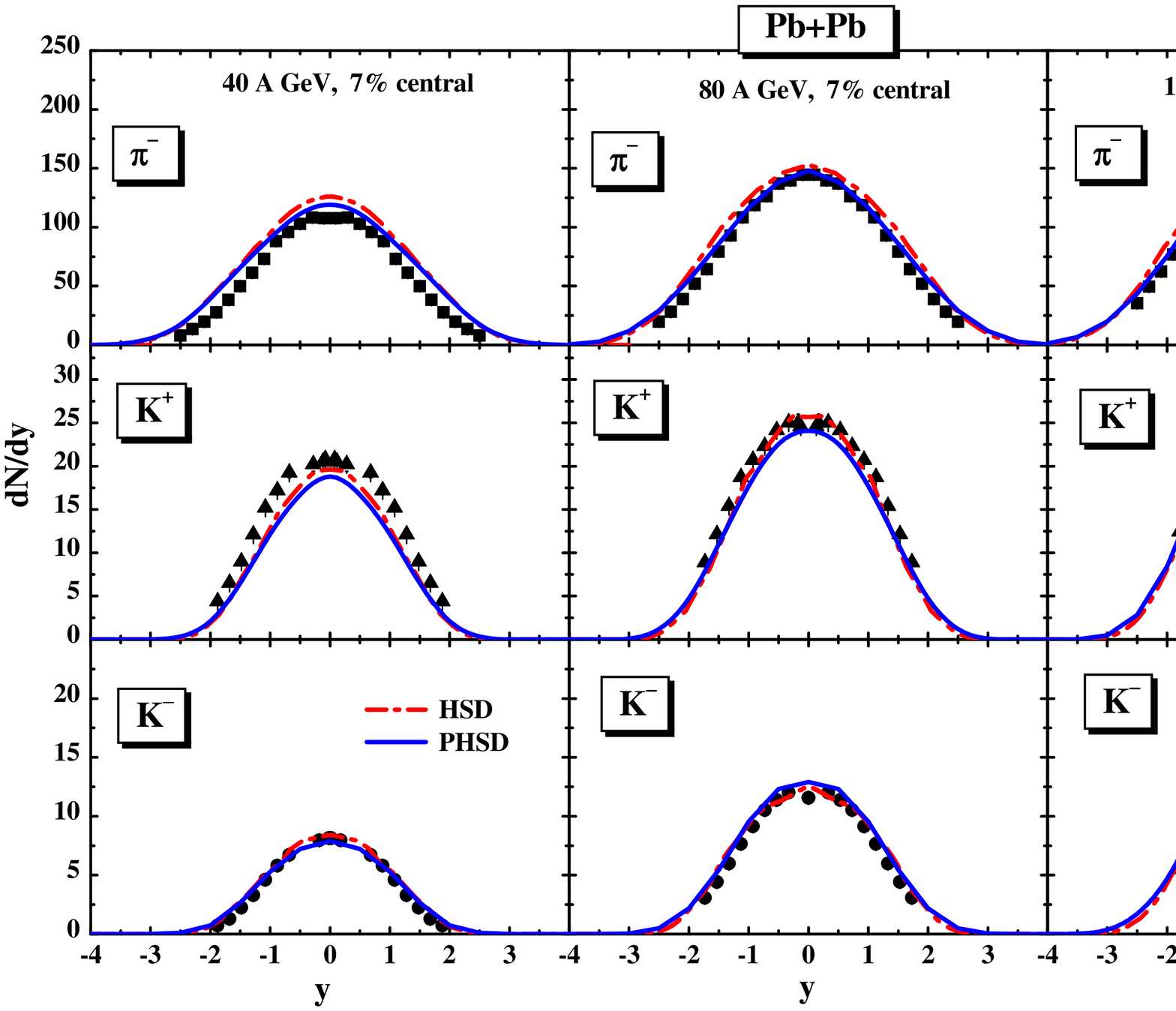,width=11cm}} \caption{The
rapidity distribution of $\pi^-$ (upper part), $K^+$ (middle part)
and $K^-$ (lower part)  for 7\% or 5\% central Pb+Pb collisions at
40, 80 and 158 A$\cdot$GeV from PHSD (solid blue lines) in
comparison to the distribution from HSD (dashed red lines) and the
experimental data from the NA49 Collaboration
\protect\cite{NA49a,NA49ai}. The figures are taken from Ref.
\protect\cite{CasBrat}.} \label{fig13}
\end{figure}
\begin{figure} \hspace{-0.3cm}
 {\psfig{figure=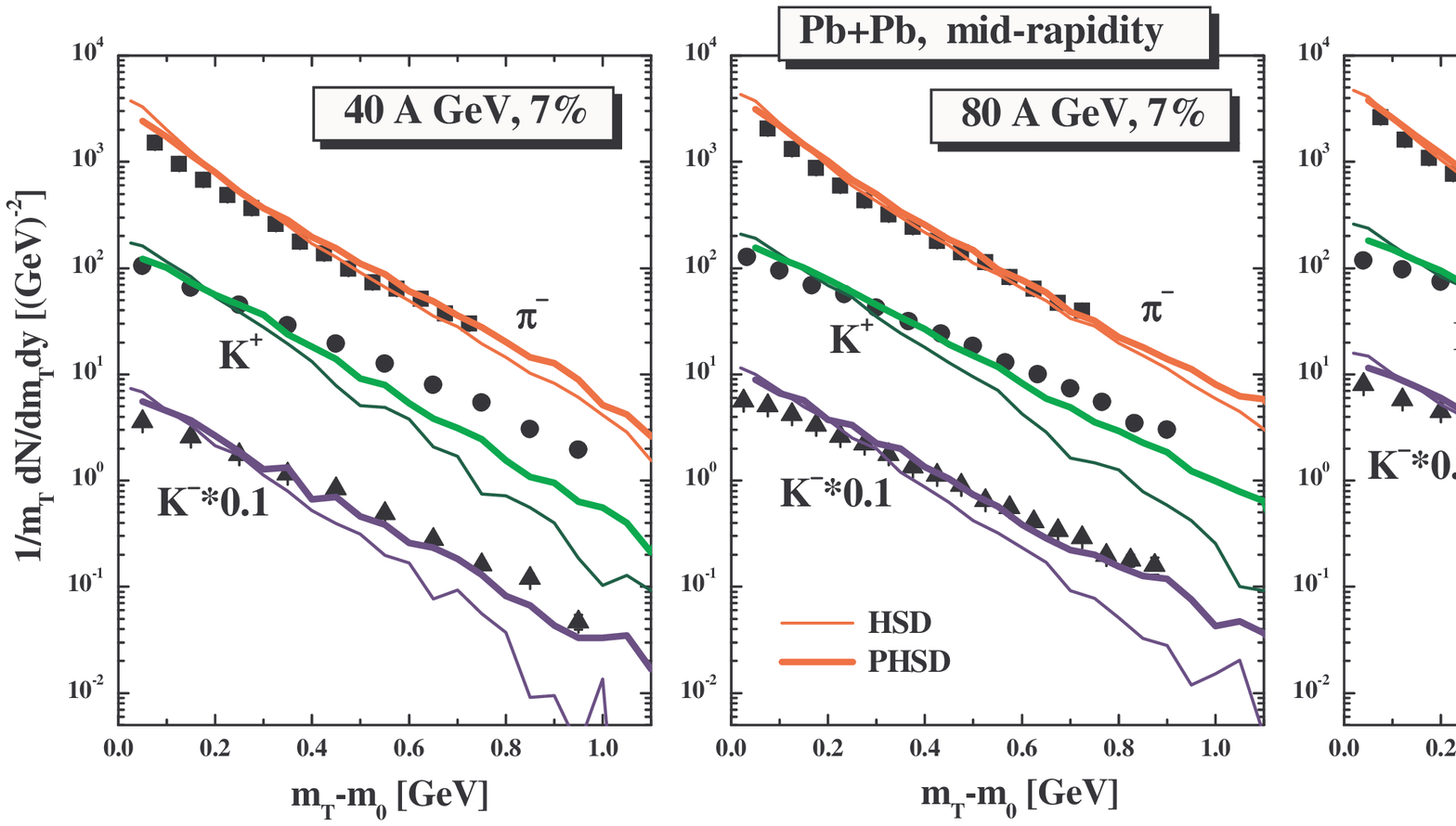,width=12.6cm}} \caption{The
$\pi^-$, $K^+$ and $K^-$ transverse mass spectra for central Pb+Pb
collisions at 40, 80 and 158 A$\cdot$GeV from PHSD (thick solid
lines) in comparison to the distributions from HSD (thin solid
lines) and the experimental data from the NA49 Collaboration
\protect\cite{NA49a}.  The figures are taken from Ref.
\protect\cite{CasBrat}. } \label{fig14} \end{figure}

Since PHSD is essentially fixed by lQCD data at $\mu_q$ = 0 in
thermal equilibrium in the partonic phase and by HSD in the hadronic
phase,  it is of interest how the PHSD approach compares to the HSD
model (without explicit interacting partonic degrees-of-freedom) as
well as to experimental data from the SPS, RHIC or LHC
collaborations. We start with proton rapidity distributions at the
SPS that demonstrate the amount of initial baryon stopping and thus
control the energy transfer in relativistic nucleus-nucleus
collisions. Since we find the HSD results for the proton rapidity
distribution $dN/dy$ to be identical with the PHSD results (within
statistics) we will only compare PHSD calculations to data of the
NA49 Collaboration. Accordingly, in Fig. \ref{fig15g} the proton
rapidity distributions from PHSD are compared to the data from Ref.
\cite{NA49yp} for 7\% central Pb+Pb collisions at 40 and 80
A$\cdot$GeV (l.h.s.). The r.h.s. of Fig. \ref{fig15g} shows the
net-proton $dN/dy$ from PHSD for 158 A$\cdot$GeV Pb+Pb collisions
for different centrality bins (bin 0 -- 0-5\%; bin 1 -- 5-12\%; bin
2 -- 12.5-23.5\%; bin 3 -- 23.5-33.5\%; bin 4 -- 33.5-43.5\% and bin
5 -- 43.5-78.5\% central events) in comparison to the  experimental
data from Ref. \cite{NA49yp2}. In fact, the PHSD results demonstrate
that the baryon stopping is reasonably reproduced in Pb+Pb
collisions as a function of bombarding energy and centrality of the
reaction at the SPS energies.
We note additionally that at SPS energies
the antiprotons from HSD are about the same as from
PHSD as well as $\Lambda + \Sigma^0$ and even $\Xi^-$ baryons, however,
the antibaryons  with antistrangeness ${\bar \Lambda} + {\bar \Sigma}^0$
and  ${\bar \Xi}^+$ are more abundant in PHSD than in HSD due to a large
contribution from hadronization. For further details on the baryon/antibaryon sector
in HSD and PHSD we refer the reader to Ref. \cite{CasBrat}.

Since the energy is dominantly transferred to mesons,
which asymptotically appear mostly as pions and kaons, we continue
with pion and $K^\pm$ rapidity distributions for 7\% central Pb+Pb
collisions at 40 and 80 A$\cdot$GeV and 5\% central collisions at
158 A$\cdot$GeV since here rather complete data sets are available
from the experimental side \cite{NA49a}. The results from PHSD
(solid blue lines)  are compared in Fig. \ref{fig13} with the
corresponding results from HSD (dashed red lines) and the
experimental data for the same centralities in comparison to the
rapidity spectrum from HSD (dashed red lines) and the experimental
data from the NA49 Collaboration \cite{NA49a}. The actual deviations
between the PHSD and HSD spectra are very moderate; the $\pi^-$
rapidity distribution is slightly squeezed in width (in PHSD) and
shows a more pronounced peak at midrapidity (at 158 A$\cdot$GeV)
more in line with the data. Nevertheless, it becomes clear from Fig.
\ref{fig13} that the energy transfer - reflected in the light meson
spectra - is rather well described by PHSD, which thus passes
another test. Fig. \ref{fig13} demonstrates that the longitudinal
motion is rather well understood within the transport approaches and
dominated by initial string formation and decay. Actually, there is
no sizeable sensitivity of the rapidity spectra to an intermediate
partonic phase. But what about the transverse degrees-of-freedom?

\begin{figure}
\centerline{\psfig{figure=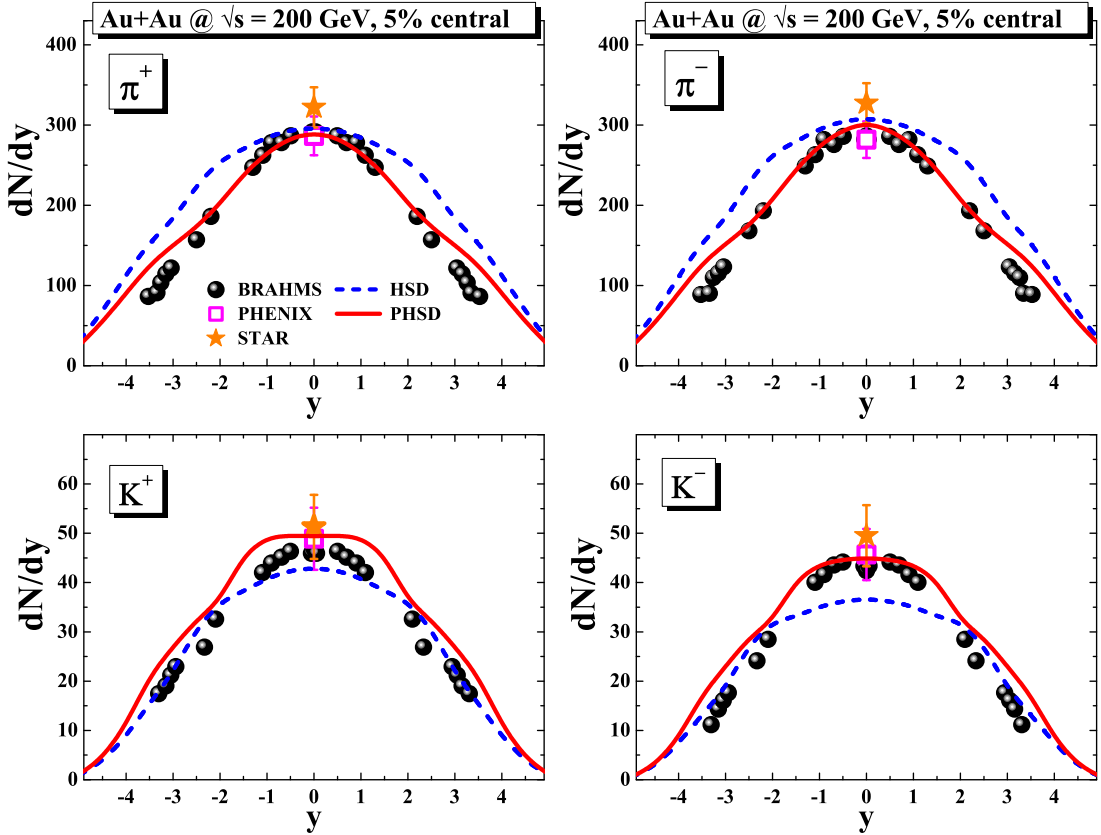,width=15cm}} \caption{The
rapidity distribution of $\pi^+$ (upper part, l.h.s.), $K^+$ (lower
part, l.h.s.), $\pi^-$ (upper part, r.h.s.) and $K^-$ (lower part,
r.h.s.) for  5\% central Au+Au collisions at $\sqrt{s}$ = 200 GeV
from PHSD (solid lines) in comparison to the distribution from HSD
(dashed lines) and the experimental data from the RHIC
Collaborations \protect\cite{PHENIX2,Adams:2003xp}. The figure is
taken from Ref. \protect\cite{BrCa11}.} \label{fig13b}
\end{figure}

 The answer to this question is offered in Fig. \ref{fig14}
where we show the transverse mass spectra of $\pi^-$, $K^+$  and
$K^-$ mesons for 7\% central Pb+Pb collisions at 40 and 80
A$\cdot$GeV and 5\% central collisions at 158 A$\cdot$GeV in
comparison to the data of the NA49 Collaboration \cite{NA49a}.  Here
the slope of the $\pi^-$ spectra is only slightly enhanced in PHSD
(thick solid lines) relative to HSD (thin solid lines)  which
demonstrates that the pion transverse mass spectra also show no sizeable
sensitivity to the partonic phase. However, the $K^\pm$ transverse
mass spectra are substantially hardened with respect to the HSD
calculations at all bombarding energies - i.e. PHSD is more in line
with the data - and thus suggest that partonic effects are better
visible in the strangeness degrees-of-freedom. The hardening of the
kaon spectra can be traced back to parton-parton scattering as well
as a larger collective acceleration of the partons in the transverse
direction due to the presence of repulsive fields for the partons.
 The enhancement of the spectral slope for kaons and
anti-kaons in PHSD (due to collective partonic flow) shows up much
clearer for the kaons due to their significantly larger mass
(relative to pions). We recall that in Refs.
\cite{Bratkovskaya:2003ie,brat04} the underestimation of the $K^\pm$
slope by HSD (and also UrQMD) had been suggested to be a signature
for missing partonic degrees-of-freedom. In fact, the PHSD
calculations support this early suggestion.

\begin{figure}
\centerline{\psfig{figure=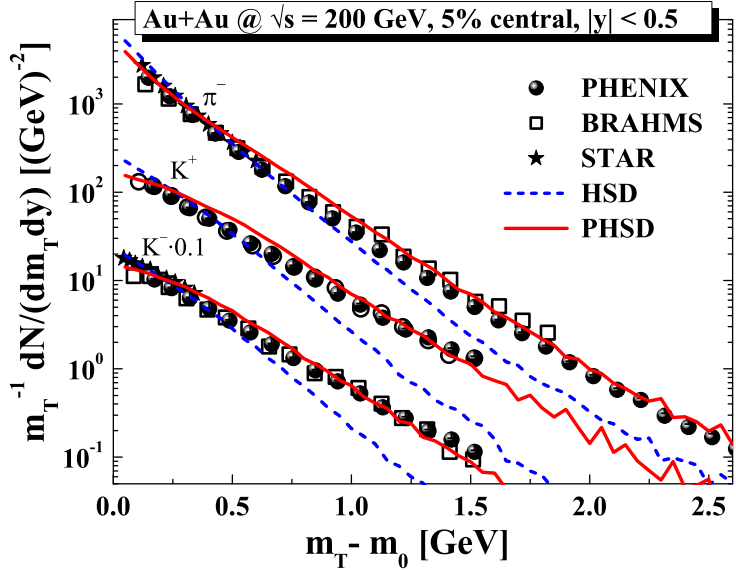,width=10.3cm}} \caption{
 The $\pi^-$, $K^+$ and $K^-$
transverse mass spectra for 5\% central Au+Au collisions at
$\sqrt{s}$ = 200 GeV from PHSD (solid  lines) in comparison to the
distributions from HSD (dashed  lines) and the experimental data
from the BRAHMS, PHENIX and STAR Collaborations
\protect\cite{PHENIX2,Adams:2003xp}.  The figure is taken from Ref. \protect\cite{BrCa11}.}
 \label{fig14b}
\end{figure}

 We continue with rapidity spectra from PHSD (solid red
lines) for charged pions and kaons in 5\% central Au+Au collisions
at $\sqrt{s_{NN}}$ = 200 GeV which are compared in Fig. \ref{fig13b}
to the data from the RHIC Collaborations \cite{PHENIX2,Adams:2003xp}
as well as to results from HSD (dashed blue lines). We find the
rapidity distributions of the charged mesons to be slightly narrower
than those from HSD and actually closer to the experimental data.
Also note that there is slightly more production of $K^\pm$ mesons
in PHSD than in HSD while the number of charged pions is slightly
lower. The actual deviations between the PHSD and HSD spectra are
not dramatic but more clearly visible than at SPS energies (cf.
Figs. 8,9). Nevertheless, it becomes clear from Fig. \ref{fig13b}
that the energy transfer in the nucleus-nucleus collision from
initial nucleons to produced hadrons - reflected dominantly in the
light meson spectra - is rather well described by PHSD also at the top RHIC energy.

 Independent information on the active degrees-of-freedom
is provided again by transverse mass spectra of the hadrons
especially in central collisions. The PHSD results for the top RHIC
energy are displayed in Fig.  \ref{fig14b}  where we show the
transverse mass spectra of $\pi^-$, $K^+$  and $K^-$ mesons for 5\%
central Au+Au collisions at $\sqrt{s}$ = 200 GeV in comparison to
the data of the RHIC Collaborations \cite{PHENIX2,Adams:2003xp}.
Here the slope of the $\pi^-$ spectra is slightly enhanced in PHSD
(solid red lines) relative to HSD (dashed blue lines)  which
demonstrates that the pion transverse mass spectra also show some
sensitivity to the partonic phase (contrary to the SPS energy
regime). The $K^\pm$ transverse mass spectra are substantially
hardened with respect to the HSD calculations - i.e. PHSD is more in
line with the data - and thus suggest that partonic effects are
better visible in the strangeness degrees-of-freedom. The hardening
of the kaon spectra can be traced back also to parton-parton
scattering as well as a larger collective acceleration of the
partons in the transverse direction due to the presence of the
repulsive scalar mean-field for the partons.

\begin{figure}
  \includegraphics[width=0.49\textwidth]{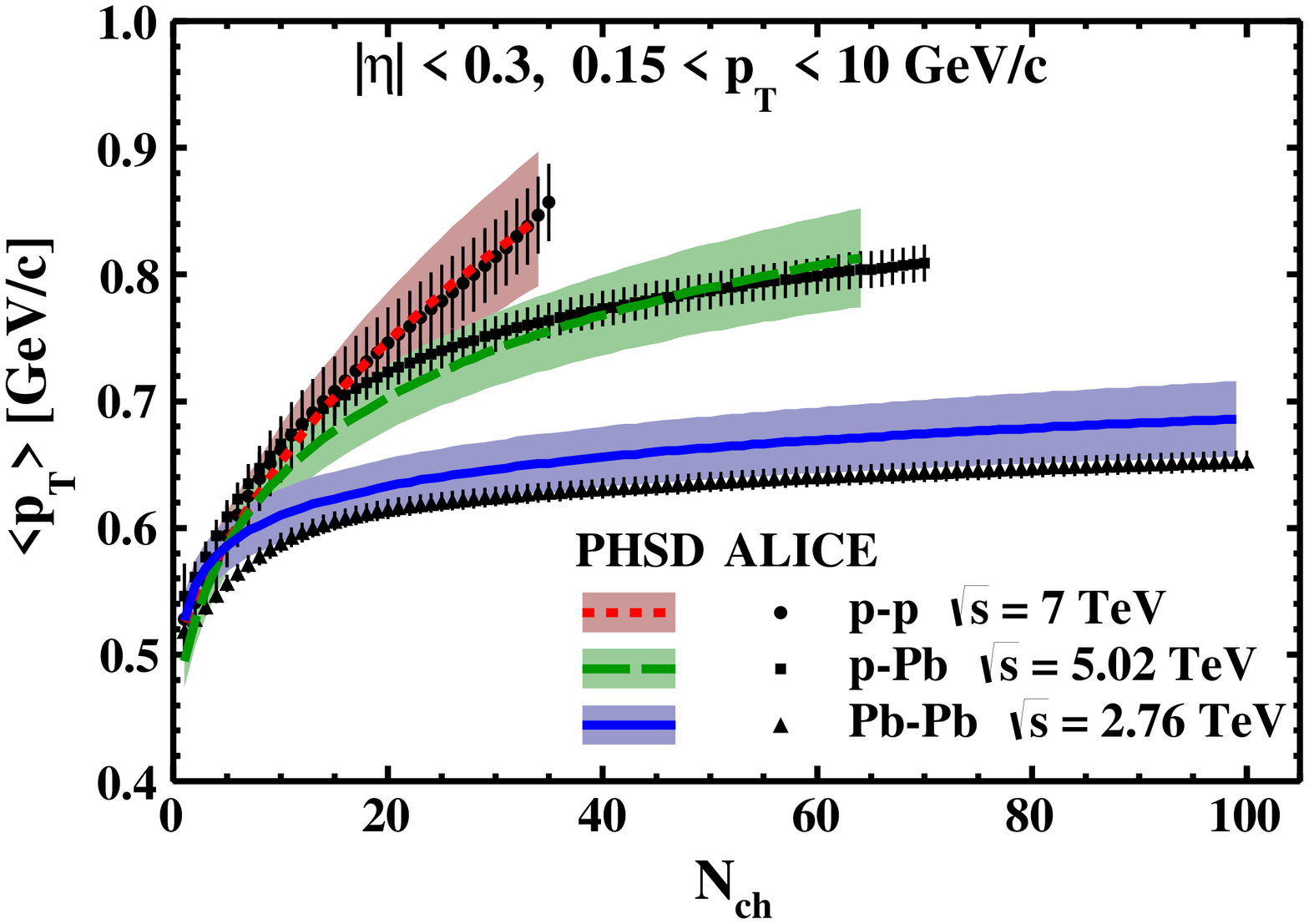} \hspace{0.1cm} \includegraphics[width=0.49\textwidth]{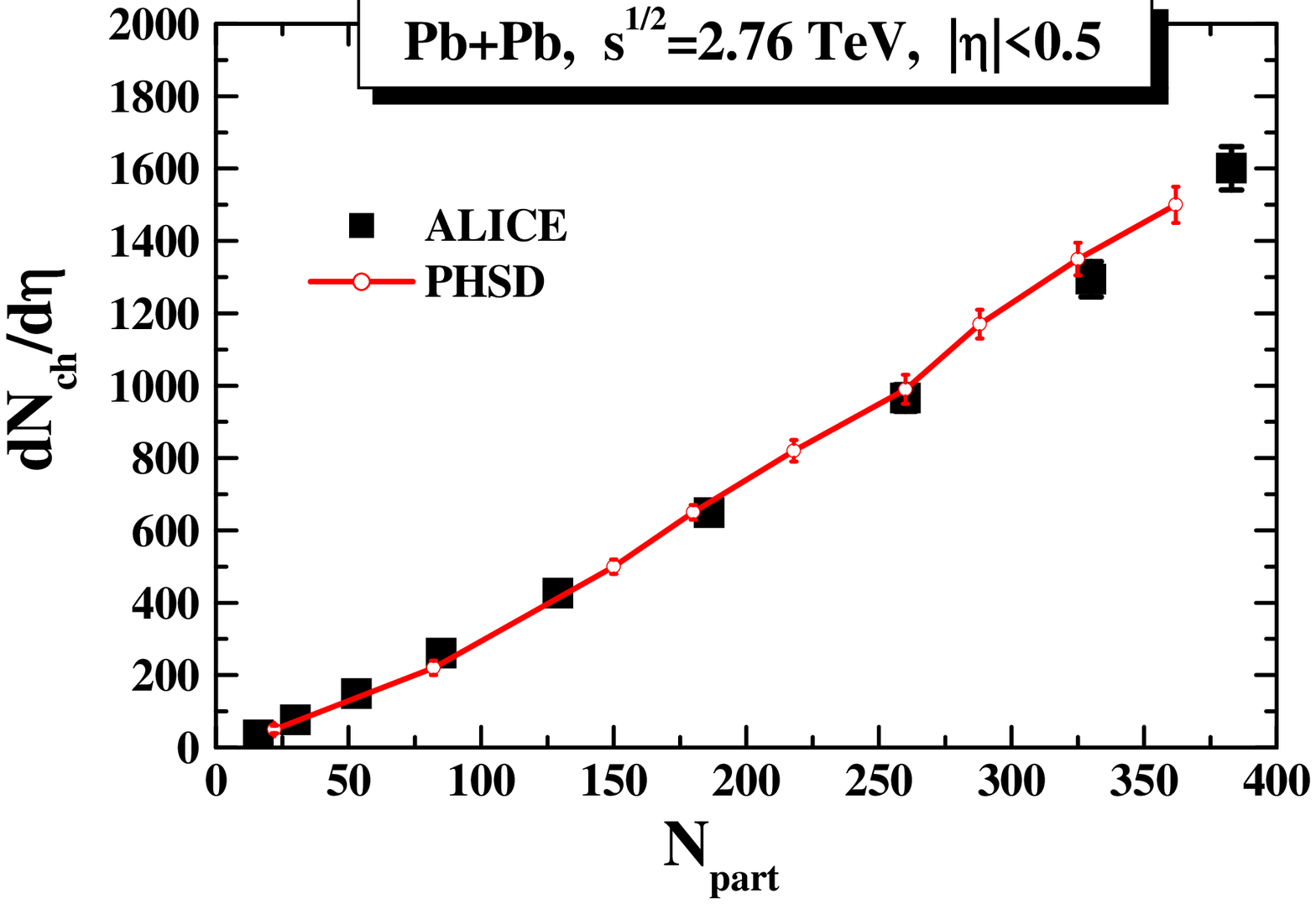}
\caption{ (l.h.s.) Mean $p_T$ results for p-p, p-Pb and Pb-Pb
collisions from   the PHSD transport approach in comparison to the ALICE experimental
  data from Ref.~\protect\cite{Abelev:2013bla} at midrapidity. Note the different invariant
  energies for p-p, p-Pb and Pb-Pb collisions. (r.h.s.) {P}seudo-rapidity distribution of charged
{hadrons} $d N_{ch}/d \eta$ at midrapidity as a function of the
number of participants $N_{part}$ from PHSD (solid line) in
comparison to the data from the ALICE Collaboration
\protect\cite{ALICE:2011ab} for Pb+Pb at $\sqrt{s_{NN}}$= 2.76 TeV.
The figures are taken from Ref. \protect\cite{Konchakovski:2014fya}.
} \label{fig:meanpt}
\end{figure}

 We, finally, come to the presently highest laboratory
energies  for Pb+Pb collisions at the LHC, however, recall that the
PHSD approach had to be properly upgraded to LHC energies with
respect to a more recent PYTHIA 6.4
implementation~\cite{Konchakovski:2014wqa}. The transition between
the different PYTHIA regions in energy is smooth with respect to
$\sqrt{s_{NN}}$ of the individual collisions such that PHSD
preserves all results at lower bombarding energies where PYTHIA 6.4
does not work sufficiently well. In PYTHIA 6.4 we use the Innsbruck
pp tune (390) which allows to describe reasonably the p-p collisions
at $\sqrt{s_{NN}}=$ 7 TeV in the framework of the PHSD transport
approach (cf.\ Fig.~1 in Ref.~\cite{Konchakovski:2014wqa}). The
overall agreement with LHC experimental data for the distribution in
the charged particle multiplicity $N_{ch}$, the charged particle
pseudorapidity distribution, the transverse momentum $p_T$ spectra
and the correlation of the average $p_T$ with the number of charged
particles $N_{ch}$ is satisfactory. Also a variety of observables
from p-Pb collisions at $\sqrt{s_{NN}}= $ 5.02 TeV compare quite
well with the experimental observations~\cite{Konchakovski:2014wqa}.

\begin{figure}
 \hspace{0.3cm} \includegraphics[width=0.44\textwidth]{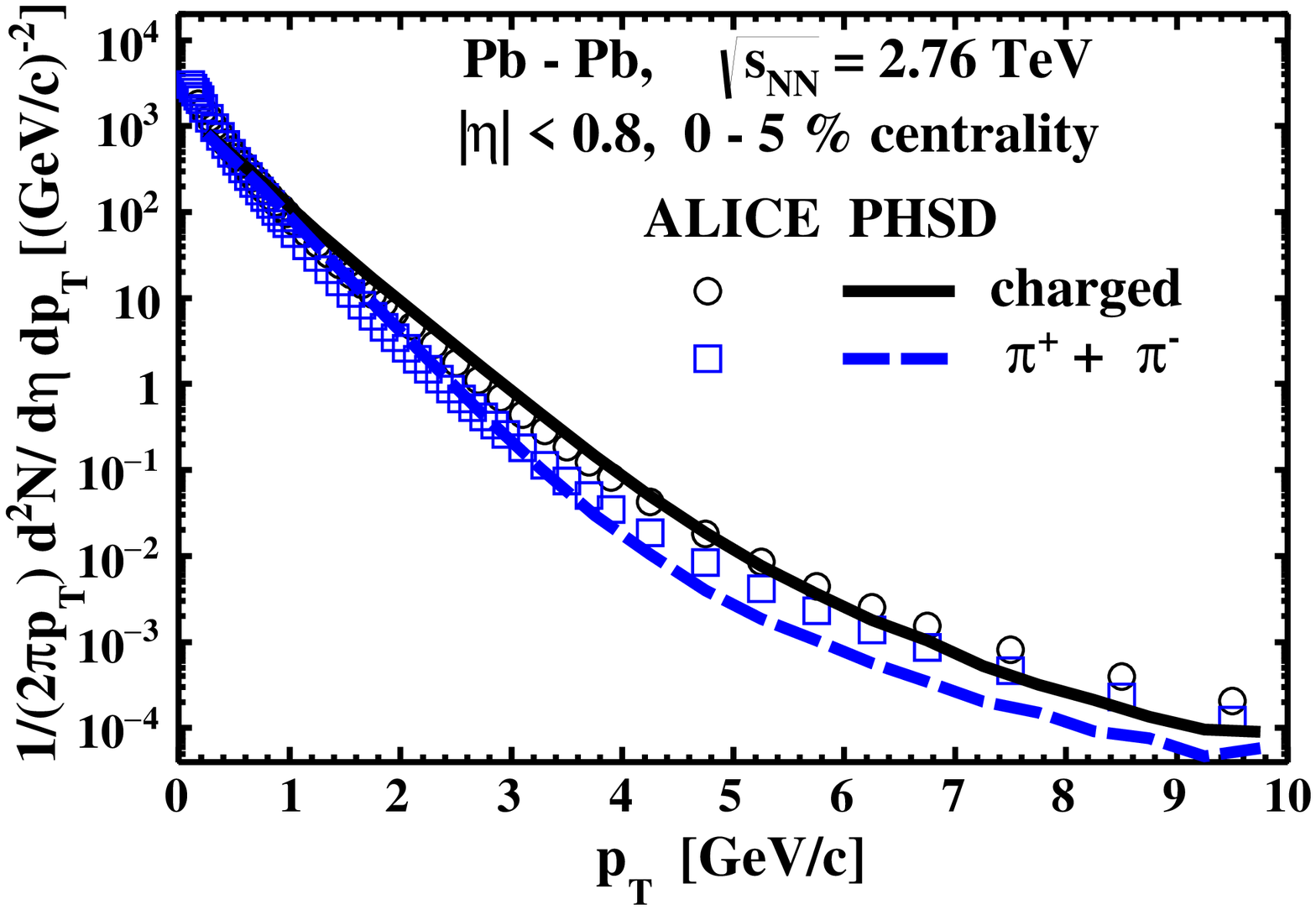} \hspace{1cm} \includegraphics[width=0.44\textwidth]{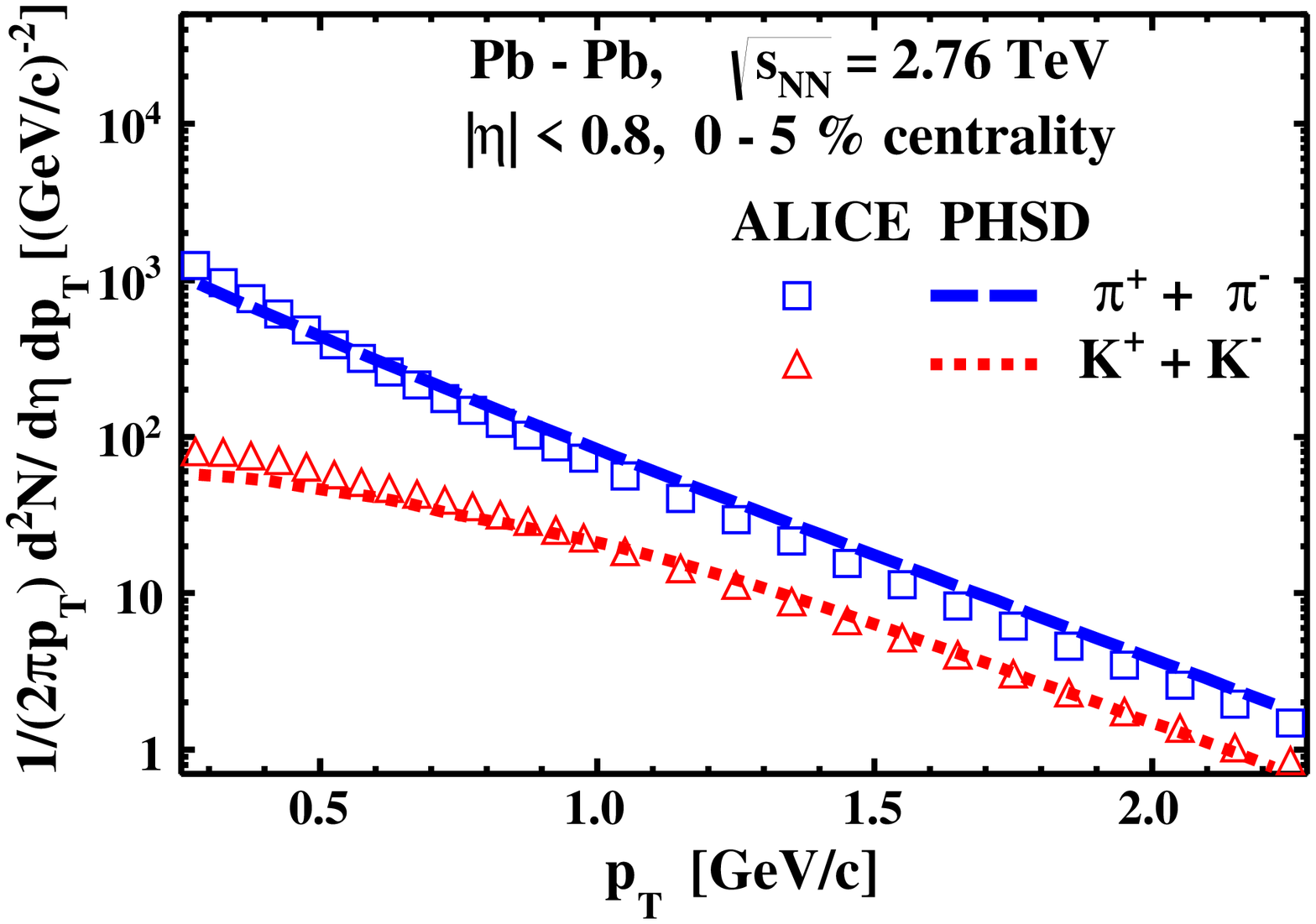}
\caption{(l.h.s.) Transverse momentum spectra from PHSD in
comparison to the
  results of the ALICE Collaboration for all charged
  particles~\protect\cite{Abelev:2012hxa, Abelev:2013ala} (solid line) as well
  as for charged pions~\protect\cite{Abelev:2014laa} (dashed line). (r.h.s.) Transverse momentum
  spectra from PHSD for $p_T \leq $ 2 GeV/c in comparison to the
  results of the ALICE Collaboration \protect\cite{Abelev:2012hxa, Abelev:2013ala,Abelev:2014laa}
  for pions and  kaons. The figures are taken from Ref. \protect\cite{Konchakovski:2014fya}.}
\label{fig:pt}
\end{figure}

One might ask whether the PHSD approach still works at LHC energies
for nucleus-nucleus (Pb+Pb) collisions although the invariant energy
is higher by about a factor of 13.8 compared to the top RHIC energy.
In Fig.~\ref{fig:meanpt} (l.h.s.) we compare the average $p_T$ (at
midrapidity) as a function of charged multiplicity $N_{ch}$ in p+p
reactions at $\sqrt{s_{NN}}=$ 7 TeV, p+Pb collisions at
$\sqrt{s_{NN}}=$ 5.02 TeV and Pb+Pb collisions at $\sqrt{s_{NN}}=$
2.76 TeV from the PHSD to the experimental data from
Ref.~\cite{Abelev:2013bla}. Note that for low multiplicities
($N_{ch} < 5$) the mean $p_T$ is almost independent on energy (see
also Ref.~\cite{Abelev:2013bla}) which in PHSD can be traced back to
the fact that (for the acceptance $|\eta| \leq$ 0.3, 0.15 $\leq p_T
\leq $ 10 GeV/c) only events with one or two binary collisions
$N_{bin}$ are selected for all systems. Actually, the correlation
$<p_T>(N_{ch})$ only weakly depends on $\sqrt{s_{NN}}$ for $pp$
reactions at these LHC energies, however, when plotting
$p_T(N_{ch})$ on an event-by-event basis, large fluctuations in
$p_T$ or $N_{ch}$ are obtained within PHSD. The same holds true for
p+Pb and Pb+Pb reactions where a fixed $N_{ch}$ can be obtained by
reactions with a varying number of binary collisions $N_{bin}$. Each
of these binary reactions then has a low $N_{ch}$ and $<p_T>$,
respectively. The ensemble average finally leads to the average
correlation shown in Fig.~\ref{fig:meanpt} (l.h.s.). Nevertheless,
the agreement between data and calculations (within the statistical
accuracy) is encouraging. {Note again that only very peripheral
Pb+Pb collisions are probed for $N_{ch} < $ 100.}

In order to shed some light on the centrality dependence of charged
particle production we display in Fig.~\ref{fig:meanpt} (r.h.s.) the
results for the pseudo-rapidity distribution $dN_c/d\eta$ at
midrapidity from the default PHSD calculations in comparison to the
ALICE data as a function of the number of participants $N_{part}$
that has been determined dynamically in the PHSD calculations. A
quite acceptable agreement is seen, suggesting that the bulk parton
dynamics is not much different at top RHIC and LHC energies.

 We continue with the transverse momentum spectra for
central Pb+Pb reactions at $\sqrt{s_{NN}}=$~2.76 TeV (0-5\%
centrality) which are compared in Fig.~\ref{fig:pt} with results
from the ALICE Collaboration for all charged
particles~\cite{Abelev:2012hxa, Abelev:2013ala} (PHSD: black solid
line) as well as for charged pions~\cite{Abelev:2014laa} (PHSD:
dashed blue line). Note that except for the upgrade in the PYTHIA
version no additional parameters or changes have been introduced in
the PHSD. In this respect the approximate reproduction of the
midrapidity $p_T$ spectra for central collisions over 7 orders of
magnitude in Fig.~\ref{fig:pt} (l.h.s.) is quite remarkable.  A
closer look at the low momentum spectra is offered in Fig.
~\ref{fig:pt} (r.h.s.) where the PHSD spectra for pions and kaons
are compared to  results of the ALICE Collaboration
\cite{Abelev:2012hxa, Abelev:2013ala,Abelev:2014laa} (symbols).

In summarizing, the partonic phase in PHSD at the top RHIC energy
and at LHC leads to a narrowing of the longitudinal momentum
distribution, a reduction of pion production, a slight enhancement
of kaon production  and to a hardening of their transverse mass
spectra relative to HSD (closer to the data). These effects are
clearly visible especially in the transverse degrees-of-freedom and
are more pronounced than at SPS energies due to the larger space-time
region of the partonic phase.

\subsubsection*{Collective flow}

\begin{figure} \phantom{a} \hspace{-0.7cm}
  \includegraphics[width=0.54\textwidth]{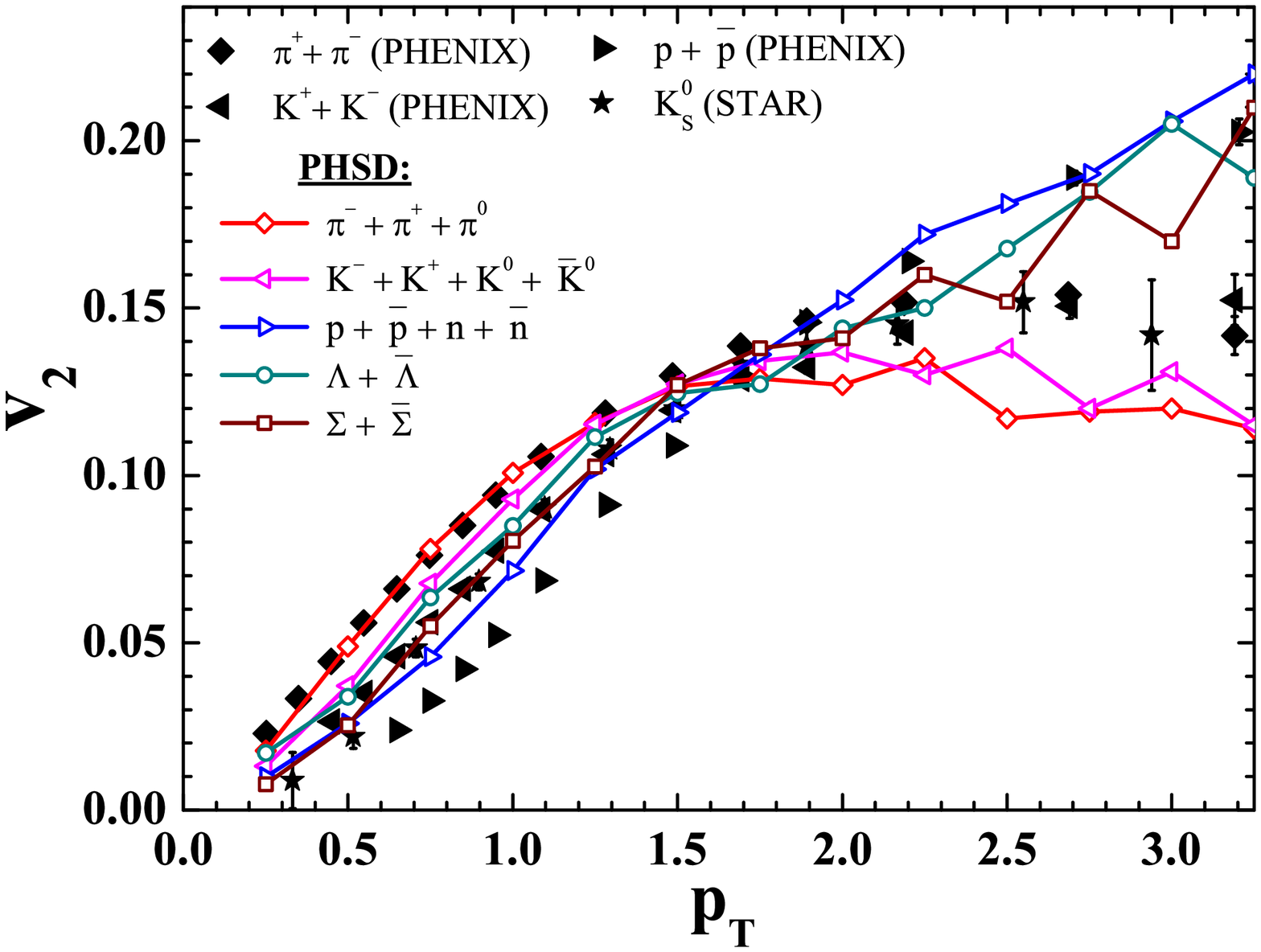} \phantom{a} \hspace{-0.7cm}
  \includegraphics[width=0.51\textwidth]{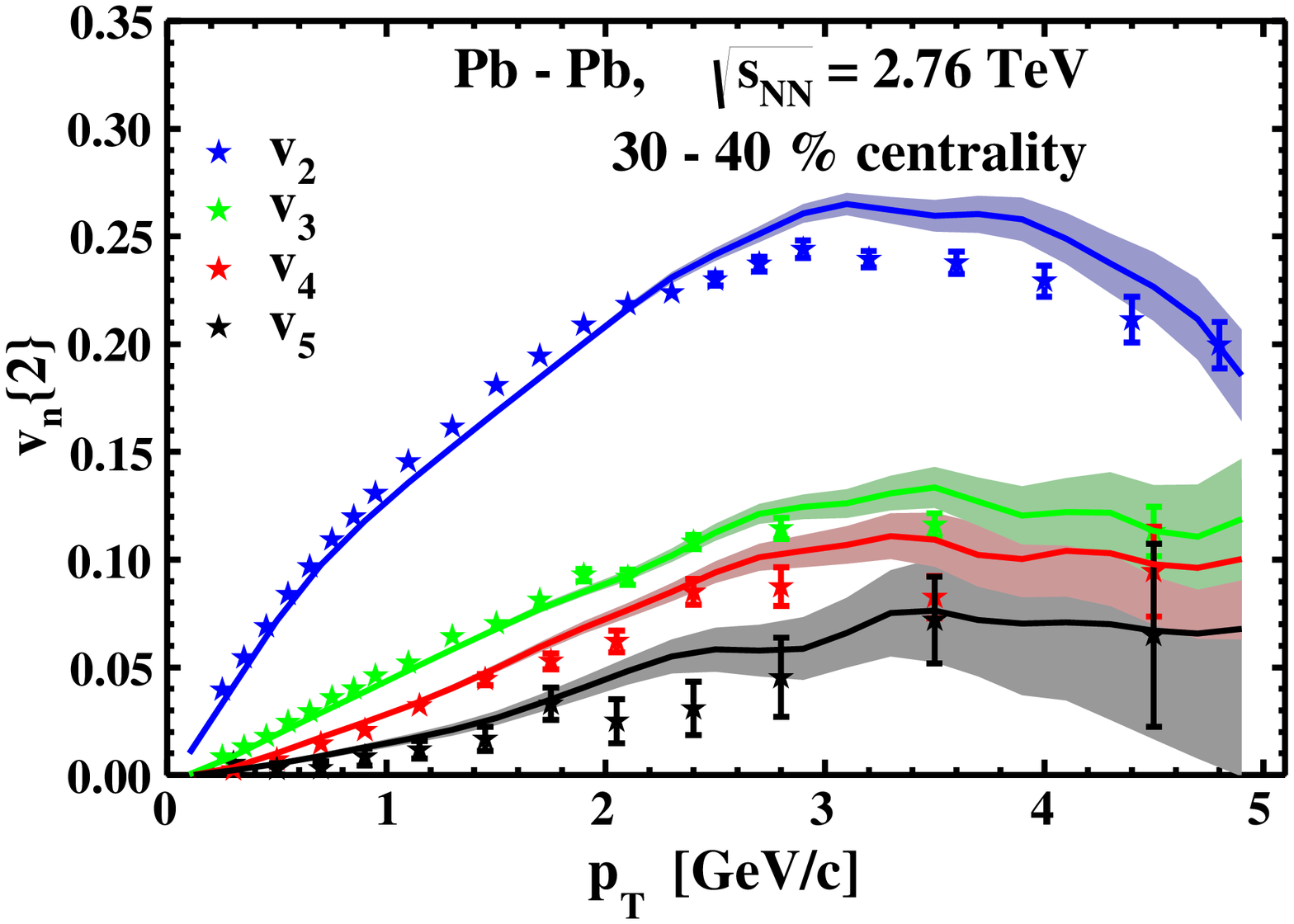}
\caption{({ l.h.s.}) The hadron elliptic flow $v_2$ for inclusive
Au+Au collisions as a function of the transverse momentum $p_T$ (in
GeV) for different hadrons in comparison to the data from the STAR
\protect\cite{STAR6,STAR6i} and PHENIX Collaborations
\protect\cite{SCALING2} within the same rapidity cuts. ({ r.h.s.})
The flow coefficients $v_2$, $v_3$, $v_4$ and $v_5$ of all
  charged particles as a function of $p_T$ for the centralities 30-40\% in case of Pb-Pb collisions at $\sqrt{s_{NN}}$
  = 2.76 TeV. The ALICE data
  have been adopted from Ref.~\protect\cite{ALICE:2011ab}. The figures are taken from Refs. \protect\cite{BrCa11,Konchakovski:2014fya}. }
 \label{fig22} \label{fig:vn}
\end{figure}

Of additional interest are the collective properties of the
strongly interacting system which are explored experimentally by the azimuthal
momentum distribution of particles in a fixed rapidity interval.
The azimuthal momentum distribution of the emitted particles is
commonly expressed in the form of a Fourier series as
\be \label{vns} E\frac{d^3N}{d^3p}=  \frac{d^2N}{2\pi
p_Tdp_Tdy}\left(\! 1\!+\! \sum^\infty_{n=1} 2v_n(p_T) \cos
[n(\psi-\Psi_n)]\! \right)\!, \ee
where $v_n$ is the magnitude of the $n'$th order harmonic term
relative to the angle of the initial-state spatial plane of symmetry
$\Psi_n$ and $p=(E,\vec{p})$ is the four-momentum of the particle
under consideration. We here focus on the coefficients $v_2$, $v_3$
and $v_4$ which implies that we have to perform event-by-event
calculations in order to catch the initial fluctuations in the shape
of the interaction zone and the event plane $\Psi_{EP}$; e. g., we
calculate the triangular flow $v_3$ with respect to $\Psi_3$ as
$v_3\{\Psi_3\} = \langle
\cos(3[\psi-\Psi_3])\rangle/\rm{Res}(\Psi_3)$.  The event plane
angle $\Psi_3$ and its resolution $\rm{Res}(\Psi_3)$ are evaluated
as described in Ref.~\cite{{Adare:2011tg}} via the two-sub-events
method~\cite{Poskanzer:1998yz,Bilandzic:2010jr}.

We here briefly summarize the main results. Fig. \ref{fig22}
(l.h.s.) shows the final hadron $v_2$ versus the transverse momentum
$p_T$ for different particle species at the top RHIC energy in
comparison to the data from the STAR \cite{STAR6,STAR6i} and PHENIX
Collaborations \cite{SCALING2}. We observe a mass separation in
$p_T$ as well as a separation in mesons and baryons for $p_T >$ 2
GeV roughly in line with data. The elliptic flow of mesons is
slightly underestimated for $p_T >$ 2 GeV in PHSD which is opposite
to ideal hydrodynamics which overestimates $v_2$ at high transverse
momenta. On the other hand, the proton (and antiproton) elliptic
flow is slightly overestimated at low  $p_T < $ 1.5 GeV. We note in
passing that also the momentum integrated results for $v_2$ as a
function of the number of participating nucleons $N_{part}$ from
PHSD compare well to the data from Ref. \cite{PHOBOS4}. In contrast,
the HSD results clearly underestimate the elliptic flow as pointed
out before \cite{brat03}. The relative enhancement of $v_2$ in PHSD
with respect to HSD can be traced back to the high interaction rate
in the partonic phase and to the repulsive scalar mean-field for
partons; the PHSD calculations without mean-fields only give a small
enhancement for the elliptic flow relative to HSD.

Stepping up in energy of the collision to $\sqrt{s_{NN}}=2.76$~TeV
reached at the LHC, the PHSD results for the flow coefficients
$v_2$, $v_3$, $v_4$ and $v_5$ of all charged particles are shown in
Fig.~\ref{fig:vn} as a function of $p_T$ for the centralities
30-40\% in Pb+Pb collisions (r.h.s.) in comparison to the ALICE data
from Ref.~\cite{ALICE:2011ab}. The PHSD results for $v_2(p_T),
v_3(p_T)$ and $v_4(p_T)$ describe the data reasonably up to about
3.5~GeV/c, whereas at higher transverse momenta the statistics of
the present calculations is insufficient to draw robust conclusions.
This also holds for the flow coefficient $v_5$ which still is in
line with the data within error bars. It is quite remarkable that
the collective behavior is reproduced in the PHSD  approach not only
for the semi-central collisions ($30-40$ \%) but also for $0-5$\%
central collisions, which are sensitive to the initial state
fluctuations (see Ref.~\cite{Konchakovski:2014fya}).

These tests indicate that the 'soft' physics at LHC in central A-A
reactions is very similar to the top RHIC energy regime although the
invariant energy is higher by more than an order of magnitude.
Furthermore, the PHSD approach seems to work from lower SPS energies
up to LHC energies for p-p, p-A as well as A-A collisions, i.e. over
a range of more than two orders in $\sqrt{s_{NN}}$. Note that for
even lower bombarding energies the PHSD approach merges to the HSD
model which has been successfully tested from the SIS to the SPS
energy regime in the past \cite{Cass99,Cass00,GEISS}. Since the bulk dynamics
is well described in PHSD in comparison to experimental data in a wide dynamical range
we may continue with the electromagnetic emissivity of the reactions
which (in principle) does not employ any new parameter.

%%%%%%%%%%%%%%%%%%%%%%%%%%%%%%%%%
\section{Implementation of photon and dilepton production in transport approaches}
\label{section_phsd}
%%%%%%%%%%%%%%%%%%%%%%%%%%%%%%%%%

\subsection{Photon sources in relativistic heavy-ion collisions}

The {\em inclusive} photon yield as produced in $p+p$, $p+A$ and
$A+A$ collisions is divided into ``{\em decay} photons" and ``{\em
direct} photons". The {\it decay photons } -- which constitute the
major part of the {\em inclusive} photon spectrum -- stem from the
photonic decays of hadrons (mesons and baryons) that are produced in
the reaction. These decays occur predominantly at later times and
outside of the active reaction zone and therefore carry limited
information on the initial high-energy state. Consequently, it is
attempted to separate the {\em decay} photons from the inclusive
yield (preferably by experimental methods) and to study the
remaining ``{\em direct} photons". One usually uses the ``cocktail"
method to estimate the contribution of the photon decays to the
spectra and to the elliptic flow $v_2$, which relies (among others)
on the $m_T$-scaling assumption and on the photon emission only by
the finally produced hadrons with momentum distributions of the
final states. Depending on the particular experimental set-up,
different definitions of the {\em decay} photons are applied by the
various collaborations: all groups subtract the decays of $\pi^0$-
and $\eta$-mesons, however, some groups also subtract the decays of
the less abundant and short-living particles $\eta^\prime$,
$\omega$, $\phi$, $a_1$ and the $\Delta$-resonance. Indeed, the
determination of the latter contributions (in particular from $a_1$
and $\Delta$) by experimental methods is questionable, because of
the photon emission during the multiple absorption and regeneration
in the initial interaction phase. Therefore, a theoretical
understanding of the {\em decay} photon contributions to the
inclusive spectrum is important. Especially for analyzing
simultaneously various measurements at different energies and within
different experimental settings a theoretical analysis is mandatory
which accounts for the different experimental acceptance cuts (from
various collaborations) and allows for comparing spectra at
different centralities and bombarding energies, ultimately bridging
the gap from p-p to central heavy-ion collisions.

Within the PHSD we calculate the photon production from the
following hadronic decays: $$ \pi^0 \to \gamma+ \gamma,
\mbox{\hspace{0.4cm}} \eta \to \gamma + \gamma,
\mbox{\hspace{0.4cm}} \eta^\prime \to \rho + \gamma,
\mbox{\hspace{0.4cm}} \omega  \to  \pi^0 + \gamma,
\mbox{\hspace{0.4cm}} \phi \to \eta + \gamma, \mbox{\hspace{0.4cm}}
 a_1  \to  \pi + \gamma, \mbox{\hspace{0.4cm}} \Delta  \to
\gamma+N, $$
where the parent hadrons may be produced in baryon-baryon ($BB$),
meson-baryon ($mB$) or meson-meson ($mm$) collisions in the course
of the heavy-ion collision or may stem from hadronization. The decay
probabilities are calculated according to the corresponding
branching ratios taken from the latest compilation by the Particle
Data Group~\cite{PDG}. The broad resonances -- including the $a_1,
\rho, \omega$ mesons -- in the initial or final state are treated in
PHSD in line with their (in-medium) spectral functions and the
differential photon or dilepton yield is integrated in time (see
below).

Let us briefly describe the evaluation of the photon production in
the decays of the $\Delta$-resonance as an important example. The $\Delta \to N\gamma$ width
depends on the resonance mass $M_\Delta$, which is distributed
according to the $\Delta$ spectral function. Starting from the
pioneering work of Jones and Scadron \cite{Jones73}, a series of
models \cite{Wolf90,Krivor02,ZatWolf03} provided the mass-dependent
electromagnetic decay width of the $\Delta$-resonance in relation to the total
width of the baryon. We employ the model of Ref.~\cite{Krivor02} in
the present calculations where the spectral function of the $\Delta$-resonance
is assumed to be of relativistic Breit-Wigner form. Furthermore, we
adopt the "Moniz" parametrization \cite{Monitz} for the shape of
the $\Delta$-spectral function, i.e. the dependence of the width on
the mass $\Gamma^{tot}(M_\Delta)$.

The {\it direct photons} are obtained by subtraction of the
decay-photon contributions from the inclusive (total) spectra
measured experimentally. So far, the following contributions to the
{\em direct} photons have been identified:
\begin{itemize}
\item The photons at large transverse
momentum $p_T$, so called {\it prompt} or {\it pQCD} photons, are
produced in the initial hard $N+N$ collisions and stem from jet
fragmentation; these contributions are well described by
perturbative QCD (pQCD). The latter, however, might be modified in
$A+A$ contrary to $p+p$ reactions due to a modification of the
parton distributions (initial state effect) or the parton energy
loss in the medium (final state effect). In  $A+A$ collisions at
large $p_T$ there may also arise contributions from the induced
jet-$\gamma$-conversion in the QGP and the jet-medium photons from
the scattering of hard partons with thermalized partons
$q_{hard}+q(g)_{QGP} \to \gamma + q(g)$; however,  these
contributions are subleading. As noted above the prompt photons are
well modeled by perturbative QCD calculations.
\item After the subtraction of the {\it prompt} photons from the {\it direct}
photon spectra, there is a significant remaining photon yield for
$p_T<3$~GeV, which is denoted as {\it thermal} photons.
These low-$p_T$ photons can be emitted by various partonic and
hadronic sources as listed below:
\begin{enumerate}
\item
Photons that are radiated by  quarks in the interaction with
antiquarks and gluons, $$ q+\bar{q}\rightarrow g+\gamma,
\hspace{2cm} q/\bar{q}+g\rightarrow q/\bar{q}+\gamma. $$ In
addition, photon production in the bremsstrahlung reactions
$q+q/g\to q+q/g+\gamma$ is possible~\cite{Haglin:1992fy}. The
implementation of the photon production by the quark and gluon
interactions in the PHSD is based on the off-shell cross sections
for the interaction of the massive dynamical quasi-particles as
described in Ref.~\cite{olena2010,Linnyk:2013hta}. The photon
production rates in a thermal medium -- calculated  within the DQPM
effective model for QCD -- are within a factor of 2 similar to the
rates obtained by the resummed pQCD approach from
Ref.~\cite{Arnold:2001ms} (see Section~\ref{sect:partonic}). Since
the quark-gluon-plasma produced in the heavy-ion collisions is
strongly-interacting, the Landau-Migdal-Pomeranchuk (LPM) coherence
effect can be important, too (cf. Section~\ref{sect:LPM}).
\item
All colliding hadronic charges (meson, baryons) can also radiate
photons by the bremsstrahlung processes:
$$
m+m\to m+m+\gamma\label{mmBr} \mbox{  \hspace{1cm} }  m+B\to
m+B+\gamma .
$$
These processes have been studied within the HSD/PHSD in
Refs.~\cite{Linnyk:2013hta,Bratkovskaya:2008iq,Linnyk:2013wma,Linnyk:2015tha}
in continuation of earlier work at lower energies \cite{Bauer,PR90}.
The implementation of photon bremsstrahlung from hadronic reactions
in transport approaches has been based until recently in the 'soft
photon' approximation (SPA). The soft-photon
approximation~\cite{gale87a,gale87b,gale87c} relies on the
assumption that the radiation from internal lines is negligible and
the strong interaction vertex is on-shell which is valid only at low
energy (and $p_T$) of the produced photon. Since the relatively high
transverse momenta of the {\em direct} photons ($p_T=0.5-1.5$~GeV)
are most important for a potential understanding of the ``direct
photon puzzle" we have departed from the SPA in the
PHSD~\cite{Linnyk:2015tha}. The PHSD results presented in this
review have been obtained employing microscopic  one-boson-exchange
(OBE) calculations instead (cf. Section~\ref{sect:brems}).
\item Additionally, the photons can be produced in binary
meson+meson and meson+baryon collisions. We consider within the PHSD
the {\em direct} photon production in the following $2\to2$
scattering processes:
$$
\pi + \pi \rightarrow \rho + \gamma , \mbox{\hspace{1cm}} \pi + \rho
\rightarrow \pi + \gamma , \hspace{1cm}
 V+N \to \gamma+N, \hspace{1cm} \mbox{where } V=\rho, \
\phi, \  \omega, \mbox{ and }N=n,p,
$$
accounting for all possible charge combinations (cf.
 Subsection~\ref{sect:22}).  Further mesonic $2\to2$ reactions with the
 allowed quantum numbers for photon production, such as $\pi+\omega$, $\rho+\omega$, $V+\eta$ etc.,
 can also contribute~\cite{Holt:2015cda}, but are discarded in the actual calculations
 that focus on the leading channels.
\end{enumerate}
\end{itemize}

\subsection{Photon production by dynamical quasiparticles in the QGP}
\label{sect:partonic}

\begin{figure} \centering
\includegraphics[width=0.59\textwidth]{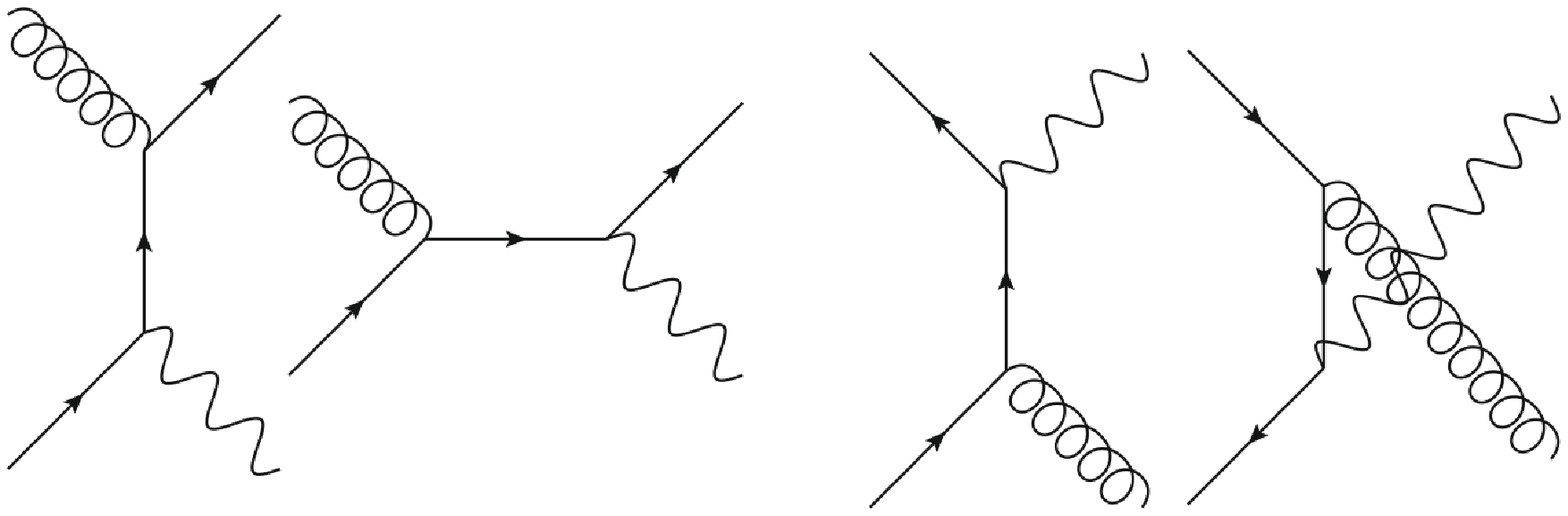}
\caption{Feynman diagrams for the leading partonic sources of {\em
thermal} photons ($q({\bar q})+g \rightarrow q({\bar q})+\gamma$ and
$q+{\bar q} \rightarrow g+\gamma$) included in the PHSD
calculations. The propagators and strong coupling are employed from
the DQPM.} \label{figol1}
\end{figure}

We start with the description of photon production in the
interactions of quarks and gluons in the quark-gluon plasma, which
dominantly proceeds through the quark-antiquark annihilation and the gluon
Compton scattering processes:
$$
q+{\bar q}  \rightarrow  g+\gamma \mbox{  \hspace{1cm} }
q({\bar q})+g  \rightarrow q({\bar q})+\gamma ,
$$
that are diagrammatically  presented in Fig.~\ref{figol1}.

In the strongly interacting QGP the gluon and quark propagators (in
PHSD) differ significantly from the non-interacting propagators such
that bare production amplitudes can no longer be
used~\cite{Cassing:2008nn,Cassing:2007nb}. The off-shell quarks and
gluons have finite masses and widths, which parametrize the resummed
interaction of the QGP constituents. The perturbative QCD results
for the cross sections of the processes in Fig.~\ref{figol1} have to
be generalized in order to include the finite masses for fermions
and gluons as well as their broad spectral functions. In
Ref.~\cite{Marzani:2008uh}, the influence of the gluon
off-shellness (fixed to $m_g^2=|\vec{k_g}|^2$) on the photon
production was studied but the quark masses had been neglected and
the spectral functions were assumed to be $\delta$-functions
(quasi-particle approximation). On the other hand, in
Ref.~\cite{Wong:1998pq} a finite quark mass has been incorporated in the
elementary cross sections for both the quark annihilation and the
gluon-Compton scattering processes (though the gluon was taken to be
massless and the quasiparticle approximation maintained). However, the
formulae from Ref.~\cite{Wong:1998pq} still could not consistently
describe the photon production by the effective dressed quarks and
gluons within the PHSD transport approach, because they did not
account for the finite width. Furthermore, in all the previous calculations the
masses of quarks and antiquarks were assumed to be equal, which is not the
case for the scattering of off-shell  particles with continuous mass distributions.

The evaluation of the cross sections for {\it dilepton} production by
off-shell partons, taking into account finite masses for quarks,
antiquarks (with generally $m_q\ne m_{\bar q}$) and gluons $m_g$ as
well as their finite spectral width (by integrating over the mass
distributions) has been carried out in Refs.~\cite{olena2010,Linnyk:2004mt}.
In order to obtain the cross sections for the {\em real} photon
production, we use the relation between the real photon production
cross section and the cross section for dilepton
production~\cite{PHENIXlast}:
\begin{equation}
\label{realvirtual} \frac{d\sigma (\gamma)}{dt}=\lim_{M\to 0}
\frac{3\pi}{\alpha} \frac{M^2}{L(M)} \frac{d^2 \sigma (e^+e^-)}{dM^2
dt},
\end{equation}
where $M^2$ is the invariant mass squared of the lepton pair (virtual photon),
while the kinematical factor
$L(M)$ is given by
\begin{equation}
L(M)=\sqrt{1-\frac{4 m_e^2}{M^2}}(1+\frac{2m_e^2}{M^2})
\end{equation}
with $m_e$ denoting the lepton mass.
We employ the DQPM parametrization for the effective quark and
gluon propagators in the calculation of ${d^2 \sigma
(e^+e^-)}/{dM^2 dt}$ thus going beyond the leading twist approximation~\cite{Linnyk:2006mv}. We
refer the reader to Ref.~\cite{olena2010} for details of the
calculations and provide only the necessary  steps here.
We briefly summarize the differences of our 'effective'
approach from the standard pQCD:
\begin{itemize}
\item
We take into account full off-shell kinematics, i.e. the transverse
motion and virtuality of the partons,
\item
quark and gluon lines in the diagrams in Fig.~\ref{figol1} and in the
leading-order diagram $q+\bar q\to\gamma^*$ (which is relevant only
for dilepton production) are
dressed with non-perturbative spectral functions and self-energies:
the cross sections are derived for arbitrary masses of all external
parton lines and integrated over these virtualities weighted with
spectral functions (see e.g.
Refs~\cite{Bratkovskaya:2008iq,Linnyk:2004mt} for an introduction to
the method); the internal lines are dressed with self energies.
\item
Strong vertices are modified compared to pQCD by replacing the perturbative
coupling (that runs with the momentum transfer) with the
running coupling $\alpha_S(T)$ that depends on the temperature $T$ of the medium
according to the parametrization of lattice data  in Ref.~\cite{Cassing06}, while
the temperature $T$ is related to the local energy density $\epsilon({\bf r};t)$ by
the lQCD equation of state. Note that close to $T_c$ the effective
coupling  $\alpha_S(T)$ increases with  decreasing temperature  much faster than the pQCD prediction.
\item
Due to the broad widths of quarks and gluons in the
sQGP~\cite{Cassing:2007nb} -- which is the consequence of their high
interaction rate -- there are non-vanishing contributions also from
the  decays of virtual quarks ($q\to q+g+l^+l^-$)
and gluons ($g\to q + \bar q +l^+l^-$), which are forbidden
kinematically in pQCD. However, we  presently discard these processes in PHSD.
\end{itemize}

Due to the factorization~\cite{McLerran:1984ay}, the dilepton
emission from the QGP created in the heavy-ion collision is given by
the convolution of the elementary subprocess cross sections
(describing quark/gluon interactions with the emission of
dileptons) with the structure functions that characterize the
properties and evolution of the plasma (encoded in the distribution
of the quarks and gluons with different momenta and virtualities):
\bea  \frac{d ^3 \sigma ^{\mbox{\small QGP}} }{dM^2dx_Fdq_T^2} =
\sum _{abc} \! \int \! d \hat s \! \int_0^\infty \! \! d m^{i}_1 \!
\int_0^\infty \! \! d m^{i}_2 \! \int_0^\infty \!\! d \mu ^f \,
F_{ab}(\hat s,m^{i}_1,m^{i}_2)\mbox{A}_c (\mu^f) \frac{d^3 \hat
\sigma _{abc} (\hat s,m^{i}_1,m^{i}_2,\mu^{f}) }{dM^2dx_Fdq_T^2}, \
\label{factorization} \eea
where $M^2$ is the  invariant mass of the dilepton pair, $m^{i}_1$ and
$m^{i}_2$ are the masses of the
incoming partons, $\mu^f$ is the mass of the outgoing
parton, while the indices $a,b,c$ denote quark, antiquark or gluon such that
all the elementary reactions are covered.  Here, $\mbox{A}(\mu^f)$
is the spectral function for the parton in the final state.
The structure function $F_{ab}$ is a two-particle correlator that
depends on the invariant energy $\sqrt{\hat s}$ of the partonic
subprocess as well as on the virtualities of the incoming partons.
Since we work in a 2PI-like approximation, the partons in the sQGP are
characterized by single-particle distributions and we
assume that the plasma structure function is given by
\be F_{ab}(\hat s,m_{1},m_{2}) = A_a (m_{1}) A_b (m_{2}) \frac{d
N_{ab}}{d s}. \ee
In this context, the quantity $d N_{q \bar q}/d s$ has the meaning
of the differential multiplicity of $q+\bar q$ collisions in the plasma
as a function of the invariant energy squared $s$ of these collisions in the interval $ds$.
Analogously, $d N_{g q}/ds$ denotes the differential multiplicity of
$g+q$ collisions.

The off-shell partonic cross sections $\hat \sigma _{abc} (\hat
s,m^{i1},m^{i2},\mu^{f})$ for the different processes have been
derived in Ref.~\cite{olena2010}. We sketch the derivation for the
example of the Gluon-Compton scattering diagram: We start from the
formula for the unpolarized cross section, \be \label{O1}
d\sigma = \frac{ \bar{
\Sigma | M_{i\to f} |^2 } \varepsilon_1 \varepsilon_2 \Pi \frac{d^3
p_f}{(2 \pi)^3} } {\sqrt{(p_1 p_2)^2 - m_1 ^2 m_2 ^2}} \ (2 \pi) ^4
\delta (p_1+p_2-\Sigma p_f), \ee
where the incoming quark and antiquark momenta are $p_1$ and $p_2$ and
their masses $m_1$ and $m_2$, respectively. In Eq. (\ref{O1}) $p_f$ are the momenta of
the outgoing particles, i.e. of the electron (muon) and positron
(anti-muon) and gluon. We define the momenta of the internal quark --
exchanged in the two relevant diagrams (see Fig. \ref{figol1}) -- by $p_3 \equiv
p_1 - q$, $\bar p_3 \equiv p_1 - p_2 - p_3 $ and its mass by $m_3$.
The final gluon momentum is denoted by $k$ and its mass by $\mu$. Then the
matrix element of the process $q +\bar q \to g + \gamma ^*$ is given by
\be M_{i\to f}=M_a+M_b, \ee where
\bea \label{O2} M_a  =   - e_q e g_s T_{i j}^l \frac{\epsilon _{\nu} (q)
\epsilon_{\sigma l}(k)}{p_3^2-m_3^2}   u_i (p_1, m_1) \left[
\gamma^{\nu} (\hat p _3 + m_3) \gamma^{\sigma} \right] v_j ( p_2,
m_2), \nnl M_b  =  - e_q e g_s T_{i j}^l \frac{ \epsilon_{\sigma l}
(k) \epsilon_{\nu} (q) }{ \bar {p_3} ^2 -m_3^2 } u_i(p_1,m_1) \left[
\gamma^\eta (\hat{\bar{p_3}} +m_3) \gamma^\nu \right] v_j (p_2,m_2) .
 \eea
In Eq. (\ref{O2}) $e$ is the electron charge, $e_q$ is the quark fractional charge while
$T_{i j}^l$ is the generator of the SU(3) color group (that gives
 the color factor in the cross section). Furthermore, $\epsilon _{\nu} (q)$
is the polarization vector for the virtual photon with momentum $q$,
$\epsilon_{\sigma l} (k)$ is the polarization vector for the gluon
of momentum $k$ and color $l$; $u_i (p, m)$ is a Dirac spinor for
the quark with momentum $p$, mass $m$ and color $i$ and $v_i (p,m)$
is the  spinor for the anti-quark.

The squared matrix element --  summed over all spin polarizations as well as over the
color degrees-of-freedom -- can be decomposed in the
following sums:
\be \sum |M|^2 =  \sum M_a^*M_a + \sum M_b^* M_b +\sum M_a^* M_b +
\sum M_b^*M_b, \ee
where the star denotes complex conjugation.
The spinors for quark states with mass $m_i$ contribute to the
expression for the average matrix element only in the combinations
$\sum \bar u (p, m_i) u (p, m_i)=(\hat p + m_i)$ and the correlation
functions between the states with different masses do not enter
$|M|^2$. Thus we find:
\bea \sum M_a ^* M_b & = & - \frac{e_q^2 e^2 g_s^2 \mbox{Tr}\{T^2\}
}{(p_3^2-m_3^2)(\bar{p_3}^2 -m_3^2)} \left[ \mbox{Tr}\left\{
(\hat{p_2}-m_2) \gamma_\sigma (\hat{p_3} + m_3) \gamma_\nu
(\hat{p_1}+m_1) \gamma^\sigma (\hat{\bar{p_3}}+m_3) \gamma^\nu
\right\} \right. \nnl & & \left . \phantom{- \frac{e_q^2 e^2 g_s^2
\mbox{Tr}\{T^2\}
}{(p_3^2-m_3^2)(\bar{p_3}^2 -m_3^2)}} %
- \frac{1}{M^2} \mbox{Tr}\left\{ (\hat{p_2} - m_2) \gamma _\sigma
(\hat{p_3}+m_3) \hat{q} (\hat{p_1}+m_1) \gamma^\sigma
(\hat{\bar{p_3}}+m_3) \hat{q} \right\} \right. \nnl & & \left .
\phantom{- \frac{e_q^2 e^2 g_s^2 \mbox{Tr}\{T^2\}
}{(p_3^2-m_3^2)(\bar{p_3}^2 -m_3^2)}} %
-\frac{A}{k^2} \mbox{Tr}\left\{ (\hat{p_2} - m_2) \hat{k}
(\hat{p_3}+m_3) \gamma_\nu (\hat{p_1}+m_1) \hat{k}
(\hat{\bar{p_3}}+m_3) \gamma^\nu \right\} \right. \nnl & & \left .
\phantom{- \frac{e_q^2 e^2 g_s^2 \mbox{Tr}\{T^2\}
}{(p_3^2-m_3^2)(\bar{p_3}^2 -m_3^2)}} %
+\frac{A}{k^2 M^2} \mbox{Tr}\left\{ (\hat{p_2} - m_2) \hat{k}
(\hat{p_3}+m_3) \hat{q} (\hat{p_1}+m_1) \hat{k}
(\hat{\bar{p_3}}+m_3) \hat{q} \right\}
 \right], \label{M1} \eea
 \bea \sum |M_a|^2 & = & - \frac{e_q^2 e^2 g_s^2 \mbox{Tr}\{T^2\}
}{(p_3^2-m_3^2)^2} \left[ \mbox{Tr}\left\{ \gamma_\sigma (\hat{p_3}
+ m_3) \gamma_\nu (\hat{p_1}+m_1) \gamma^\nu (\hat{p_3} + m_3)
\gamma^\sigma  (\hat{p_2}-m_2) \right\} \right. \nnl & & \left .
\phantom{ - \frac{e_q^2 e^2 g_s^2 \mbox{Tr}\{T^2\}
}{(p_3^2-m_3^2)^2}} %
- \frac{1}{M^2} \mbox{Tr}\left\{  \gamma_\sigma (\hat{p_3} + m_3)
\hat{q}(\hat{p_1}+m_1) \hat{q} (\hat{p_3} + m_3) \gamma^\sigma
(\hat{p_2}-m_2) \right\} \right. \nnl & & \left . \phantom{-
\frac{e_q^2 e^2 g_s^2 \mbox{Tr}\{T^2\}
}{(p_3^2-m_3^2)^2}} %
-\frac{A}{k^2} \mbox{Tr}\left\{ \hat{k} (\hat{p_3} + m_3) \gamma_\nu
(\hat{p_1}+m_1) \gamma^\nu (\hat{p_3} + m_3) \hat{k} (\hat{p_2}-m_2)
\right\} \right. \nnl & & \left . \phantom{- \frac{e_q^2 e^2 g_s^2
\mbox{Tr}\{T^2\}
}{(p_3^2-m_3^2)^2}} %
+\frac{A}{k^2 M^2} \mbox{Tr}\left\{ \hat{k} (\hat{p_3} + m_3)
\hat{q} (\hat{p_1}+m_1) \hat{q} (\hat{p_3} + m_3) \hat{k}
(\hat{p_2}-m_2) \right\}
 \right] . \label{M2} \eea
%\end{widetext}
Note that by the transformation $\{ p_3 \to \bar{p_3}, p_1\to p_2,
p_2 \to p_1, m_1 \to -m_2, m_2 \to -m_1 \}$ we readily obtain $\sum
M_b ^* M_a$ from $\sum M_a ^* M_b$ and $\sum |M_b|^2 $ from $\sum
|M_a|^2 $. In equations (\ref{M1}) and (\ref{M2}) the factor $A$ sets the
gauge.
For instance, in the generalized renormalizable gauge we have $A=(1-\lambda)
k^2 / (k^2-\lambda \mu^2)$ and specifically in Feynman gauge $\lambda=0$.
We used the feynpar.m~\cite{feynpar.m} package of the Mathematica
program~\cite{Mathematica} to evaluate the traces of the products of
the gamma matrices.
The resulting cross sections are given in Ref.~\cite{olena2010}.
Since the final formulae for the cross sections are quite lengthy,
we do not repeat them here. But we note that it is seen from the
explicit results that the quark off-shellness leads to higher twist
corrections ($\sim m_q^2/s,m_q^2/t,m_q^2/u$). These corrections are
small in hard hadron scattering at high center-of-mass energy
$\sqrt{s}>10$~GeV but become substantial for  photon production in
the sQGP, where the characteristic $\sqrt{s}$ of parton collisions
is of the order of a few GeV.

We take $d^2\sigma (e^+e^-)/dM^2 dt$ from Ref.~\cite{olena2010} and
use relation (\ref{realvirtual}) to implement the real photon
production in the off-shell quark and gluon interactions into the
PHSD transport approach. In each interaction of $q+\bar q$ or
$q/\bar q+g$ the photon production probability and the elliptic
flow of the produced photon are recorded differentially in
transverse momentum $p_T$, rapidity $y$ and interaction time $t$.

\subsection{Thermal rates and the Landau-Migdal-Pomeranchuk effect}
\label{sect:LPM}

Using the cross sections for photon radiation by dressed quarks and
gluons in the processes $q \bar q \to g \gamma$ and $qg\to q \gamma$
from Ref.~\cite{olena2010} we can calculate the differential rate of
photons from a thermalized strongly interacting QGP. Fig.~\ref{amy}
presents the invariant rate of photons produced from a QGP   at the
temperature $T=200$~MeV (red solid line) in comparison to the
leading-order Log-resummed perturbative QCD rate (blue solid line)
from Arnold, Moore and Yaffe (AMY rate, taken from
Ref.~\protect{\cite{Arnold:2001ms}}). One observes a qualitative
agreement between the results of both approaches although the
degrees-of-freedom and their couplings are different. We mention
that photon rates calculated recently at the NLO in  perturbative
QCD~\cite{Ghiglieri:2013gia,Vujanovic:2014xva,Ghiglieri:2014vua}
also are approximately in line with those presented in
Fig.~\ref{amy} (l.h.s.).

\begin{figure}
\includegraphics[width=0.46\textwidth]{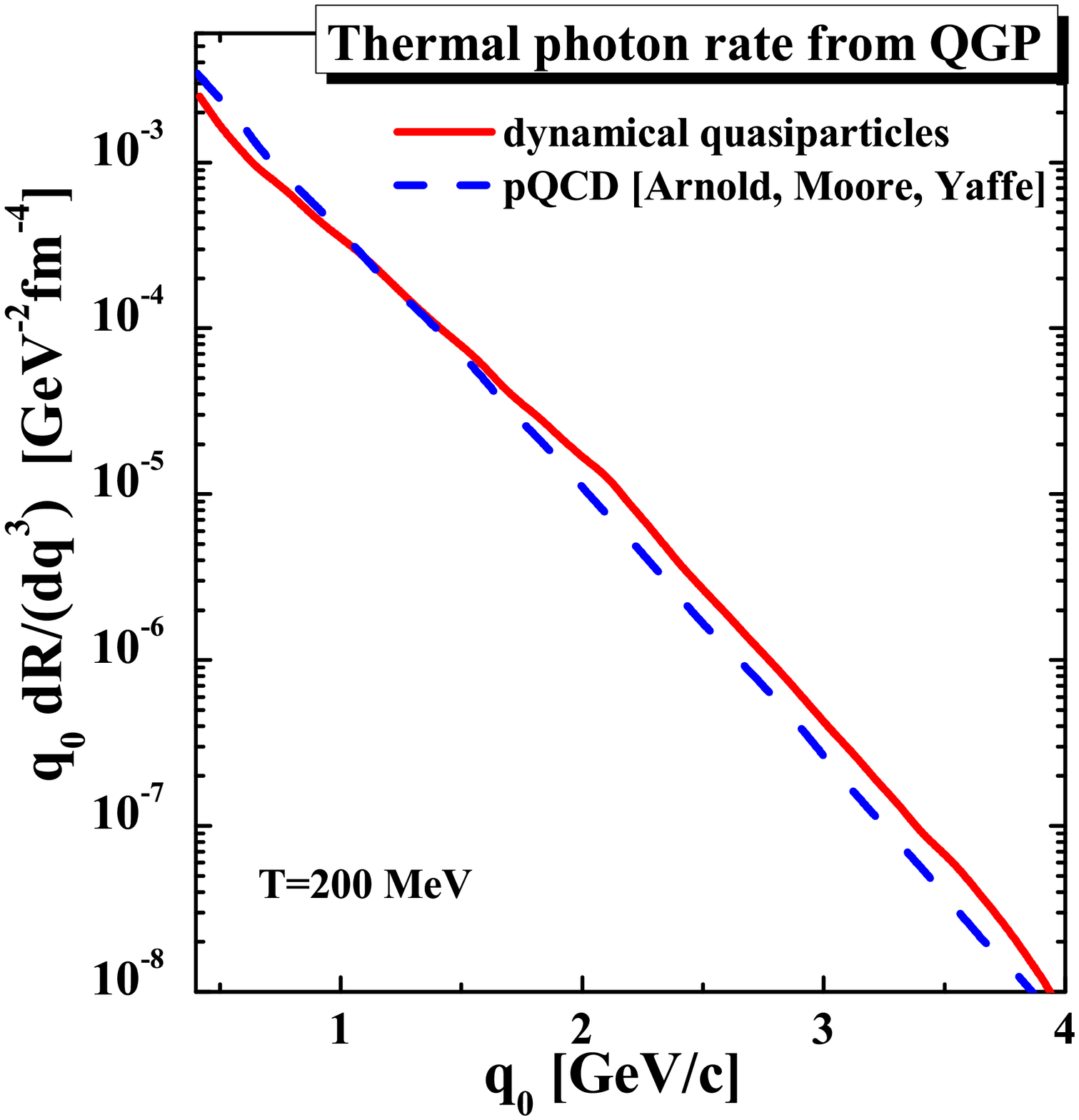} \hspace{1cm} \includegraphics[width=0.42\textwidth]{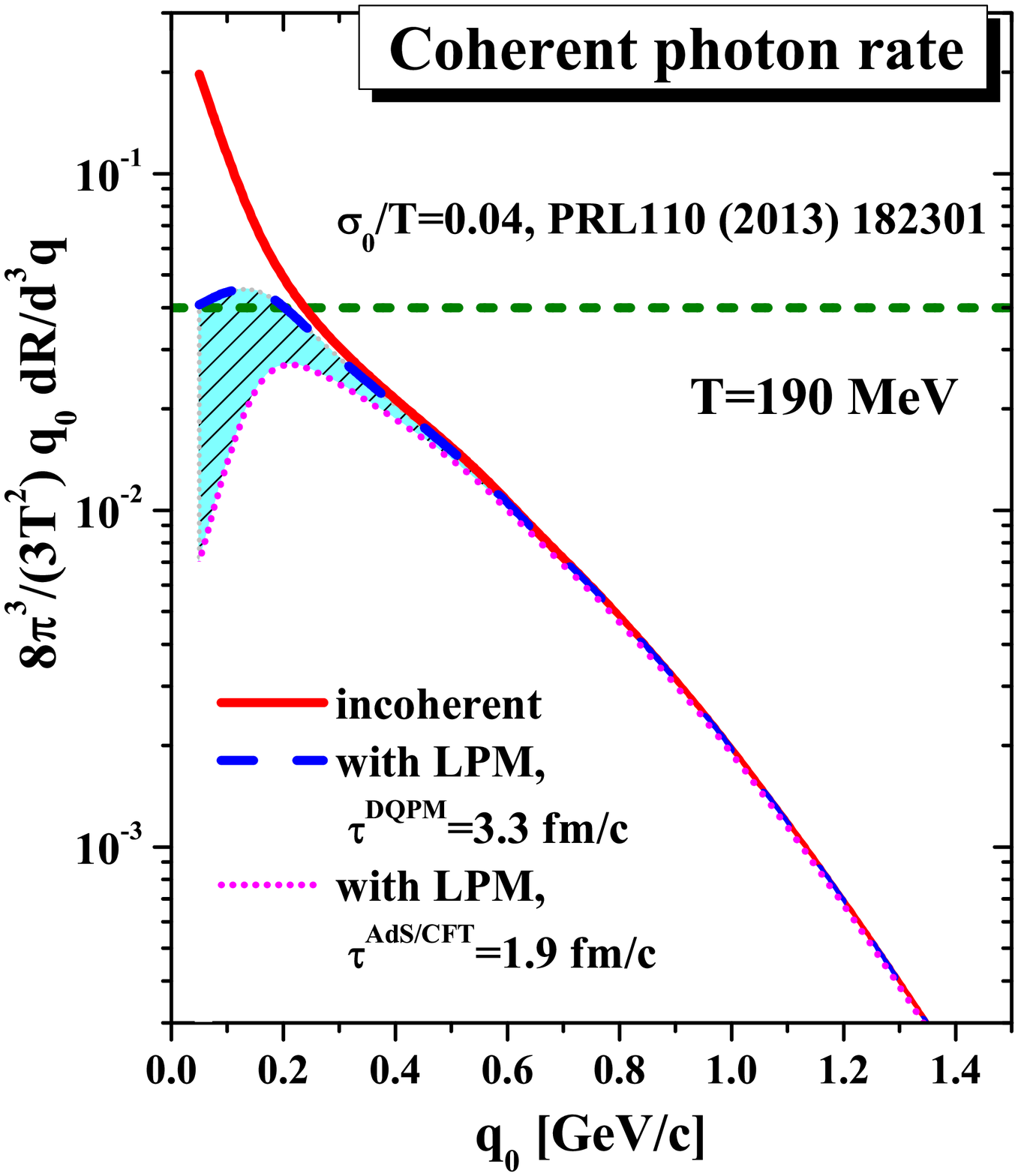}
\caption{   (l.h.s.) Invariant rate of photons produced from the
strongly-interacting quark-gluon plasma (at temperature $T$= 200
MeV) consisting of massive broad quasi-particle quarks and gluons
(red solid line). The leading-order pQCD rate (blue dashed line)
from Ref.~\protect{\cite{Arnold:2001ms}} (AMY-rate) is shown for
comparison. (r.h.s.) Incoherent invariant photon production rate
from the strongly-interacting quark-gluon plasma (at temperature
$T$= 190 MeV) consisting of massive broad quasi-particle quarks and
gluons (red solid line) scaled by $8\pi^3/(3 T^2)$ in order to match
the electric conductivity for $q_0\to0$ (cf.
(\protect{\ref{ratecond}})). The blue dashed line and the magenta
dotted line show the coherent rates for two assumptions on the
average time between the collisions $\tau$, i.e. from the DQPM model
(upper, dashed line) and from the AdS/CFT correspondence as a lower
limit (dotted line). The figures are taken from Ref.
\protect\cite{Linnyk:2015tha}. \label{LPM} \label{amy} }
\end{figure}

The radiation of photons by charged particles  is modified in the
medium compared to the vacuum. One of such medium effects is caused
by the absence of well-defined incoming and outgoing asymptotic
states due to the multiple scattering of particles in a strongly
interacting environment. If the subsequent scatterings occur within
the time necessary for photon radiation $\tau_\gamma \sim 1/q_0$,
then the amplitudes for the emission of photons before and after the
charged particle  scattering have to be summed coherently. The
effect of this destructive interference on the photon spectrum  by
electrons transversing a dense medium was first studied by Landau
and Pomeranchuk in Ref.~\cite{Landau:1953um,Landau:1953gr} and
Migdal in Ref.~\cite{Migdal:1956tc}. Accordingly, the
Landau-Pomeranchuk-Migdal (LPM) effect modifies the spectrum of
photons produced in the medium in comparison to the incoherent sum
of emissions in quasi-free scatterings, leading especially to a
suppression of the low energy photons because the formation time of
the photon $\tau_\gamma$ is proportional to the inverse photon
energy $1/q_0$. In particular, the LPM effect regularizes the
$1/q_0$ divergence of the quasi-free bremsstrahlung spectra. The LPM
suppression and the induced thermal mass of the medium quanta (the
dielectric effect) together ensure that the photon spectrum is
finite in the limit $q_0\to 0$.

The importance of the LPM effect for the case of dilepton and photon
production from QCD systems was shown in
Refs.~\cite{Cleymans:1992kb,Cleymans:1992kb2,Cleymans:1992je,Knoll:1993ic}
long ago.
The magnitude of the LPM suppression is governed by the average time
between the collisions $\tau$, which in turn is given by the inverse
scattering length $a$ or by the inverse average spectral width of
the particles~$\gamma$:
\be \tau=\frac{1}{a} \sim \frac{1}{\gamma}. \ee
The LPM suppression is more pronounced in case of small $\tau$, i.e.
for high reaction rates $\gamma$. Thus we expect it to be important
for the emission of photons from the strongly-interacting
quark-gluon plasma (sQGP) as created in the early phase of the
heavy-ion collision. Indeed, it was shown in
Refs.~\cite{BrCa11,Cassing:2013iz} in the scope of the DQPM  that
the average collision time of partons is as short as $\tau\approx
2\!-3$~fm/c for temperatures in the range $T\!=\!1\!\!-\!2 \ T_c$,
where $T_c \approx 158$~MeV is the deconfinement transition
temperature. In comparison, the average time between pion collisions
in a thermalized pion gas at temperatures $T<T_c$ is above 10~fm/c
\cite{Cleymans:1992kb,Cleymans:1992kb2}.

Let us now quantify the magnitude of the LPM effect on the spectrum
of photons radiated from the QGP as calculated within the PHSD. The
coherent photon production rate - taking into account the LPM effect
- differs from the incoherent cross section by a suppression factor,
which generally depends on the photon energy, temperature and the
interaction strength of the constituents. The coherent photon
emission rate has been  derived in
Ref.~\cite{Cleymans:1992kb,Cleymans:1992kb2} for an elastically
interacting pion gas in the soft photon approximation for the photon
radiation amplitudes. The authors of
Ref.~~\cite{Cleymans:1992kb,Cleymans:1992kb2} used the same method
for the calculation of the photon emission over the whole trajectory
of the charged particle as  adopted in the original work by Migdal
in Ref.~\cite{Migdal:1956tc}. After averaging over the times between
collisions $\tau$, assuming an exponential distribution,
\be \frac{d W}{d \tau} = a e^{-\tau a}, \label{scattimes} \ee
the coherent photon emission rate was found to be
\be \frac{dR}{dq^3}=N\frac{2 \alpha_{_{EM}}}{(2 \pi)^2} \left< v^2
\frac{(1-\cos^2\Theta)}{a^2+q_0^2(1-v\cos \Theta)^2} \right> ,
\label{redl} \ee
where the brackets $<.>$ stand for an average over the velocities of
the scattering particles after the scattering. In Eq. (\ref{redl})
the velocities are characterized by their absolute values $v$ and
scattering angles $\cos \Theta$ in the center-of-mass frame with
respect to the incoming (pion) momenta, while $N$ is the number of
scatterings and $\alpha_{EM}\approx 1/137$.  A realistic
parametrization of the data was used for the pion elastic scattering
cross section (cf. Section~\ref{sect:brems}) but the scattering was
assumed to be isotropic. We recall that the incoherent rate is
obtained from Eq. (\ref{redl}) in the limit $a=0$.

An analytical form of the coherence factor was obtained in
Ref.~\cite{Knoll:1993ic} in the model of hard scattering centers,
using a quantum mechanical approach to coherently sum the photon
amplitudes from all the scatterings. In the thermal medium the
spacial distribution of the scattering centers is assumed random.
Consequently, the function (\ref{scattimes}) naturally arises in
this model for the distribution of times between collisions by a
direct calculation of the two-particle correlation function. The
quenching factor in the dipole limit ($\vec q=0$) was found to be
\be (G(q_0 \tau))^2 = \left( \frac{(q_0 \tau)^2}{1+(q_0 \tau)^2}
\right)^2. \label{lenk} \ee
Although formula (\ref{lenk}) was obtained in a simple model, it is
useful because it correctly captures the dependence of the LPM
suppression on the average strength of the interaction given solely
by the mean-free-time between collisions $\tau$ in the assumption of
isotropic collisions.

We recall that the perturbative interaction of quarks and gluons is dominated by
small angle scattering due to the massless particle exchange in the
$t$-channel diagrams. In this case the coherence factor for the
quark system in the limit of small scattering angles was obtained in
Ref.~\cite{Cleymans:1992je}. However, the elastic scattering of
dressed quarks in the PHSD is not dominated by the $t\to0$ pole as
in the perturbative case since the gluon mass (of order 1 GeV) acts
as a regulator in the amplitude. Accordingly, the angular
distribution for quark-quark scattering is closer to an isotropic
distribution for low or moderate $\sqrt{s}$ in accordance with the
model assumptions of Ref.~\cite{Knoll:1993ic} such that the
expression (\ref{lenk}) should apply as an estimate of the LPM
suppression for the photon emission within the PHSD.

In Fig.~\ref{LPM} (r.h.s.)  we show the photon emission rate in a QGP at the
temperature $T$=190 MeV as calculated in the PHSD as an incoherent
sum of the photon emission in quark and gluon scatterings (red solid
line) which diverges for $q_0 \rightarrow 0$. The blue dashed line gives the same rate with the quenching
factor (\ref{lenk}) applied using $\tau(T)=1/\Gamma(T)\approx 3.3
$~fm/c from the DQPM (for $T$= 190 MeV).  We observe that the
suppression -- in comparison to the incoherent rate -- is visible only for
photon energies  $q_0<0.4$~GeV. For an estimate of the upper limit
on the LPM suppression we employ the relaxation time approximation
for the ratio of the shear viscosity over entropy density $\eta/s$
which gives $\eta/s \approx $0.14 at $T$=190 MeV in the
DQPM~\cite{Ozvenchuk:2012fn,Ozvenchuk:2012kh}. The lowest bound as
conjectured within the AdS/CFT correspondence is $\eta/s=1/(4 \pi)
\approx 0.08$. In the relaxation time approximation this corresponds
to a lower value of $\tau \approx 1.9$~fm/c. The coherent photon
rate in this case is given by the (lowest) magenta dotted line and even shows a
peak in the photon rate for $q_0 \approx$ 0.2 GeV.

In order to further clarify the strength of the LPM suppression of
the photon emission in the sQGP, we use the knowledge of the
electric conductivity $\sigma_0(T)$ of the sQGP from the
DQPM~\cite{Cassing:2013iz} which is roughly in line with more recent
results from lattice QCD (cf. Fig. 6, r.h.s.). We recall that the photon emission
rate from a thermal medium is controlled by
$\sigma_0$ via the relation ~\cite{Yin:2013kya},
\be \frac{\sigma_0}{T}=\frac{8 \pi^3}{3 T^2} \lim _ {q_0\to 0}
\left( q_0 \frac{d R}{ d^3 q} \right), \label{ratecond} \ee
where $T$ is the temperature of the system, $q_0$ is the photon
energy and $\vec q$ is the photon momentum. Using the number for
$\sigma_0/T$  from the PHSD at the temperature of $T=190$~MeV from
Ref.~\cite{Cassing:2013iz} (or Fig. 6, r.h.s.), we obtain a limiting
value for the scaled photon emission rate of $0.04$ for $q_0\to0$
according to formula (\ref{ratecond}) (green short dashed line in
Fig.~\ref{LPM}, r.h.s.). The blue dashed line in Fig.~\ref{LPM} --
the estimate of the rate based on  formula (\ref{lenk}) and the DQPM
average spectral width of the quarks/antiquarks -- indeed approaches
the limiting value of 0.04 as given by the kinetic calculations of
the electric conductivity.

Taking into account some uncertainty in the determination of $\tau$
and the expression (\ref{lenk}), we conclude from Fig.~\ref{LPM} (r.h.s.) and
analogous calculations at different temperatures that the LPM effect
influences the photon production from the QGP for photon energies
below $q_0 \approx 0.4$~GeV, but is negligible for higher photon
energies. We note in passing that the suppression of the photon
spectrum in the hadronic phase is much smaller due to the lower
interaction rate, i.e. a longer interaction time $\tau$ and thus a
lower LPM suppression factor at the same photon energy.

\subsection{Bremsstrahlung $m+m\to m+m+\gamma$ beyond the soft-photon approximation}
\label{sect:brems}

We briefly sketch the description of the photon
bremsstrahlung in meson+meson scattering beyond the
soft-photon approximation~\cite{Low:1958sn}. Since pions are the
dominant meson species in the heavy-ion collisions \cite{Linnyk:2013hta}, we concentrate
here on the description of the bremsstrahlung photon production in
pion+pion collisions.
In order to calculate the differential cross sections for the photon
production in the processes of the type $\pi+\pi\to \pi+\pi+\gamma$
we use the one-boson exchange (OBE) model as  originally applied
in Ref.~\cite{Eggers:1995jq} to the dilepton bremsstrahlung in pion+pion
collisions, later on in Ref.~\cite{Liu:2007zzw} to the low-energy
photon bremsstrahlung in pion+pion and kaon+kaon collisions.
The calculations are based on  a covariant microscopic
effective theory with the interaction Lagrangian,
\be L_{int}= g_{\sigma} \sigma \partial _\mu \vec \pi \partial ^\mu
\vec \pi + g _\rho \vec \rho ^\mu \cdot (\vec \pi \times
\partial _\mu \vec \pi) + g_f f_{\mu \nu} \partial ^\mu \vec \pi
\cdot
\partial ^\nu \vec \pi, \label{Lagrangian} \ee
as suggested in Refs.~\cite{Haglin:1992fy,Eggers:1995jq}.
Within this model the interaction of pions is described by the
exchange of scalar, vector and tensor resonances: $\sigma$, $\rho$
and $f_2(1270)$, respectively. Additionally,  form factors are
incorporated in the vertices in the $t$- and $u$-channels to account for
the composite structure of the mesons and thus to
effectively suppress the high momentum transfers,
\be h_{\alpha}(k^2)=\frac{m_\alpha^2-m_\pi^2}{m_\alpha^2-k^2}, \ee
where $m_\alpha=m_\sigma$ or $m_\rho$ or $m_f$ is the mass of the
exchanged meson and $k^2$ is the momentum transfer squared.

The cross section for elastic $\pi+\pi\to\pi+\pi$ scattering is given by
\be \frac{d \sigma_{el} (s)}{dt} = \frac{|M_{el}|^2}{16 \pi s
(s-4m_\pi^2)}, \label{el1} \ee
where the matrix element $|M|^2$ is calculated by coherently
summing up the Born diagrams of the $\sigma$-, $\rho$- and $f_2$-meson exchange
in $t$, $s$ and $u$  channels
(the $u$-channel diagrams are needed only in case of identical
pions), %%
\bea |M_{el}|^2&=& |M^s(\sigma)\!+\! M^t(\sigma)\! +\! M^u(\sigma)
+\!M^s(\rho)\!+\!M^t(\rho)\!+\!M^u(\rho)\! +\!M^s(f)\!+\!M^t(f)\!
+\!M^u(f)|^2  . \ \ \ \ \ \ \ \
 \eea
Let us define the four-momenta of the incoming pions as
$p_a=(E_a,\vec p_a)$ and $p_b=(E_b,\vec p_b)$, the momenta of the
outgoing pions as $p_1=(E_1,\vec p_1)$ and $p_2=(E_2,\vec p_2)$ and
the four-momentum of the exchanged resonance ($\sigma$, $\rho$ or
$f_2$) as $k$.
The propagators of the massive and broad scalar and vector particles
are used to describe the exchange of the $\sigma$ and $\rho$ mesons
(see e.g. Ref.~\cite{Eggers:1995jq}). The resonance $f_2$ is a
spin-2 particle, for which the full momentum-dependent propagator
has been derived in Ref.~\cite{vanDam:1970vg}. The polarization sum is
\bea P_{\mu \nu \alpha\beta}  & = & \frac{1}{2} (g_{\mu\alpha}
g_{\nu\beta} + g_{\mu\beta} g_{\nu\alpha} - g_{\mu\nu}
g_{\alpha\beta}  ) \!\!\!\!\!\!\!\!\! - \frac{1}{2}
(g_{\mu\alpha}\frac{k_\nu k_\beta}{m_f^2} +  g_{\mu\beta}\frac{k_\nu
k_\alpha}{m_f^2} + g_{\nu\alpha}\frac{k_\mu k_\beta}{m_f^2} +
g_{\nu\beta}\frac{k_\mu k_\alpha}{m_f^2}  ) \nnl &&
\!\!\!\!\!\!\!\!\! + \frac{2}{3} (\frac{1}{2} g_{\mu \nu} +
\frac{k_\mu k_\nu}{m_f^2} )(\frac{1}{2} g_{\alpha \beta} +
\frac{k_\alpha k_\beta}{m_f^2} ). \eea
Following the example of the dilepton production study in
Ref.~\cite{Eggers:1995jq}, we use the same propagator for the $f_2$
resonance while additionally accounting for its finite width by
adding an imaginary part to the self-energy in accordance with its lifetime.

%\begin{figure}
%\includegraphics[width=0.43\textwidth]{pipi.eps} \hspace{0.5cm} \includegraphics[width=0.54\textwidth]{angular.eps}
%\caption{ {\it (l.h.s.)} Cross section of pion+pion elastic
%scattering within the OBE effective models are compared to the
%experimental data
%from~\protect{\cite{Srinivasan:1975tj,Protopopescu:1973sh}}. {\it
%(r.h.s.)} Angular differential cross section for pion+pion elastic
%scattering within two effective models: the exchange of  two mesonic
%resonances, scalar $\sigma$ and vector $\rho$ (blue dashed line),
%and the exchange of  three resonances $\sigma$, $\rho$ and the
%tensor resonance $f_2(1270)$ of the particle data booklet~\cite{PDG}
%(red solid line with symbols and gray dashed line). The red solid
%line with star symbols shows the model with the full momentum
%dependence of the $f_2$ propagator, while the grey dashed line is
%obtained neglecting the momentum dependence of the $f_2$ propagator.
%The green dashed line shows the constant and isotropic
%$\sigma_{el}=10$~mb for comparison. \label{pipi}}
%\end{figure}

As a result, the following expressions are obtained for the matrix
elements in case of elastic $\pi+\pi$ scattering diagrams (we give here
explicitly the $t$- and $s$-channel results, the $u$-channels can be
easily obtained by the crossing relations):
\bea M^t(\sigma) & = &  \frac{-g_\sigma^2 h_\sigma^2(t) \left( 2
m_\pi^2-t \right)^2}{t-m_\sigma^2+im_\sigma \Gamma_\sigma},
\hspace{1cm} M^s(\sigma)  =  \frac{-g_\sigma^2 \left(s - 2 m_\pi^2
\right)^2}{s-m_\sigma^2+im_\sigma \Gamma_\sigma}, \nnl M^t(\rho) & =
& \frac{-g_\rho^2 h_\rho^2(t) \left(
s-u\right)^2}{t-m_\rho^2+im_\rho \Gamma_\rho}, \hspace{1.5cm}
M^s(\rho) =  \frac{g_\rho^2 \left( u-t\right)^2}{s-m_\rho^2+im_\rho
\Gamma_\rho}, \label{elasticformulas} \\ M^t(f) & = & \frac{g_f^2
h_f^2(t) }{t-m_f^2+im_f \Gamma_f} \frac{1}{2} \left( \frac{2}{3} (2
m_\pi^2-t)^2-(s-2 m_\pi^2)^2 - (2m\pi^2-u)^2 \right), \nonumber \eea
\bea  M^s(f) & =  & \frac{g_f^2 }{s-m_f^2+im_f \Gamma_f} \frac{1}{2}
\left( \frac{2}{3} (s-2 m_\pi^2)^2-(2 m_\pi^2-t)^2 - (2m\pi^2-u)^2
\right), \nonumber
\eea
where the Mandelstamm variables are defined as
$s=(p_a+p_b)^2=(p_1+p_b)^2$, $t=(p_a-p_1)^2=(p_b-p_2)^2$,
$u=(p_a-p_2)^2=(p_b-p_1)^2$. We point out that the formulae
(\ref{elasticformulas}) are compact, because the masses of all pions
were assumed to be equal to $m_\pi$ and the energy-momentum
conservation $p_a+p_b=p_1+p_2$ has been used. These conditions are
not satisfied for the off-shell $\pi+\pi\to\pi+\pi$ subprocess,
which  we encounter in the subsequent calculation of the
bremsstrahlung photon production $\pi+\pi\to\pi+\pi+\gamma$. For the
actual calculation we obtained the off-shell generalizations
$M(p_a,p_b,p_1,p_2)$ of the formulae (\ref{elasticformulas}), which
are too lengthy to be presented here explicitly.

A reduced version of the model with the exchange of only two
resonances -- the scalar $\sigma$ and the vector $\rho$ meson -- was
used by the authors of Ref.~\cite{Liu:2007zzw} to calculate the rate
of the photon production from the $\pi+\pi\to\pi+\pi+\gamma$ process
at low transverse momenta of the photons ($p_T<0.4$~GeV). This
approximation is suitable at low $p_T$ because the photon rate in
this kinematical region is dominated by  pion collisions of low
center-of-mass energy $\sqrt{s}$, for which the contribution of the
$f_2$-exchange is small. However, relatively high transverse momenta
of photons $p_T=1-2$~GeV are of interest for our goal of clarifying
the ``puzzling" high elliptic flow of {\em direct} photons.
Therefore, we use the OBE model with three mesons as interaction
carriers (including the tensor particle $f_2(1270)$) in the PHSD
calculations.

\begin{figure}
\hspace{0.1cm}
\includegraphics[width=0.45\textwidth]{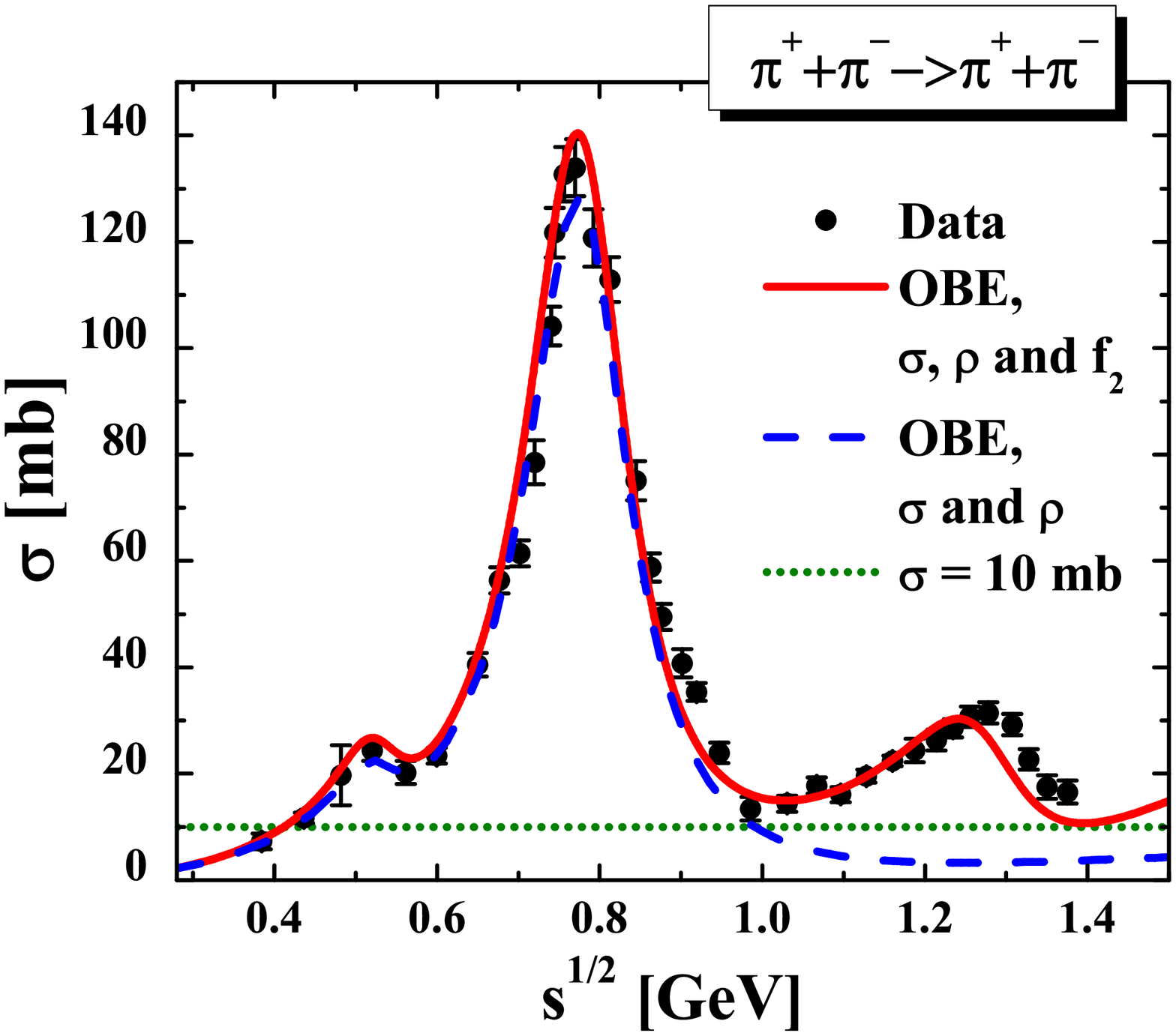} \hspace{0.5cm}
\includegraphics[width=0.48\textwidth]{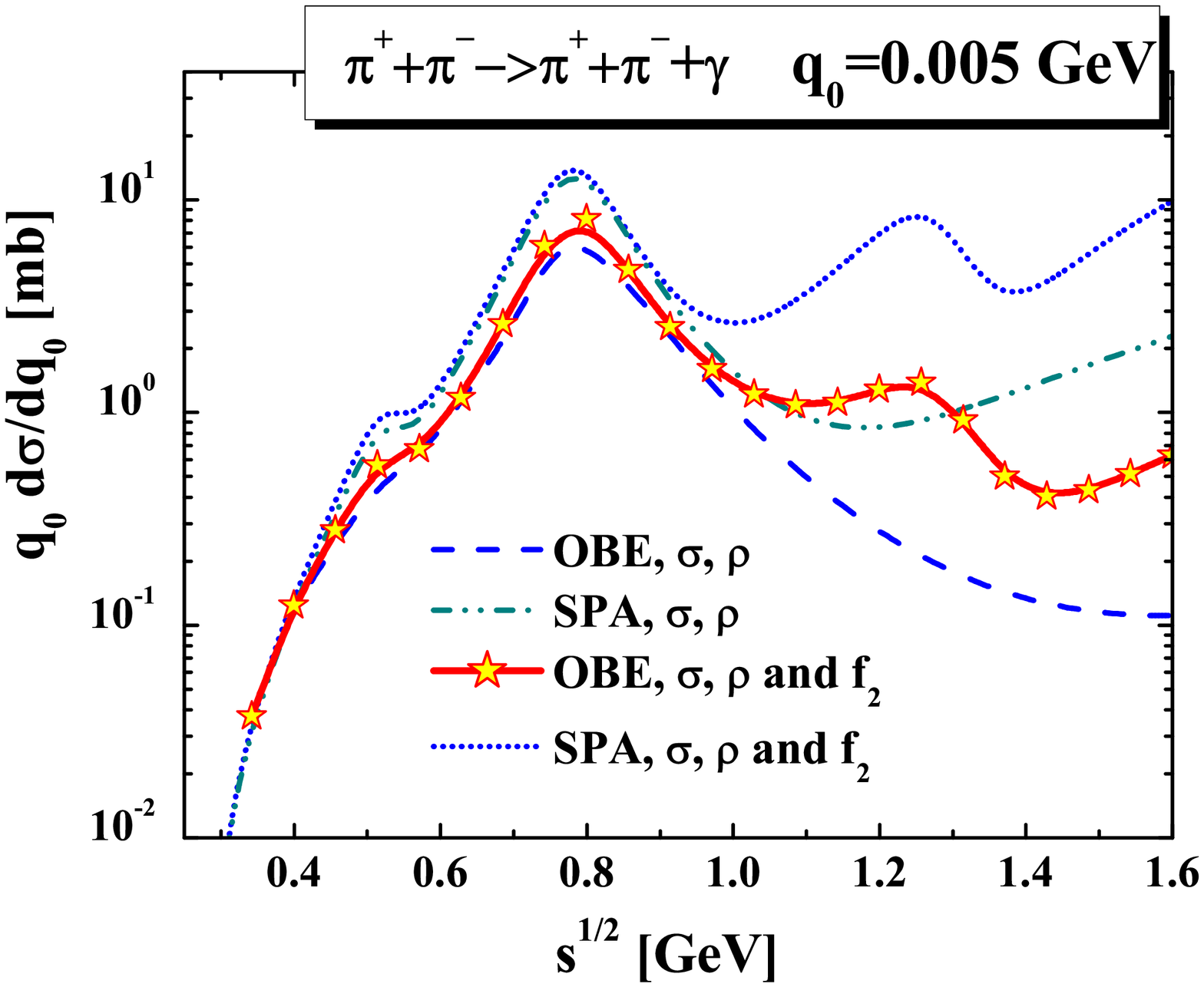}
\caption{ { (l.h.s.)} Cross section for pion+pion elastic scattering
within the OBE effective models in comparison to the experimental
data from
Refs.~\protect{\cite{Srinivasan:1975tj,Protopopescu:1973sh}}: the
exchange of  two mesonic resonances, scalar $\sigma$ and vector
$\rho$ (blue dashed line), and the exchange of  three resonances
$\sigma$, $\rho$ and the tensor resonance $f_2(1270)$ of the
particle data booklet~\cite{PDG} (red solid line). The green dashed
line shows the constant and isotropic $\sigma_{el}=10$~mb for
orientation. { (r.h.s.)}  Cross section for the production of a
photon with energy $q_0=0.005$~GeV in the process
$\pi+\pi->\pi+\pi+\gamma$ within the following models: the exact OBE
cross section within the effective model taking into account scalar,
vector and tensor interactions via the exchange of $\sigma$, $\rho$
and $f_2(1270)$-mesons (red line with star symbols), the soft photon
approximation to this model (blue dotted line); the OBE result
within the model taking into account only the scalar and vector
interactions via the exchange of $\sigma$ and $\rho$ mesons (blue
dashed line), and the soft photon approximation to this model (cyan
dash-dot-dotted line). The figures are taken from Ref.
\protect\cite{Linnyk:2015tha}. \label{VariantsSPA}  \label{pipi}}
\end{figure}

Phenomenological coupling constants, masses and widths of the three
interaction-carriers -- entering the Lagrangian (\ref{Lagrangian}) --
have to be fixed to the integrated energy-dependent cross section for
 pion+pion elastic scattering $\sigma_{el} (\sqrt{s})$, which is known
experimentally as a function of $\sqrt{s})$. We present in Fig.~\ref{pipi} (l.h.s.) the integrated cross
section for $\pi+\pi$ elastic scattering in two versions of
the OBE model described above: taking into account the 2 resonances
$\sigma$, $\rho$ (dashed blue line) and taking into account the 3
resonances $\sigma$, $\rho$ and $f_2$ (solid red line). Fitting the
parameters of both variants of the OBE model (with two- or
three-resonance exchange) to the data
from Refs.~\cite{Srinivasan:1975tj,Protopopescu:1973sh} we find the  best-fit parameters: $g_\sigma
m_\sigma = 2.0$, $m_\sigma= 0.525$~GeV, $\Gamma_\sigma=0.100$~GeV,
$g_\rho=6.15$, $m_\rho=0.775$~GeV, $\Gamma_\rho=0.15$~GeV, $g_f
m_f=8.0$, $m_f=1.274$~GeV, $\Gamma_f=0.18$~GeV. %%
%% ! update
%%
The values of the masses and widths suggest an identification of
the $\rho$-resonance to the $\rho$-meson and of the particle $f_2$
to the $f_2(1270)$ in the particle data book~\cite{PDG}.
One sees in Fig.~\ref{pipi} (l.h.s.) that the tensor particle $f_2$ is
important for the description of the pion interaction at higher
collision energies $\sqrt{s}>1$~GeV. Neglecting the contribution of
the $f_2$ leads to an underestimation of the $\pi+\pi$ elastic
scattering cross section by an order of magnitude around
$\sqrt{s}=1.2-1.3$~GeV. Later data on the $\pi+\pi$ interaction at
$\sqrt{s}$ above 1~GeV -- extracted in Ref.~\cite{Aston:1990wg} from
the measurement of the $K+p\to\Lambda+\pi+\pi$ reaction -- also
point to the importance of the tensor interaction in the resonance
region of the $f_2(1270)$.

Within  the OBE model  for the covariant
interactions of pions (described above), we can also calculate the emission of photons from
the colliding pions by gauge coupling to the external hadron lines.
The Feynman diagrams for the photon production in the process
$\pi+\pi\to\pi+\pi+\gamma$ are shown in Fig.~\ref{diagramsOBE}. For
identical pions, e.g. $\pi^+ + \pi^+$, the $u$-channel diagrams have
to be added, which are obtained from the $t$-channel diagrams by
exchanging the outgoing pions. The applicability of this method is
not limited to low photon energies but is restricted only
by the applicability of the effective model to the description of
the pion-pion (elastic) scattering.

Let us again denote the four-momenta of the incoming pions by $p_a$
and $p_b$, the momenta of the outgoing pions by $p_1$ and $p_2$,
and the photon momentum by $q=(q_0,\vec q)$. The
cross section for photon production in the process
\be \pi(p_a) + \pi(p_b) \to \pi(p_1) + \pi(p_2) +\gamma(q) \ee
then is given by
\be d \sigma^{\gamma} = \frac{1}{2\sqrt{s(s-4m_\pi^2)}}
|M(\gamma)|^2 d R _3, \label{r3} \ee
where $d R_3$ is the three-particle phase space, which depends on
the momenta of the outgoing pions and of the photon,
\bea dR_3 & = & \frac{d^3p_1}{(2\pi)^3 2 E_1} \frac{d^3p_2}{(2\pi)^3
2 E_2} \frac{d^3q}{(2\pi)^3 2 q_0}
 (2\pi)^4 \delta^4\left(
p_a+p_b-p_1-p_2-q\right).
 \eea
The cross section (\ref{r3}) will be integrated over the final pion
momenta to obtain the differential photon spectrum $d\sigma/d^3q$.
The $\delta$-function allows to perform four
integrations analytically and the remaining two are done numerically.

The matrix element $M$ in (\ref{r3}) is a coherent sum of the
diagrams presented in Fig.~\ref{diagramsOBE} -- i.e. of the photon
attached to each pion line $\pi_a$, $\pi_b$, $\pi_1$ and $\pi_2$ --
and of contact terms, which account for the emission from the
vertices and the internal lines:
\bea |M(\gamma)|^2 = M^*_\mu(\gamma) M^\mu(\gamma) = \left|
M_a^\mu+M_b^\mu + M_1^\mu+M_2^\mu + M_c^\mu \right|^2.
 \eea
The complex matrix elements for the photon emission from each of the
pion lines $M_i^\mu$ are calculated as sums of the three meson
exchanges ($\sigma$, $\rho$, $f_2$). For instance:
\bea M^\mu _1 = e J_1^\mu \left[ M^s_{el}(p_a,p_b,p_1+q,p_2)
 + M^t_{el}(p_a,p_b,p_1+q,p_2)  +
M^u_{el}(p_a,p_b,p_1+q,p_2) \right] \eea
with
\bea J_{a,b}^\mu= -Q_{a,b}\frac{(2 p_{a,b}-q)^\mu}{2 p_{a,b}\cdot
q}, \hspace{1cm} J_{1,2}^\mu = Q_{1,2}\frac{(2 p_{1,2}-q)^\mu}{2
p_{1,2}\cdot q}, \label{charges2} \eea
where $Q_i$ are the charges of the pions in terms of the electron
charge $e$. The matrix elements for the pion elastic subprocess
$M_{el}(p_a,p_b,p_1+q,p_2)$ are the off-shell generalizations of the
formulae (\ref{elasticformulas}).

The contact term $M_c^\mu$ is taken from Ref.~\cite{Liu:2007zzw},
 Eq. (14), where it was derived by demanding the gauge invariance of
the resulting cross section. Indeed, the gauge invariance of the
result has to be restored~\cite{Haglin:1989ga} in calculations
within effective models. In the present work, we have used the
contact terms in order to cancel the gauge-dependent parts in the
matrix element as  in  Ref.~\cite{Liu:2007zzw}. Alternatively, one can take
into account additional diagrams with the emission of photons from
the internal lines (see Refs~\cite{Eggers:1995jq}) but this method
does not always eliminate the need for contact terms (see
Ref.~\cite{Haglin:1989ga}). We have verified that $q_\mu M^\mu
(\gamma) = 0$ in our calculations and thus
the resulting cross sections are gauge invariant.
%Comparing our results to calculations with a different gauge-fixing
%method will allow to quantify the uncertainty of the effective model
%applied (work in progress).

\begin{figure}
\hspace{1cm} \includegraphics[width=0.9\textwidth]{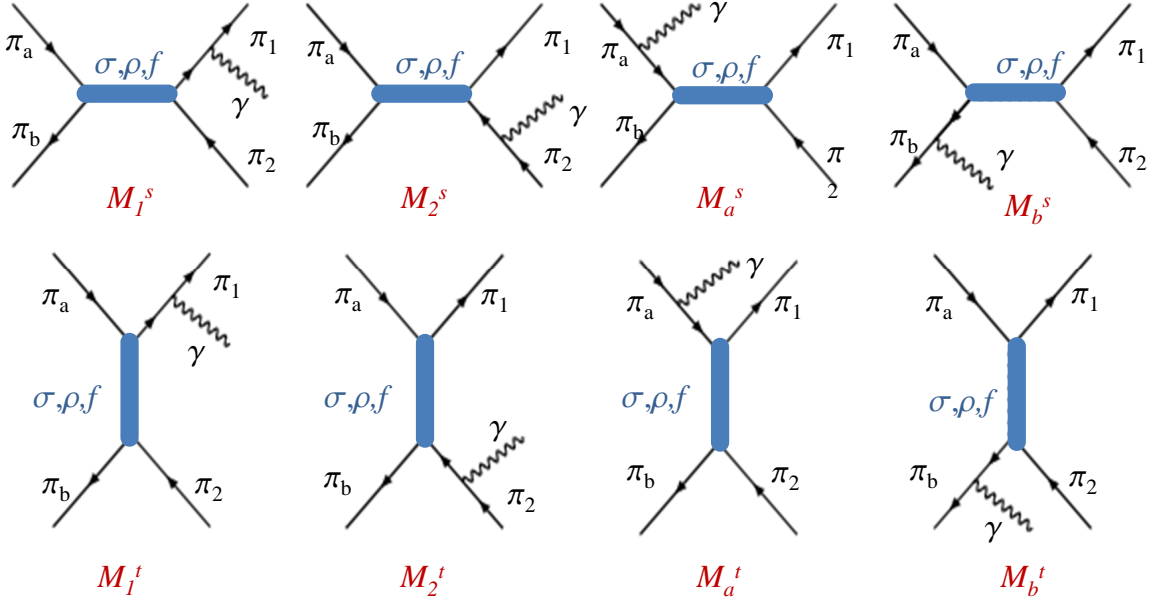}
\caption{  Feynman diagrams for photon production in the reaction
$\pi+\pi\to\pi+\pi+\gamma$ in the one-boson exchange (OBE) model.
The time goes from left to right. For  identical pions, e.g. $\pi^+
+ \pi^+$, the $u$-channel diagrams have to be added.
\label{diagramsOBE} }
\end{figure}

%Within the soft-photon approximation (SPA)~\cite{gale87a,gale87b,gale87c} one assumes
%that the strong interaction vertex is on-shell.
The soft photon
approximation is based on the first-order expansion in the Low
theorem~\cite{Low:1958sn} and is valid at low photon energy and low
$\sqrt{s_{mm}}$ of the meson+meson collision, as has been studied in
detail for the production of dileptons in Ref.~\cite{Eggers:1995jq}.
In this case the strong interaction part and the electromagnetic
part can be separated, i.e. the soft-photon cross section for the
reaction $m_1 + m_2 \to m_1 + m_2 + \gamma$ can be written as
\be  q_0 \frac{d\sigma^{\gamma}(s)}{d^3q} = \frac{\alpha_{_{EM}}}{4
\pi^2} \!\!\!\! \int\limits_{-\lambda(s,m_a^2,m_b^2)/s}^0
\!\!\!\!\!\!\!\!\!\! |\epsilon \cdot J(q,t)|^2 \frac{d\sigma_{el}
(s)}{dt} dt, \label{int_spa_formula} \ee
where $\alpha_{_{EM}}$ is the fine structure constant, $t$ is the
momentum transfer squared  in the $\pi+\pi\to\pi+\pi$ subprocess
and $\epsilon$ is the photon polarization. In (\ref{int_spa_formula})  $J^\mu$ is the
electromagnetic current
\be J^\mu=-Q_a\frac{p_a^\mu}{(p_a\cdot q)}
-Q_b\frac{p_b^\mu}{(p_b\cdot q)} +Q_1\frac{p_1^\mu}{(p_1\cdot q)}
+Q_2\frac{p_2^\mu}{(p_2\cdot q)}. \nonumber \ee
The polarization sum
\be |\epsilon \cdot J|^2 = \left\{ \sum_{pol \ \lambda } J\cdot
\epsilon_\lambda  J\cdot \epsilon_\lambda \right\} \ee
depends on the photon momentum $q$, the charges of the pions $Q_i$
as well as on the invariant kinematic variables, including $t$. For
the case of equal-mass particle scattering
($m_a=m_b=m_1=m_2=m_\pi$) one obtains~\cite{Haglin:1992fy}:
\bea |\epsilon\cdot J|^2 & = & \frac{1}{q_0^2}\left\{
-(Q_a^2+Q_b^2+Q_1^2+Q_2^2)
-2(Q_aQ_b\!+\!Q_1Q_2)\frac{s-2m_\pi^2}{\sqrt{s(s-4m_\pi^2)}} \ln
\left(\frac{\sqrt{s}+\sqrt{s-4m_\pi^2}}{\sqrt{s}-\sqrt{s-4m_\pi^2}}
\right)
\right. \nnl
&&  \hspace{0.8cm} \left.
+2(Q_aQ_1\!+\!Q_bQ_2)\frac{2m_\pi^2-t}{\sqrt{t(t-4m_\pi^2)}} \ln
\left(\!\!
\frac{\sqrt{-t+4m_\pi^2}+\sqrt{-t}}{\sqrt{-t+4m_\pi^2}-\sqrt{-t}}
\right)
 \right. \nnl
&&  \hspace{0.8cm} \left.
+2(Q_aQ_2\!+\!Q_bQ_3)\frac{s-2m_\pi^2+t}{\sqrt{(s+t)(s+t-4m_\pi^2)}}
 \ln
\left(\frac{\sqrt{s+t}+\sqrt{s+t-4m_\pi^2}}{\sqrt{s+t}-\sqrt{s+t-4m_\pi^2}}
\right) \right\}. \label{charges1} \eea
In Eq. (\ref{int_spa_formula}), $d \sigma_{el}(s)/dt$ is the on-shell
differential elastic $\pi+\pi$ cross section, which is a function of
the invariant energy $\sqrt{s}$ and the pion scattering angle via the four-momentum
transfer squared $t$.

The expression (\ref{int_spa_formula}) is considerably simpler in
comparison to the ``full" OBE formula (\ref{r3}) due to the
factorization of the diagrams from Fig.~\ref{diagramsOBE} into an
electromagnetic part and an elastic $\pi+\pi\to\pi+\pi$
subprocess, for the cross section of which the photon $q$-dependence is
omitted. This corresponds to neglecting the off-shellness of the
pion  emitting the photon, e.g. for the pion $a$:
\be p_a-q\approx p_a. \ee
Consequently, the sub-process invariant energy $\sqrt{s_2}$ is also
approximated by the total invariant energy $\sqrt{s}$ of the process
$\pi+\pi\to\pi+\pi+\gamma$:
\be s_2\equiv (p_a+p_b-q)^2 \approx (p_a+p_b)^2 = s, \ee
and the limits of integration over $t$ are also taken as for the
on-shell case, i.e. from $-\lambda(s,m_a^2,m_b^2)/s$ to 0, while the
actual integration over the full 3-particle phase space $R_3$ in the exact
treatment (\ref{r3}) involves different limits for $t$.

In Fig.~\ref{pipi} (r.h.s.) we show the resulting cross sections for
the photon production in the process $\pi+\pi\to\pi+\pi+\gamma$
within the following models: the ``full" OBE taking into account
scalar, vector and tensor interactions via the exchange of $\sigma$,
$\rho$ and $f_2(1270)$-mesons (red line with star symbols), the soft
photon approximation (\ref{int_spa_formula}) to this model (blue
dotted line); the OBE result employing  only the scalar and vector
interactions via the exchange of $\sigma$ and $\rho$ mesons (blue
dashed line), and the soft photon approximation to this model  (cyan
dash-dot-dotted line). For the very low energy of the photon of
$q_0=5$~MeV the SPA agrees with the ``exact" cross section very well
in the region of $\sqrt{s}<0.9$~GeV (see Fig.~\ref{pipi}, r.h.s.).
However, the discrepancy to the OBE result is increasing rapidly
with growing $\sqrt{s}$;  the calculations for the higher photon
energy of $q_0=0.5$~GeV show an even larger discrepancy between the
SPA and the exact OBE result (cf. Ref.~\cite{Linnyk:2015tha}).

\begin{figure}[t]
 \hspace{0.4cm}
 \includegraphics[width=0.43\textwidth]{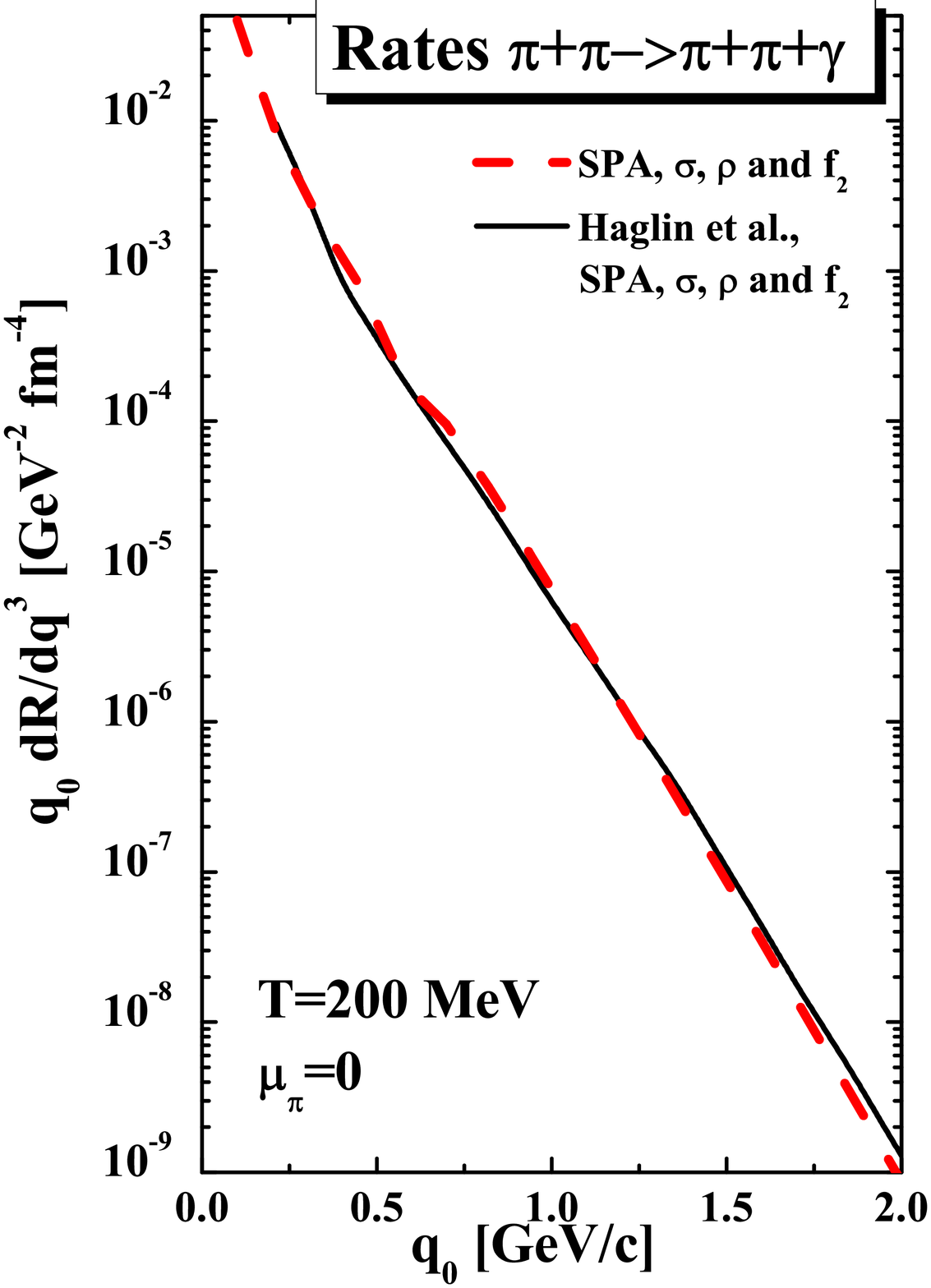} \hspace{0.4cm}
\includegraphics[width=0.43\textwidth]{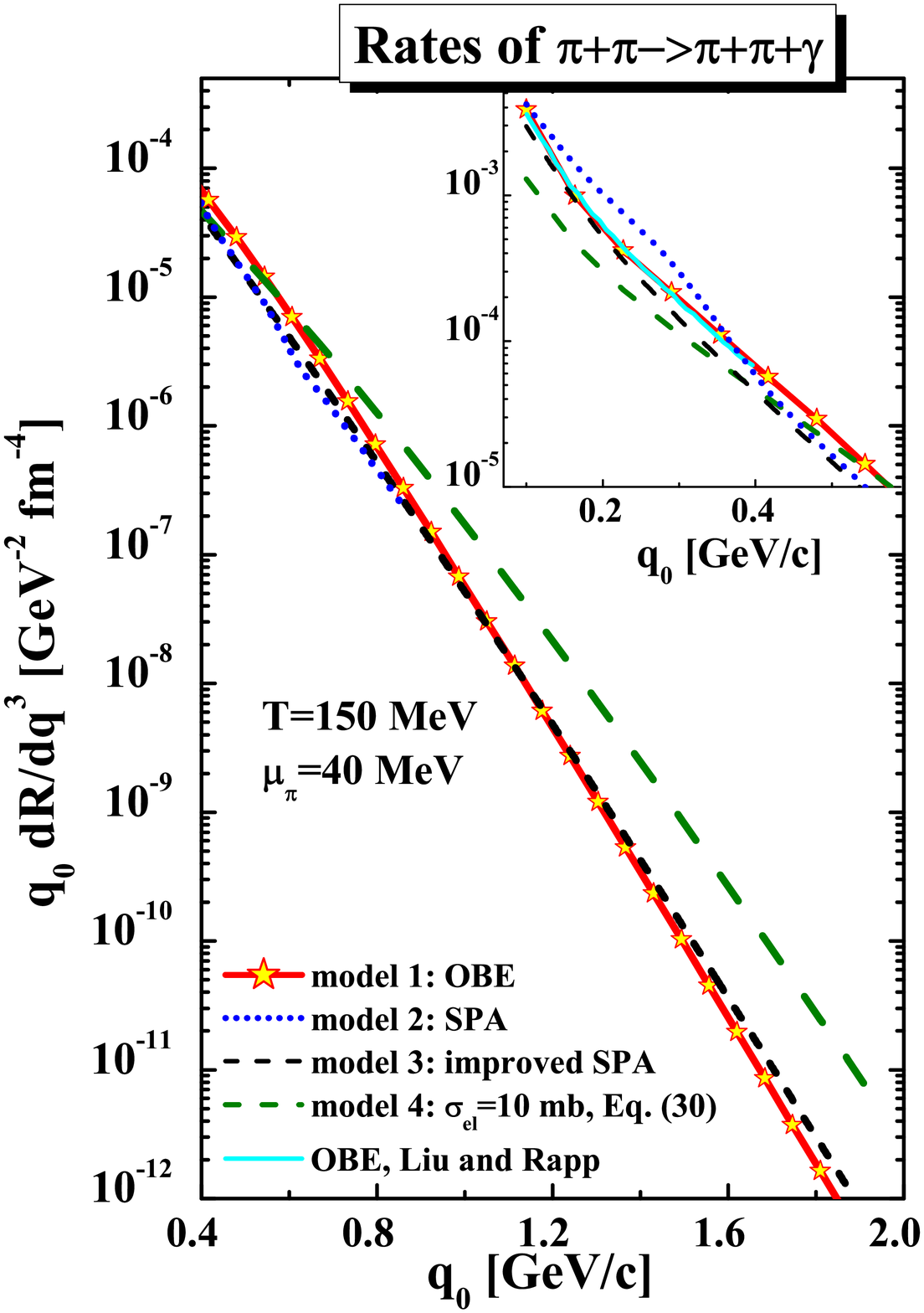}
\caption{ { (l.h.s.)} Invariant rate of the bremsstrahlung-photon
production from an equilibrated pion gas at a temperature of
$T=200$~MeV and pion chemical potential $\mu_\pi=0$ as calculated in
the OBE model with three resonance exchange within the soft-photon
approximation (red dashed line). The black solid line from
Ref.~\protect{\cite{Haglin:2003sh}} is shown for comparison and
validation of our calculations. { (r.h.s.)}
 Invariant rate of
bremsstrahlung photons produced from an equilibrated pion gas at
$T=150$~MeV and $\mu_\pi=40$~MeV versus the photon energy $q_0$. The
in-let shows the same quantity for the range of photon energies
$q_0=0.1-0.4$~GeV. The calculations have been performed within the
following models: (1) OBE model beyond the soft-photon approximation
(red solid line with star symbols); (2) OBE model within the soft
photon approximation (blue dotted line); (3) OBE model within the
improved soft photon approximation (black short-dashed line) -- the
invariant energy $\sqrt{s_2}$ of the on-shell $\pi+\pi$ elastic
process is not equal to the total invariant energy of the process
$\sqrt{s}$: $s_2=s-q_0 \sqrt{s}$; (4) the soft photon approximation
with the constant isotropic elastic cross section of
$\sigma_{el}=10$. The cyan solid line from Liu and
Rapp~\protect{\cite{Liu:2007zzw}} is shown for comparison. The
figures are taken from Ref.~\protect\cite{Linnyk:2015tha}.
\label{rates}}
\end{figure}

Using the cross section for the $\pi+\pi\to\pi+\pi+\gamma$ reaction
according to Eq. (\ref{int_spa_formula}) as a function of the photon
energy $q_0$ and the collision energy $\sqrt{s}$,  the yield
$dN/d^3q$ and the invariant rate $q_0 dR/d^3q$ for bremsstrahlung
photon production from an equilibrated pion gas can be evaluated in
a straight forward manner.
Within kinetic theory, the rate of photon production in the
collisions of particles $a$ and $b$ in a thermalized medium (number
of photons produced per unit space-time volume $d^4x$) is an
integral over the three-momenta of the incoming particles (in the classical limit):
\bea q_0 \frac{dN}{dx^4 d^3q}  = g \int \! ds \int \! \frac{d^3
p_a}{(2\pi)^3} \int \! \frac{d^3 p_b}{(2\pi)^3} \, e^{-(E_a+E_b)/T}
\  v_{rel} \, \, q_0 \, \frac{d\sigma^\gamma}{d^3q}
\delta(s-(p_a+p_b)^2), \ \ \label{kinetic} \eea
where $T$ is the temperature, $v_{rel}$ is the relative velocity
given by
\be v_{rel}=\frac{\sqrt{(p_a\cdot p_b)^2-m_a^2 m_b^2}}{E_a E_b} ,\ee
and $g=(2s_a+1)(2s_b+1)$ is the spin degeneracy factor. Integrating
the expression (\ref{kinetic}) over the particle momenta one
obtains~\cite{Haglin:1992fy}:
\be q_0  \frac{dN}{d^4xd^3q}= \frac{T^6 g}{16 \pi^4} \int
\limits_{z_{min}}^\infty \!\! dz
\frac{\lambda(z^2T^2,m_a^2,m_b^2)}{T^4} K_1(z)
q_0\frac{d\sigma^\gamma}{d^3q} , \ee
where $ z_{min} = (m_a+m_b)/T $, $z=\sqrt{s}/T$, and $K_1(z)$ is the
modified Bessel function.

The expression (\ref{kinetic}) can be generalized to account for
quantum effects such as Bose enhancement or Pauli blocking
(depending on the particle type) by integrating additionally over
the momenta of the final particles and changing the Boltzmann
distributions to  Fermi or Bose distribution functions $f_i(T)$:
\bea q_0 \frac{dN}{dx^4 d^3q}  = g \! \int \! ds \! \int \!
\frac{d^3 p_a}{(2\pi)^3} \! \int \! \frac{d^3 p_b}{(2\pi)^3} \! \int
\! \frac{d^3 p_1}{(2\pi)^3} \! \int \! \frac{d^3 p_1}{(2\pi)^3}
f_a(T) f_b(T) (1 \pm f_1(T)) (1 \pm f_2(T))
\nnl
  \times v_{rel} \, \, q_0 \, \frac{d\sigma^\gamma}{d^3q} \,
\delta(s-\!(p_a+p_b)^2) , \ \  \label{kinetic2} \eea
where the (-) sign has to be used in case of fermions.
In the current Subsection we calculate the thermal rates according to
formula (\ref{kinetic}). However, within the PHSD transport approach
for the heavy-ion collisions in Section~\ref{sect:results}  the
effects of the quantum statistics will be taken into account (although of subleading importance).

In Fig.~\ref{rates} (l.h.s.) the rates are presented for a temperature
$T=200$~MeV and pion chemical potential $\mu_\pi=0$ for the  OBE
model with three resonance exchanges adopting the soft-photon
approximation (red dashed line). We confirm the results from
Haglin \cite{Haglin:2003sh} (black solid line) calculated within the
same assumptions (SPA, three resonances) but with a slightly
different  parameter set of the Lagrangian. It is, however,
questionable that the SPA is applicable at high photon energies.

%\begin{figure} \centering
%\includegraphics[width=0.5\textwidth]{XsectE05.eps}
%\caption{(Color on-line) Cross section for the production of photons
%with energy $q_0=0.5$~GeV in the process $\pi+\pi->\pi+\pi+\gamma$
%within the following models: the ``exact´´ OBE cross section within
%the effective model taking into account the scalar, vector and
%tensor interactions via the exchange of $\sigma$, $\rho$ and
%$f_2(1270)$-mesons gives the black line with star symbols, the soft
%photon approximation to this model is shown by the red solid line.
%\label{XsectE05}}
%\end{figure}

We note that the accuracy of the SPA approximation can be
significantly improved and the region of its applicability can be
extended by slightly modifying the formula (\ref{int_spa_formula})
-- i.e. by evaluating the on-shell elastic cross section at the
invariant energy $\sqrt{s_2}$ of the sub-process. The latter is
kinematically fixed to \be s_2=s-q_0 \sqrt{s} \ne s .\ee
Thus the modified SPA formula is
\be q_0 \frac{d\sigma^{\gamma}(s)}{d^3q} = \frac{\alpha_{_{EM}}}{4
\pi} \!\!\!\!\! \int\limits_{-\lambda(s_2,m_a^2,m_b^2)/s_2}^0 \!\!
\!\!\!\!\!\!\!\!\!\! |\epsilon \cdot J(q,t)|^2 \frac{d\sigma_{el}
(s_2)}{dt} dt. \label{int_spa_formula2} \ee
In the following, we will denote the approximation (\ref{int_spa_formula2})
as ``improved SPA" and will show below that it provides a fairly good
description of the exact photon production rates.

It is instructive to compare the photon production rates beyond
the soft photon approximation for the
$\pi+\pi\to\pi+\pi+\gamma$ reaction  to the rates from the
exact OBE expression (\ref{r3}).
We present the calculated invariant rate $q_0 d R/dq^3$ of
bremsstrahlung photons produced from an equilibrated pion gas at
$T=150$~MeV and $\mu_\pi=40$~MeV in Fig.~\ref{rates} (r.h.s.). The results of
the following models are compared:
\begin{itemize}
\item model 1 (red solid line): exact rates within the one-boson exchange model (OBE) {\em
beyond} the soft-photon approximation  -- i.e. using the formula
(\ref{r3}) for the photon production cross section $q_0\
d\sigma^\gamma/d^3q$;
\item
model 2 (blue dotted line): result within the soft photon approximation
-- i.e. using the formula (\ref{int_spa_formula}) -- while using the
elastic $\pi+\pi$ cross section calculated within the OBE model as
given by  equations (\ref{el1})-(\ref{elasticformulas});
\item
model 3 (black short-dashed line): results of the {\em improved} soft
photon approximation
 -- i.e. using the formula (\ref{int_spa_formula2})
in stead of (\ref{int_spa_formula}) -- and the same pion elastic
scattering cross section as in the model 2;
\item
model 4 (upper green dashed line): soft photon approximation using a
constant isotropic elastic cross section of $\sigma_{el}=10$~mb and
assuming for the pion charges $Q_a=Q_1=1$, $Q_b=Q_2=0$. In this case the elastic cross section
does not depend on $\sqrt{s}$ and thus there is no difference
between the SPA and improved SPA.
\end{itemize}
The rate of bremsstrahlung photons at low transverse momenta
$p_T<0.4$~GeV has been calculated before by Liu and Rapp in
Ref.~\protect{\cite{Liu:2007zzw}} within the one-boson exchange
model with the exchange of two resonances for the same system. This
previous result is shown for comparison by the cyan dashed line and
is confirmed by our present calculations. The agreement is expected,
since our calculations differ only in the inclusion of the
$f_2$-meson exchange, which is important for larger $\sqrt{s}$ and
does not play an important role for the production of low transverse
momentum photons, which is dominated by low $\sqrt{s}$ of the
$\pi+\pi$ collisions.

On the other hand, the SPA (model 2) deviates from the exact OBE
result (model 1) even at low $q_0$  because the former directly
follows the $\sqrt{s}$ dependence of the elastic $\pi-\pi$ cross
section. Since the formula (\ref{int_spa_formula}) does not account
for the off-shellness of the emitting pion, it overestimates the
high-$\sqrt{s}$ regime of the elastic cross section in line with the
findings of Refs.~\cite{Haglin:1992fy,Eggers:1995jq}. We note that
the  OBE model presented here is constrained by the pion scattering data
only up to  $\sqrt{s_{\pi\pi}}=1.4$~GeV and generally cannot be
extended to larger $\sqrt{s}$. Thus the SPA scenario "model 2" is not
reliable for large $q_0$ (approximately for $q_0>0.8$~GeV). This is
not the case for the improved SPA (model 3).

%\begin{figure}
%\begin{center}
%\includegraphics[width=\textwidth]{dNdS.eps}
%\caption{  The distribution in the invariant collision energy
%$\sqrt{s}$ for the elastic scattering of mesons on mesons (left
%panel), mesons on baryons (middle panel) and for the elastic
%scattering of partons (right panel) in the course of a $Au+Au$
%collision at $\sqrt{s_{NN}}=200$~GeV for $b=7$~fm as calculated
%within the PHSD approach. Here $V\equiv (\rho, \omega, \phi)$; $K$
%denotes all the strange mesons $K\equiv (K, \bar K, K^*, \bar K^*)$
%and $B$ stands for the baryons $B=(p, n, \Delta, \dots)$.
%}\label{dNdS}
%\end{center}
%\end{figure}

One can see in Fig.~\ref{rates} (r.h.s.) that the {\em improved} SPA
(\ref{int_spa_formula2}) gives a very good approximation to the
exact result at higher photon energies of up to  $q_0 \approx
2$~GeV. This is because the $\sqrt{s_2}$ of the subprocess is always
below $\sqrt{s}$, and the OBE model for the elastic cross section is
sufficiently realistic in this region of $\sqrt{s_2}$. In
comparison, the constant cross-section approximation overestimates
the exact rates for $q_0>1$~GeV and underestimates for
$q_0<0.4$~GeV. This model corresponds to the approximation used
previously in the transport calculations in
Refs.~\cite{Linnyk:2013hta,Bratkovskaya:2008iq,Linnyk:2013wma} for
an estimate of the photon bremsstrahlung in meson+meson collisions.
In the following we will report on results based on the exact OBE
cross section $d \sigma^\gamma/d^3q$ for $\pi+\pi$ bremsstrahlung.
The bremsstrahlung photon production in collisions of other meson
types is treated in analogy to the $\pi+\pi$ collisions by means of
mass-scaled cross sections.

We note that another important source of  photons is the
bremsstrahlung in {\it meson+  baryon} collisions. As we have shown
above, the SPA gives a good approximation to the exact rates, if we
use the correct invariant energy in the hadronic subprocess
$s_2=s-q_0 \sqrt{s}$ and a realistic model for the differential
cross section of the subprocess, i.e. for the elastic scattering of
mesons on baryons. The cross sections for the meson+baryon elastic
scatterings (implemented within the PHSD transport approach) have
been previously adjusted to the data differentially in energy and
angular distribution.  Thus we evaluate the photon production in the
processes $m+B\to m+B+\gamma$ in the PHSD by using realistic elastic
scattering cross sections taken at the correct invariant energy
$\sqrt{s_2}$ in the scope of the improved SPA.

\subsection{Binary meson+meson and meson+nucleon reactions}
\label{sect:22}

We calculate the cross sections for the processes $\pi \pi
\rightarrow \rho \gamma , \pi \rho \rightarrow \pi \gamma $ as in
Ref.~\cite{Bratkovskaya:2008iq}, i.e. the total cross section
$\sigma_{\pi\pi \rightarrow \rho\gamma }(s,\rho_N)$ is obtained by
folding the vacuum cross section $\sigma_{\pi\pi \rightarrow
\rho\gamma }^{0}(s,M)$ with the (in-medium) spectral function of the
$\rho$ meson:
\begin{eqnarray}
\sigma_{\pi\pi \rightarrow \rho\gamma }(s,\rho_N) =
\label{xs_pipiVVtot} \int\limits_{M_{min}}^{M_{max}} dM \
\sigma_{\pi\pi \rightarrow \rho\gamma }^{0}(s,M) \ A(M,\rho_N)\
P(s).
\end{eqnarray}
Here $A(M,\rho_N)$ denotes the meson spectral function for given
total width $\Gamma_V^*$:
\begin{eqnarray}  \label{spfunV}
A_V(M,\rho_N) =  C_1 \ \frac{2}{\pi} \ \frac{M^2 \
\Gamma_V^*(M,\rho_N)} {(M^2-M_{0}^{*^2}(\rho_N))^2 + (M
{\Gamma_V^*(M,\rho_N)})^2},
\end{eqnarray}
with the normalization condition for any $\rho_N$, $
\int_{M_{min}}^{M_{lim}} A_V(M,\rho_N) \ dM =1$, where
$M_{lim}=2$~GeV is chosen as an upper limit for the numerical
integration while the lower limit of the vacuum $\rho$ spectral
function corresponds to the $2\pi$ decay threshold $M_{min}=2 m_\pi$ in vacuum
and $2 m_e$ in medium. $M_0^*$ is the pole mass of the vector meson
spectral function which is $M_0^*(\rho_N=0)=M_0$ in vacuum, however,
might be shifted in the medium (e.g. for the dropping mass
scenario). Furthermore, the vector meson width is the sum of the
vacuum total decay width and collisional width:
\begin{eqnarray}
\Gamma^*_V(M,\rho_N)=\Gamma_V(M) + \Gamma_{coll}(M,\rho_N) .
\label{gammas}
\end{eqnarray}
In Eq. (\ref{xs_pipiVVtot}) the function $P(S)$ accounts for the
fraction of the available part of the full spectral function
$A(M,\rho_N)$ at given energy $\sqrt{s}$, integrated over the mass
$M$ up to $M_{max}=\sqrt{s}$, with respect to the total phase space.

The cross section $\sigma_{\pi\pi \rightarrow \rho\gamma }^{0}(s,M)$
is taken in line with the model by Kapusta et al. \cite{Kapusta:1991qp} with
the $\rho$-meson mass considered as a dynamical variable, i.e
$m_\rho\to M$:
\begin{eqnarray}
%\hspace{3mm}
 \frac{d \sigma}{dt}\left( \pi^\pm \pi^0 \to \rho ^\pm
\gamma \right) & = & -\frac{\alpha g_\rho^2}{16sp_{CM}^2} \left[
\frac{(s-2M^2)(t-m_\pi^2)^2}{M^2(s-M^2)^2}
+\frac{m_\pi^2}{M^2}-\frac{9}{2}
+\frac{(s-6M^2)(t-m_\pi^2)}{M^2(s-M^2)}  \right. \nnl &&
\hspace{2cm}
 +\frac{4 (M^2-4m_\pi^2)s}{(s-M^2)^2}
\left. +\frac{4(M^2-4m_\pi^2)}{t-m_\pi^2}\left(
\frac{s}{s-M^2}+\frac{m_\pi^2}{t-m_\pi^2} \right) \right].
\end{eqnarray}
The photon production in the $\pi+\rho$ interaction is calculated
analogously (cf. Ref.~\cite{Bratkovskaya:2008iq} for details).
%\end{itemize}

We recall that the PHSD and HSD are off-shell transport approaches and thus allow to
study the effect of the modification of the vector-meson spectral functions in the medium.
In particular the photon production in secondary meson interactions
is sensitive to the properties of the vector mesons at finite density and temperature~\cite{Bratkovskaya:2008iq,Song:1994zs,Song:1993ae}. In this
respect, we stress here that the yields and the in-medium spectral
functions of vector mesons in PHSD have been independently
constrained by the comparison to the data on dilepton mass-spectra
(see Refs.~\cite{Linnyk:2011hz,Linnyk:2011vx,Bratkovskaya:2008bf}
and Section 7, respectively).

We have incorporated into the PHSD approach additionally the $2\to2$
processes  $V + N \rightarrow N + \gamma$, where $V$ stands for a
vector meson while $N$ denotes a proton or
neutron~\cite{Linnyk:2015tha}. These processes are the baryonic
counterparts to the mesonic $2\to2$ reactions
$\pi+\rho/\pi\to\gamma+\pi/\rho$. We consider the interaction
of nucleons with the mesons $V=\rho$, $\phi$, $\omega$, taking into
account the various possible charge combinations, e.g. $\rho^0+p\to
\gamma+p$, $\rho^-+p\to\gamma+n$, $\rho^+ + n\to \gamma +p$, etc.
In order to evaluate the probabilities for photon production in the
collisions of vector mesons with  nucleons, we use the inverse photoproduction
processes $\gamma+N\to\rho+N$, $\gamma+N\to\phi+N$,
$\gamma+N\to\omega+N$ (controlled by data) and employ detailed
balance to obtain the differential cross sections for the processes
$\rho+N\to\gamma+N$, $\phi+N\to\gamma+N$, $\omega+N\to\gamma+N$, i.e.
\bea \label{O5} \sigma(NV\to \gamma N) & = & \frac{g_\gamma}{g_V} \frac{
p^{*2}_{N\gamma}}{p^{*2}_{NV}} \  \sigma(\gamma N \to NV),
 \eea
where $g_\gamma=2$ and $g_V=3$ are the spin degeneracy factors of
the photon and the vector meson $V$. In Eq. (\ref{O5})
$p^*_{N\gamma}$ is the center-of-mass momentum in the $N+\gamma$
system and $p^*_{NV}$ is the center-of-mass momentum in the $N+V$
system.

The cross sections for the exclusive photo-production of $\rho$,
$\phi$ and $\omega$ vector mesons on the nucleon have been measured
by the Aachen-Berlin-Bonn-Hamburg-Heidelberg-Munich (ABBHHM)
Collaboration and published in Ref.~\cite{ABBHHM:1968aa}. In the
same work also parametrizations for these cross section have been
given that are based on the vector-meson-dominance model with a
non-relativistic Breit-Wigner (BW) spectral function for the
$\rho$-meson. Later, the fits have been updated in Ref.
\cite{Effenberger:1999ay} using  relativistic BW spectral functions
for  $\rho$, $\omega$ and $\phi$ mesons.
The total cross sections -- fitted in Ref.~\cite{Effenberger:1999ay}
to the data from Ref.~\cite{ABBHHM:1968aa} -- are given by
\be \sigma(\gamma N\to VN) =\frac{1}{p^*_{N\gamma} s} \int d\mu
|M_V|^2 p^*_{NV} A_V(\mu), \label{cros} \ee
where the mass $\mu$ of the vector meson is distributed according to the spectral
function $A_V(\mu)$:
\be A_V (\mu) = \frac{2}{\pi} \frac{\mu ^2 \Gamma (\mu)}{
(\mu^2-M_i^2)^2 + \mu^2 \Gamma^2(\mu) }, \ee
with $M_i$ denoting the pole mass of the meson. The matrix elements
for the reactions $\gamma + N \to V + N$  are parametrized as
 \bea
|M_\rho|^2&\!=\!&0.16 \mbox{ mb GeV}^2, \nnl
|M_\omega|^2&\!=\!&\frac{0.08 \ p_{VN}^{*2}}{2(\sqrt{s}-1.73
\mbox{\, GeV})^2+p_{VN}^{*2}} \mbox{ mb GeV}^2, \nnl
 |M_\phi|^2&\!=\!&0.004 \mbox{ mb
GeV}^2.
 \label{fits} \eea
The cross sections (\ref{cros}) with the parameters (\ref{fits}) are
consistent with the dynamics of vector mesons in the PHSD/HSD, where
also relativistic BW spectral functions for vector mesons are used
and propagated off-shell.

For the angular distribution of the $\rho$-meson production in the
process $\gamma+N\to N+\rho$, we follow  the suggestion of
Ref.~\cite{Effenberger:1999ay},
\be \frac{d \sigma}{dt} \sim \exp (B t), \label{param} \ee
with the photon-energy dependent parameter $B$ (fitted to the data):
$B=5.7$ for $q_0 \le 1.8$~GeV,
$B=5.43$ for $1.8 < q_0 \le 2.5$~GeV,
$B=6.92$ for $2.5 < q_0 \le 3.5$~GeV,
$B=8.1$ for $3.5 < q_0 \le 4.5$~GeV,
$B=7.9$ for $q_0 > 4.5$~GeV.
The data in Ref.~\cite{Effenberger:1999ay} have shown that the cross
section is dominated by the $t\approx 0$ region in line with the vector dominance model (VDM) where the
process $\gamma + N \to V+N$ is described by the incident photon
coupling to the vector meson of helicity $\pm1$, which consequently
is scattered elastically by the nucleon.
% (cf. Refs.~\cite{Klopot:2013laa,Sakurai:1960ju,Chew:1957tf}).

%------------------------------------------------------------------------

\subsection{Dilepton sources}

Dileptons ($e^+e^-$, $\mu^+\mu^-$ pairs or virtual photons) can be
emitted from all stages of the heavy-ion reactions as well as real
photons. One of the advantages of dileptons -- compared to photons
-- is an additional 'degree-of-freedom': the invariant mass $M$
which allows to disentangle various sources. There are the following
production sources of dileptons
in $p+p, p+A$ and $A+A$ collisions:\\
1) Hadronic sources:\\
(i) at low invariant masses ($M < 1$ GeV$c$) -- the  Dalitz decays
of mesons and  baryons $(\pi^0,\eta,\Delta, ...)$ and the direct decay of
vector mesons  $(\rho, \omega, \phi)$ as well as hadronic bremsstrahlung~\cite{Bratkovskaya:2008iq}; \\
(ii) at intermediate masses (1 GeV$ < M < 3$ GeV) -- leptons from
correlated $D+\bar D$ pairs~\cite{Linnyk:2011hz}, radiation from
multi-meson reactions ($\pi+\pi, \ \pi+\rho, \ \pi+\omega, \
\rho+\rho, \ \pi+a_1, ... $)
denoted by ``$4\pi$" contributions~\cite{Song:1994zs,Li:1998ma,vanHees:2006ng,vanHees:2007th}; \\
(iii) at high invariant masses ($M > 3$ GeV) -- the direct decay of
vector mesons  $(J/\Psi, \Psi^\prime)$~\cite{Linnyk:2009nx} and
initial 'hard'
Drell-Yan annihilation to dileptons ($q+\bar q \to l^+ +l^-$, where $l=e,\mu$)~\cite{Linnyk:2006mv}.\\
2) 'thermal' QGP dileptons radiated from the partonic interactions
in heavy-ion collisions that contribute dominantly to the
intermediate masses. The leading processes are the 'thermal' $q\bar
q$ annihilation ($q+\bar q \to l^+ +l^-$, \ \ $q+\bar q \to g+ l^+
+l^-$) and Compton scattering ($q(\bar q) + g \to q(\bar q) + l^+
+l^-$) in the QGP~\cite{Kapusta:1991qp}.

The dilepton production by a (baryonic or mesonic) resonance $R$
decay can be schematically presented in the following way:
\begin{eqnarray}
 BB &\to&R X   \label{chBBR} \\
 mB &\to&R X \label{chmBR} \\
       R & \to & e^+e^- X, \label{chRd} \\
       R & \to & m X, \ m\to e^+e^- X, \label{chRMd} \\
       R & \to & R^\prime X, \ R^\prime \to e^+e^- X, \label{chRprd}
\end{eqnarray}
i.e. in a first step a resonance $R$ might be produced in
baryon-baryon ($BB$) or meson-baryon ($mB$) collisions
(\ref{chBBR}), (\ref{chmBR})  or be formed in the hadronization
process. Then this resonance can couple to dileptons directly
(\ref{chRd}) (e.g., Dalitz decay of the $\Delta$ resonance: $\Delta
\to e^+e^-N$) or decays to a meson $m$ (+ baryon) or in
(\ref{chRMd})  produce dileptons via direct decays ($\rho, \omega$)
or Dalitz decays ($\pi^0, \eta, \omega$). The resonance $R$ might
also decay into another resonance $R^\prime$ (\ref{chRprd}) which
later produces dileptons via Dalitz decay.

The electromagnetic part of all conventional dilepton sources  --
$\pi^0, \eta, \omega$  Dalitz decays, direct decay of vector mesons
$\rho, \omega$ and $\phi$ -- are  described in detail in
Ref.~\cite{BCM00SIS} -- where dilepton production in $pp$ and $pd$
reactions has been studied. Actual modifications -- relative to
Ref.~\cite{BCM00SIS} -- are related to the Dalitz decay of baryonic
resonances and especially the strength of the $pp$ and $pn$
bremsstrahlung since calculations by Kaptari and K\"ampfer in 2006
\cite{Kaptari:2005qz} indicated that the latter channels might have
been severely underestimated in  previous studies on dilepton
production at SIS energies. For the results reported here we adopt
the parametrizations from Ernst et al. \cite{Ernst:1997yy} (Eqs. (9)
to (13)) for the Dalitz decays of the baryonic resonances which are
also incorporated in the PLUTO simulation program of the HADES
Collaboration.
%\cite{PLUTO}.
For the bremsstrahlung channels in $pp$ and $pn$
reactions we adopt the results from the OBE model calculations by Kaptari and K\"ampfer in Ref.
\cite{Kaptari:2005qz}.

%------------------------------------------------------------------------
\subsection{Vector-meson spectral functions}

In order to explore the influence of in-medium effects on the
vector-meson spectral functions we incorporate the effect of
collisional broadening (as in Refs. \cite{BratKo99,GKC97a,GKC97b}),
i.e. the vector meson width has been implemented as:
\begin{eqnarray}
\Gamma^*_V(M,|\vec p|,\rho_N)=\Gamma_V(M) + \Gamma_{coll}(M,|\vec
p|,\rho_N) . \label{gammas2}
\end{eqnarray}
Here $\Gamma_V(M)$ is the total width of the vector mesons
($V=\rho,\omega$) in the vacuum. For the $\rho$ meson we use
\begin{eqnarray}
\Gamma_\rho(M) &\simeq& \Gamma_{\rho\to\pi\pi}(M) =  \Gamma_0
\left(M_0\over M\right)^2 \left(q\over q_0\right)^3 \ F(M) ,
\label{Widthrho} \\
&& q = {(M^2-4m_\pi^2)^{1/2}\over 2}, \
  \ q_0 = {(M_0^2-4m_\pi^2)^{1/2}\over 2}. \nonumber
\end{eqnarray}
In Eqs. (\ref{Widthrho}) $M_0$ is the vacuum pole mass of the vector
meson spectral function, $F(M)$ is a formfactor taken from Ref.
\cite{Rapp} as
\begin{eqnarray}
F(M)={\left(2\Lambda^2 +M_0^2 \over  2\Lambda^2 + M^2 \right)^2}
\label{Frapp}\end{eqnarray} with a cut-off parameter
$\Lambda=3.1$~GeV. This formfactor was introduced in Ref.
\cite{Rapp} in order to describe the $e^+e^-$ experimental data with
better accuracy.
 For the $\omega$ meson a constant total vacuum width is
used: $\Gamma_\omega\equiv \Gamma_\omega(M_0)$, since the $\omega$
is a narrow resonance in vacuum.

The collisional width in Eq. (\ref{gammas2}) is approximated as
\begin{eqnarray}
\Gamma_{coll}(M,|\vec p|,\rho_N) = \gamma \ \rho_N < v \
\sigma_{VN}^{tot} > \approx  \ \alpha_{coll} \ \frac{\rho_N}{\rho_0}
.... \label{dgamma}
\end{eqnarray}
Here $v=|{\vec p}|/E; \ {\vec p}, \ E$ are the velocity, 3-momentum
and energy of the vector meson in the rest frame of the nucleon
current and $\gamma^2=1/(1-v^2)$. Furthermore, $\rho_N$ is the
nuclear density and $\sigma_{VN}^{tot}$ the meson-nucleon total
cross section. The parameter $\alpha_{coll}$ is determined
dynamically within the transport calculation by recording
the $\rho$ collision rate as a function of the baryon density $\rho_N$.

In order to explore the observable consequences of vector-meson mass
shifts at finite nuclear density -- as suggested by the CBELSA-TAPS
data \cite{tapselsa} for the $\omega$ meson -- the in-medium vector
meson pole masses are modeled (optionally) according to the Hatsuda
and Lee \cite{H&L92} or Brown/Rho scaling \cite{BrownRho,BrownRho2}
as
\begin{eqnarray}
\label{Brown} M_0^*(\rho_N)= \frac{M_0} {\left(1 + \alpha {\rho_N /
\rho_0}\right)},
\end{eqnarray}
where $\rho_N$ is the nuclear density at the resonance decay
position $\vec r$; $\rho_0 = 0.16 \ {\rm fm}^{-3}$ is the normal
nuclear density and $\alpha \simeq 0.16$ for the $\rho$ and $\alpha
\simeq 0.12$ for the $\omega$ meson \cite{Metag07}. The
parametrization (\ref{Brown}) may be employed also at much higher
collision energies (e.g. FAIR and SPS) and one does not have to
introduce a cut-off density in order to avoid negative pole masses.
Note that the effective mass (\ref{Brown}) is uniquely fixed by the
'customary' expression $M_0^*(\rho_N) \approx M_0 (1 - \alpha
\rho_N/\rho_0)$ in the low density regime.

\begin{figure}
%{{\psfig{figure=f_pi.eps,width=0.45\textwidth}}}
\hspace{1.3cm} {\psfig{figure=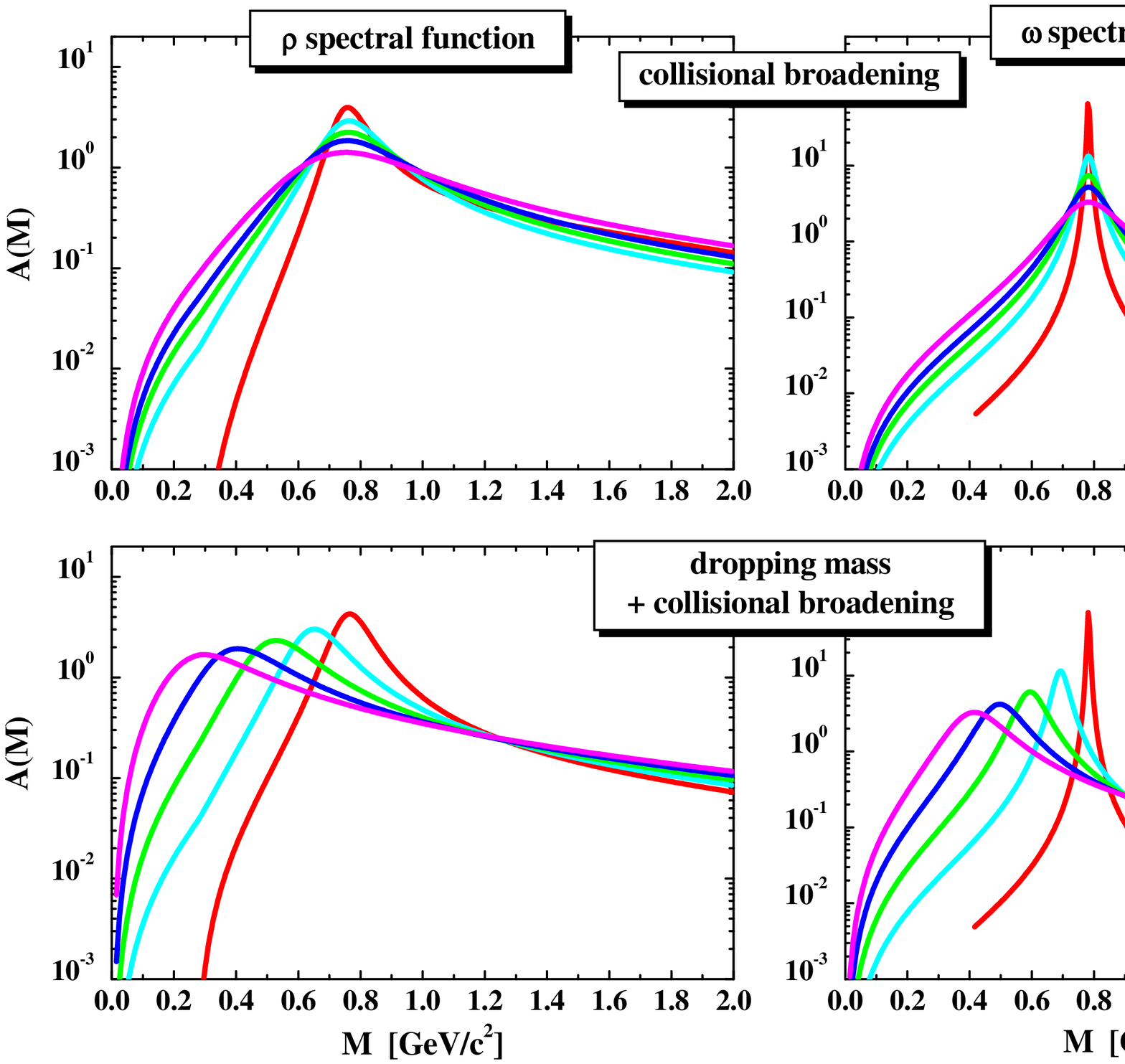,width=0.6\textwidth}} \caption{
%{\it (l.h.s.)} The ratio $R(M)=\sigma(e^+e^-\to \pi^+\pi^-)/
%\sigma(e^+e^-\to \mu^+\mu^-)$. The solid (red) line corresponds to
%the result with the formfactor $F(M)$ (\ref{Frapp}) while the dashed
%(blue) line is without $F(M)$. The compilation of the experimental
%data is taken from Ref. \protect\cite{expFPI}. {\it (r.h.s.)}
The spectral functions for the $\rho$ and $\omega$ meson
 in the case of the 'collisional broadening' scenario (upper part)
and the 'dropping mass + collisional broadening' scenario (lower
part) for nuclear densities of 0,1,2,3,5$\times\rho_0$ as employed
in the transport calculations (see text for details). The figures
are taken from Ref. \protect\cite{Bratkovskaya:2007jk}.} \label{Fpi}
\label{Fig00}
\end{figure}

The resulting spectral functions for the $\rho$ and $\omega$ meson
are displayed in Fig. \ref{Fig00} for the case of 'collisional
broadening' (upper part) as well as for the 'dropping mass +
collisional broadening' scenario (lower part) for densities of
0,1,2,3,5 $\times \rho_0$.  Note that in vacuum the hadronic widths
vanish for the $\rho$ below the two-pion mass and for the $\omega$
below the three-pion mass. With increasing nuclear density $\rho_N$
elastic and inleastic interactions of the vector mesons shift
strength to low invariant masses. In the 'collisional broadening'
scenario we find a dominant enhancement of strength below the pole
mass for the $\rho$ meson while the $\omega$ meson spectral function
is drastically enhanced in the low- and high-mass region with
density (on expense of the pole-mass regime). In the 'dropping mass
+ collisional broadening' scenario both vector mesons dominantly
show a shift of strength to low invariant masses with increasing
$\rho_N$.
Qualitatively similar pictures are obtained for the $\phi$
meson but quantitatively smaller effects are seen due to the lower
effect of mass shifts and a substantially reduced $\phi N$ cross
section which is a consequence of the $s\bar{s}$ substructure of the
$\phi$ meson. Since the $\phi$ dynamics turn out to be of minor
importance for the dilepton spectra to be discussed below we discard
an explicit representation.
The 'family' of spectral functions shown in Fig. \ref{Fig00} allows for a
sufficient flexibility with respect to the possible scenarios
outlined above. A comparison to dilepton data is expected to provide
further constraints on the possible realizations.

%------------------------------------------------------------------------
\subsection{Off-shell propagation and the time-integration method}

\begin{figure} \hspace{2.5cm}
{\psfig{figure=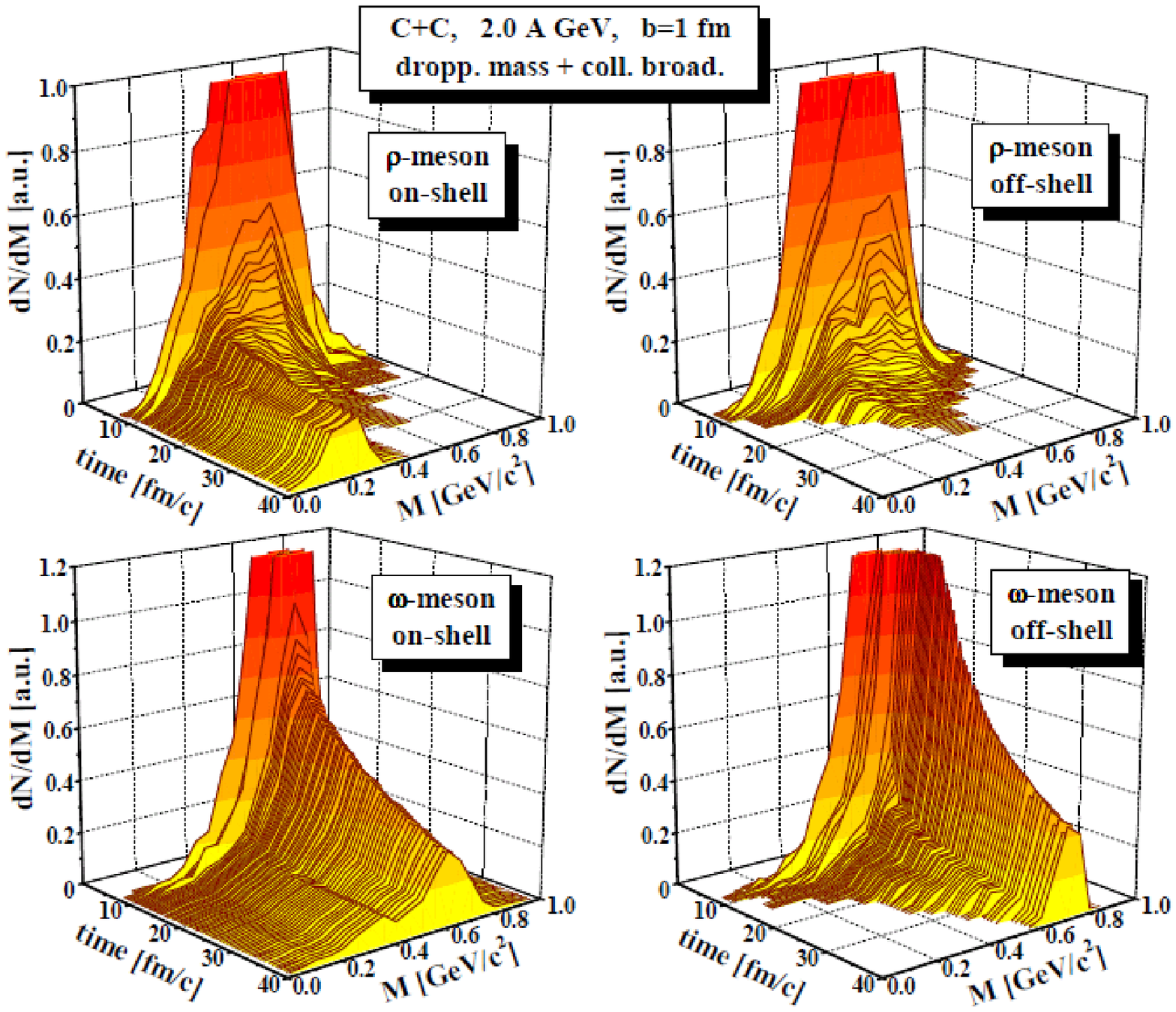,width=10cm}} \caption{Time evolution of
the mass distribution of $\rho$ (upper part) and $\omega$ (lower
part) mesons for  central $C+C$ collisions (b=1 fm) at 2 A GeV for
the dropping mass + collisional broadening scenario. The l.h.s. of
Fig. \protect\ref{Fig3Donoff} correspond to the calculations with
on-shell dynamics whereas the r.h.s. show the off-shell results. The
figures are taken from Ref. \protect\cite{Bratkovskaya:2007jk}.}
\label{Fig3Donoff} \vspace*{1mm}
\end{figure}

The propagation of  broad resonances in the off-shell transport
approach has been described in  Section 2.7 above. In order to
demonstrate the importance of off-shell transport dynamics we
present in Fig. \ref{Fig3Donoff} the time evolution of the mass
distribution of $\rho$ (upper part) and $\omega$ (lower part) mesons
for  central C+C collisions (b=1 fm) at 2 A GeV for the dropping
mass + collisional broadening scenario (as an example). The l.h.s.
of Fig. \ref{Fig3Donoff} corresponds to the calculations with
on-shell propagation whereas the r.h.s. show the results for the
off-shell dynamics. As seen from Fig. \ref{Fig3Donoff} the initial
$\rho$ and $\omega$ mass distributions are quite broad even for a
small system such as $C+C$ where, however, the baryon density at 2 A
GeV may reach (in some local cells) up to $2 \rho_0$. The number of
vector mesons decreases with time due to their decays and the
absorption by baryons ($\rho N \rightarrow \pi N$ or $\rho N
\rightarrow \pi\pi N$). Most of the $\rho$ mesons decay/disappear
already inside the ``fireball" for density  $\rho_N > 0$. Due to the
``fireball" expansion the baryon density drops quite fast, so some
amount of $\rho$ (and $\omega$) mesons reach the very low density
zone or even the 'vacuum'. Since for the off-shell case (r.h.s. of
Fig. \ref{Fig3Donoff}) the $\rho$ and $\omega$ spectral functions
change dynamically by propagation in the dense medium according to
Eqs. (\ref{eomr}) and (\ref{eomp}) they regain the vacuum shape for
$\rho_N\to 0$. This does not happen for the on-shell treatment
(l.h.s. of Fig. \ref{Fig3Donoff}); the $\rho$ spectral function does
not change its shape by propagation but only by explicit collisions
with other particles. Indeed, there is a number of $\rho$'s which
survive the decay or absorption and leave the ``fireball" with
masses below $2m_\pi$.

Accordingly, the approximate on-shell propagation leads to the
appearance of $\rho$ mesons in the vacuum with $M\le 2m_\pi$, which
can not decay to two pions; thus they live practically 'forever'
since the probability to decay to other channels is very small.
Indeed, such $\rho$'s will continuously shine  low mass dileptons
which leads to an apparent 'enhancement/divergence' of the dilepton
yield at low masses (note, that the dilepton yield is additionally
enhanced by a factor $\sim 1/M^3$). The same statements are valid
for the $\omega$ mesons (cf.  lower part of Fig.  \ref{Fig3Donoff}):
since the $\omega$ meson is a long living resonance, a larger amount of
$\omega$'s survives with an in-medium like spectral function  in the
vacuum (in case of on-shell dynamics). Such $\omega$'s with $M <
3m_\pi$ can decay only to $\pi \gamma$ or electromagnetically (if $M
< m_\pi $). Since such phenomena appearing in on-shell transport
descriptions (including an explicit vector-meson propagation)
contradict basic physical principles, an off-shell treatment is
mandatory.

%------------------------------------------------------------------------
%%%\subsection{Time integration method}

Since the dilepton production is a very rare process (e.g.  the
branching ratio for the vector meson decay is $\sim 10^{-5}$), a
perturbative method is used in the transport calculation in order to
increase statistics. In the PHSD approach (in this report as well as
in earlier investigations
\cite{Linnyk:2011hz,Bratkovskaya:2008bf,Bratkovskaya:2007jk,Brat97,CBRW97,Linnyk:2009nx})
we use the time integration  (or 'shining') method first introduced
by Li and Ko in Ref. \cite{LiKo95}. The main idea of this method is
that dileptons can be emitted during the full lifetime of the
resonance $R$ before its strong decay into hadrons or absorption by
the surrounding medium. For example, the $\rho^0$ decay (with
invariant mass $M$) to $e^+ e^-$ during the propagation through the
medium from the production time $t=0$ up to the final (``death")
time $t_F$ -- which might correspond to an absorption by baryons or
to reactions with other hadrons as well as the strong decay into two
pions -- is calculated as
\begin{equation}
{d N^{\rho\to e^+ e^-}\over d M} = \sum_{t=0}^{t_F} \Gamma^{\rho^0
\to e^+ e^-} (M) \cdot {\Delta t \over \gamma (\hbar c)} \cdot
{1\over \Delta M} \label{Nrho}
\end{equation}
in the mass bin $\Delta M$ and time step $\Delta t$ (in fm$/c$). In
(\ref{Nrho}) $\gamma$ is the Lorentz factor of the $\rho$-meson with
respect to the calculational frame. The electromagnetic decay width
is defined as
\begin{equation}
\Gamma^{\rho^0 \to e^+ e^-}(M) = C_\rho {{M_0^*}^4 \over M^3},
\label{gamrn}
\end{equation}
where $C_\rho= {\Gamma^{\rho\to e^+e^-}(M_0) / M_0}$. Here $M_0$ is
the vacuum pole mass, $M_0^*$ is the in-medium pole mass which is
equal to the vacuum pole mass for the collisional broadening
scenario, however, is shifted for the dropping mass scenario
according to Eq. (\ref{Brown}). The time integration method allows
to account for the full in-medium dynamics of vector mesons from
production (``birth") up to their ``death". In case of the $\rho$
propagation in the vacuum only the 2 pion-decay channel contributes
and the default results are regained after time integration.
%We note that by calculating the dilepton emission only at
%the strong decay vertex (e.g. as in Refs. \cite{bleicher2,Vogel07})
%the dilepton rate (as well as the density dependence of the dilepton
%emission) is underestimated since a sizeable part of the emission
%history (e.g. before the absorption point by baryons or before the
%decay to pions) is lost.

%------------------------------------------------------------------------
\subsection{$e^+e^-$ bremsstrahlung in $p+p$ and $p+n$ reactions}

The soft-photon approximation (SPA) \cite{gale87a,gale87b,gale87c}
has been discussed in detail in Section~\ref{sect:brems} in case of
meson-meson collisions. In spite of the general limitation of the
SPA it has been widely used for the calculation of the
bremsstrahlung dilepton spectra by different transport groups
\cite{Cass99,Wolf90,Ernst:1997yy,Xiong:1990bg}. Note, at those early
times the applicability of the SPA for an estimate of dilepton
radiation from $NN$ collisions at 1-2 GeV bombarding energies has
been supported by independent One-Boson-Exchange (OBE) model
calculations by Sch\"afer et al.  \cite{Schaefer89} and later on by
Shyam et al. \cite{Shyam03}. In these models the effective
parameters have been adjusted to describe elastic $NN$ scattering at
intermediate energies. The models have then been applied to
bremsstrahlung processes including the interference of different
diagrams for the dilepton emission from all charged hadrons.  As
shown in Ref. \cite{Schaefer89} the $pp$ bremsstrahlung is much
smaller than $pn$ bremsstrahlung due to a destructive interference
of amplitudes from the initial and final radiation. We note that
gauge invariant results have been obtained in Refs.
\cite{Schaefer89,Shyam03} by `gauging' the phenomenological form
factors at the meson-baryon vertices. However, there are different
schemes to introduce gauge invariance in OBE models - as stressed by
Kondratyuk and Scholten \cite{Scholten} - which lead to sizeably
different cross sections. One also has to point out that already in
1997 an independent study by de Jong et al. \cite{deJong97} - based
on a full $T$-matrix approach - has indicated that the validity of
the SPA for $e^+e^-$ bremsstrahlung at intermediate energies of 1-2
GeV may be very questionable. However, in  Ref. \cite{deJong97} only
the $pp$ reaction has been considered; indeed, the bremsstrahlung in
the full T-matrix approach is larger by a factor of about 3 than the
corresponding SPA calculations (and OBE results).

In 2005 new covariant OBE calculations for dilepton
bremsstrahlung have been performed by Kaptari and K\"ampfer
\cite{Kaptari:2005qz}. The effective parameters for $NN$ scattering
have been taken similar to Ref. \cite{Shyam03}, however, the restoration
of gauge invariance has been realized in a different way. As
mentioned above, there are several prescriptions for restoring gauge
invariance in effective theories including momentum-dependent form
factors in interactions with charged hadrons \cite{Scholten} and the
actual results depend on the prescription employed.
The scheme in Ref. \cite{Kaptari:2005qz} is to include explicitly
the vertex form factors into the Ward-Takahashi identity for the
full meson-exchange propagators, which is different from the method
used in Refs. \cite{Schaefer89,Shyam03}.
In this review we will report on results based on the bremsstrahlung calculations from
Kaptari and K\"ampfer in Ref. \cite{Kaptari:2005qz}.

%%%%%%%%%%%%%%%%%%%%%%%%%%%%%%%%%%%%%%%%%%%%%%%%%%%%%%%%%%%%%%%%%%%%%%%%%%

% part of the review in PPNP 2015
\section{Results on photon production in $p+A$ and $A+A$ collisions}
\label{sect:results}

\begin{figure} \hspace{0.02cm}
\includegraphics[width=0.56\textwidth]{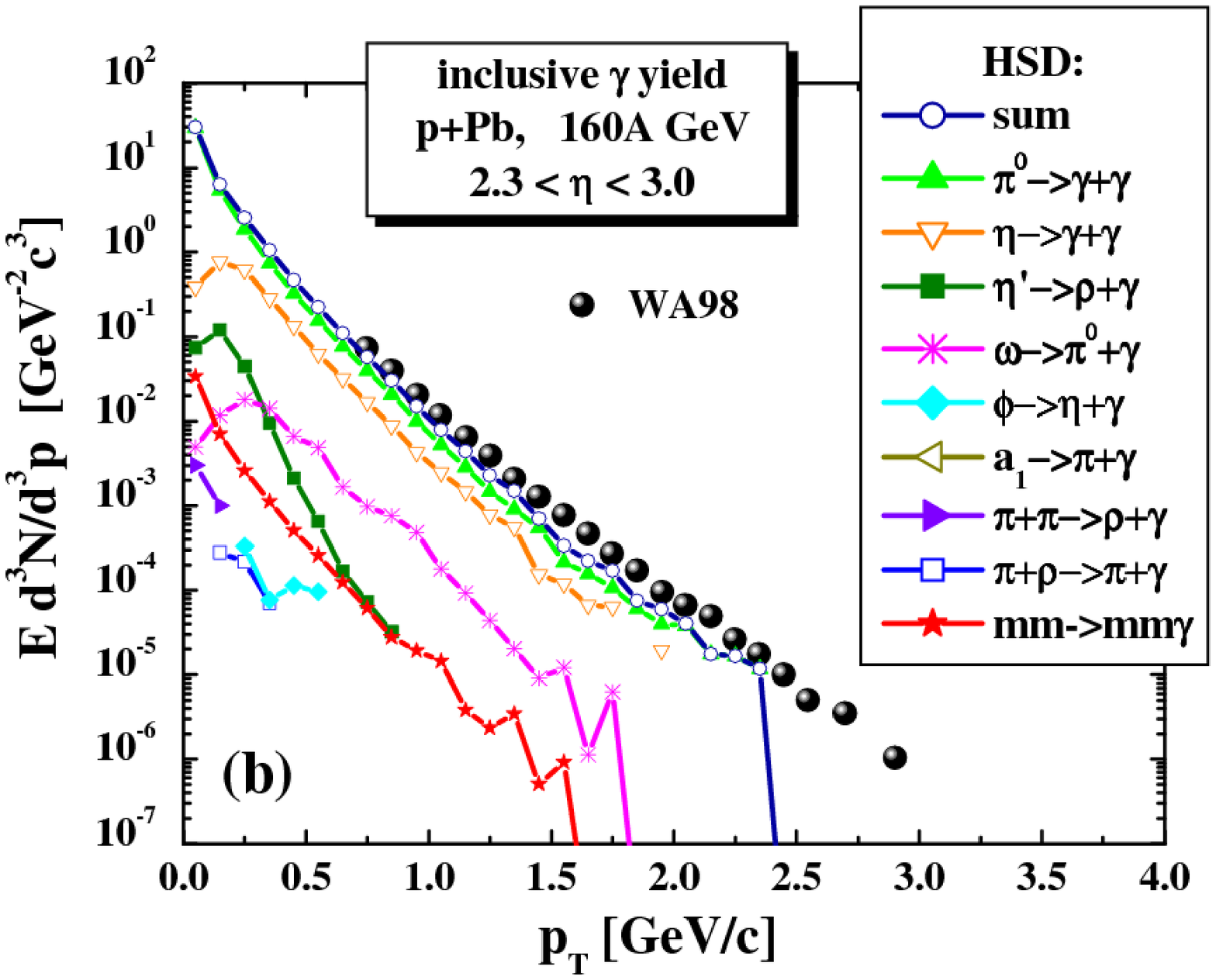}
\includegraphics[width=0.4\textwidth]{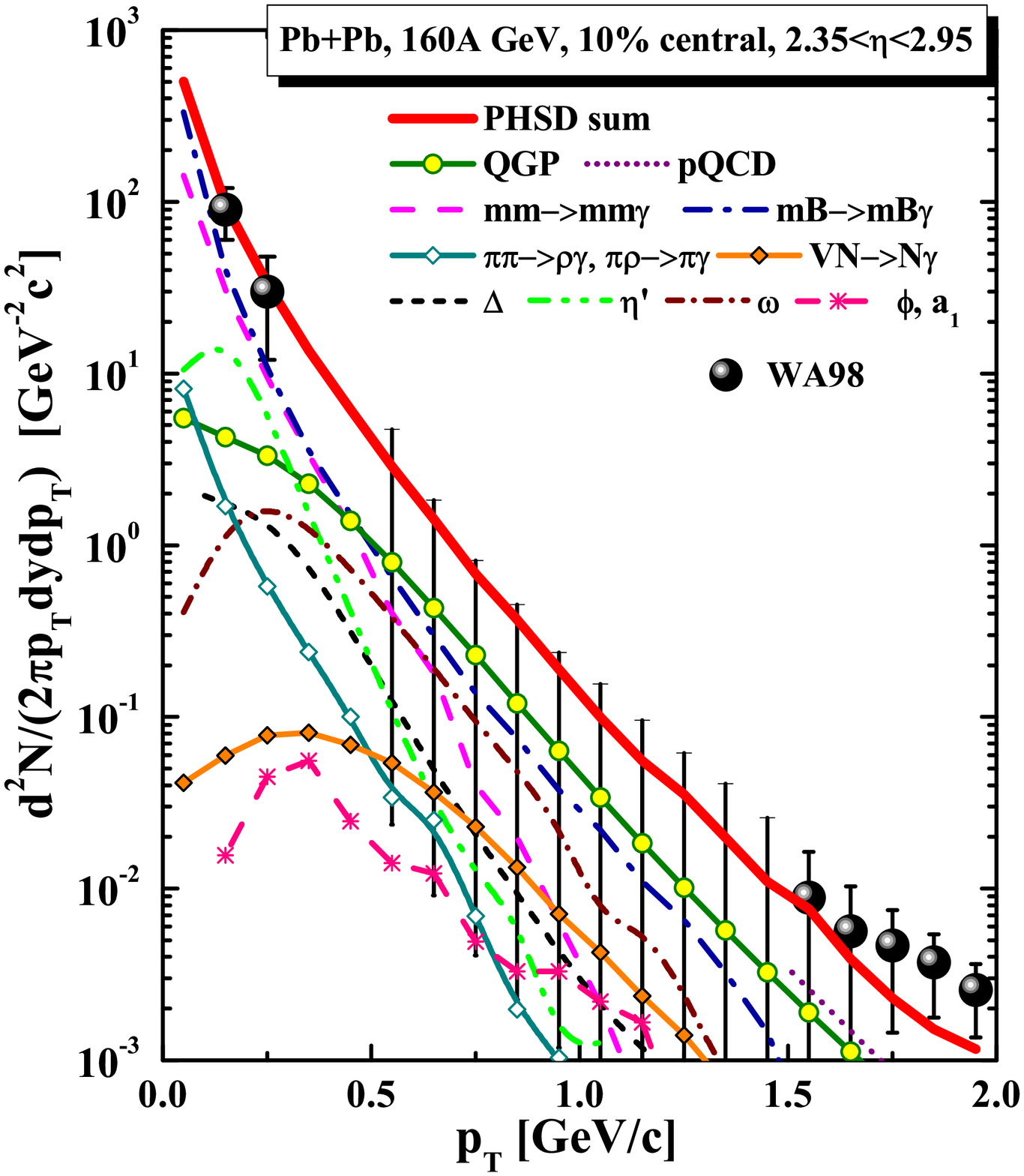}
\caption{(l.h.s.) Comparison of the PHSD/HSD calculations for {\em
inclusive} photons for p+Pb collisions at 160 GeV to the data of the
WA98 Collaboration from
Refs.~\protect\cite{Aggarwal:2000th,Aggarwal:2000ps}. See legend for
the contribution from the individual channels. (r.h.s.) Comparison
of the PHSD calculations for {\em direct} photons from Pb+Pb at 158
A GeV collisions to the data of the WA98 Collaboration from
Refs.~\protect\cite{Aggarwal:2000th,Aggarwal:2000ps}. In comparison
to the original HSD study~\protect\cite{Bratkovskaya:2008iq}: (i)
the meson+baryon bremsstrahlung (blue dash-dotted line), $\Delta$
decays (black short-dashed line) and the photons from QGP (green
line with round symbols) are added (ii) and the meson+meson
bremsstrahlung is now calculated beyond the SPA (magenta dashed
line). The black line with diamond symbols labeled as ``other"
includes: $\omega$, $\eta'$, $\phi$ an d $a_1$-meson decays, binary
channels $\pi+\rho/\pi\to\pi/\rho+\gamma$ and $N+V\to N+\gamma$. The
figures are taken from Refs.
\protect\cite{Bratkovskaya:2008iq,Linnyk:2015tha}. \label{WA98} }
\end{figure}

Direct photons are expected to provide a powerful probe of the
quark-gluon plasma (QGP) as created in ultra-relativistic nuclear
collisions. The photons interact only electromagnetically and thus
escape to the detector almost undistorted through the dense and
strongly interacting medium. Thus the photon transverse-momentum
spectra and their azimuthal asymmetry carry  information on the
properties of the matter under extreme conditions, existing in the
first few fm/c of the collisional evolution. On the other hand, the
measured photons provide a time-integrated picture of the heavy-ion
collision dynamics and are emitted from every moving electric charge
-- partons or hadrons. Therefore, a multitude of photon sources has
to be differentiated in order to access the signal of interest. The
dominant contributions to the {\em inclusive} photon production are
the decays of mesons, dominantly pions, $\eta$- and $\omega$-mesons.
Experimental collaborations subtract the ``{\em decay} photons" from
the inclusive photon spectrum using a cocktail
calculation~\cite{PHENIX1,Wilde:2012wc} and obtain the ``direct"
photons.

In particular the {\em direct} photons at transverse momenta $p_T<$
3 GeV/c are expected to be dominated by  "thermal" sources, i.e. the
radiation from the strongly interacting Quark-Gluon-Plasma
(sQGP)~\cite{Shuryak:1978ij,Shuryak:1978ij2} and the secondary
meson+meson and meson+baryon interactions in the hadronic
phase~\cite{Song:1994zs,Li:1998ma}. These partonic and hadronic
channels have been studied within PHSD in detail in
Refs.~\cite{Linnyk:2013hta,Linnyk:2013wma} at
% Linnyk:2013zfaCassing:2014vna
Relativistic-Heavy-Ion-Collider energies and it was found that the
partonic channels constitute up to  half of the observed {\em
direct} photon spectrum for central collisions. Other theoretical
calculations also find a significant or even dominant contribution
of the photons produced in the QGP to the {\em direct} photon
spectrum
\cite{Chatterjee:2005de,Liu:2009kq,Dion:2011vd,Dion:2011pp,Chatterjee:2013naa,Shen:2013vja}.

The low-$p_T$ {\em direct} photons probe not only the
temperature~\cite{PHENIX1,Wilde:2012wc,Shen:2013vja} of the produced
QCD-matter, but also its (transport) properties, for instance, the
shear viscosity $\eta$ (cf. Section 4.4). Using the {\em direct}
photon elliptic flow $v_2$ (a measure of the azimuthal asymmetry in
the photon distribution) as a viscosimeter was first suggested by
Dusling in Ref.~\cite{Dusling:2009bc}; this idea was later supported
by the calculations in Refs.~\cite{Dion:2011vd,Dion:2011pp,
Shen:2013vja,Shen:2013cca}. It was also suggested that the photon
spectra and $v_2$ are sensitive to the collective directed flow of
the system~\cite{vanHees:2014ida,Shen:2014cga}, to the equation of
state~\cite{vanHees:2014ida,Goloviznin:2012dy}, to the possible
production of a
Glasma~\cite{McLerran:2014hza,Monnai:2014kqa,Liu:2012ax}, to the
rate of chemical equilibration in the
QGP~\cite{Ozvenchuk:2012fn,Ozvenchuk:2012kh,Almasi:2014pya} and to
the asymmetry induced by the strong magnetic field (flash) in the
very early stage of the heavy-ion
collision~\cite{Bzdak:2012fr,Tuchin:2014pka,Tuchin:2012mf}.

However, the observation by the PHENIX Collaboration~\cite{PHENIX1}
that the elliptic flow $v_2(p_T)$ of {\em direct} photons produced
in minimum bias Au+Au collisions at $\sqrt{s_{NN}}=200$~GeV is
comparable to that of the produced pions was a surprise and in
contrast to the theoretical expectations and predictions. Indeed,
the photons produced by partonic interactions in the quark-gluon
plasma phase have not been expected to show considerable flow
because they are dominated by the emission in the initial phase
before the elliptic flow fully develops. We here report about the
studies within the PHSD approach on this issue and compare to other
models in context of the available data from the different
collaborations.

\begin{figure}
\hspace{0.5cm}
\includegraphics[width=0.47\textwidth]{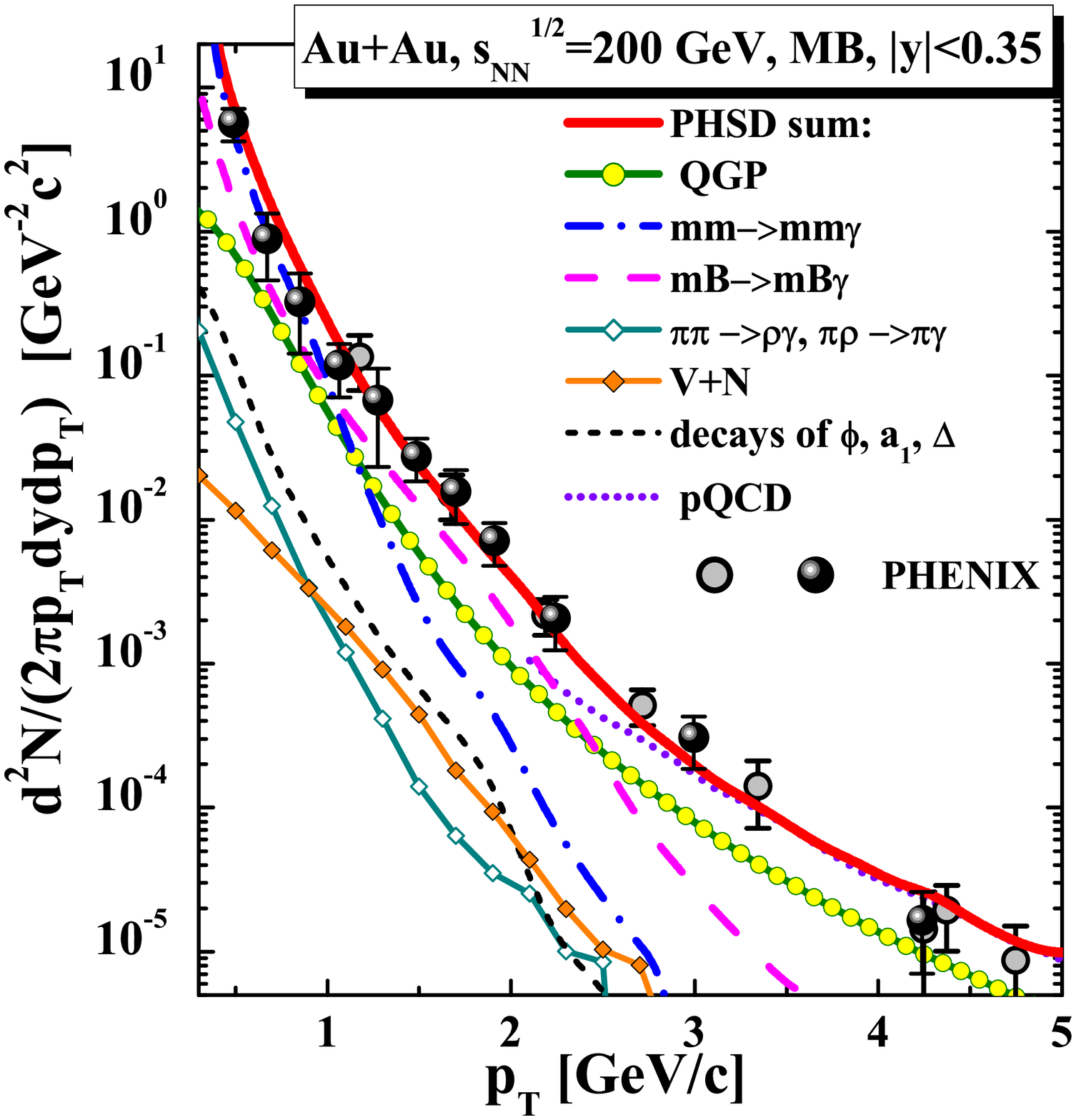}
\includegraphics[width=0.45\textwidth]{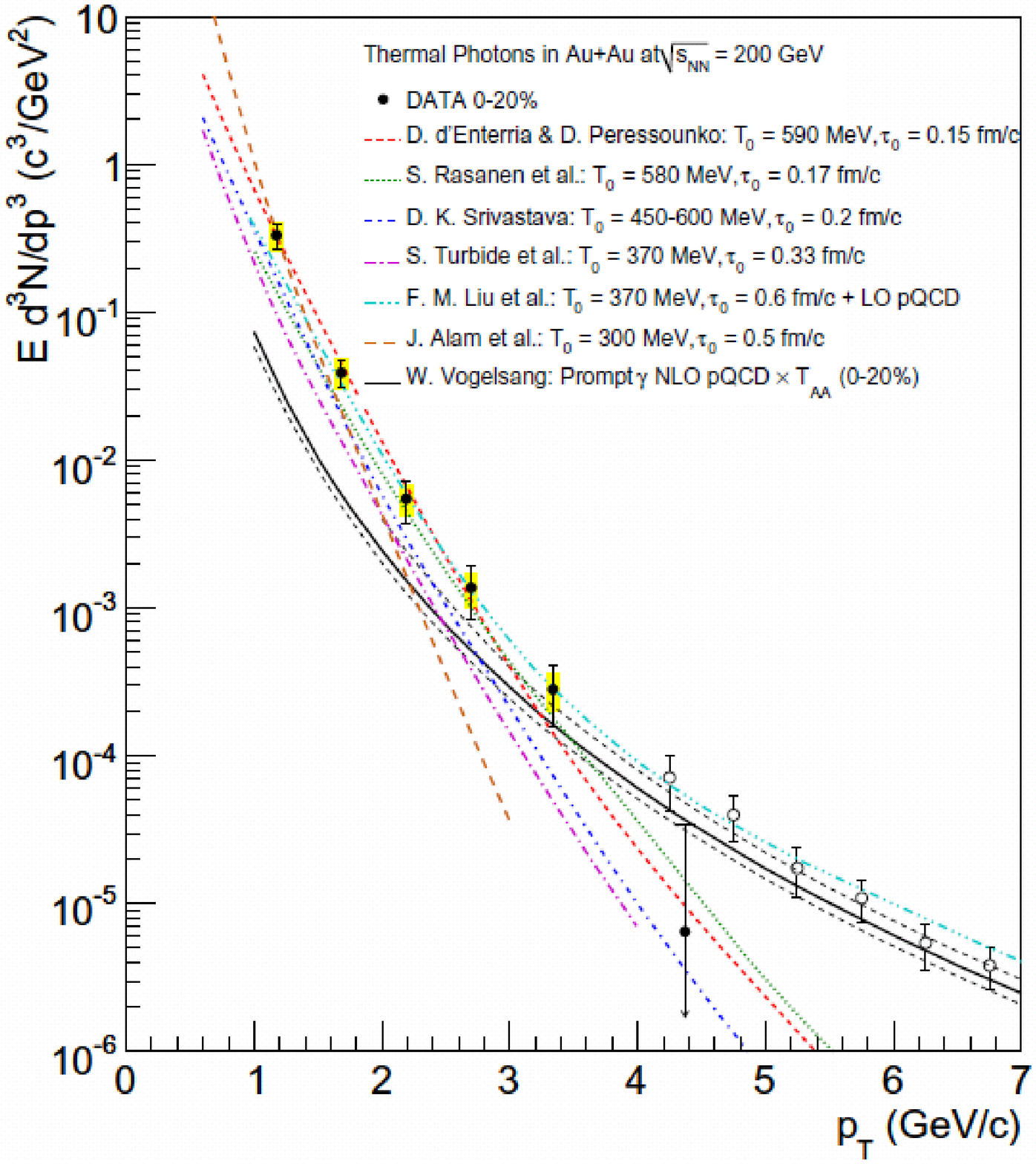}
\caption{
  ({ l.h.s.}) PHSD results for the spectrum of {\em direct} photons
produced in 0-40\% most central Au+Au collisions at
$\sqrt{s_{NN}}=200$~GeV as a function of the transverse momentum
$p_T$ at midrapidity $|y|< 0.35$. The data of the PHENIX
Collaboration are adopted from Refs.
\protect{\cite{Adare:2008ab,Adare:2014fwh}}. For the individual
lines see the legend in the figure. The figure is taken from Ref.
\protect\cite{Linnyk:2015tha}. ({ r.h.s.}) A compilation of various
predictions for the {\em direct} photon yield in hydrodynamical
models (see legend) in comparison to the data of the PHENIX
collaboration. The figure is taken from Ref.~\protect\cite{Axel}. }
  \label{spectrarhic}
\end{figure}

\subsection{Direct photon spectra from SPS to LHC energies}

We start with the system p+Pb at 160 GeV, i.e. at the top SPS
energy.  Fig.~\ref{WA98} (l.h.s.) shows the comparison of the
HSD/PHSD calculations to the data of the WA98 Collaboration from
Ref.~\cite{Aggarwal:2000th,Aggarwal:2000ps}  in the pseudorapidity
interval $2.3 < \eta < 3.0$. In this case almost the entire photon
spectrum is described by the contribution from pion and $\eta$
decays while the contribution from the heavier mesons is not
leading. The successful description of these data by PHSD is
dominantly due to the fact that the meson production itself is
described very well in p+A reactions~\cite{Bratkovskaya:2008iq}.

We continue with Pb+Pb collisions at $\sqrt{s_{NN}}$ = 17.3 GeV.
Fig.~\ref{WA98} (r.h.s.) shows the comparison of the PHSD
calculations~\cite{Linnyk:2015tha} for the {\em direct} photon
$p_T$-spectrum to the data of the WA98 Collaboration from
Ref.~\cite{Aggarwal:2000th,Aggarwal:2000ps} for 10\% centrality in
the pseudorapidity interval $2.35 < \eta < 2.95$. In addition to the
sources, which had been incorporated in the original HSD study in
2008, the meson+baryon bremsstrahlung, $VN\to N \gamma$, $\Delta\to
N\gamma$ decay and the QGP channels are added. Compared to the
earlier results of Ref.~\cite{Bratkovskaya:2008iq}, the description
of the data is further improved and the conclusions remain
unchanged: the bremsstrahlung contributions are essential for
describing the data at low $p_T$.  This interpretation is shared by
the authors of
Refs.~\cite{Liu:2007zzw,Haglin:2003sh,Dusling:2009ej}, who also
stressed the importance of the meson+meson bremsstrahlung in view of
the WA98 data using  hydrodynamical or fireball models. Note that
the photon contribution from the QGP is practically negligible at
this bombarding energy for low $p_T$ and reaches at most 25\% at
$p_T>0.5$~GeV.

We now step on to the top RHIC energy of $\sqrt{s_{NN}}$ = 200 GeV
and report on PHSD  results for the differential photon spectra for
the system Au+Au.  The  {\em direct} photon spectrum -- as a sum of
partonic as well as hadronic sources -- in 0-40\% central Au+Au
collisions  is presented in Fig.~\ref{spectrarhic} (l.h.s.) as a
function of the transverse momentum $p_T$ at midrapidity $|y|<
0.35$. While the 'hard' $p_T$ spectra are dominated by the 'prompt'
(pQCD) photons, the 'soft' spectra are filled by the 'thermal'
sources: the QGP gives up to $~50\%$ of the {\em direct} photon
yield below 2 GeV/$c$, a sizeable contribution stems from hadronic
sources such as meson-meson ($mm$) and meson-Baryon ($mB$)
bremsstrahlung while the contribution from binary $mm$ reactions is
of subleading order. Thus, according to the  PHSD results the $mm$
and $mB$ bremsstrahlung turn out to be an important source of {\em
direct} photons also at the top RHIC energy. We note, that the
bremsstrahlung channels are not included in the $mm$ binary 'HG'
rate by Turbide et al. in Ref. \cite{Turbide:2003si} used in the
hydro calculations addressed above. We stress that $mm$ and $mB$
bremsstrahlung  can not be subtracted experimentally from the photon
spectra and have to be included in theoretical models.

The right panel of Fig.~\ref{spectrarhic} shows a compilation of
various predictions for the {\em direct} photon yield in
hydrodynamical models (see legend) in comparison to the data of the
PHENIX collaboration from Ref.~\cite{Axel}.  The NLO pQCD
calculations for the prompt photon production from Vogelsang have
been added to the {\em thermal} photon spectra.  The actual results
for the {\em direct} photon spectra depend on the initial
temperature $T_0$ (varying by about a factor of 2) and the hydro
starting time $\tau_0$ which are fitted differently to final
hadronic spectra, respectively. All these models only give a very
low elliptic flow for the {\em direct} photons.

\begin{figure}
\hspace{0.1cm} \includegraphics[width=0.47\textwidth]{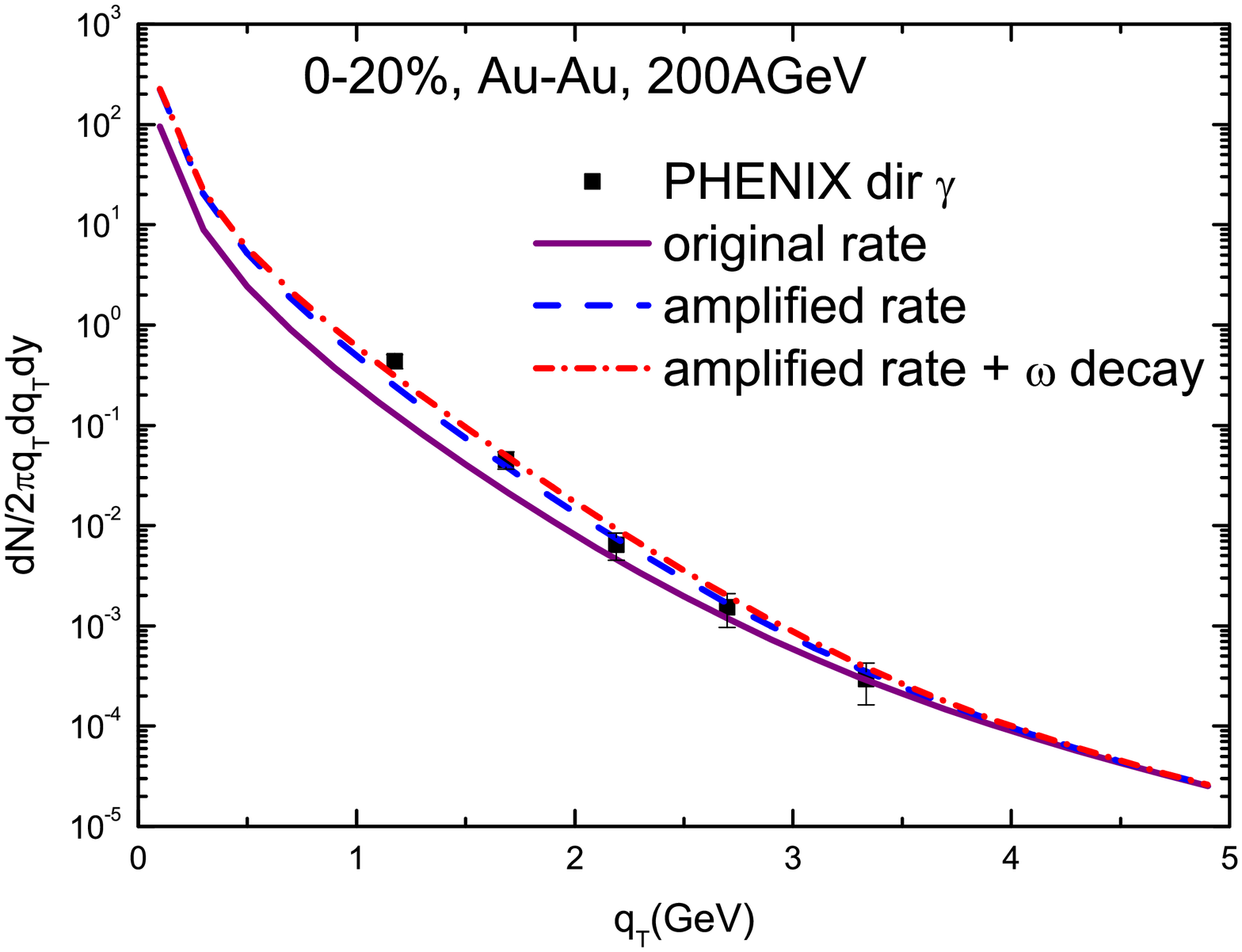}
\includegraphics[width=0.49\textwidth]{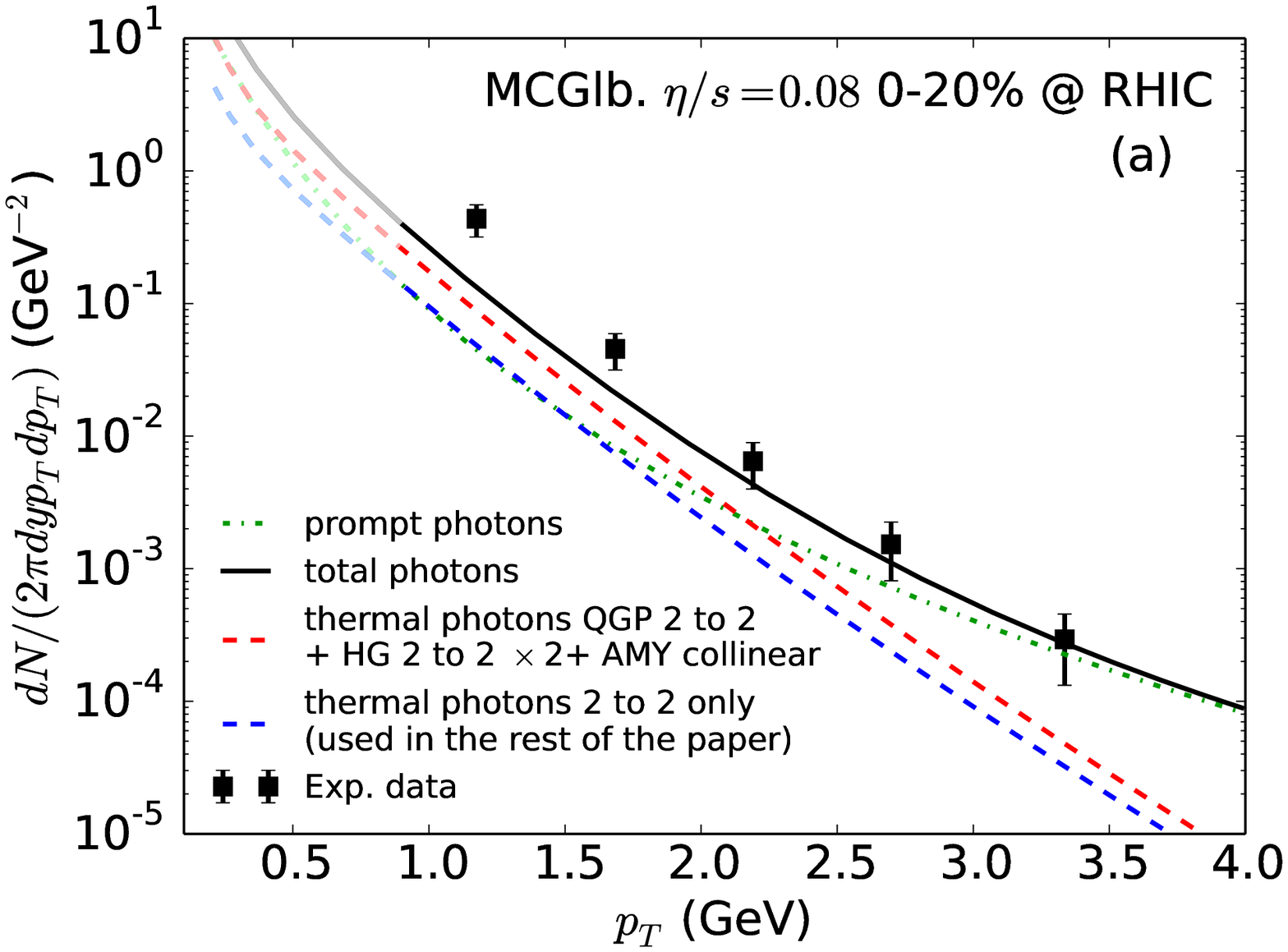}
\caption{ ({ l.h.s.}) {\em Direct} photon spectra
 from the model of van Hees et al. \protect\cite{vanHees:2014ida} at RHIC when adding $\omega
  \rightarrow\pi^0 + \gamma$ decays at thermal freeze-out to a scenario
  with amplified rates at temperatures close to the pseudo-critical transition temperature $T_c$ (dash-dotted line),
  compared to the amplified rate (dashed line) and default-rate (solid line) scenarios.
  The figure is taken from Ref.\protect\cite{vanHees:2014ida}. \label{fig_rhic-omg}
 ({ r.h.s.}) Calculated photon spectra in the viscous hydrodynamical model from Shen et al.~\protect\cite{Shen:2013vja}
  in comparison to the data from the PHENIX Collaboration \protect\cite{Adare:2008ab}.}
  \label{RHIChyd}
\end{figure}

As an example for more recent calculations we show in
Fig.~\ref{fig_rhic-omg} (l.h.s.) the results from the model of van
Hees et al.~\cite{vanHees:2014ida} which is describing the PHENIX
data~\cite{Adare:2008ab} with a good accuracy. The calculations of
Ref.~\cite{vanHees:2014ida} are based on a hydrodynamical model for
the ``fireball" evolution with the hypothesis that the rates of
photon production are amplified for temperatures close to the
hadronization transition and adding to the thermal spectra
(calculated with the amplified rates) the photon contribution from
final-state $\omega$-mesons  at thermal freeze-out. The spectra
presented on the right hand side of Fig.~\ref{RHIChyd} have been
calculated by Shen et al.~\cite{Shen:2013vja} using a viscous
hydrodynamical evolution and taking into account viscous effects in
the photon rates. In this approach -- that reproduces the final
hadron spectra and hadron $v_2$ -- the data are underestimated
considerably. The discrepancy becomes enhanced  when an alternative
scenario of a gluon-dominated initial state is considered since the
gluons do not carry electric charge.

\begin{figure}
\phantom{a} \hspace{-0.7cm}
\includegraphics[width=0.8\textwidth]{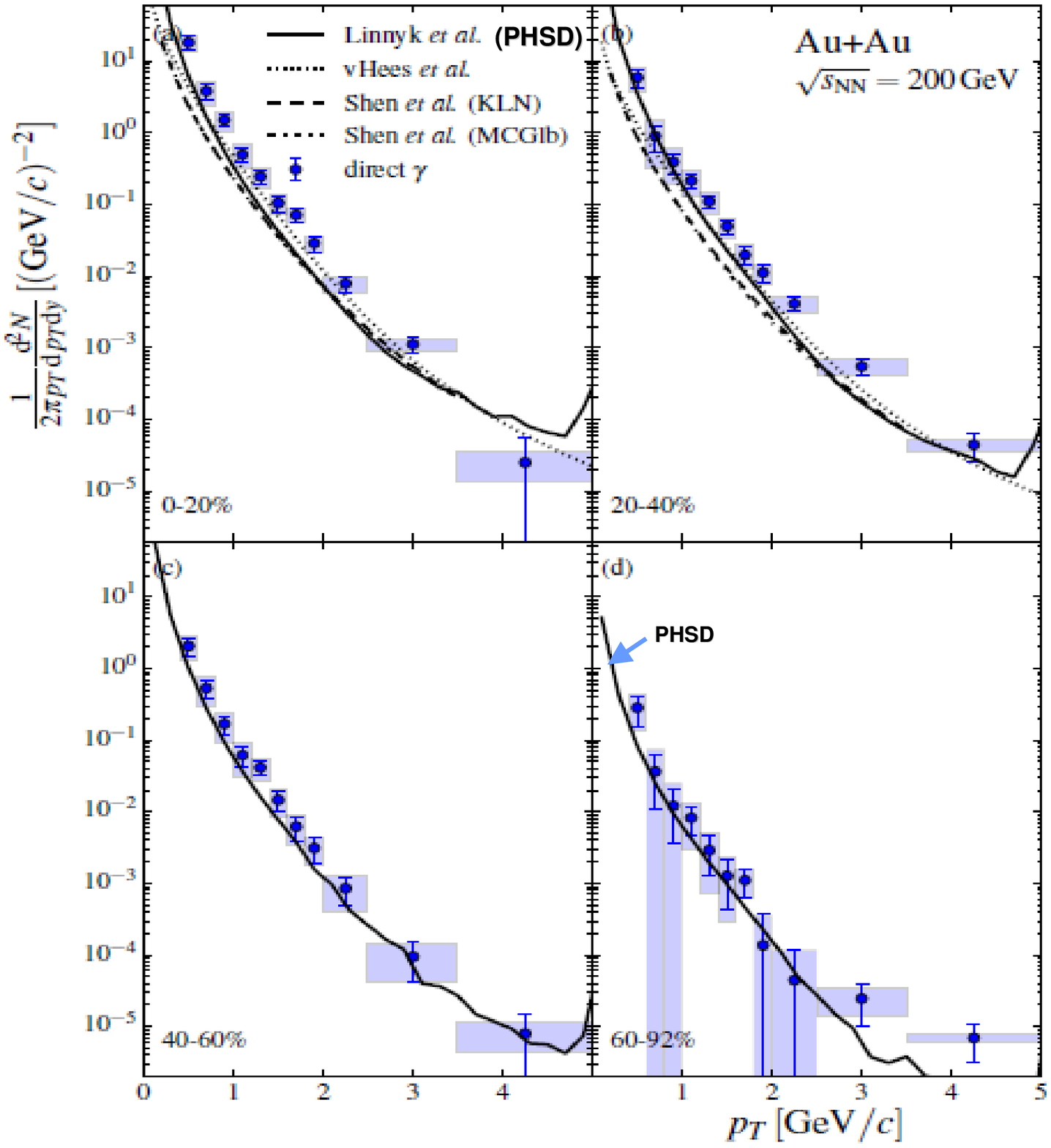} \phantom{a} \hspace{-4cm}
\includegraphics[width=0.4\textwidth]{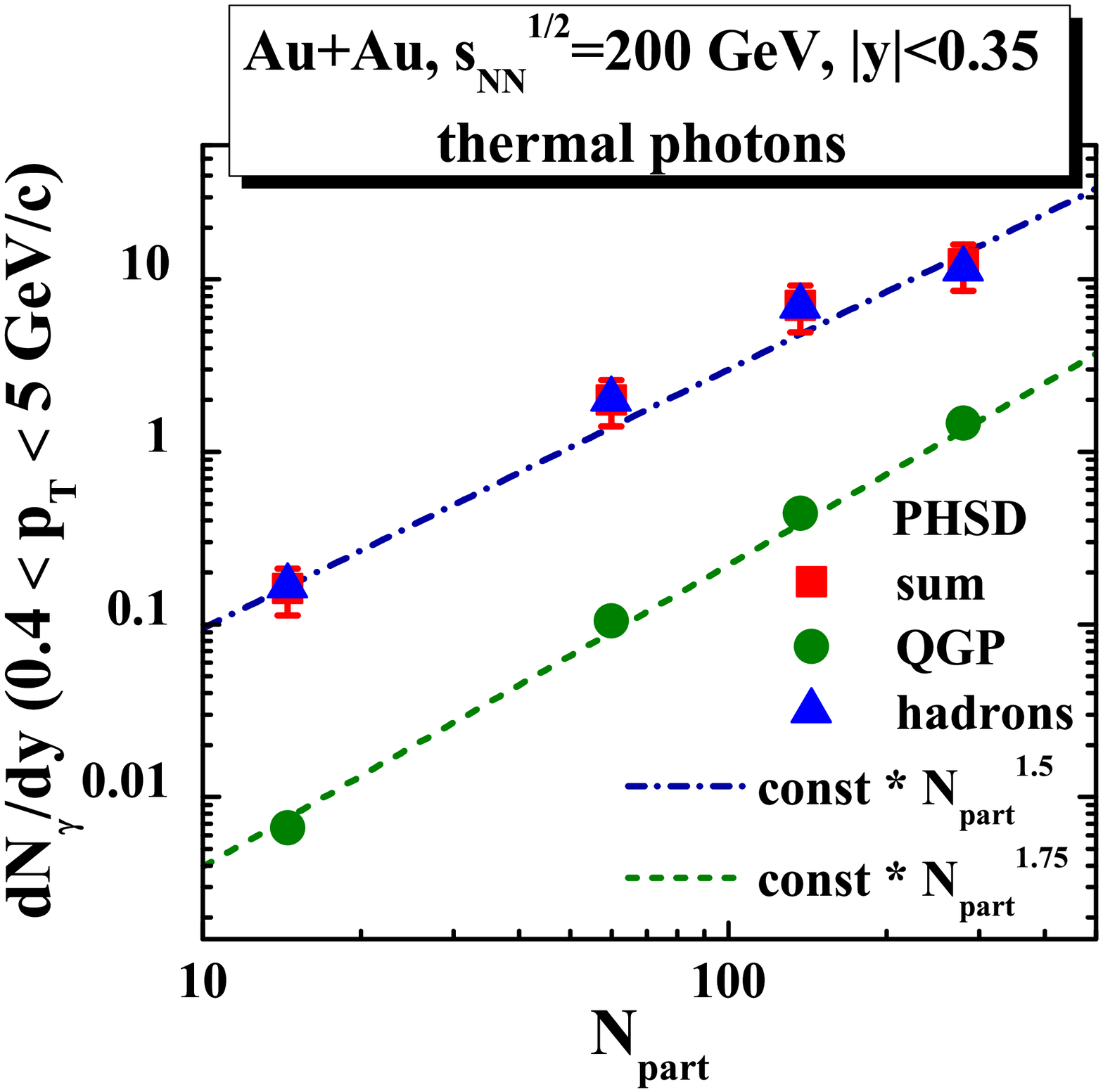}
 \caption{({\it l.h.s}) Centrality dependence of the {\em direct} photon
$p_T$-spectra for 0-20\%, 20-40\%, 40-60\%, 60-92\% central Au+Au
collisions at $\sqrt{s}=200$ GeV: model predictions vs. the PHENIX
data  \protect\cite{Adare:2014fwh}. The PHSD predictions are denoted
by 'Linnyk et al.' (solid lines). The figure is taken from Ref.
\protect\cite{Mizuno:2014via}. ({\it r.h.s}) The scaling of the
integrated {\em thermal} photon yield from PHSD as a function of the
number of participating nucleons in Au+Au collisions at
$\sqrt{s_{NN}}=200$ GeV for the hadronic channels (upper symbols)
and partonic channels (lower symbols). The figure is taken from
Ref.~\protect\cite{Linnyk:2013wma}. } \label{fig:N3PHSD} \label{222}
\end{figure}

\subsubsection*{Photon sources: QGP vs. HG}

The question: "what dominates the photon spectra - {\it QGP
radiation or hadronic contributions}" can be addressed
experimentally by investigating the centrality dependence of the
photon yield since the QGP contribution is expected to decrease when
going from central to peripheral collisions where the hadronic
channels are dominant. Fig. \ref{fig:N3PHSD} (l.h.s.) shows the
centrality dependence of the {\em direct} photon $p_T$-spectra  for
0-20\%, 20-40\%, 40-60\%, 60-92\% central Au+Au collisions at
$\sqrt{s}=200$ GeV. The solid dots stand for the recent PHENIX data
\cite{Adare:2014fwh,Mizuno:2014via} whereas the lines indicate the
model predictions: solid line - PHSD (denoted as 'Linnyk et al.')
\cite{Linnyk:2013hta,Linnyk:2013wma,Linnyk:2015tha}, dashed and
dashed-dotted lines ('Shen et al. (KLN)' and 'Shen et al.' (MCGib)')
are the results from viscous (2+1)D VISH2+1 \cite{Shen:2014cga} and
(3+1)D MUSIC \cite{Dion:2011vd,Dion:2011pp} hydro models whereas the
dotted line ('vHees et al.') stands for the results of the expanding
fireball model \cite{vanHees:2011vb}. As seen from Fig.
\ref{fig:N3PHSD} (l.h.s.) for the central collisions the models
deviate up to a factor of 2 from the data and each other due to the
different dynamics and sources included (as discussed above);  for
the (semi-)peripheral collisions the PHSD results - dominated by
$mm$ and $mB$ bremsstrahlung - are consistent with the data which
favor these hadronic sources. Presently, no results from the other
models for peripheral reactions are known.

\begin{figure} \centering
\includegraphics[width=0.95\textwidth]{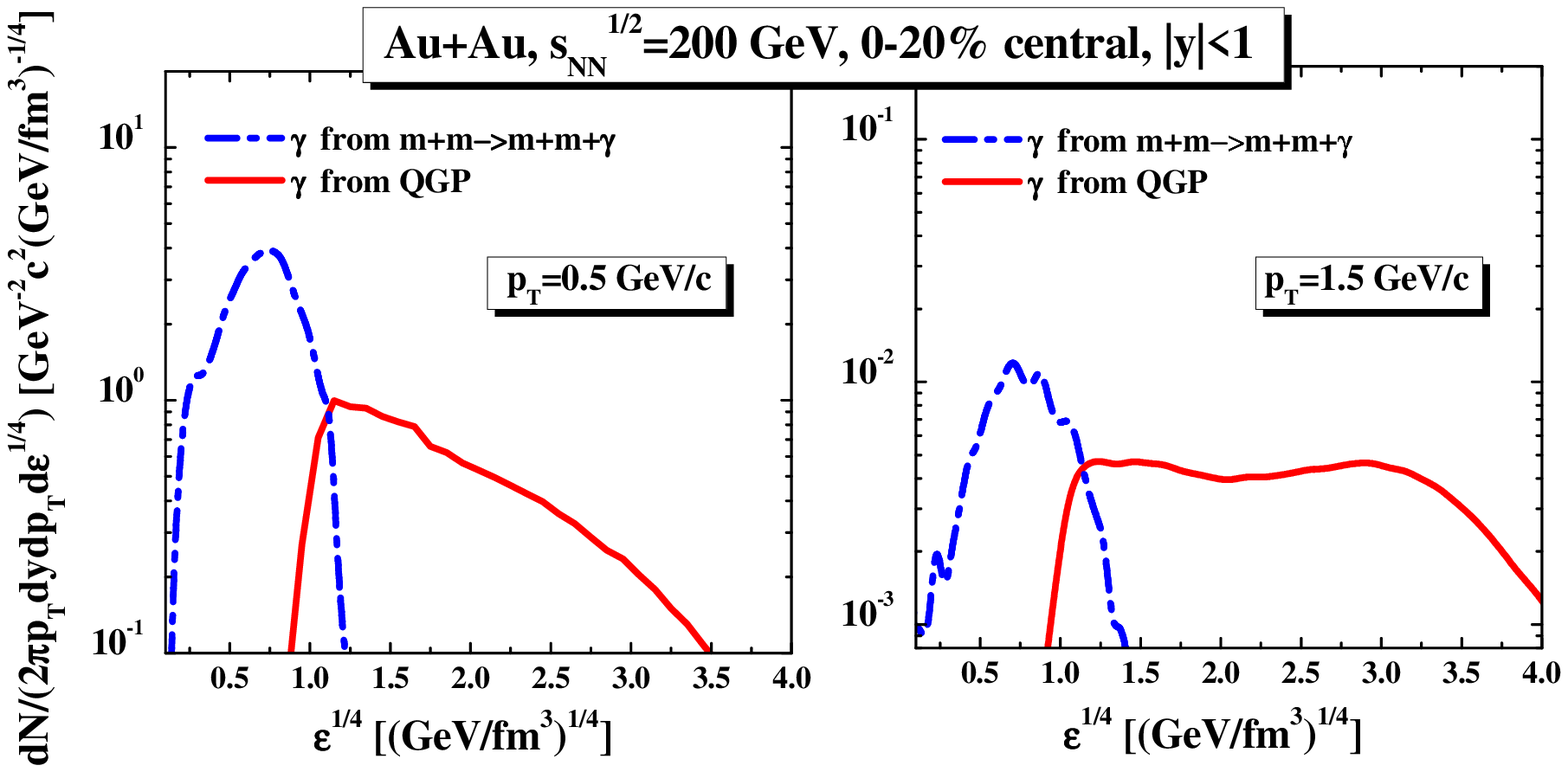}
\caption{ (l.h.s.) Photon yield  with a transverse momentum
$p_T=0.5$~GeV/c at midrapidity produced in 0-20 \% most central
$Au+Au$ collisions as a function of the approximate local
``temperature" (the fourth-root of the energy density) from the PHSD
from meson-meson bremsstrahlung (dash-dotted lines) and gluon
Compton scattering (solid lines). \label{Evol05} (r.h.s.) Same as in
the left panel for photons with a transverse momentum
$p_T=1.5$~GeV/c.
 \label{Evol15}  }
\end{figure}

The centrality dependence of the {\em direct} photon yield,
integrated over different $p_T$ ranges, has been measured by the
PHENIX Collaboration, too  \cite{Adare:2014fwh,Mizuno:2014via}. It
has been found that the midrapidity 'thermal' photon yield scales
with the number of participants as $dN/dy \sim N_{part}^\alpha$ with
$\alpha =1.48\pm 0.08$ and only very slightly depends on the
selected $p_T$ range (which is still in the 'soft' sector, i.e. $<
1.4$ GeV/$c$). Note that the 'prompt' photon contribution (which
scales as the $pp$ 'prompt' yield times the number of binary
collisions in $A+A$) has been subtracted from the data. The PHSD
predictions \cite{Linnyk:2013hta,Linnyk:2013wma,Linnyk:2015tha} for
the minimum bias Au+Au collisions give $\alpha (total) \approx 1.5$
(cf. Fig. \ref{222}, r.h.s.) which is dominated by hadronic
contributions while the QGP channels scale with $\alpha (QGP) \sim
1.75$.  A similar finding has been obtained by the viscous (2+1)D
VISH2+1 and (3+1)D MUSIC hydro models \cite{Shen:2013vja}:
$\alpha(HG) \sim 1.46, \ \ \alpha(QGP) \sim 2, \ \ \alpha(total)
\sim 1.7$. Thus, the QGP photons show a centrality dependence
significantly stronger than that of hadron gas (HG) photons.

Next, let us investigate  the photon production across the phase
transition in the heavy-ion collision to check whether the observed
yield of {\em direct} photons is produced dominantly in some
particular region of the energy-density or in some particular phase
of matter. Fig.~\ref{Evol05} shows the yield of photons produced at
midrapidity in 0-20 \% most central $Au+Au$ collisions at
$\sqrt{s_{NN}}$ = 200 GeV  as a function of the approximate local
``temperature" (i.e. the fourth-root of the energy density) from the
PHSD. The left panel of Fig.~\ref{Evol05} presents the calculations
for  photons with a transverse momentum $p_T=0.5$~GeV/c, while the
right panel corresponds to  photons with a transverse momentum
$p_T=1.5$~GeV/c. We observe that the early, hot state does not
dominate the photon production in the QGP contrary to expectations
of the static thermal fireball model, where photon production is
roughly proportional to a power of the temperature ($\sim T^4$). The
integration over the dynamical evolution of the heavy-ion collision
leads to similar contributions of the different energy density
regions for transverse momenta in the order of 1 -- 1.5 GeV/c since
the rate decreases but the space-time volume increases. For the low
$p_T$=0.5 GeV/c, i.e. at the lower end of the experimental spectra,
the hadronic contribution is clearly larger than the partonic one.
The photon production in the hadronic phase is dominated by the
lower energies/temperatures because of the very long times over
which the produced hadrons continue to interact elastically, which
is accompanied by the photon bremsstrahlung in case of charged
hadrons.

%\subsection{Spectra: LHC energy regime}
%\label{sect:lhc}

\begin{figure}[t]
\begin{minipage}{0.48\linewidth}
\includegraphics[width=\textwidth]{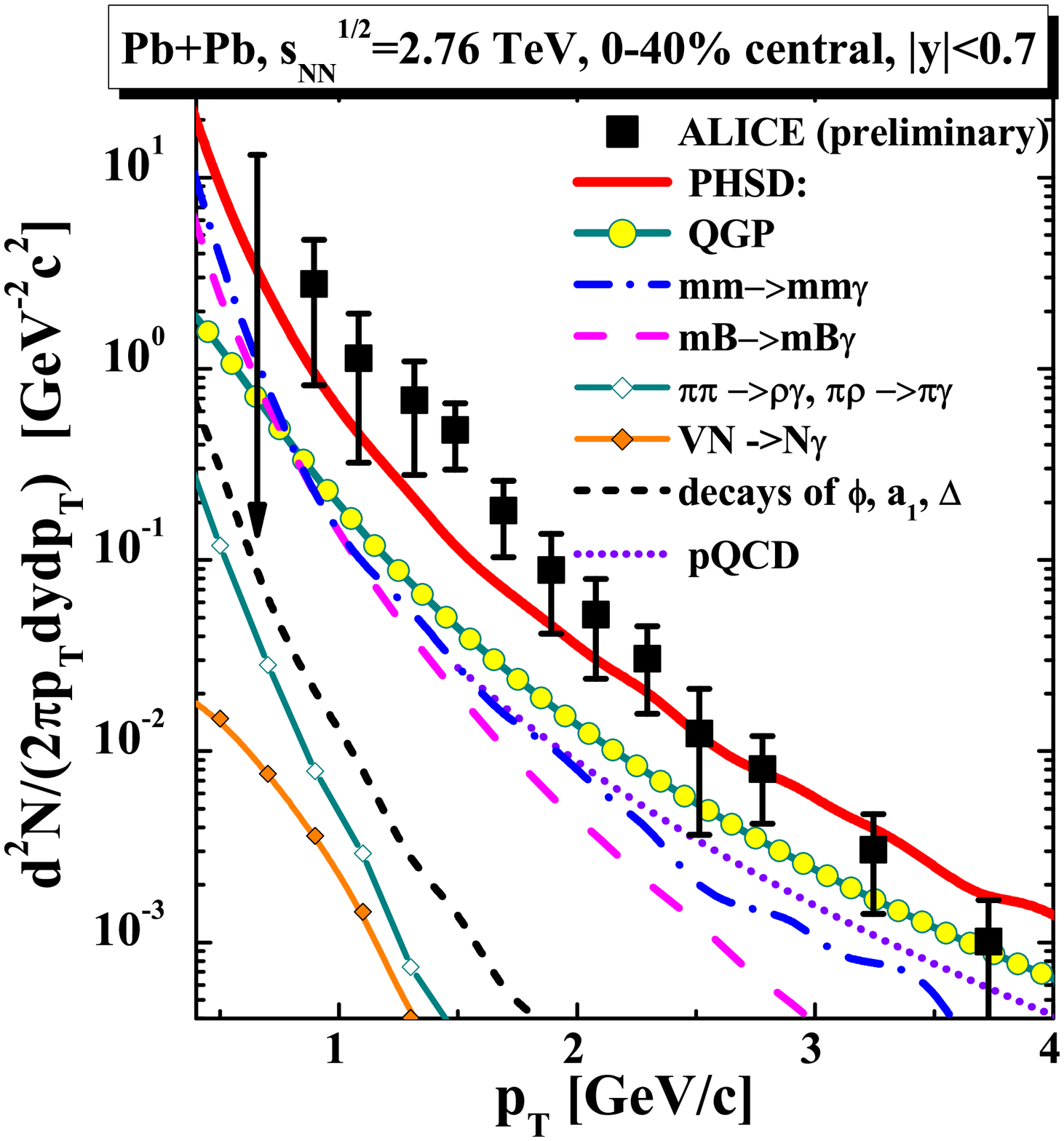}
\end{minipage}
\hspace{0.6cm}
\begin{minipage}{0.38\linewidth}
\includegraphics[width=\textwidth]{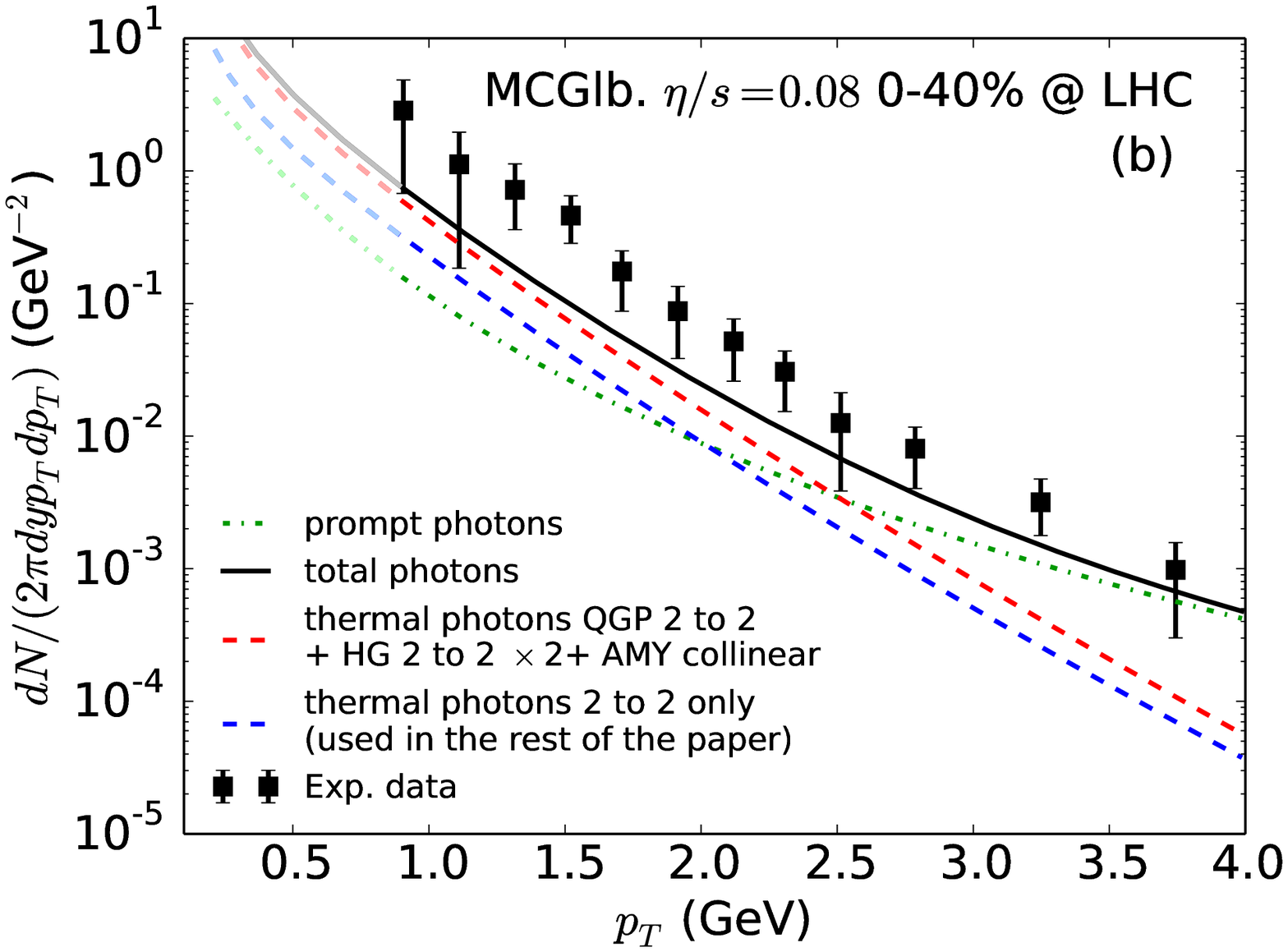}
\includegraphics[width=\textwidth]{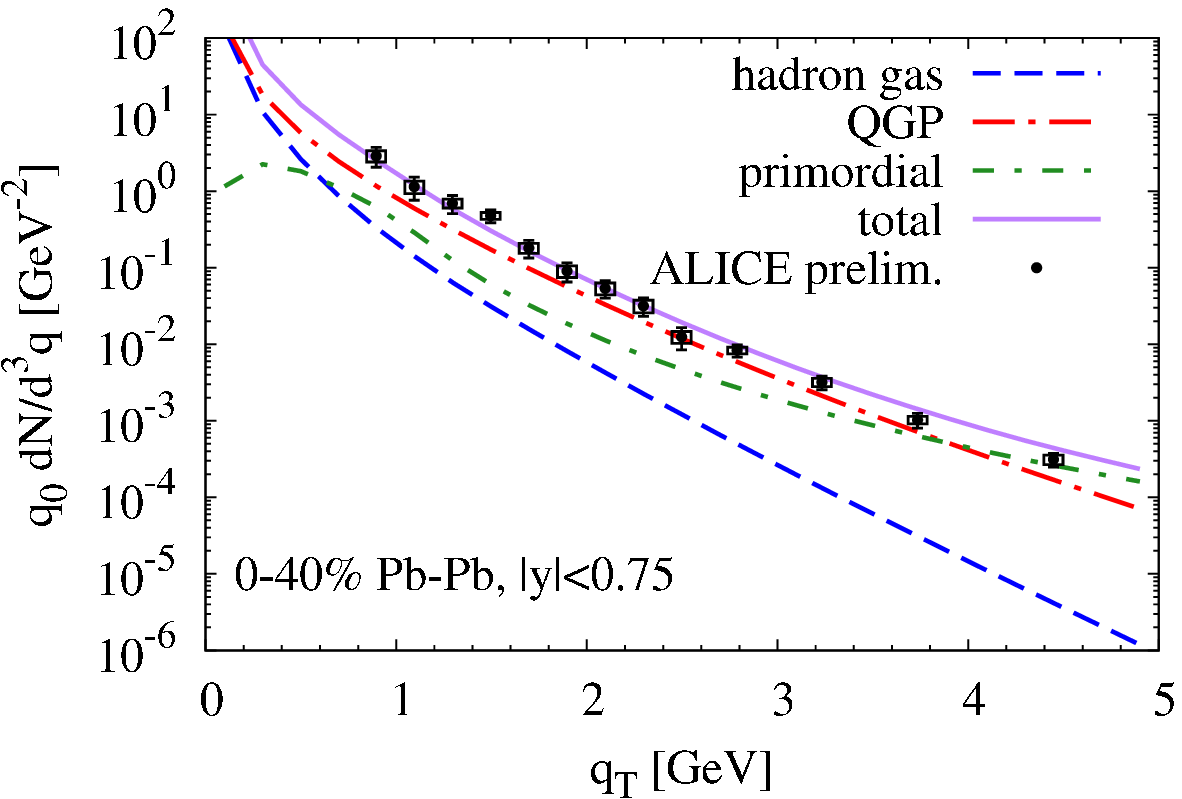}
\end{minipage}
\caption{ ({ l.h.s.}) The yield of {\em direct} photons at
midrapidity in Pb+Pb collisions at the invariant energy
$\sqrt{s_{NN}}=2.76$~TeV for 0-40\% centrality as predicted within
the PHSD in comparison to the preliminary data from the ALICE
Collaboration~\protect\cite{Wilde:2012wc}. The figure is taken from
Ref. \protect\cite{Linnyk:2015tha}. \label{spectralhc} ({ r.h.s.,
upper panel:}) Photon spectra in $0{-}40\%$ centrality Pb+Pb
collisions at the LHC as calculated within the viscous hydrodynamics
by Shen et al. \protect\cite{Shen:2013cca}. The Pb+Pb data are from
the ALICE Collaboration~\protect\cite{Wilde:2012wc}. The figure is
taken from Ref.~\protect\cite{Shen:2013cca}. ({ r.h.s., lower
panel:}) The same observable as calculated in the upper panel with
an ideal hydrodynamical evolution and amplified photon rates around
the transition temperature by van Hees et al.
\protect\cite{vanHees:2014ida}. The figure is taken from Ref.
\protect\cite{vanHees:2014ida}. }
\end{figure}

\begin{figure}[htb]
\hspace{2.5cm} \includegraphics[width=0.6\textwidth]{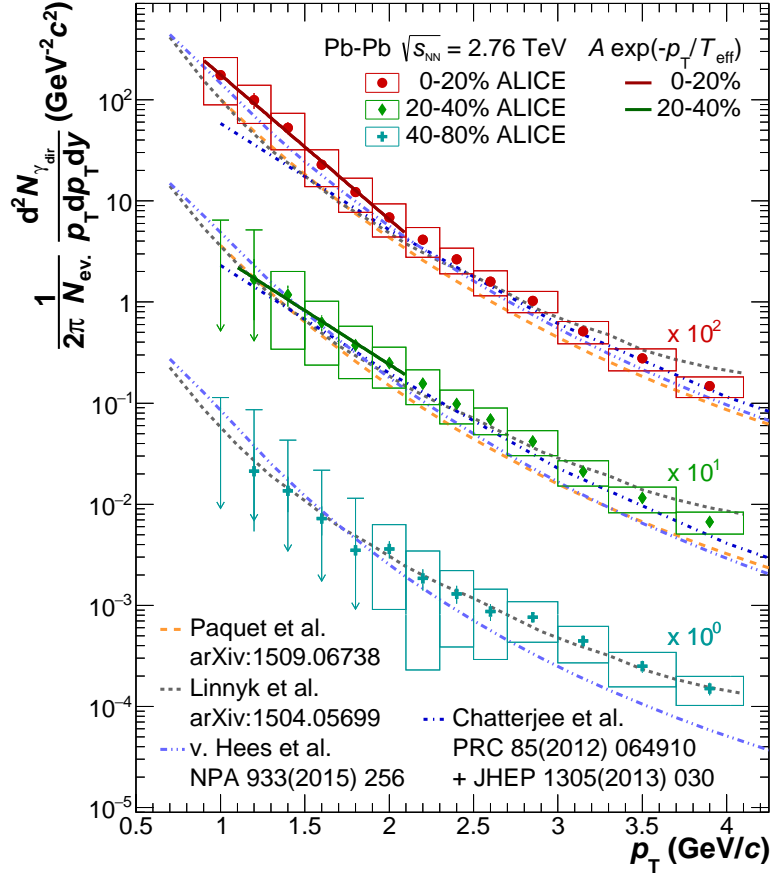}
 \caption{Centrality dependence of the {\em direct} photon
$p_T$-spectra for 0-20\%, 20-40\%, 40-80\% central Pb+Pb collisions
at $\sqrt{s_{NN}}$=2.76 TeV: model predictions vs. the ALICE data
\protect\cite{Adam:2015lda}. The PHSD predictions are denoted by
'Linnyk et al.' (dotted lines). The figure is taken from Ref.
\protect\cite{Adam:2015lda}. \label{fig:ALICE}}
 \end{figure}

We now increase the invariant collision energy $\sqrt{s_{NN}}$ by a
factor of 13.8.  In Fig.~\ref{spectralhc} (l.h.s.) we show the
 {\em direct} photon yield from PHSD in Pb+Pb collisions at the
invariant energy $\sqrt{s_{NN}}=2.76$~TeV for 0-40\% centrality in
comparison  to the preliminary data of the ALICE Collaboration from
Ref.~\cite{Wilde:2012wc}. We find a rather good overall agreement
with the data within about a factor of 2 in the range of transverse
momenta $p_T$ from 1 to 4 GeV. On the other hand, the calculations
tend to underestimate the preliminary data in the low-$p_T$
region~\cite{Linnyk:2015nea}.  However, the significance of the
comparison is not robust until the final data will be available. We,
furthermore, present in Fig. ~\ref{spectralhc} (r.h.s.) the photon
spectra for $0{-}40\%$ centrality Pb+Pb collisions at the LHC as
calculated within the viscous hydrodynamics by Shen et al.
~\protect\cite{Shen:2013cca} in comparison to the Pb+Pb data from
the ALICE Collaboration~\cite{Wilde:2012wc} (upper right panel). In
the right bottom panel we show the same observable as calculated in
the ideal hydrodynamical model with amplified photon rates around
the transition temperature by van Hees et al.
\cite{vanHees:2014ida}. Similar to RHIC energies the viscous hydro
calculations \cite{Shen:2013cca} underestimate the measured photon
yield for $p_T <$ 2 GeV/c while the model of van Hees et al.
\cite{vanHees:2014ida} with amplified rates at $T_c$ performs
better.

An actual overview on the current situation with respect to the {\em
direct} photon yields at different centralities has been provided by
the ALICE Collaboration in Ref. \cite{Adam:2015lda} and is displayed
in Fig. \ref{fig:ALICE}. The figure shows the centrality dependence
of the {\em direct} photon $p_T$-spectra for 0-20\%, 20-40\%,
40-80\% central Pb+Pb collisions at $\sqrt{s_{NN}}$=2.76 TeV in
comparison to various model predictions. The PHSD predictions are
denoted by 'Linnyk et al.' (dotted lines) and are compatible with
the measurements within the error bars. This roughly holds also for
the other models.

In conclusion, we have found that from SPS to LHC energies the
radiation from the sQGP constitutes less than half of the observed
number of {\em direct} photons for central reactions in the PHSD.
The hydrodynamical and fireball models predict a larger fraction of
the QGP photons to the total yield and are substantially lower in
the hadronic contributions.  The radiation from hadrons and their
interaction -- which are not measured separately so far -- give a
considerable contribution in the PHSD especially at low transverse
momentum. The dominant hadronic sources are the meson decays, the
meson-meson bremsstrahlung and the meson-baryon bremsstrahlung.
While the first (e.g. the decays of $\omega$, $\eta$', $\phi$ and
$a_1$ mesons) can be subtracted from the photon spectra once the
mesonic yields are determined independently by experiment, the
reactions $\pi+\rho\to\pi+\gamma$, $\pi+\pi\to \rho+\gamma$, $V+N\to
N+\gamma$, $\Delta \to N+\gamma$ as well as the meson-meson and
meson-baryon bremsstrahlung can be 'separated' from the partonic
sources only with the assistance of theoretical models (and
corresponding uncertainties).

\subsection{Elliptic flow of {\em direct} photons}
\label{sect:v2}

We recall that the azimuthal momentum distribution of the photons is
expressed in the form of a Fourier series as,
\be E\frac{d^3N}{d^3p}=  \frac{d^2N}{2\pi
p_Tdp_Tdy}\left(\! 1\!+\! \sum^\infty_{n=1} 2v_n(p_T) \cos
[n(\psi-\Psi_n)]\! \right)\!, \ee
where $v_n$ is the magnitude of the $n'$th order harmonic term
relative to the angle of the initial-state fluctuating spatial plane
of symmetry $\Psi_n$ and $p=(E,\vec{p})$ is the four-momentum of the
photon. We here focus on the coefficients $v_2$ and $v_3$ which
implies that we have to perform event by event calculations in order
to catch the initial fluctuations in the shape of the interaction
zone and the event plane $\Psi_{EP}$. We calculate the triangular
flow $v_3$ with respect to $\Psi_3$ as $v_3\{\Psi_3\} = \langle
\cos(3[\psi-\Psi_3])\rangle/\rm{Res}(\Psi_3)$. The event plane angle
$\Psi_3$ and its resolution $\rm{Res}(\Psi_3)$ are evaluated as
described in Ref.~\cite{{Adare:2011tg}} via hadron-hadron
correlations by the two-sub-events
method~\cite{Poskanzer:1998yz,Bilandzic:2010jr}.

We note again that the second flow coefficient $v_2$ carries information
on the interaction strength in the system -- and thus on the state
of matter and its properties -- at the space-time point, from which
the measured particles are emitted. The elliptic flow $v_2$ reflects
the azimuthal asymmetry in the momentum distribution of the produced
particles ($p_x$ vs $p_y$), which is correlated with the
geometrical azimuthal asymmetry of the initial reaction region. If
the produced system is a weakly-interacting gas, then the initial
spatial asymmetry is not effectively transferred into the final
distribution of the momenta. On the other hand, if the produced matter
has the properties of a liquid, then the initial geometrical
configuration is reflected in the final particle momentum
distribution.

More than a decade ago, the WA98 Collaboration has measured the
elliptic flow $v_2$ of photons produced in $Pb+Pb$ collisions at the
beam energy of $E_{beam}=158$~AGeV~\cite{Aggarwal:2004zh}, and it
was found that the $v_2(\gamma^{incl})$ of the
low-transverse-momentum inclusive photons was about equal to the
$v_2({\pi})$ of pions within the experimental uncertainties. This
observation lead to the conclusion that either (Scenario a:) the
contribution of the {\em direct} photons to the inclusive ones is
negligible in comparison to the {\em decay} photons, i.e. dominantly
the $\pi^0$ decay products, or (Scenario 2:) the elliptic flow of
the {\em direct} photons is comparable in magnitude to the
$v_2(\gamma^{incl})$, $v_2(\gamma^{decay})$ and $v_2({\pi})$.
However, in view of the {\em direct} photon spectrum from WA98,
which we described in Section 6.1, there is a significant finite
yield of {\em direct} photons at low transverse momentum. Thus the
scenario 1 can be ruled out. Furthermore, the observed {\em direct}
photons of low $p_T$ must have a significant elliptic anisotropy
$v_2$ of the same order of magnitude as the hadronic flow since they
dominantly stem from hadronic sources.  Thus the
interpretation~\cite{Bratkovskaya:2008iq,Liu:2007zzw} of the
low-$p_T$ {\em direct} photon yield measured by WA98 -- as
dominantly produced by the bremsstrahlung process in the mesonic
collisions $\pi+\pi\to\pi+\pi+\gamma$ -- is in accord also with the
data on the photon elliptic flow $v_2(\gamma^{incl})$ at the top SPS
energy.

Let us note that the same conclusions apply also to the most recent
studies of the photon elliptic flow at RHIC and LHC. The PHENIX and
ALICE Collaborations have measured the inclusive photon $v_2$ and
found that at low transverse momenta it is comparable to the
$v_2(p_T)$ of {\em decay} photons as calculated in cocktail
simulations based on the known mesonic $v_2(p_T)$. Therefore, either
(a) the yield of the {\em direct} photons to the inclusive ones is
not statistically significant in comparison to the {\em decay}
photons or (b) the elliptic flow of the {\em direct} photons must be
as large as $v_2(\gamma^{decay})$ and $v_2(\gamma^{incl})$.

\begin{figure}
\centering
\includegraphics[width=0.95\linewidth]{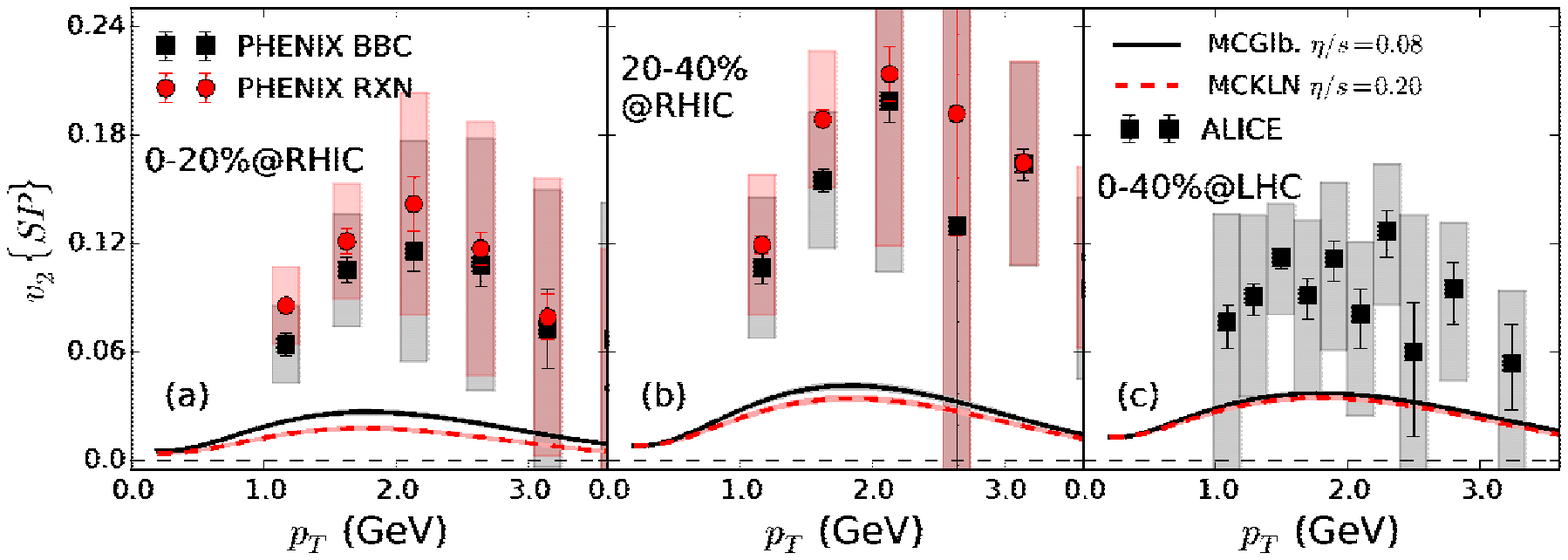}
  \caption{Comparison of {\em direct} photon (prompt + thermal (QGP+HG)) elliptic flow
      from event-by-event viscous hydrodynamics with recent experimental data from (a)
      0-20\% and (b) 20-40\% central  Au+Au collisions at RHIC
       \protect\cite{PHENIX1} and (c) from 0-40\% central Pb+Pb collisions
     at the LHC \protect\cite{Lohner:2012ct}. The solid black (dashed red) lines correspond to MCGlb
     (MCKLN) initial conditions evolved with a shear viscosity $\eta/s{\,=\,}0.08$ (0.2),
       respectively. The figure is taken from Ref.~\protect\cite{Shen:2013cca}.  }
  \label{fig2}
\end{figure}

\subsubsection*{The direct photon $v_2$ ``puzzle"}

The recent observation by the PHENIX Collaboration \cite{PHENIX1}
that the elliptic flow $v_2(p_T) $ of {\em direct} photons produced
in minimum bias Au+Au collisions at $\sqrt{s_{NN}}=200$~GeV is
comparable to that of the produced pions was a surprise and in
contrast to the theoretical expectations and predictions
\cite{Chatterjee:2005de,Liu:2009kq,Dion:2011vd,Dion:2011pp,Chatterjee:2013naa}.
Indeed, the photons produced by partonic interactions in the
quark-gluon plasma phase have not been expected to show a
considerable flow because - in a hydrodynamical picture - they are
dominated by the emission at high temperatures, i.e. in the initial
phase before the elliptic flow fully develops. Since the {\em
direct} photon $v_2(\gamma^{dir})$ is a 'weighted average' ($w_i$)
of the elliptic flow of individual contributions $i$,
\be \label{dir2}  v_2 (\gamma^{dir}) = \sum _i  v_2 (\gamma^{i}) w_i
(p_T) =  \frac{\sum _i  v_2 (\gamma^{i}) N_i (p_T)}{\sum_i N_i
(p_T)}, \ee
a large QGP contribution gives a smaller $v_2(\gamma^{dir})$.

A sizable photon $v_2$ has been observed also by the ALICE
Collaboration \cite{Wilde:2012wc,Lohner:2012ct} at the LHC. None of
the theoretical models could describe simultaneously the photon
spectra and $v_2$ which may be noted as a ``puzzle" for theory (cf.
Fig. \ref{fig2} in case of viscuous hydro calculations by Shen et
al. in Ref. \cite{Shen:2013cca}). Moreover, the PHENIX and ALICE
Collaborations have reported recently the observation of non-zero
triangular flow $v_3$ (see Refs. \cite{Ruan:2014kia,BockQM14}).
Thus, the consistent description of the photon experimental data
remains a challenge for theory and has stimulated a couple of new
ideas and developments that are briefly outlined in the following.

\subsubsection*{Developments in hydrodynamical models}

The following developments in the hydrodynamical modeling of the
heavy-ion collision evolution and the photon rates were stimulated
by the puzzling disagreement between the models and the photon data (cf. Fig. \ref{fig2}).

I.) The first hydrodynamical calculations on photon spectra were
based on  ideal hydrodynamics with smooth Glauber-type initial conditions
(cf. Ref. \cite{PHENIXlast}). The influence of {\it event-by-event
(e-b-e) fluctuating initial conditions} on the photon observables
was investigated within the (2+1)D Jyv\"askyl\"a ideal hydro model
\cite{Chatterjee:2013naa} which includes the equilibrated QGP and
Hadron Gas (HG) fluids. It has been shown that 'bumpy' initial
conditions based on the Monte-Carlo Glauber model lead to a slight
increase at high $p_T$ ($>$ 3 GeV/$c$) for the yield and $v_2$ which
is, however, not sufficient to explain the experimental data -- see
the comparison of model calculations with the PHENIX data in Figs.
7,8 of Ref. \cite{Chatterjee:2013naa} and with the ALICE data in Figs.
9,10 of Ref. \cite{Chatterjee:2013naa}.

II.) The influence of {\it viscous corrections} on photon spectra
and anisotropic flow coefficients $v_n$ has been investigated in two
independent viscous hydro models: 1) (3+1)D MUSIC
\cite{Dion:2011vd,Dion:2011pp} which is based on ``bumpy" e-b-e
fluctuating initial conditions from IP-Glasma and includes viscous
QGP (with lQCD EoS) and HG fluids; 2) (2+1)D VISH2+1
\cite{Shen:2014cga} with 'bumpy' e-b-e fluctuating initial
conditions from the Monte-Carlo Glauber model and viscous QGP (with
lQCD EoS) and HG fluids. The photon rate has been modified in Refs.
\cite{Dion:2011vd,Dion:2011pp,Shen:2014cga} in order to account for
first order non-equilibrium (viscous) corrections to the standard
equilibrium rates (i.e. the thermal QGP \cite{Arnold:2001ms} and HG
\cite{Turbide:2003si} rates). It has been found that the viscous
corrections only slightly increase the high $p_T$ spectra compared
to the ideal hydro calculations while they have a large effect on
the anisotropic flow coefficients $v_n$.
% - see Fig. \ref{fig:generic}.
Interesting to note that the viscous suppression of hydrodynamic
flow for photons is much stronger than for hadrons. Also the photon
$v_n$ coefficients are more sensitive to the QGP shear viscosity
which might serve the photon flow observables as a {\em QGP
viscometer} as suggested in Ref. \cite{Shen:2014cga}.

\begin{figure}
\hspace{-0.5cm}
\includegraphics[width=0.38\textwidth]{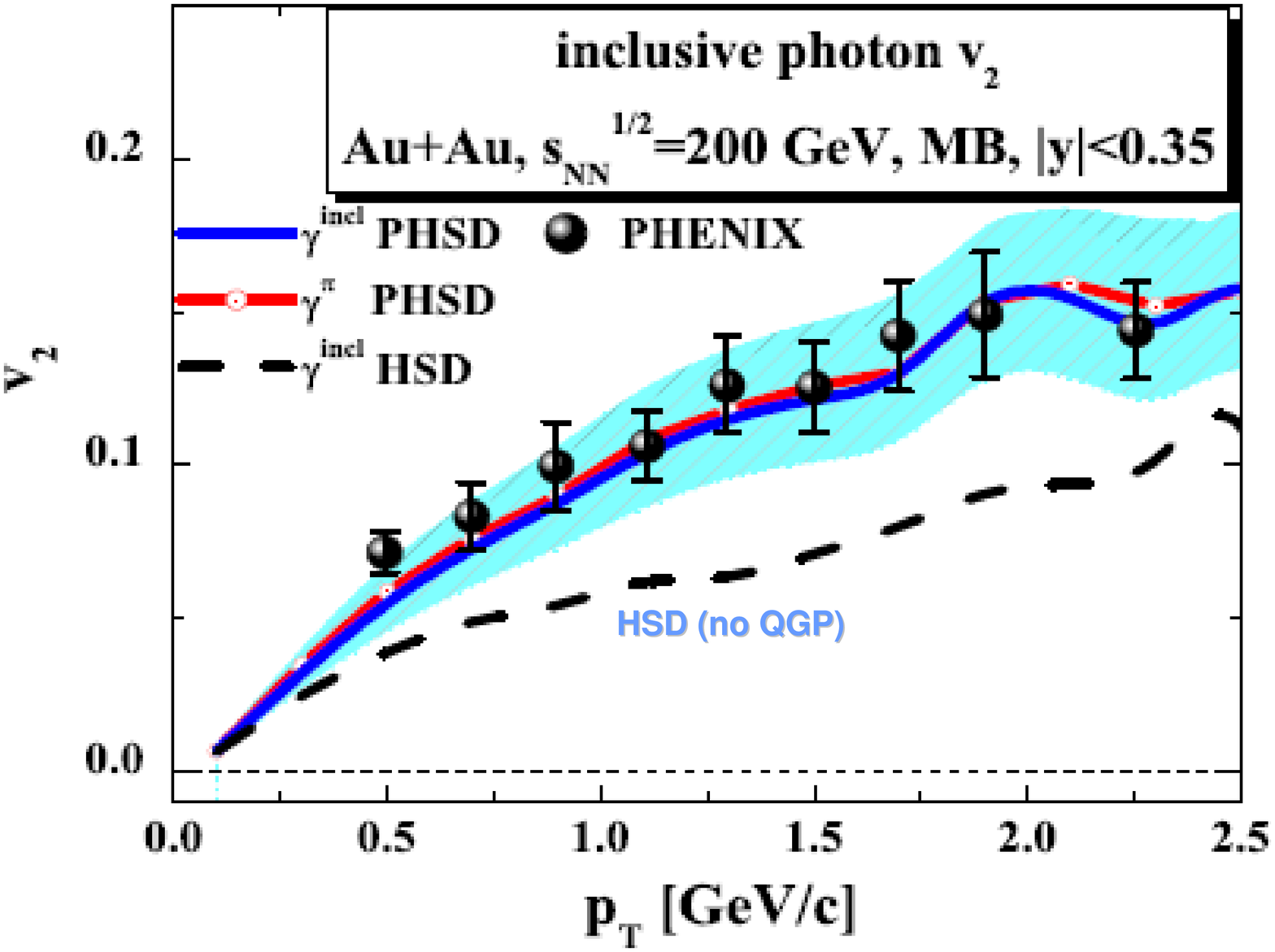}
\hspace{-0.75cm}
\includegraphics[width=0.32\textwidth]{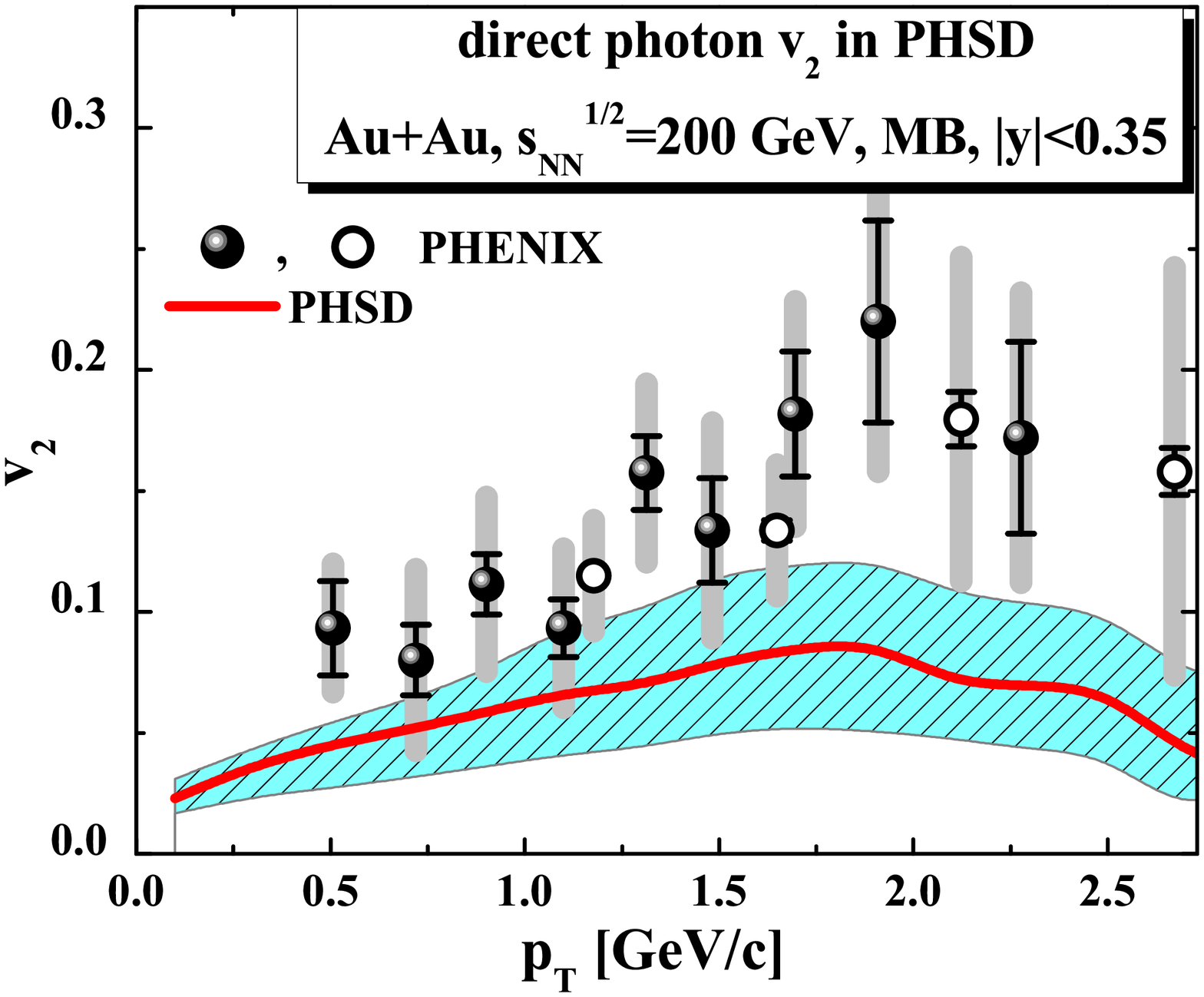}
\includegraphics[width=0.34\textwidth]{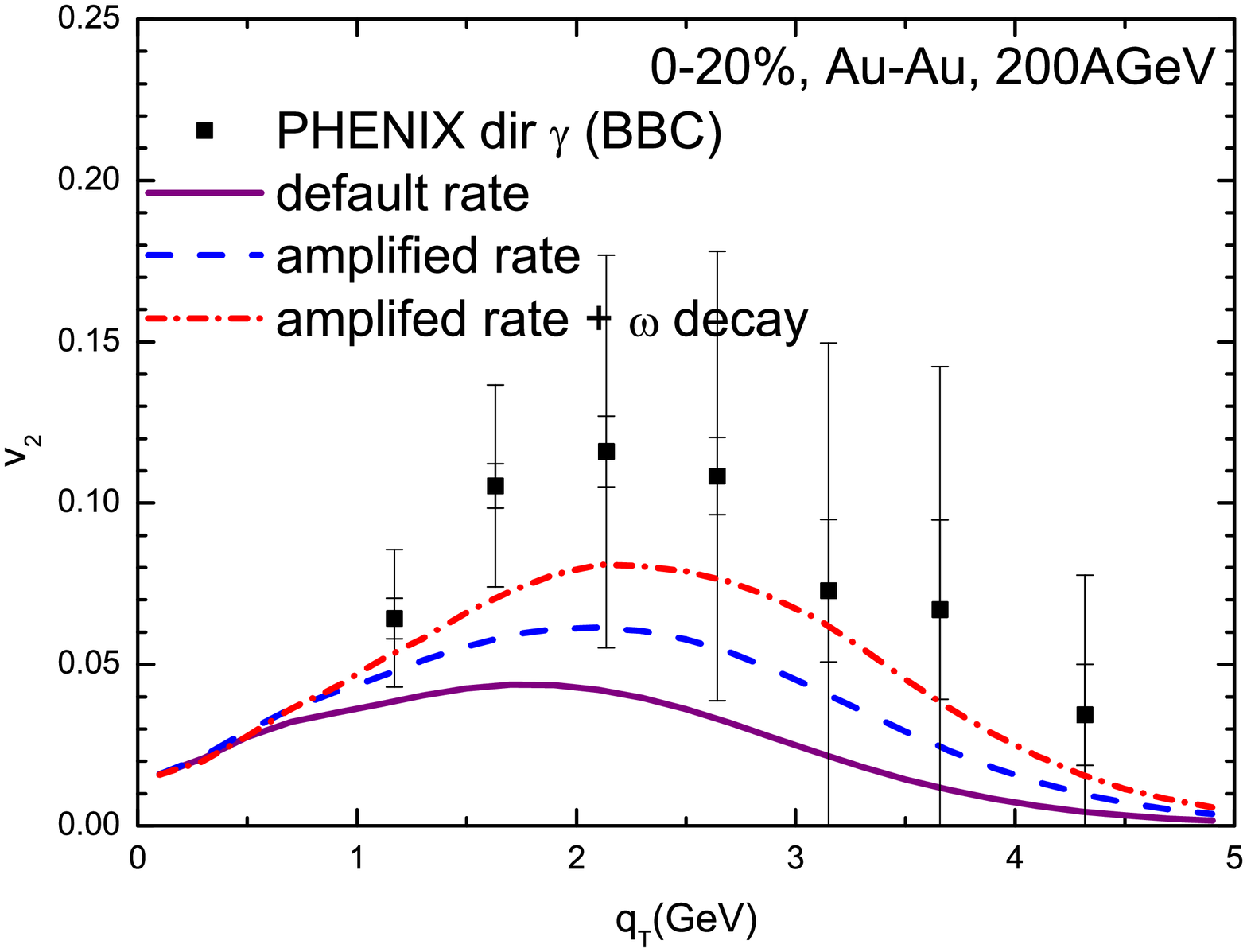}
\caption{{\em Inclusive} (left) and {\em direct} (middle) photon
elliptic flow coefficient $v_2(p_T)$ from the PHSD approach in
comparison to the PHENIX data \protect\cite{PHENIX1} for midrapidity
minimum bias Au+Au collisions at $\sqrt{s}=200$ GeV. The figures are
taken from
Ref.~\protect\cite{Linnyk:2013hta,Linnyk:2013wma,Linnyk:2015tha}. ({
r.h.s.}) {\em Direct} photon spectra $v_2$
 from the fireball model at RHIC when adding $\omega
  \rightarrow\pi^0 + \gamma$ decays at thermal freeze-out to the scenario
  with amplified rates (dash-dotted line), compared to the amplified-rate
  (dashed line) and default-rate (solid line) scenarios.
  The figure is taken from Ref.\protect\cite{vanHees:2014ida}.
 }
\label{fig:v2PHSD} \label{rhicV2}
\end{figure}

It is important to stress that the state-of-art hydro models
discussed above reproduce well the hadronic 'bulk' observables (e.g.
rapidity distributions, $p_T$ spectra and $v_2, v_3$ of hadrons).
However, in spite of definite improvements of the general dynamics
by including the fluctuating initial conditions (IP-Glasma or
MC-Glauber type) and viscous effects, the hydro models underestimate
the spectra and $v_2$ of photons at RHIC and LHC energies.

III.) Another idea, which has been checked recently within the
(2+1)D VISH2+1 viscous hydro model by Shen et al.
\cite{Shen:2014cga}, corresponds to the generation of {\it
'pre-equilibrium' flow} (see Ref. \cite{Shen:2014lpa}). The idea of
'initial' flow has been suggested in Ref. \cite{vanHees:2011vb} and
modeled as a rapid increase of bulk $v_2$ in the expanding fireball
model which leads to a substantial enhancement of photon $v_2$. In a
viscous hydro model \cite{Shen:2014lpa} the generation of
pre-equilibrium flow has been realized using a free-streaming model
to evolve the partons to 0.6 fm/c where the Landau matching takes
over to switch to viscous hydro. Such a scenario leads to a quick
development of momentum anisotropy with saturation near the critical
temperature $T_c$.  Although the pre-equilibrium flow effect
increases the photon $v_2$ slightly this is not sufficient to
reproduce the ALICE data (the same holds for the PHENIX data at RHIC
energies). Note, that the actual strength of such an effect depends
on the way of its modeling (cf. Ref. \cite{vanHees:2011vb}).
Moreover, the physical origin of such 'initial' (pre-equilibrium)
flow has to be justified/found first before robust conclusions can
be drown.

One may speculate about the possible effects on  photon observables
from further improvements of hydro models such as an inclusion of
the finite bulk viscosity as well as other transport coefficients
and their temperature dependence  etc. However, the failure of the
state-of-art viscous hydro models to describe the photon observables
is striking  although the hadronic 'bulk' dynamics is well
reproduced.

\begin{figure}[t]
\hspace{0.2cm} \includegraphics[width=0.95\textwidth]{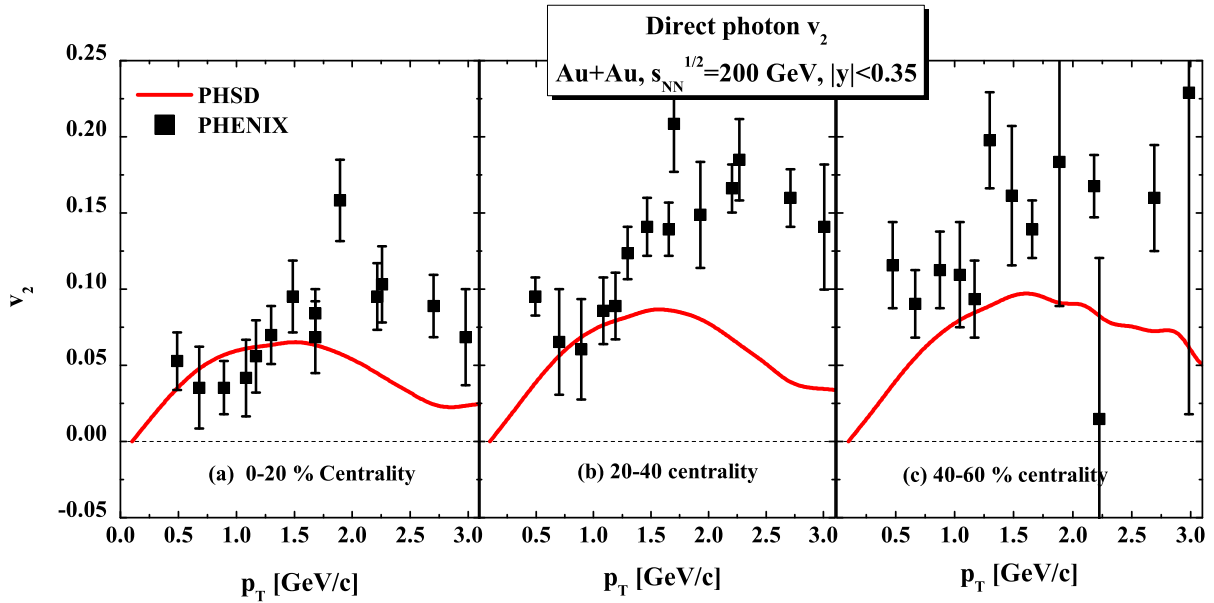}
\caption{ Centrality dependence of the {\em direct} photon $v_2$ for
Au+Au collisions at $\sqrt{s_{NN}}$ = 200 GeV for different
centralities (see legend); the data from the PHENIX
Collaboration~\protect\cite{Adare:2014fwh,Adare:2015lcd} are
compared to the earlier PHSD predictions from
Ref.~\protect\cite{Linnyk:2013wma}.
 } \label{v2sum}
\end{figure}

\subsubsection*{Photons from non-equilibrium transport}

In order to shed some light on the photon $v_2$ puzzle outlined
above, we consider the  influence of {\it non-equilibrium dynamics}
on the photon production in the following. As a 'laboratory' for
that we will employ the microscopic PHSD  transport approach that
has been derived and described in Sections 2-4, while the
implementation of photon production by the various partonic and
hadronic channels has been explained in Section 5. Since the
elliptic flow of pions (or charged hadrons) is under control in PHSD
in comparison to the data from the PHENIX, STAR and ALICE
Collaborations (cf.
Refs.~\cite{Konchakovski:2014fya,Linnyk:2013hta,PHENIX1,Adams:2004bi,Adler:2003kt});
also the spectrum of their {\em decay} photons is  predicted
reliably by the approach. This allows for a solid computation of the
{\em direct} photon yield at all energies from SPS to LHC.

In the PHSD the {\em direct} photon $v_2(\gamma^{dir})$ is
calculating by building the weighted sum of the channels, which are
not subtracted by the data-driven methods, as follows: the photons
from the quark-gluon plasma, from the initial hard parton collisions
(pQCD photons), from the decays of short-living resonances
($a_1$-meson, $\phi$-meson, $\Delta$-baryon), from the binary
meson+meson and meson+baryon channels ($\pi+\rho\to\pi+\gamma$,
$\pi+\pi\to\rho+\gamma$, $V+p/n\to n / p+\gamma$), and from the
bremsstrahlung in the elastic meson+meson and meson+baryon
collisions ($m+m\to m+m+\gamma$, $m+B\to m+B+\gamma$). The {\em
direct} photon $v_2$ is extracted by summing up the elliptic flow of
the individual channels contributing to the {\em direct} photons,
using their contributions to the spectrum as the relative
$p_T$-dependent weights, $w_i(p_T)$, cf. Eq.~\ref{dir2}.

\begin{figure}[t]
\hspace{0.5cm}\includegraphics[width=0.45\textwidth]{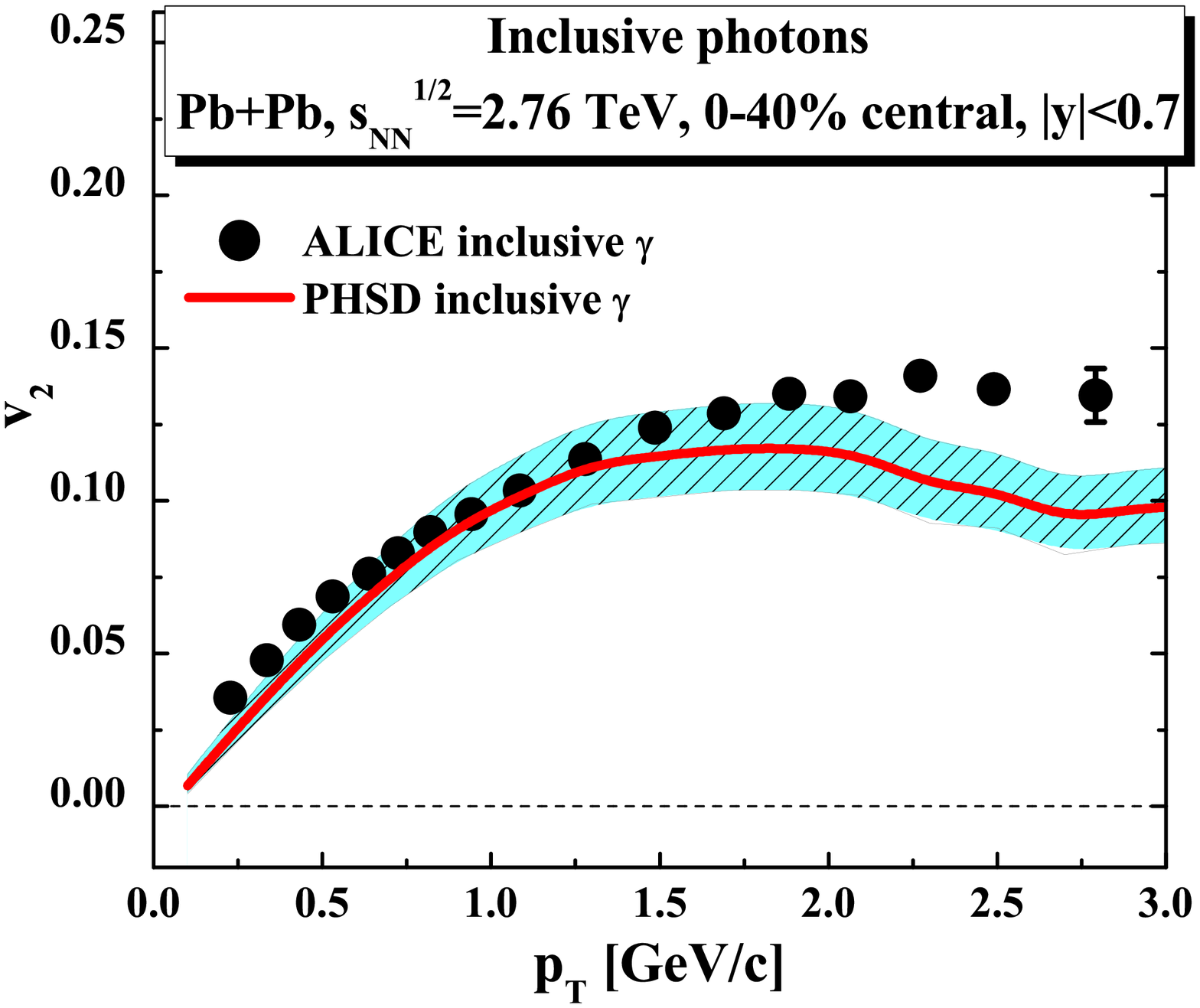}
\hspace{0.2cm}
\includegraphics[width=0.47\textwidth]{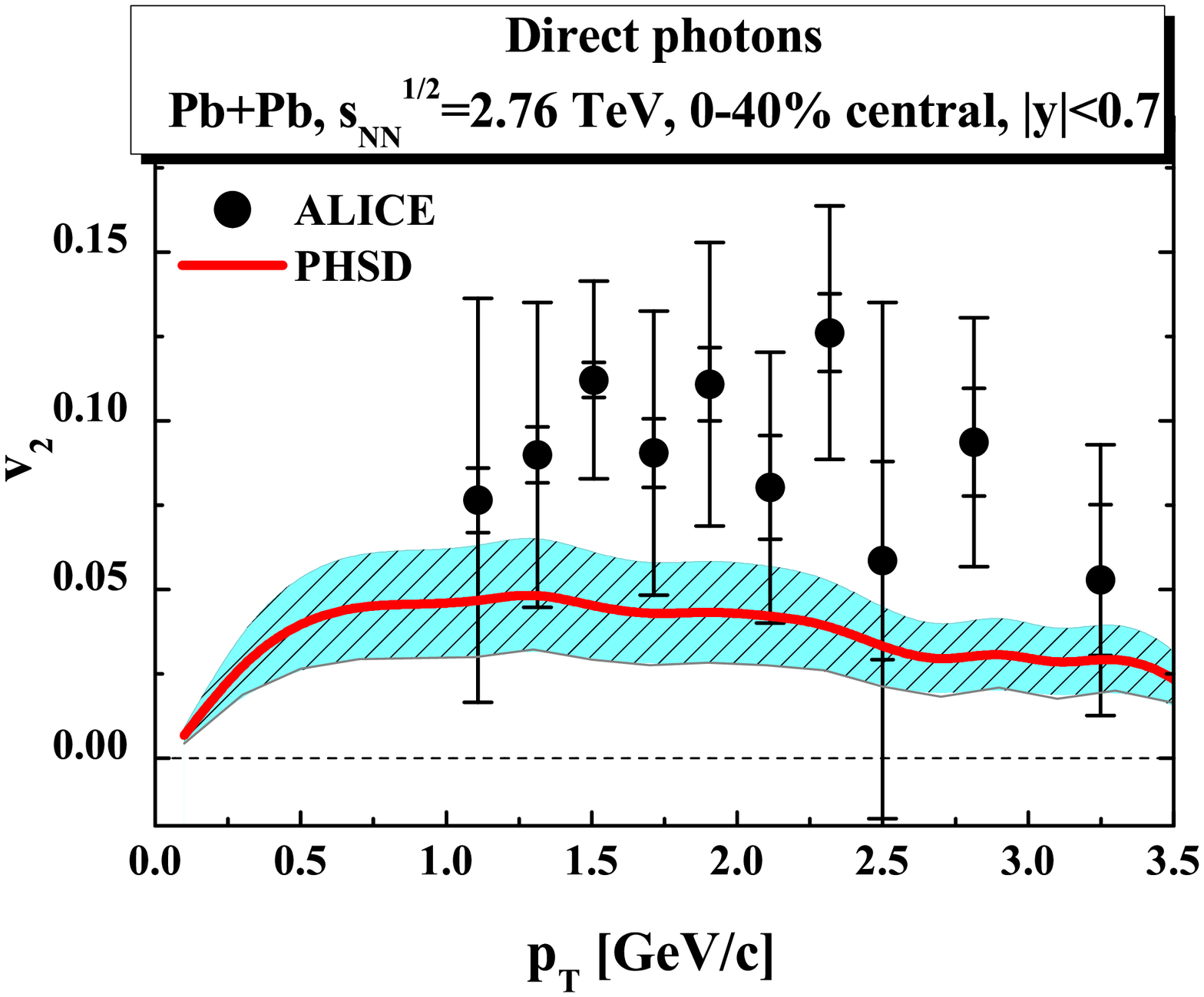}
\caption{ ({ l.h.s.}) Elliptic flow $v_2$ versus transverse momentum
$p_T$ for the {\em inclusive} photons produced in  0-40\% central
Pb+Pb collisions at $\sqrt{s_{NN}}=2.76$~TeV as calculated by the
PHSD (solid red line); the blue error band reflects the finite
statistics and the uncertainty in the modeling of the cross sections
for the individual channels. \label{inclv2lhc}  ({ r.h.s.}) Elliptic
flow $v_2$ versus transverse momentum $p_T$ for the {\em direct}
photons produced in  0-40\% central Pb+Pb collisions at
$\sqrt{s_{NN}}=2.76$~TeV as predicted by the PHSD (solid red line);
the blue error band is dominated by the uncertainty in the modeling
of the cross sections for the individual channels. The data from the
ALICE Collaboration are taken from
Ref.~\protect\cite{Morreale:2014spa}. } \label{lhcv2dir}
\end{figure}

The results for the elliptic flow $v_2(p_T)$ of {\em direct photons}
produced in $Au+Au$ collisions at the top RHIC energy are shown in
the middle panel of Fig.~\ref{rhicV2} while the elliptic flow in
the left panel  in comparison to the PHENIX data  \cite{PHENIX1}.
Since the inclusive photons dominantly stem from $\pi^0$ decay the
left panel of Fig. \ref{rhicV2} demonstrates again that the pion
$v_2$ is under control in PHSD while HSD calculations (dashed line)
fail substantially.
%In comparison to the previous results within the
%PHSD approach~\cite{Linnyk:2013hta,Linnyk:2013wma}, the elliptic
%flow in the intermediate region of the transverse momenta
%$1.0<p_T<2.0$~GeV is reduced by about 50\% due to the improvements
%in our treatment of the bremsstrahlung channels beyond the
%soft-photon approximation.
According to the PHSD calculations for the {\em direct} photon
spectra almost half of the {\em direct} photons measured by PHENIX
(in central collisions) stems from the collisions of quarks and
gluons in the deconfined medium created in the initial phase of the
collision. The photons produced in the QGP carry a very small $v_2$
and lead to an overall {\em direct} photon $v_2$ about a factor of 2
below the pion $v_2(\pi)$ even though the other channels in the sum
(\ref{dir2}) have large elliptic flow coefficients $v_2$ of the
order of $v_2(\pi)$ (cf. Fig.~7 of Ref.~\cite{Linnyk:2013hta}). This
leads to a  final elliptic flow for {\em direct} photons which is
about half of the measured $v_2$ in PHSD. The right panel of
Fig.~\ref{rhicV2} shows the photon $v_2$ from the fireball model of
van Hees et al. \cite{vanHees:2014ida} for different scenarios: the
solid line corresponds to the 'default scenario', which is
comparable to the PHSD results for $v_2$ (middle panel). The dashed
line is obtained when amplifying the production rate close to $T_c$
while the dash-dotted line additionally includes the photons from
$\omega$-decay at freeze-out. We note that in PHSD we do not find an
enhanced photon rate close to $T_c$ (cf. Fig. 27) and the
$\omega$-decay contributions are included by default. In summary, we
conclude that the PHSD results lead to a {\em direct} photon $v_2$
at RHIC which is substantially larger that that from hydro
calculations (cf. Fig. 29) but still underestimates the PHENIX data
at RHIC.

\begin{figure}[t]
\centerline{\includegraphics[width=0.45\textwidth]{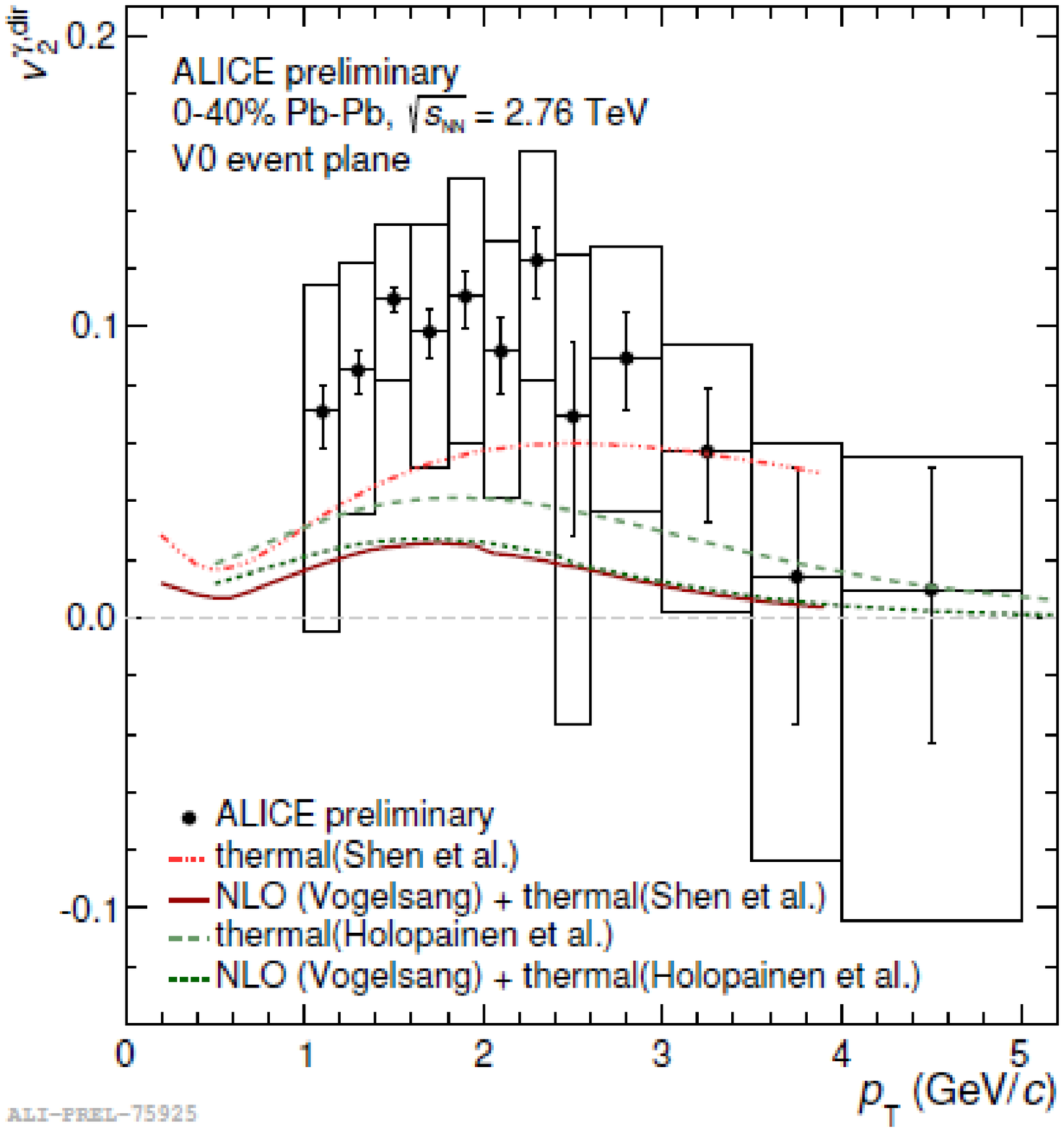}\hspace{0.4cm}
\includegraphics[width=0.49\textwidth]{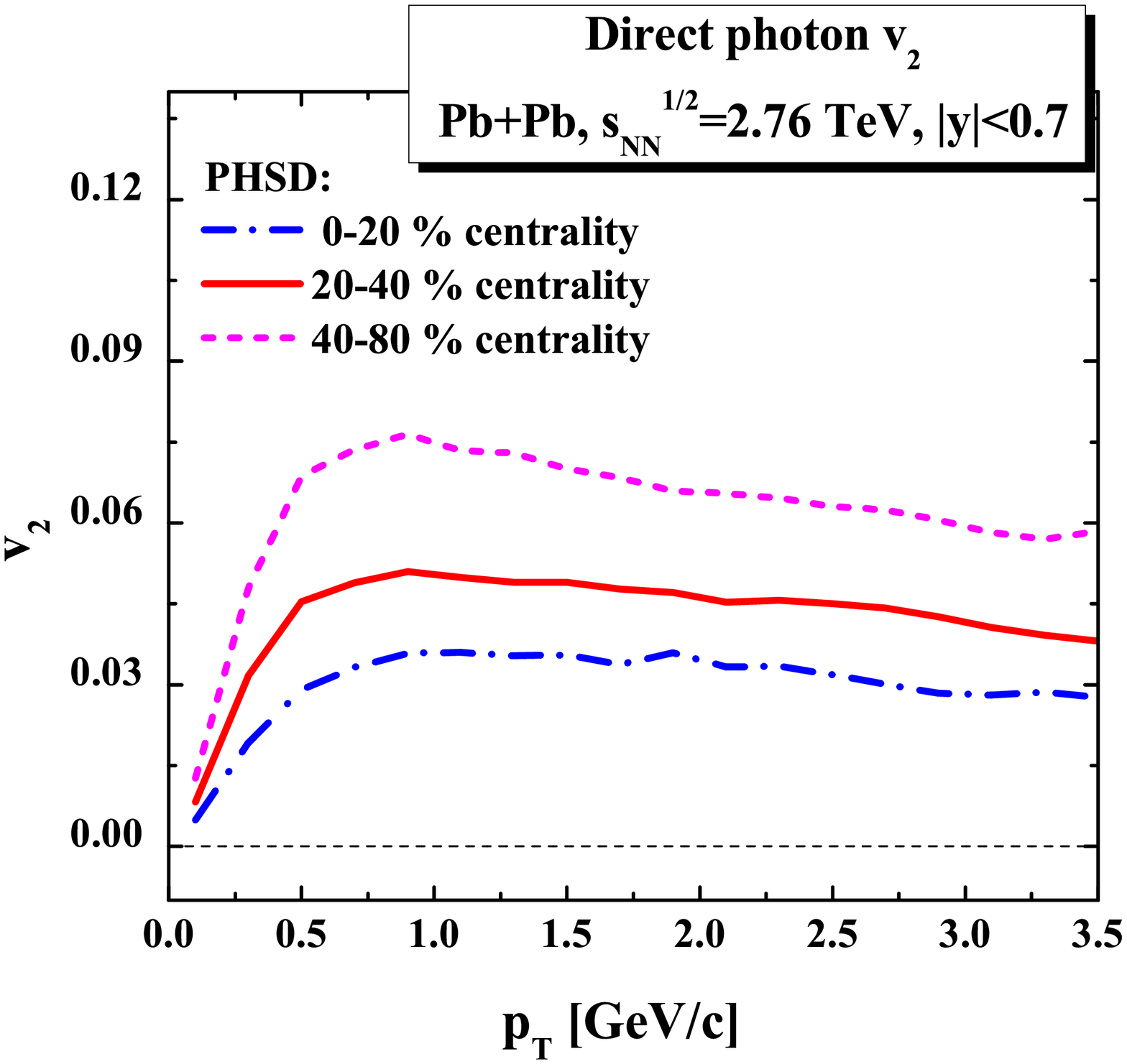}}
\caption{(l.h.s.) Preliminary data of the ALICE Collaboration for
the {\em direct} photon elliptic flow $v_2$ in comparison to
theoretical calculations from
Refs.~\protect\cite{Shen:2013cca,Gordon:1994ut,Holopainen:2011pd}.
The figure is taken from Ref.~\protect\cite{Morreale:2014spa}.
(r.h.s.) Predictions for the elliptic flow $v_2$ of {\em direct}
photons from PHSD versus transverse momentum $p_T$ produced in Pb+Pb
collisions at $\sqrt{s_{NN}}$ = 2.76 TeV for different centrality
classes (see legend). } \label{alicepho}
\end{figure}

The PHSD results are readily understood as follows: the partonic
collisions -- producing photons in the QGP -- take place throughout
the evolution of the collision. The initial high partonic collision rate falls rapidly
with time in PHSD and is followed up by a significant amount of hadronic production channels.
Thus the production of photons from the QGP is no longer
dominated by the early times. As a consequence, the elliptic flow
`picked up' by the photons from the parent parton collisions
saturates after about 5~fm/c and reaches a relatively low value of
about $0.02$, only. We note that a delayed production of charges
from the strong gluon fields
(`glasma'~\cite{McLerran:2011zz,Chiu:2012ij,Blaizot:2012qd,Gale:2012rq})
might shift the QGP photon production to somewhat later times when
the elliptic flow is built up more but this also decreases the
amount of QGP photons! However, we cannot quantitatively ans\-wer
whether the additional evolution in the pre-plasma state could
generate considerable additional $v_2$ while reproducing the photon
spectra.

A preliminary summary of the current situation  is displayed in Fig.
\ref{v2sum} where the photon elliptic flow from PHENIX is compared
to the PHSD predictions for different centrality classes 0-20\% (a),
20-40\% (b) and 40-60\% (c). Whereas the elliptic flow is roughly
described in the most central class there is an increasing tendency
to underestimate in the PHSD the strong elliptic flow especially for
peripheral collisions where some additional source might be present.
Thus the observed centrality dependence of the elliptic flow is
roughly in agreement with the interpretation that a large fraction
of the {\em direct} photons is of hadronic origin (in particular
from the bremsstrahlung in meson+meson and meson+baryon collisions);
the latter contribution becomes stronger or even dominant in more
peripheral collisions. But more precise data will be mandatory for a
robust conclusion.

We finally present the PHSD predictions/calculations for the
elliptic flow of inclusive and {\em direct} photons produced in
$Pb+Pb$ collisions at the energy of $\sqrt{s_{NN}}=2.76$~TeV at the
LHC within the acceptance of the ALICE detector. Since the pion
$v_2$ is described well within the PHSD at $\sqrt{s_{NN}}$ = 2.76
TeV this is expected also for the inclusive photon $v_2$ due to the
dominance of photons from $\pi^0$ decay. The left panel of
Fig.~\ref{lhcv2dir} presents predictions/calculations for the
elliptic flow $v_2$ versus transverse momentum $p_T$ for the {\em
inclusive} photons produced in 0-40\% central Pb+Pb collisions at
$\sqrt{s_{NN}}=2.76$~TeV (solid red line) with the blue error band
reflecting the finite statistics and the theoretical uncertainty in
the modeling of the cross sections for the individual channels. The
elliptic flow $v_2(p_T)$ of {\em direct} photons produced in 0-40\%
central Pb+Pb collisions at $\sqrt{s_{NN}}=2.76$~TeV as predicted by
the PHSD (solid red line) is shown in the right panel of
Fig.~\ref{lhcv2dir}, the blue error band is dominated by the
uncertainty in the modeling of the cross sections for the individual
channels.  As in case of the PHENIX data at RHIC  the preliminary
data of the ALICE Collaboration for the {\em direct} photon elliptic
flow $v_2$ for the same centrality  are slightly higher than the
PHSD predictions (although compatible within error bars). The
different lines in Fig. \ref{alicepho} (l.h.s.) show the {\em
direct} photon $v_2(p_T)$ from the theoretical calculations in Refs.
\protect\cite{Shen:2013cca,Gordon:1994ut,Holopainen:2011pd} (see
legend) which are similar to the PHSD predictions or even below. The
situation at the LHC energy of $\sqrt{s_{NN}}$ = 2.76 TeV is thus
comparable to the one at the top RHIC energy and the $v_2$ puzzle
remains.

We, furthermore, provide predictions for the centrality dependence
of the {\em direct} photon $v_2(p_T)$ in Pb+Pb collisions at
$\sqrt{s_{NN}}$ = 2.76 TeV in the  centrality classes 0-20\%,
20-40\% and 40-80\%  which are of relevance for the upcoming
measurements by the ALICE Collaboration at the LHC. The actual
results from PHSD are displayed in Fig. \ref{alicepho} (r.h.s.) and
show a very similar centrality dependence as in case of Au+Au
collisions at the top RHIC energy.

We note that there are other scenarios towards the solution of the
{\em direct} photon $v_2$ puzzle proposed during the 'Quark
Matter-2014 Conference': early-time magnetic field effects
\cite{Basar:2012bp,Basar:2014swa}, Glasma effects
\cite{McLerran:2014hza},  or non-perturbative effects of a
'semi-QGP' \cite{Hidaka:2014uia}. We discard an explicit discussion
of these suggestions.

%-----------------------------------------------------------------------------

\subsection{Triangular flow of direct photons}
\label{sect:v3}

\begin{figure}[t]
 \includegraphics[width=\textwidth]{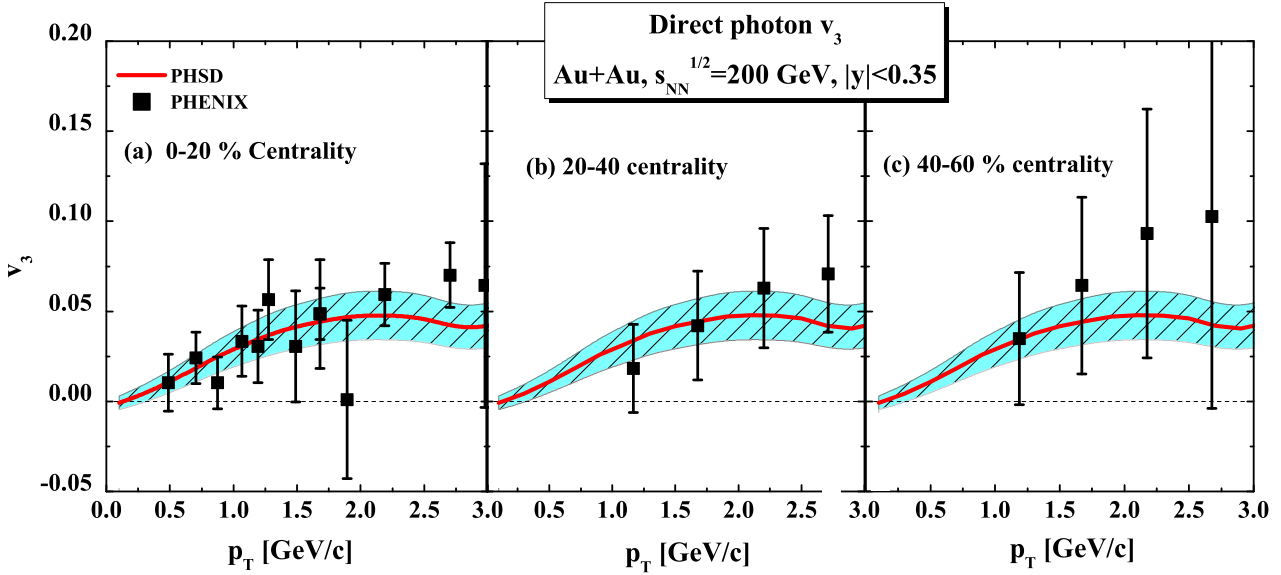}
\caption{(Color on-line) Triangular flow $v_3$ versus transverse
momentum $p_T$ for the {\em direct} photons produced in Au+Au
collisions at $\sqrt{s_{NN}}=200$~GeV in three centrality classes
(see legends).
%; the blue band reflects the uncertainty in
%the modeling of the cross sections for the individual channels.
The PHSD results are shown by the solid red lines in comparison to
the data of the PHENIX Collaboration (black symbols)
taken from Ref.~\protect{\cite{Bannier:2014bja,Adare:2015lcd}}.}
\label{v3}
\end{figure}

We have seen in the previous Subsections that the measured spectra
of {\em direct} photons could be reproduced by the PHSD calculations
at least within a factor of 2 (which is comparable with the current
accuracy of the measurements). Also, the {\em inclusive} photon
$v_2$ was well described and the elliptic flow of {\em direct}
photons was qualitatively in line with the data and attributed
essentially to hadronic sources although still underestimating the
data.

On the other hand, there exists an alternative interpretation of the
strong elliptic flow of {\em direct} photons, in which the azimuthal
asymmetry of the photons is due to the initial strong magnetic field
essentially produced by spectator charges (protons). Indeed, the
magnetic field strength in the very early reaction stage reaches up
to $eB_y \approx 5 m_\pi^2$ in semi-peripheral $Au+Au$ collisions at
$\sqrt{s_{NN}}=200$~GeV (see the calculations within the PHSD in
Ref.~\cite{Voronyuk:2011jd}; comparable estimates have been obtained
also in Refs.~\cite{Tuchin:2014pka,Tuchin:2012mf,Skokov:2009qp}).
These strong magnetic fields might influence the photon production
via the polarization of the medium, e.g. by influencing the motion
of charged quarks in the QGP, or by directly inducing a real photon
radiation via the virtual photon ($\vec B$-field) coup\-ling to a
quark loop and (multiple) gluons; the photons are then produced
azimuthal asymmetrically with positive $v_2$.

The photon production under the influence of magnetic fields has
been calculated in
Refs.~\cite{McLerran:2014hza,Bzdak:2012fr,Tuchin:2014pka,Tuchin:2012mf,
Basar:2012bp,Skokov:2013axa}. The observed spectra and elliptic flow
of {\em direct} photons could be explained using suitable
assumptions on the conductivity, bulk viscosity or degree of
chemical equilibration in the early produced matter. The common
feature of these calculations was that the {\em triangular} flow
coefficient $v_3$ of the {\em direct} photons was expected to be
very small. Indeed, the magnetic field may lead to an azimuthal
asymmetry $v_2$ but not to a triangular mode $v_3$.

Consequently, it is of interest to measure experimentally the third
flow coefficient $v_3(p_T)$ and to compare it to the calculations in
the different classes of models: (a) those attributing the large
elliptic flow and strong yield of {\em direct} photons dominantly to
hadronic sources, e.g. the PHSD transport approach; (b) the models
suggesting the large azimuthal asymmetry and additional yield of
{\em direct} photons to be caused by the early magnetic fields; (c)
the models assuming that the yield of {\em direct} photons at low
$p_T$ is dominated by partonic channels.

\begin{figure}[t]
\includegraphics[width=0.5\textwidth]{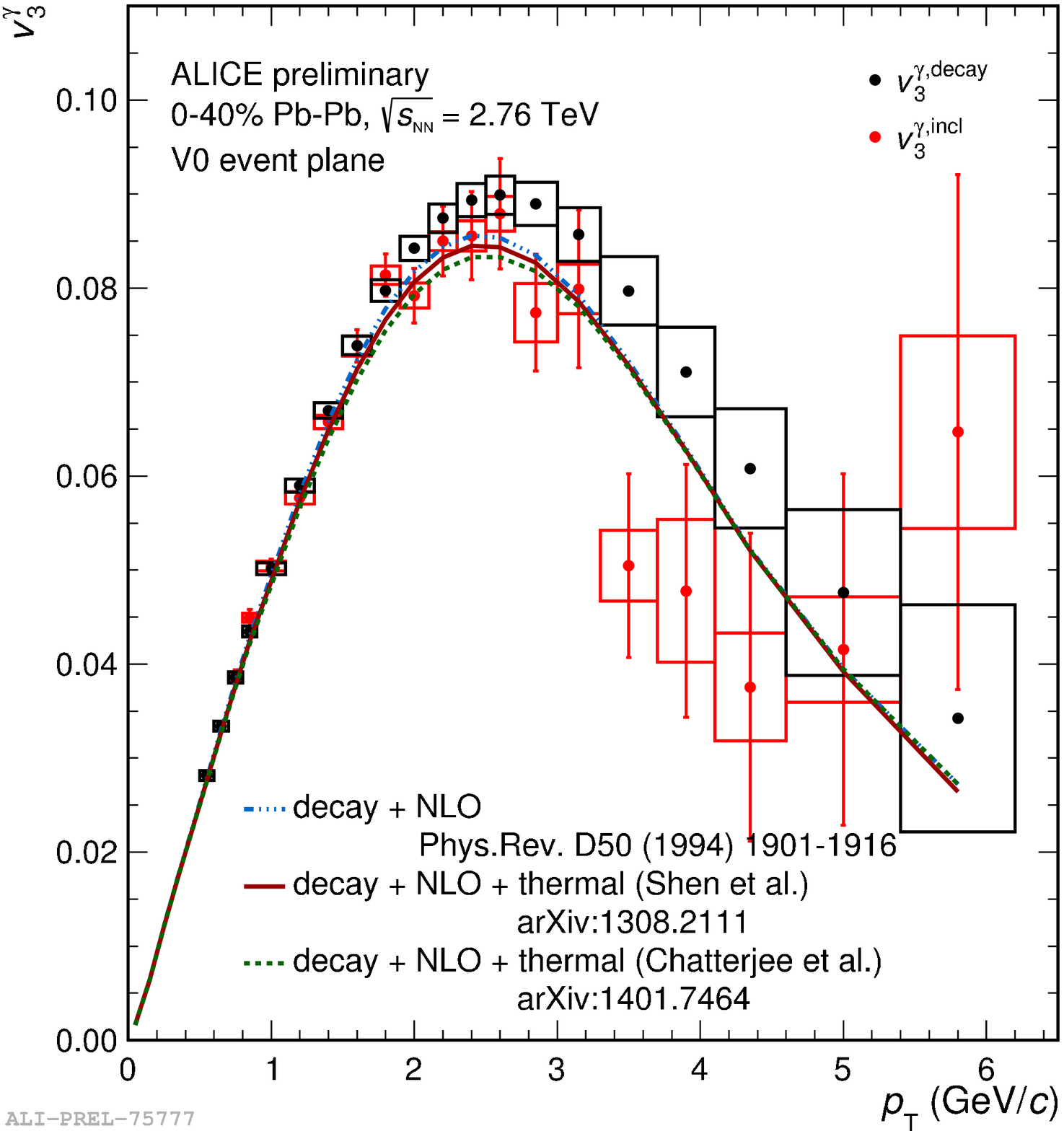}\hspace{0.2cm}
\includegraphics[width=0.49\textwidth]{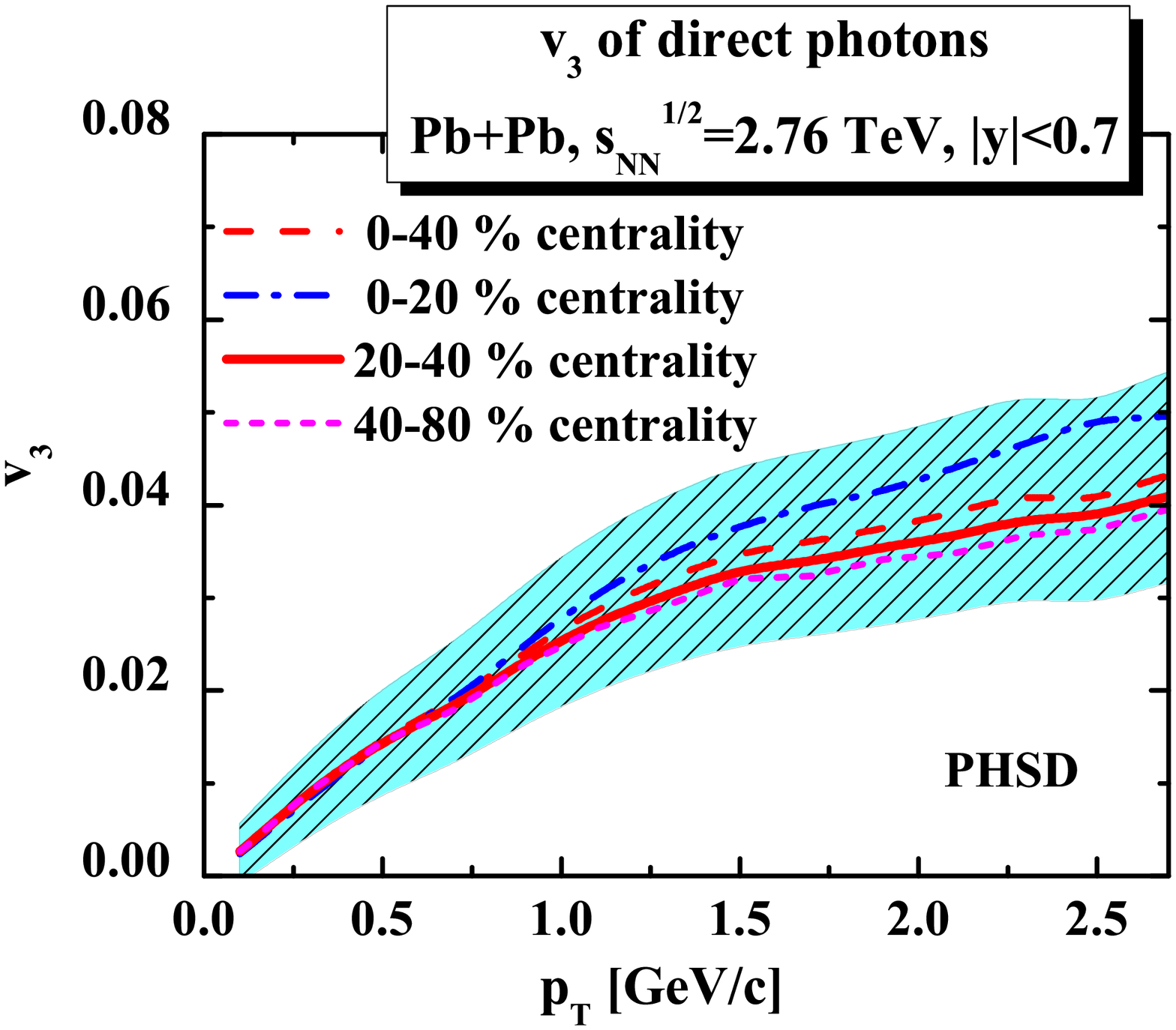}
\caption{(l.h.s.) Preliminary data of the ALICE Collaboration for
the inclusive photon $v_3(p_T)$. The lines represent contributions
of {\em decay} photons with added theoretical calculations from
Refs.~ \protect\cite{Shen:2013cca,Gordon:1994ut,Holopainen:2011pd}.
The figure is taken from Ref.~\protect\cite{Peressounko:2014nza}.
(r.h.s.) Predictions for the triangular flow $v_3$ versus transverse
momentum $p_T$ for the {\em direct} photons produced in  different
centrality classes for Pb+Pb collisions at $\sqrt{s_{NN}}=2.76$~TeV
from the PHSD (see legend); the blue band reflects the uncertainty
in the modeling of the cross sections for the individual channels
and give a measure of the present level of accuracy. The figure is
taken from Ref.~\protect\cite{Linnyk:2015tha}.
 } \label{Alicev3}
\end{figure}

In Fig.~\ref{v3} we present our results for the triangular flow
$v_3$ versus transverse momentum $p_T$ for the {\em direct} photons
produced in Au+Au collisions at $\sqrt{s_{NN}}=200$~GeV from the
PHSD (solid red lines) for 0-20\% (a), 20-40\% (b) and 40-60\% (c)
centrality.
%; the blue band reflects
%the uncertainty in the mode\-ling of the cross sections for the
%individual channels, in particular the bremsstrahlung.
The PHSD gives a positive non-zero triangular flow of {\em direct}
photons up to 6\% with very little centrality dependence on the
level of the present accuracy ($\sim 25\%$). The PHSD results are in
agreement with the data of the PHENIX Collaboration from
Refs.~\cite{Mizuno:2014via,Ruan:2014kia,Bannier:2014bja} which
suggests that the scenario (a) is at least compatible with the
measurements.

The preliminary data of the ALICE Collaboration for the $v_3$ of
inclusive photons in Fig. \ref{Alicev3} (l.h.s.) do not seem to
point towards an interpretation of the {\em direct} photons being
dominantly produced in the early stage under the influence of the
magnetic field (b), because the $v_3$ of these photons is expected
to be close to zero. Of course, the photon production in the
magnetic fields occurs on top of other channels, which may carry
finite triangular flow $v_3$. But the weighted sum of all the
channels including the magnetic-field-induced photons will give a
smaller $v_3\ne0$  than the sum without this channel. The scenario
(c) has been studied by other groups within a hydrodynamic modeling
of the collision  in Refs.~\cite{Shen:2013cca,Chatterjee:2014dqa}.
The triangular flow $v_3(p_T)$ of {\em direct} photons from
Refs.~\cite{Shen:2013cca,Chatterjee:2014dqa} is about a factor of 2
smaller than that obtained in the PHSD approach.

In Fig. \ref{Alicev3} (r.h.s.) we present predictions for the
triangular flow $v_3$ versus transverse momentum $p_T$ for the {\em
direct} photons produced in  different centrality classes for Pb+Pb
collisions at $\sqrt{s_{NN}}=2.76$~TeV from the PHSD (see legend);
the blue band reflects the uncertainty in the modeling of the cross
sections for the individual channels and give a measure of the
present level of accuracy. The centrality dependence of $v_3(p_T)$
turns out to be low and is practically constant within the accuracy
of the present PHSD calculations. An experimental  confirmation of
this expectation could further affirm the notion of large hadronic
contributions to the {\em direct} photons and in particular the
photon production via the bremsstrahlung in meson and baryon
collisions. It should be possible to differentiate between the
scenarios in the future, when  data of higher accuracy  and
information on the centrality dependence of {\em direct} photons
(especially on $v_2$ and $v_3$) will become available.

% part of the review in PPNP 2015

\section{Results on dilepton production in heavy-ion collisions}

\subsection{{SIS energies}}
 The dileptons produced in low energy heavy-ion collisions
have been measured  first  by the DLS Collaboration at Berkeley
\cite{Wilson:1997sr,Wilson:1993mp,Porter:1997rc}. The observed
dilepton yield  \cite{Porter:1997rc} in the mass range from 0.2 to
0.5 GeV in  C+C and Ca+Ca collisions at 1 A GeV was about of five
times higher than the calculations by different transport models
using the 'conventional' dilepton sources as bremsstrahlung,
$\pi^0-, \eta-, \omega-$ and $\Delta$-Dalitz decays and direct
vector mesons ($\rho, \omega, \phi$) decays
\cite{Xiong:1990bg,Wolf:1992gg,Bratkovskaya:1996bv}. Even when
including the different in-medium scenarios such as 'collisional
broadening' and 'dropping mass' for the $\rho$-meson spectral
function did not solve the ``DLS puzzle"
\cite{Ernst:1997yy,BratKo99,Bratkovskaya:1997mp,Fuchs:2005zga}.

 The recent experimental data from the HADES Collaboration
at GSI \cite{HADES07,Pachmayer:2008yn,Agakishiev:2009yf,Agakishiev:2011vf},
however, confirmed the measurement of the DLS Collaboration for  C+C
at 1.0 A GeV  \cite{Pachmayer:2008yn} as well as for the elementary
reactions \cite{HADES_pp22}.  In the mean time also
the theoretical transport approaches as well as effective models for
the elementary $NN$ reactions have been further developed.
 A possible solution of the ``DLS puzzle" from the
theoretical side has been suggested in Ref.
\cite{Bratkovskaya:2007jk} by incorporating stronger $pn$ and $pp$
bremsstrahlung contributions in line with the  updated
One-Boson-Exchange (OBE) model calculations from Kaptari and K{\"
a}mpfer \cite{Kaptari:2005qz}. As shown in Ref.
\cite{Bratkovskaya:2007jk} the results from the HSD approach  with
'enhanced' bremsstrahlung cross sections agree very well  with the
HADES  data for C+C at 1 and 2 A GeV as well as with the DLS data
for C + C and Ca + Ca at 1 A GeV, especially when including a
collisional broadening in the vector-meson spectral functions. A
similar finding has been obtained by other independent transport
groups, i.e. the IQMD \cite{Thomere:2007cj} and the Rossendorf BUU
\cite{Barz:2009yz} collaborations.

\begin{figure}[t]
%\phantom{a}\vspace*{5mm}
\hspace{.5cm}
\includegraphics*[width=7.8cm]{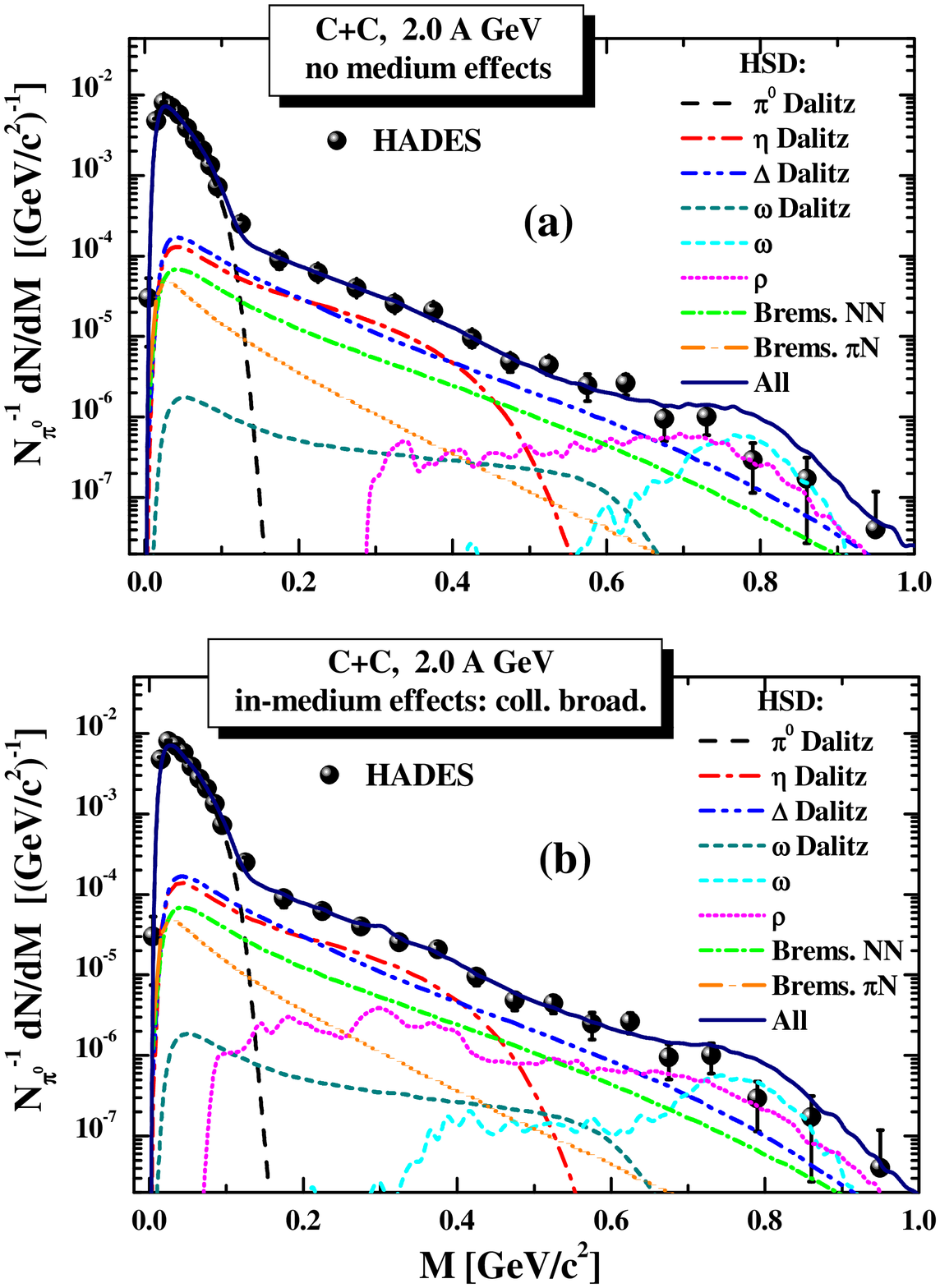}\includegraphics*[width=7.8cm]{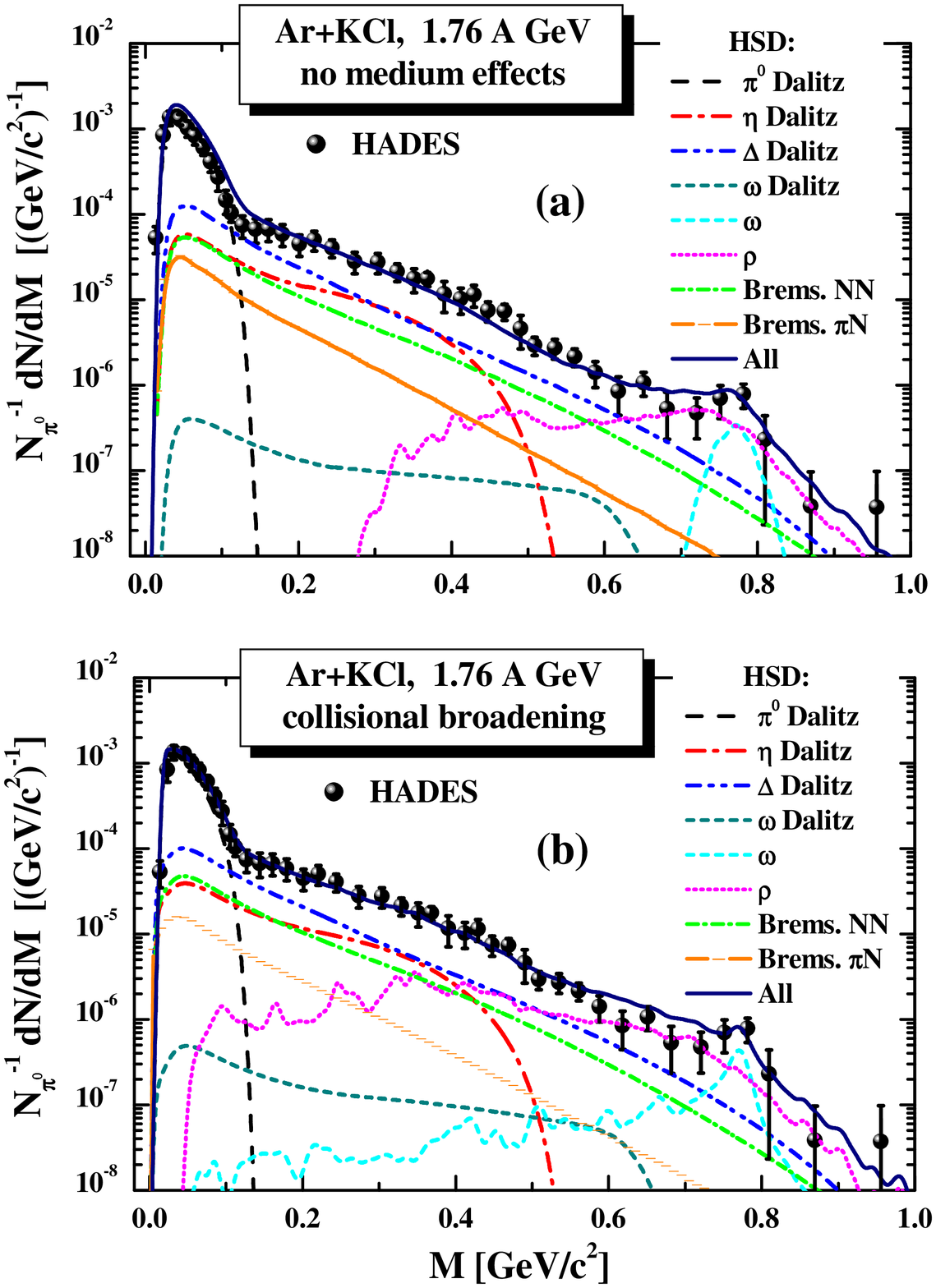}
\caption{The mass differential dilepton spectra - normalized to the
$\pi^0$ multiplicity
 - from PHSD/HSD calculations for C+C  at 2 $A$GeV (l.h.s.) and Ar+KCl
 at 1.76 $A$GeV (r.h.s.) in
comparison to the HADES data
\protect\cite{Agakishiev:2009yf,Agakishiev:2011vf}.  The upper parts (a)
shows the case of 'free' vector-meson spectral functions while the
lower parts (b) give the result for the 'collisional broadening'
scenario. The different colour lines display individual channels in
the transport calculation (see legend).  The theoretical
calculations passed through the corresponding HADES acceptance
filter and mass/momentum resolutions. The figures are taken
from Ref.~\protect\cite{BratAich}. } \label{Fig_CC20}
\end{figure}

\begin{figure}[t]
\hspace{0.8cm}
\includegraphics[width=7.3cm]{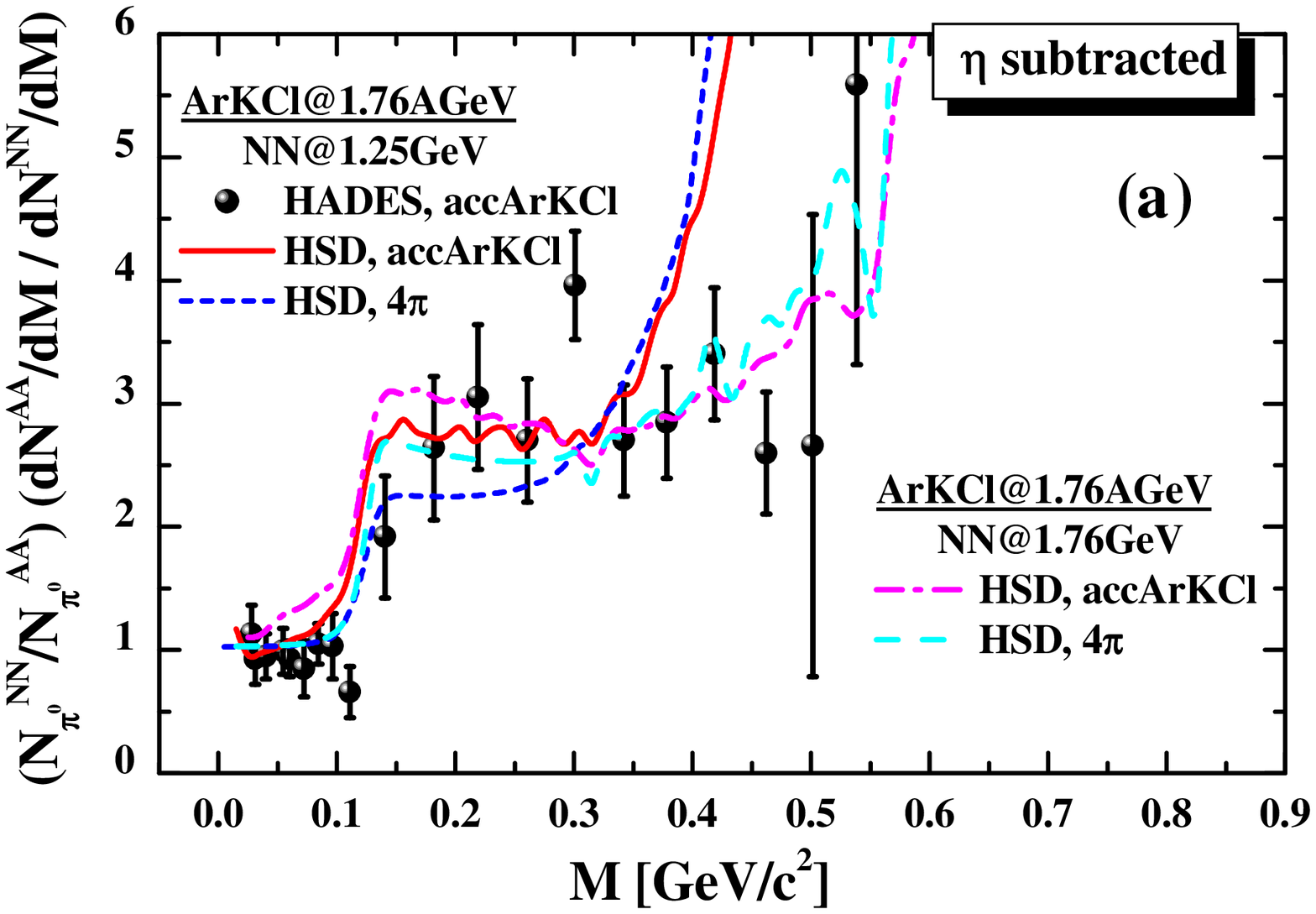}\hspace{0.3cm}
\includegraphics[width=7.3cm]{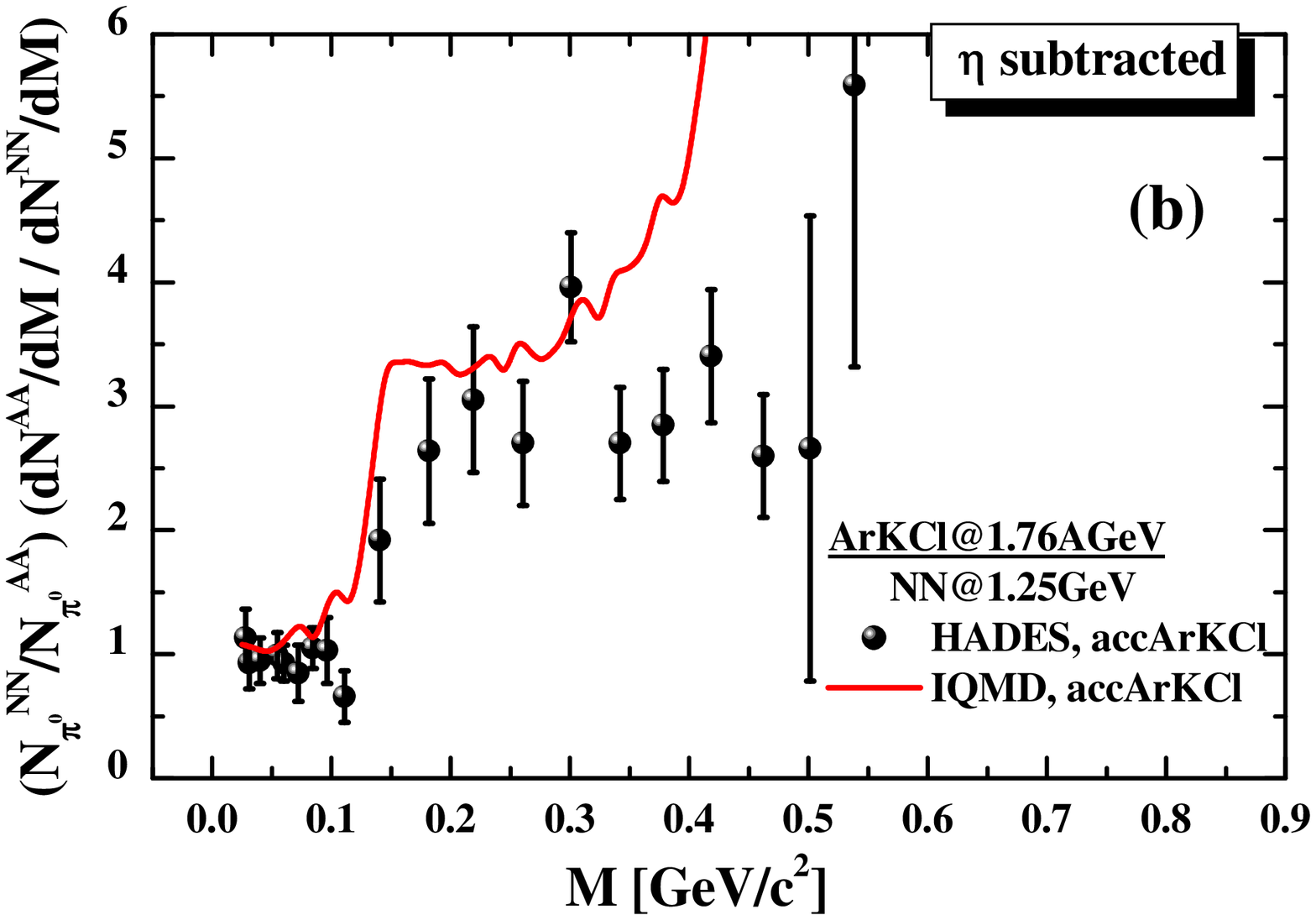}
\caption{ The PHSD/HSD (l.h.s.) and IQMD (r.h.s.) results for the
ratio of the dilepton differential spectra -- normalized to the
$\pi^0$ multiplicity and after $\eta$ Dalitz yield subtraction -- to
the isospin-averaged reference spectra $NN=(pp+pn)/2$ taken at 1.25
GeV, involving Ar+KCl experimental acceptance (solid line) and for
$4\pi$ (short dashed line). Also the PHSD/HSD results for the ratio
to the reference $NN$ spectra taken at 1.76 GeV are shown, with the
Ar+KCl experimental acceptance (dash-dotted line) and in $4\pi$
(dashed line). The figures are taken from
Ref.~\protect\cite{BratAich}.} \label{Fig_RArKNN}
\end{figure}
\begin{figure}
%\phantom{a}\vspace*{5mm}
%\includegraphics[width=8.cm]{R_AAa.eps}
%\hspace*{5mm}
%\includegraphics[width=8.cm]{R_AAb.eps}
%\vspace*{-5mm}
\hspace{0.7cm}
\includegraphics[width=0.9\textwidth]{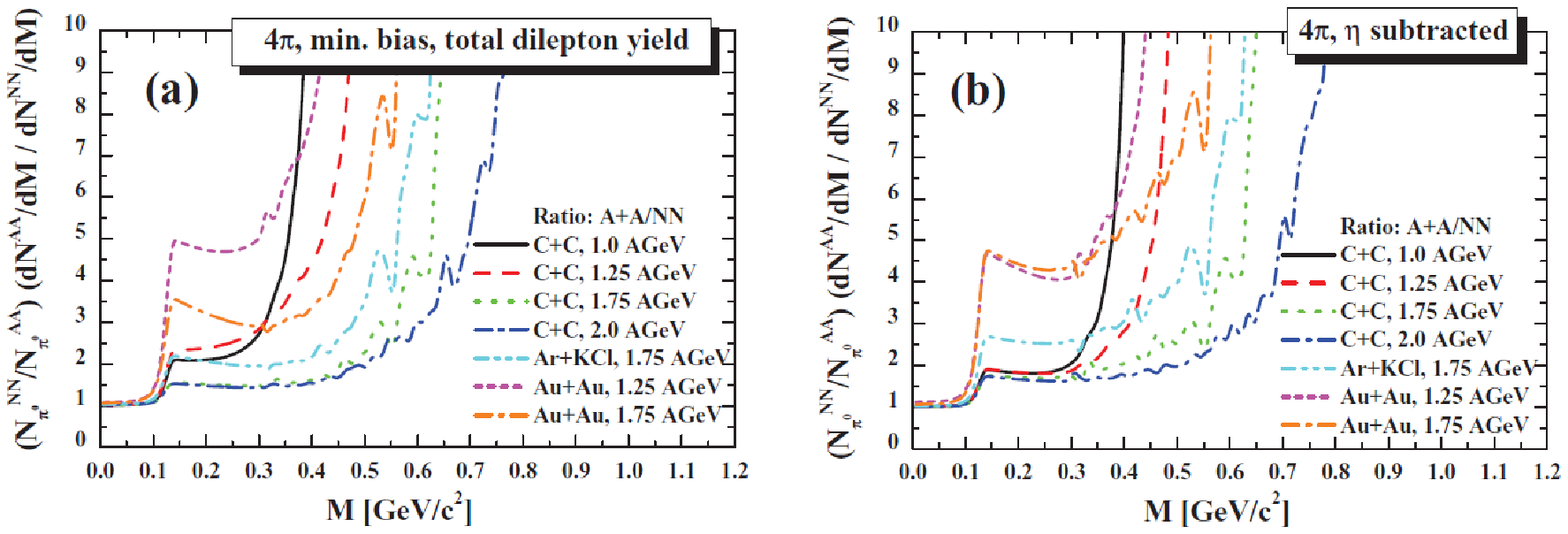}
\caption{ (l.h.s.) The ratio $(1/N_{\pi^0}^{AA}
dN^{AA}/dM)/(1/N_{\pi^0}^{NN} dN^{NN}/dM)$ of the invariant mass
differential dilepton $4\pi$ spectra - normalized to the $\pi^0$
multiplicity  - from HSD/PHSD calculations for minimum bias $A+A$
collisions: We display C+C, Ar+KCl, Au+Au collisions in comparison
to the isospin-averaged reference spectra $NN=(pn+pp)/2$  at 1.0,
1.25, 1.75, 2.0 $A$GeV. (r.h.s.) the same ratios but for the
dilepton spectra after $\eta$ Dalitz decay subtraction. The figures
are taken from Ref.~\protect\cite{BratAich}.} \label{Fig_RAA}
\end{figure}

Since all (relevant) elementary dilepton channels have been
described in Section 5 we may step on with the actual results for
A+A reactions. Note that at SIS energies of 1--2 AGeV the HSD and
PHSD results are equivalent because no partonic subsystems are
formed. Fig. \ref{Fig_CC20} (l.h.s.) shows the mass differential
dilepton spectra -- normalized to the $\pi^0$ multiplicity -- from
PHSD/HSD calculations for C+C at 2 $A$GeV in comparison to the HADES
data \cite{Agakishiev:2009yf}. The theoretical calculations passed
through the corresponding HADES acceptance filters and mass/momentum
resolutions  which leads to a smearing of the spectra at high
invariant mass and particularly in the $\omega$-resonance region.
The upper part shows  the case of 'free' vector-meson spectral
functions while the lower part presents the result for the
'collisional $\rho$ broadening' scenario. Here the difference
between the in-medium scenarios is of minor importance and partly
due to the limited mass resolution which smears out the spectra.
Fig. \ref{Fig_CC20} (r.h.s.) displays the mass differential dilepton
spectra - normalized to the $\pi^0$multiplicity - from HSD
calculations for the heavier system Ar+KCl at 1.76 $A$GeV  in
comparison to the HADES data \cite{Agakishiev:2011vf}.  The upper
part shows again the case of 'free' vector-meson spectral functions
while the lower part gives the result for the collisional broadening
of the $\rho$-meson. Also in this data set the enhancement around
the $\rho$ mass is clearly visible. For the heavier system the
'collisional broadening' scenario shows a slightly better agreement
with experiment than the 'free' result and we expect that for even
heavier systems the difference between the two approaches increases.

Some rather model independent results are  expected when comparing
the dilepton mass spectra from A+A to 1/2 (pp+pn) reactions
normalized to the $\pi^0$ multiplicity and subtracting the (known)
$\eta$-Dalitz decay contribution.
The left panel of  Fig. \ref{Fig_RArKNN} shows the PHSD/HSD results for
the ratio of the dilepton differential spectra - normalized to the
$\pi^0$ multiplicity and after $\eta$ Dalitz yield subtraction - to
the isospin-averaged reference spectra $NN=(pp+pn)/2$ taken at 1.25
GeV and employing the Ar+KCl experimental acceptance (solid line)
and in $4\pi$ (short dashed line).  We display as well the ratio of
Ar+KCl at 1.76 $A$GeV to the reference $NN$ spectrum at the same
energy, including the experimental Ar + KCl acceptance
(dash-dotted line) and in $4\pi$ (dashed line).  These results show
clearly that for invariant masses of $0 .1 \ GeV < M < 0.35$ GeV the
data as well as theory  are not a mere superposition of the
elementary spectra. The comparison also excludes that this
enhancement, observed in heavy-ion collisions, is due to acceptance
since the results with acceptance and in $4\pi$ are very similar. At
larger invariant masses theory and data do not agree because  the
bump at the invariant masses around   $ M\approx 0.5$ GeV, seen in
the experimental $pd$ reactions, is not reproduced by theory. Taking
the reference spectra at the same nominal energy the theory predicts
that this enhancement is constant up to energies of $ M\approx 0.5$
GeV. Then the Fermi motion becomes important and yields a strong
increase of the ratio. These PHSD/HSD results are confirmed by the IQMD
calculations shown in the right panel of Fig. \ref{Fig_RArKNN}.

\begin{figure}[t]
\phantom{a}\vspace*{5mm}
\centerline{\includegraphics[width=13cm]{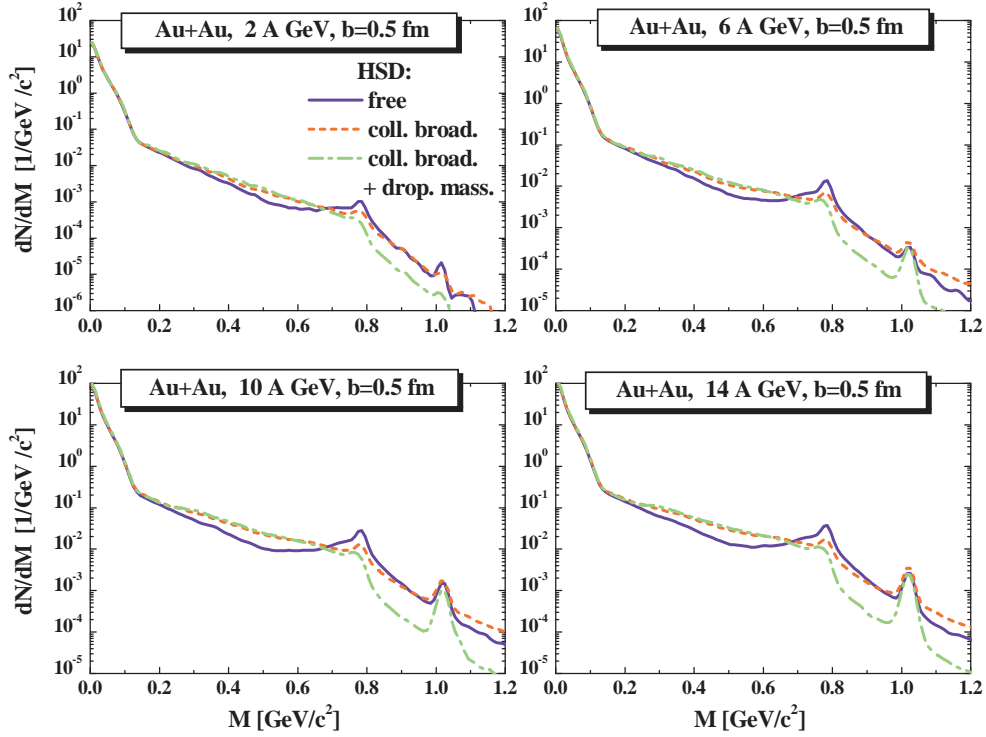}}
\vspace{0.1cm} \caption{Predictions of the HSD/PHSD transport
calculation for the mass differential dilepton spectra for very
central Au+Au collisions from 2 to 14 A GeV calculated for different
in-medium scenarios - collisional broadening and the combined
scenario (dropping mass + collisional broadening). Figure is taken
from Ref.~\protect\cite{Friman:2011zz}.} \label{Dil_Au}
\end{figure}

We note that with increasing mass A+A of the system the low mass
dilepton regime from roughly 0.15 to 0.5 GeV in the transport
calculations increases due to multiple $\Delta$-resonance production
and Dalitz decay. The dileptons from intermediate $\Delta$'s, which
are part of the reaction cycles $\Delta \to \pi N ; \pi N \to
\Delta$ and $NN\to N\Delta; N\Delta \to NN$, escape from the system
while the decay pions do not \cite{BratAich}.   With increasing
system size more generations of intermediate $\Delta$'s are created
and the dilepton yield is enhanced accordingly. In inclusive C+C
collisions there is only a moderate enhancement relative to scaled
p+p and p+n collisions due to the small size of the system while in
Ar+KCl reactions already several (3-4) reaction cycles become
visible. A similar finding has been obtained within the IQMD
transport model (cf. Figs. 25, 27, 29 in Ref. \cite{BratAich}). This
effect enhances with the system size and reaches a factor of 4.5-5.0
for Au+Au minimum bias at 1.25 A GeV as illustrated in Fig.
\ref{Fig_RAA} which presents the ratio $(1/N_{\pi^0}^{AA}
dN^{AA}/dM)/(1/N_{\pi^0}^{NN} dN^{NN}/dM)$ of the mass differential
dilepton spectra - normalized to the $\pi^0$ multiplicities  -
obtained in  HSD/PHSD calculations. Displayed are the ratios of
minimum bias C+C, Ar+KCl, Au+Au collisions and of  the
isospin-averaged reference spectra $NN=(pn+pp)/2$ at the same energy
(l.h.s.).  The right hand side depicts the same ratios but for the
dilepton spectra after $\eta$-Dalitz yield subtraction. Additionally
to the $\Delta$ regeneration, the $pN$ bremsstrahlung -- which
scales with the number of collisions and not with the number of
participants, i.e. pions -- contributes to the enhancement of the
ratio in Fig. \ref{Fig_RAA}.

Based on the study in Ref.~\cite{BratAich} this enhancement can be
attributed to two effects: i) the bremsstrahlung radiation from $pn$
and $pp$ reactions which does not scale with the pion number (i.e.
the number of participants) but rather with the number of elementary
elastic collisions; ii) the shining of dileptons from the
'intermediate' $\Delta$'s, which take part in the $\Delta \to \pi N$
and $ \pi N \to \Delta$ reaction cycle.  This cycle produces a
number of generations of $\Delta$'s during the reaction which
increases with the size of the system. At the end only one pion is
produced but each intermediate $\Delta$ has contributed to the
dilepton yield because the emitted dileptons are not absorbed
(unlike pions). This leads to an enhancement of the dilepton yield
when compared to the final number of pions.  Thus, the enhancement
confirms the predictions of transport theories that in heavy-ion
collisions several generations of $\Delta$'s are formed which decay
and are recreated by $\pi N \to \Delta$ reactions. Accordingly, the
dilepton data from $A+A$ reactions shed light on the
$\Delta$-resonance dynamics in the medium especially at SIS
energies.

\subsection{AGS energies}

In the AGS energy regime from 2 to 14 A GeV no dilepton data have
been taken so far but are foreseen in the HADES and CBM experiments
at FAIR. We thus show in Fig. \ref{Dil_Au}  the HSD/PHSD predictions
for the dilepton yields from central Au + Au collisions calculated
for different energies from 2 to 14 A GeV applying the different
in-medium scenarios: collisional broadening and the combined
approach (dropping mass + collisional broadening). One can see from
Fig. \ref{Dil_Au} that both scenarios lead to an enhancement of the
dilepton yield in the mass region $M =0.3-0.8$ GeV  by a factor of
about 2. The largest in-medium effect is, however, attributed to the
reduction of the dilepton yield between the $\omega$ and $\phi$
peaks due to the downward shift of the poles of the $\rho$ and
$\omega$ spectral functions in case of the dropping mass scenario.
However, the latter scenario is not consistent with existing
experimental data at higher energies (see Subsection 7.3), so one
has to rely most likely on the relatively modest in-medium effects
due to a collisional broadening of the vector mesons in the medium
(dashed line).

\begin{figure}[t]
%\begin{minipage} [c] {0.65\textwidth}
\hspace{0.4cm}
\includegraphics[width=0.45\textwidth]{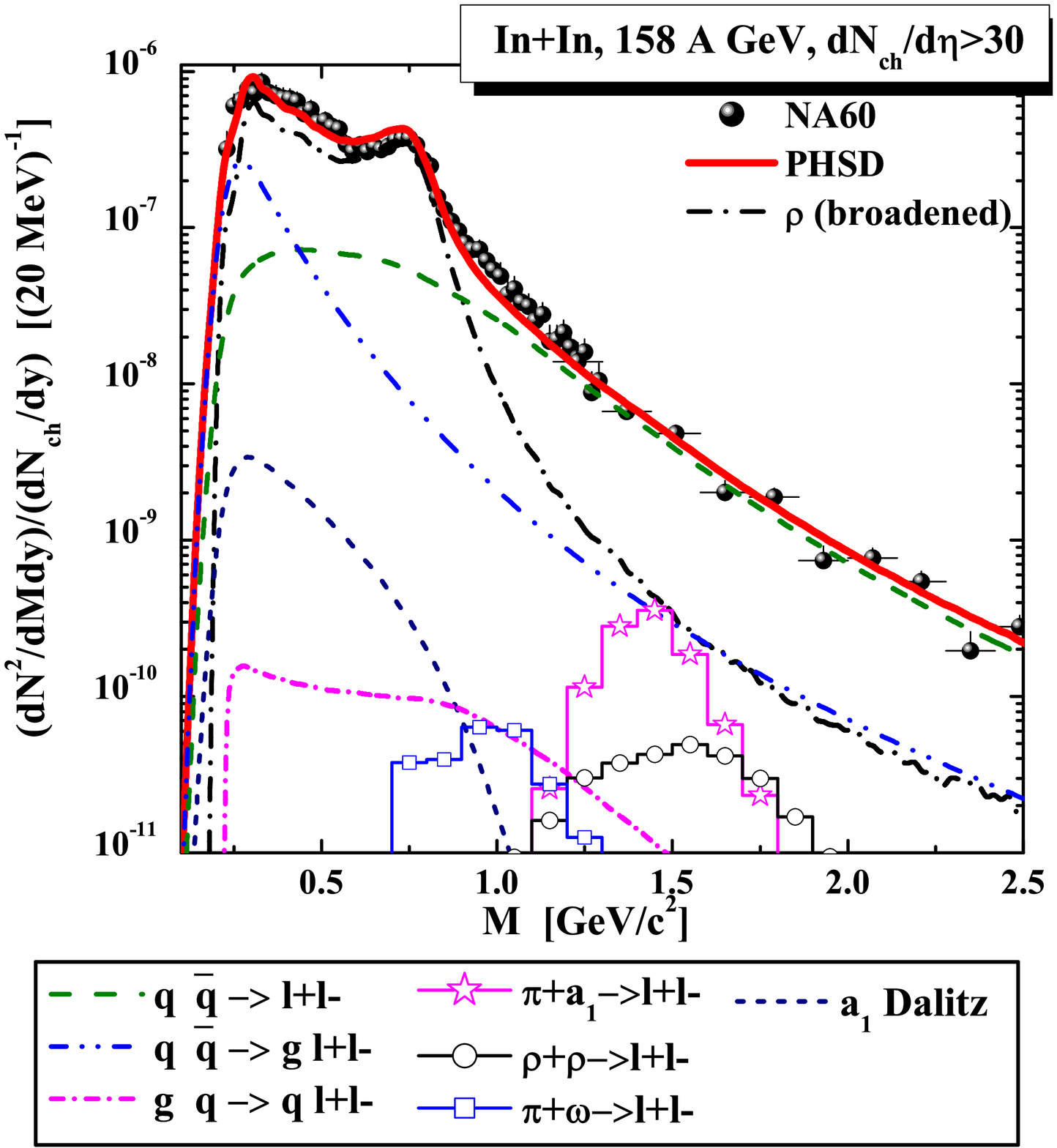} \hspace{0.4cm}
\includegraphics[width=0.46\textwidth]{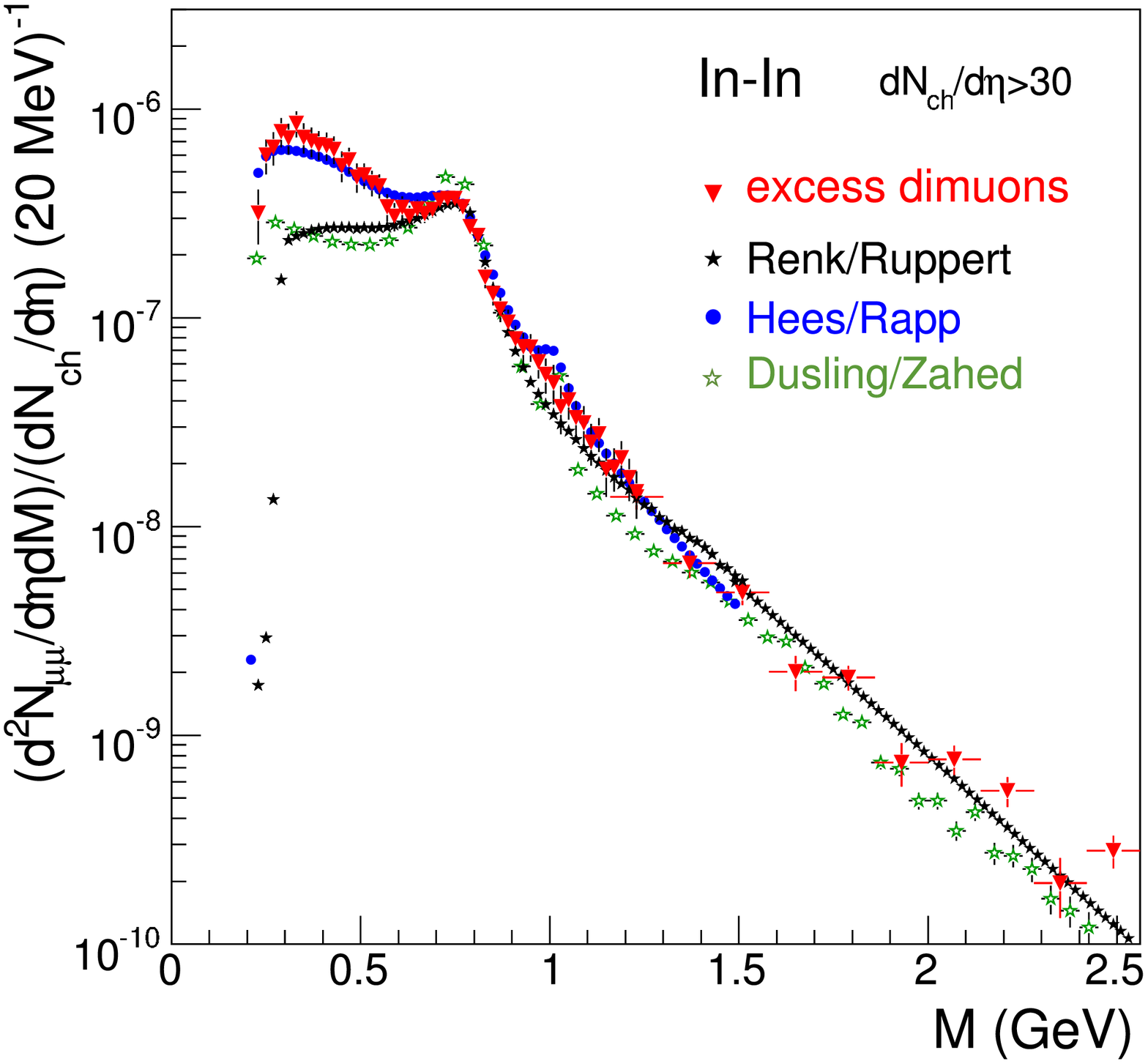}
%\end{minipage}
\caption{ ({ l.h.s.}) Acceptance corrected mass spectra of excess
dimuons from In+In at 158~A GeV integrated over $p_T$ in
$0.2<p_T<2.4$~GeV from PHSD compared to the data of
NA60~\cite{Arnaldi:2008er}. The dash-dotted line shows the dilepton
yield from the in-medium $\rho$ with a broadened spectral function,
the dashed line presents the yield from the $q+\bar q$ annihilation,
the dash-dot-dot line gives the contribution of the gluon
Bremsstrahlung process ($q\bar q\to g l^+l^-$), while the solid line
is the sum of all contributions. For the description of the other
lines, which correspond to the non-dominant channels, we refer to
the figure legend. The figure is taken from
Ref.~\protect\cite{Linnyk:2011hz}. ({ r.h.s.}) Acceptance-corrected
invariant mass spectrum of excess dimuons in In+In collisions at 158
A GeV in comparison to model results from Renk and Ruppert, van Hees
and Rapp as well as Dusling and Zahed. The figure is taken from
Refs.~\protect{\cite{NA60,NA602,NA603,NA604,Arnaldi:2008er}}. }
\label{NA60_AC}
\end{figure}
\begin{figure}[t]
\begin{center} \includegraphics[width=0.45\textwidth]{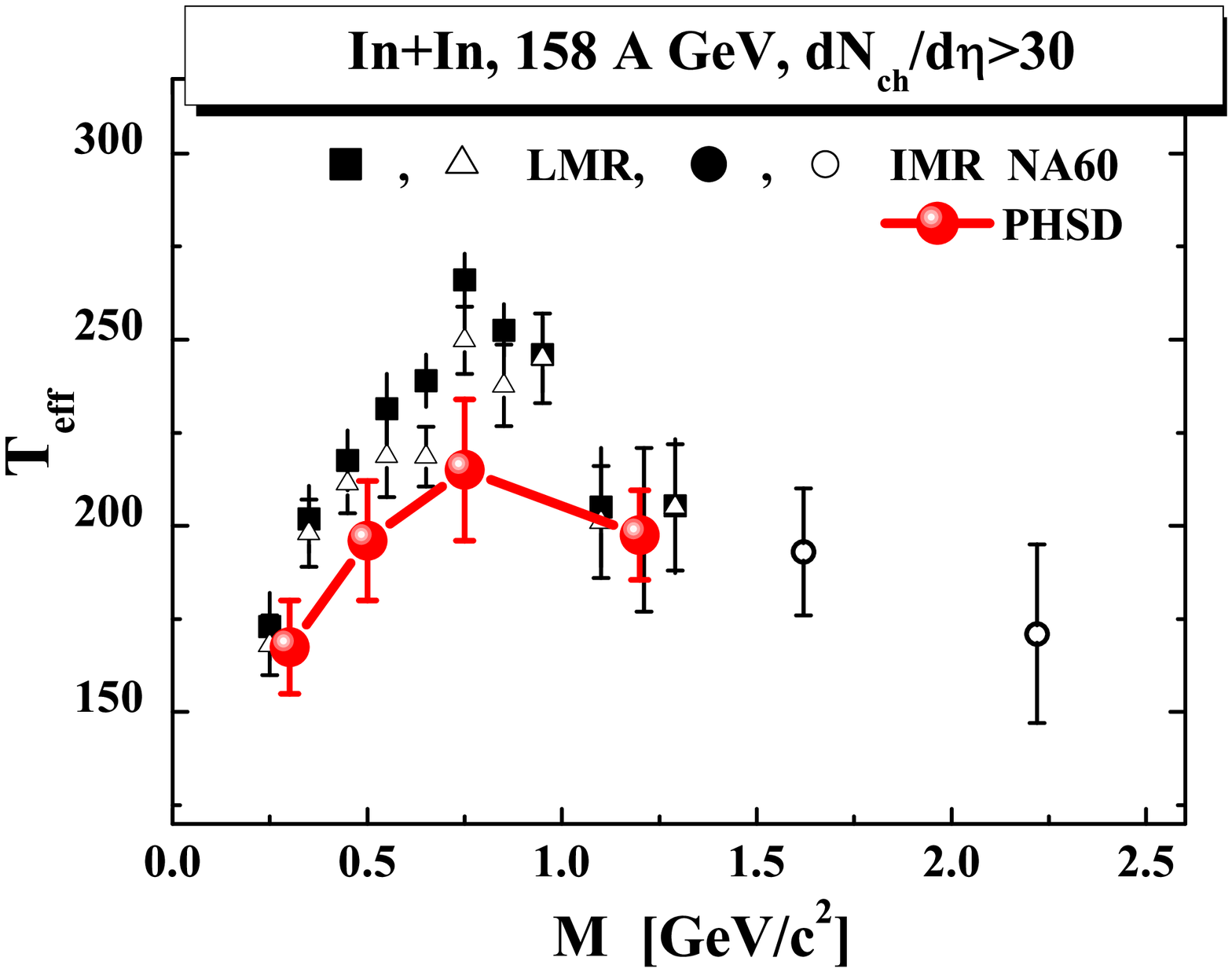}
\end{center}
\caption{The inverse slope parameter $T_{eff}$ of the dimuon yield
from In+In at 158 A GeV as a function of the dimuon invariant mass
$M$ in PHSD (solid line with full dots) compared to the data of the
NA60
Collaboration~\protect{\cite{Arnaldi:2008er,NA60,NA602,NA603,NA604}}.
The figure is taken from Ref.~\protect\cite{Linnyk:2011hz}.}
\label{NA60c}
\end{figure}

\subsection{{SPS energies}}

 We step up in energy and compare model results with
experimental data for dileptons from In+In collisions at 160 A GeV
measured by the NA60 Collaboration. In Fig.~\ref{NA60_AC} we present
PHSD results for the dilepton excess over the known hadronic sources
as produced in In+In reactions at 158~A GeV compared to the
acceptance corrected data. We find here that the spectrum at
invariant masses in the vicinity of the $\rho$-meson peak is well
reproduced by the $\rho$ meson yield, if a broadening of the meson
spectral function in the medium is assumed, while the partonic
sources account for the yield at high masses. Our analysis shows
that the contributions of the ``$4\pi$" processes (shown by the
lines with symbols) -- as first noted by the authors of
Ref.~\cite{Song:1994zs} -- are very much suppressed.

 One concludes from Fig.~\ref{NA60_AC} that the measured
spectrum for $M>1$~GeV is dominated by the {\it partonic} sources.
Indeed, the dominance of the radiation from {the} QGP over the
hadronic sources in PHSD is related to a rather long -- of the order
or 3 fm/c -- evolution in the partonic phase (in co-existence with
the space-time separated hadronic phase) on one hand (cf. Fig.~10 of
Ref.~\cite{CasBrat}) and the rather high initial energy densities
created in the collision on the other hand (cf. Fig.~6 of
Ref.~\cite{Linnyk:2008hp}). In addition, we find from
Fig.~\ref{NA60_AC} that in PHSD the partonic sources also have a
considerable contribution to the dilepton yield for $M<0.6$~GeV. The
yield from the two-to-two process $q+\bar q\to g+l^+l^-$ is
especially important close to the threshold ($ \approx 0.211$ GeV).
This conclusion from the microscopic calculation is in qualitative
agreement with the findings of an early (more schematic)
investigation in Ref. \cite{Alam:2009da}. For related results from
alternative models we refer the reader to the right panel of Fig.
\ref{NA60_AC}.

The comparison of the mass dependence of the slope parameter
evolution in PHSD and the data from NA60 is shown explicitly in
Fig.~\ref{NA60c}.  Including the partonic dilepton sources allows
to reproduce in PHSD the $m_T$-spectra as well as the finding of the
NA60 Collaboration~\cite{Arnaldi:2008er,NA60,NA602,NA603,NA604} that
the effective temperature of the dileptons (slope parameters) in the
intermediate mass range is lower than that of the dileptons in the
mass bin $0.6<M<1$~GeV, which is dominated by hadronic sources (cf.
Fig.~\ref{NA60c}).  The softening of the transverse mass spectrum
with growing invariant mass implies that the partonic channels occur
dominantly before the collective radial flow has developed. Also,
the fact that the slope in the lowest mass bin and the highest one
are approximately similar -- both in the data and in the PHSD -- can
be traced back to the two windows of the mass spectrum that in our
picture are influenced by the radiation from the sQGP: $M=2 M_\mu
-0.6$~GeV and $M>1$~GeV.  For more details we refer the reader to
Ref. \cite{Linnyk:2011hz}.

\subsection{{RHIC energies}}

Now we are coming to the top RHIC energy of
$\sqrt{s{_{NN}}}$ = 200 GeV and present the most important findings
from the PHSD study in Ref. \cite{Linnyk:2011vx}. In the left part
of Fig.~\ref{MRHIC} we show the PHSD results for the invariant mass
spectra of inclusive {dileptons in} Au+Au collisions for the
acceptance cuts on single electron transverse momenta $p_{eT}$,
pseudorapidities $\eta_e$, azimuthal angle $\phi_e$, and dilepton
pair rapidity $y$: $ p_{eT}>0.2 \mbox{ GeV}, \  |\eta_e|<0.35, \
-3\pi/16<\phi_e<5\pi/16, \ 11\pi/16 <\phi_e<19\pi/16,\ |y|<0.35.$

\begin{figure}
\hspace{0.3cm}
 \includegraphics[width=0.52\textwidth]{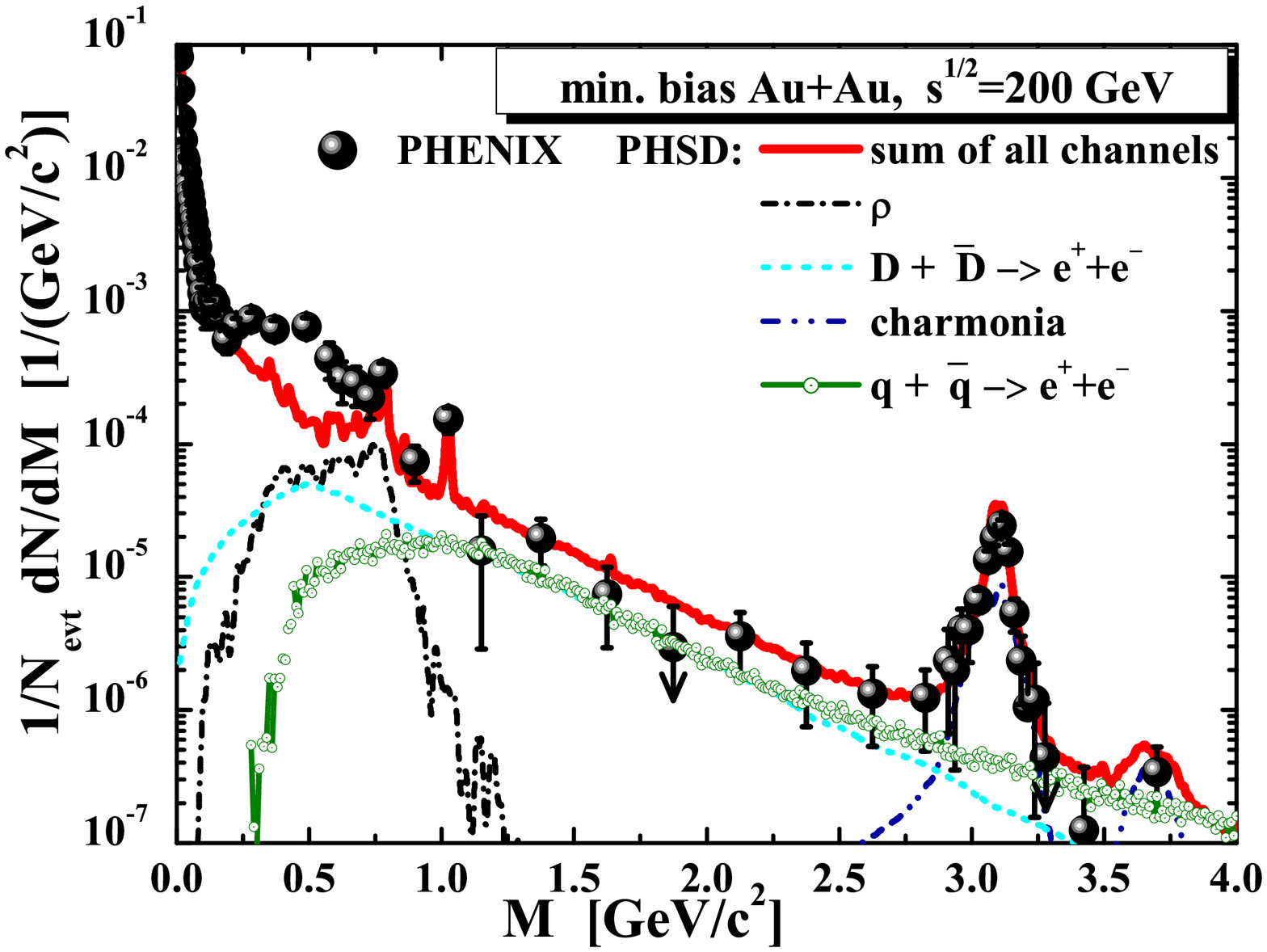}\hspace{0.5cm}
\includegraphics[width=0.42\textwidth]{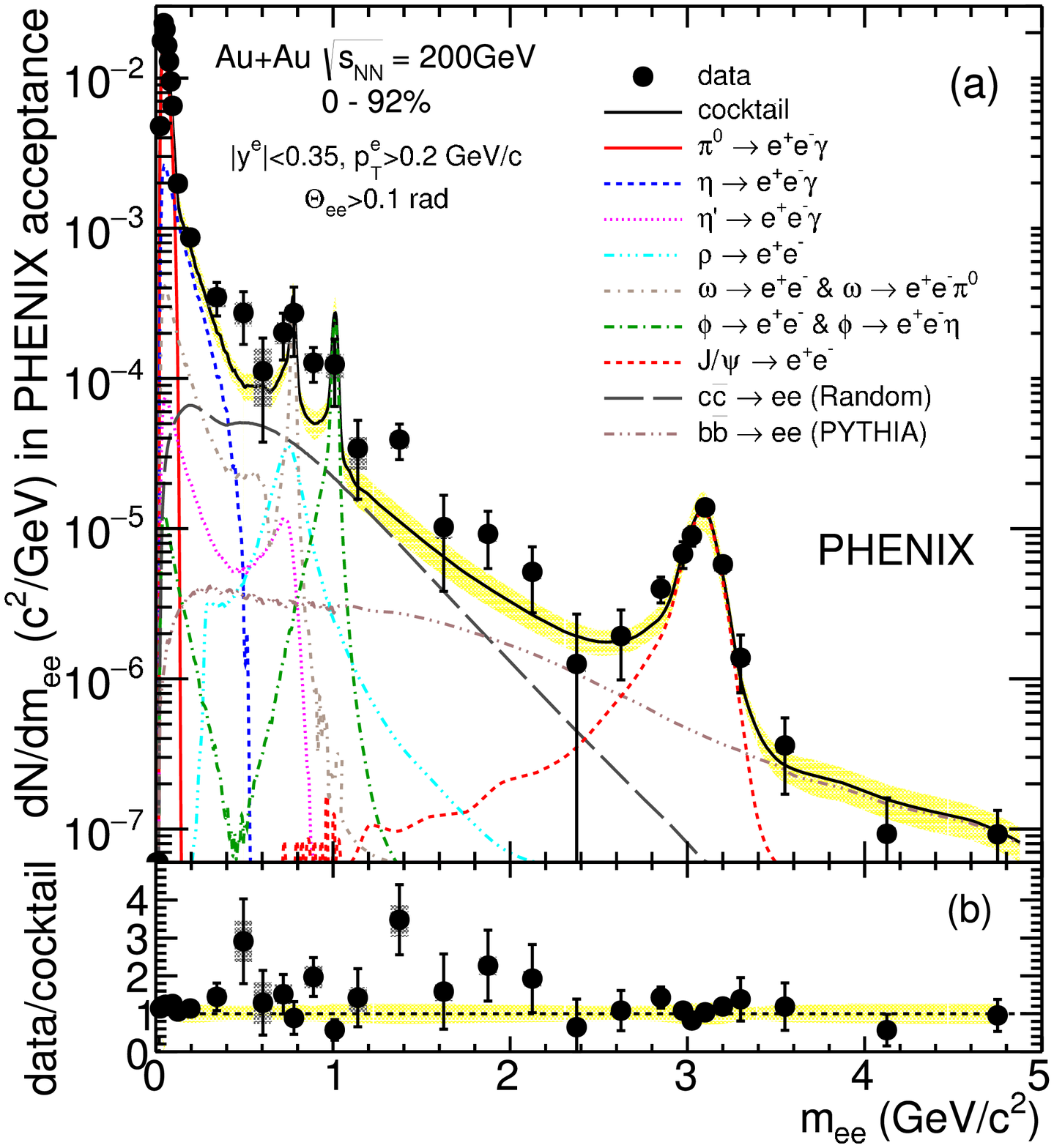}
 \caption{(l.h.s.) The PHSD results for the invariant mass spectra of
inclusive dilepton{s} in Au+Au collisions at $\sqrt{s{_{NN}}}$ = 200
GeV within the PHENIX acceptance cuts in comparison to the data from
the PHENIX
Collaboration~\protect{\cite{PHENIXlast,Toia:2005vr,Toia:2006zh,Afanasiev:2007xw}}
based on the data from 2004. The different lines indicate the
contributions from different channels as specified in the figure,
which is taken from Ref.~\protect\cite{Linnyk:2011vx}. (r.h.s.) New
data of the PHENIX Collaboration measured in 2010 with the
Hadron-Blind Detector compared to the cocktail of hadron decays. The
figure is taken from Ref.~\protect\cite{Adare:2015ila}. }
\label{MRHIC}
\end{figure}

\begin{figure} \centering
    \includegraphics[width=0.467\textwidth]{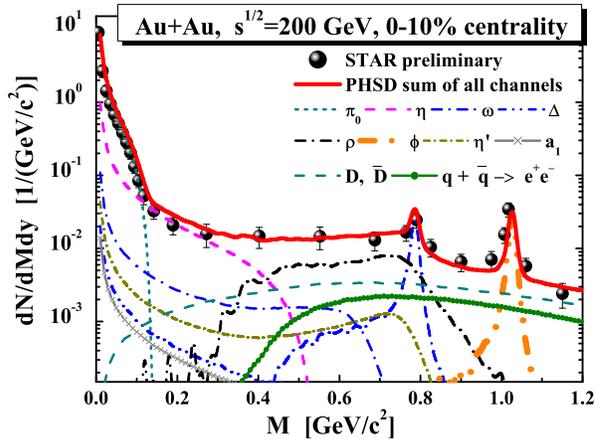}\hspace{0.3cm}
    \includegraphics[width=0.46\textwidth]{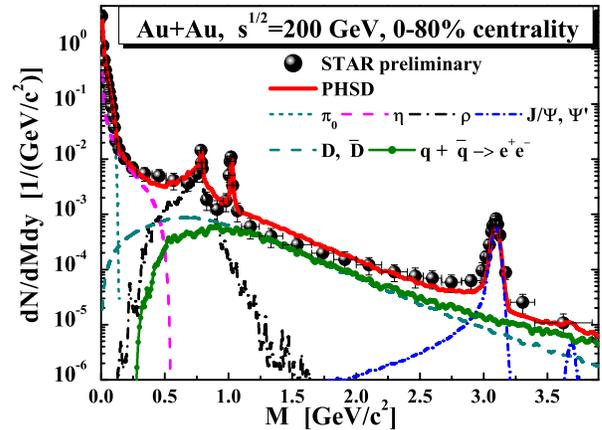}
\caption{The PHSD results for the {invariant mass} spectra of
{dileptons in Au+Au} collisions at $\sqrt{s{_{NN}}}$ = 200 GeV for
$M \! = \! 0 \! - \! 1.2$~GeV (left panel) and for
$M\!=\!0\!\,-\,\!4$~GeV (right panel) for 0 - 10 \% or 0 - 80\%
centrality within the cuts of the STAR experiment. The  data of the
STAR Collaboration are adopted from Ref. \protect\cite{Zhao:2011wa}.
The figures are taken from Ref.~\protect\cite{Linnyk:2011vx}. }
\label{MSTAR}
\end{figure}

\begin{figure}
\hspace{2cm}
\includegraphics*[width=0.75\textwidth]{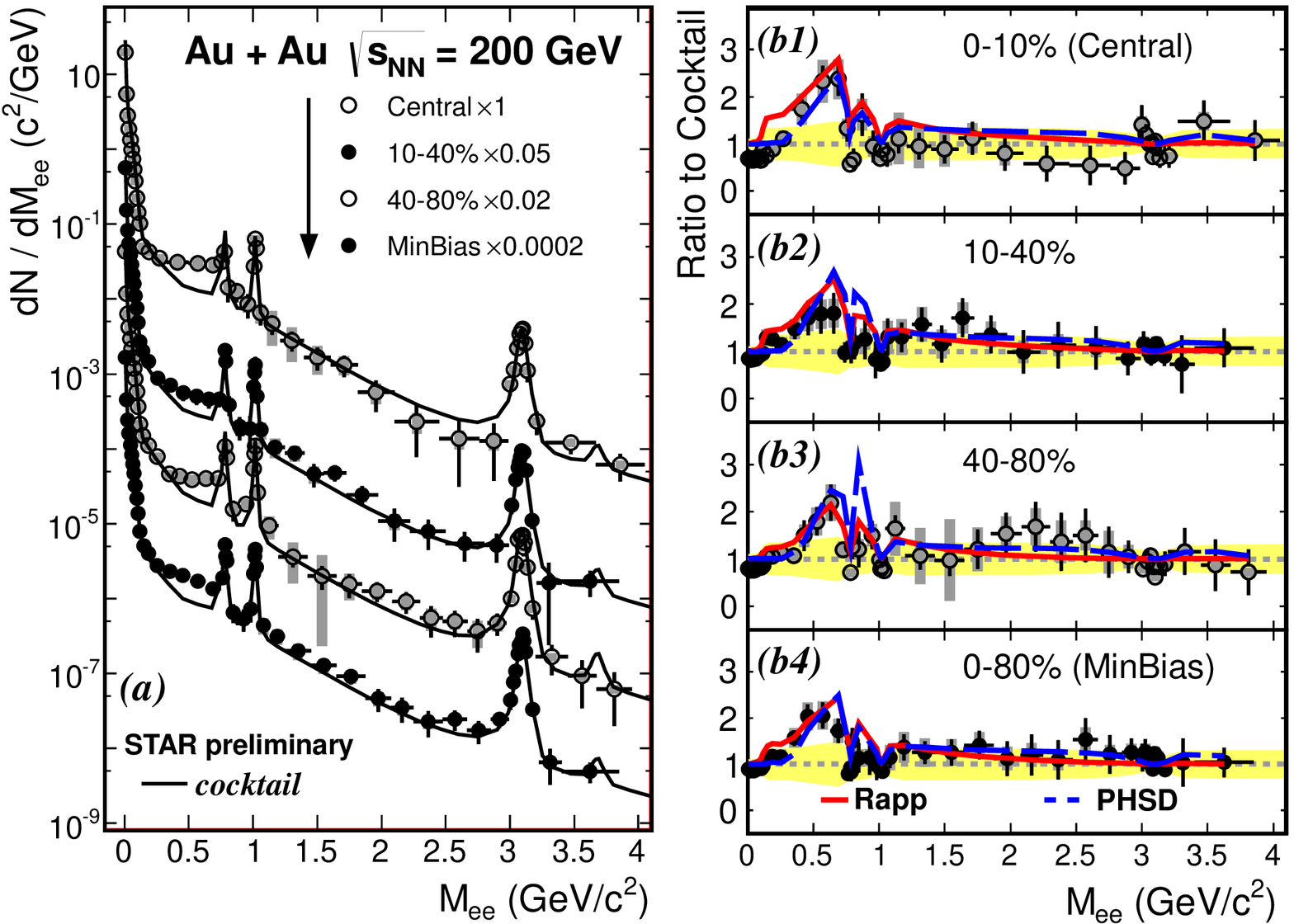}
\caption{Centrality dependence of the midrapidity dilepton yields
(left) and its ratios (right) to the 'cocktail' for 0-10\%, 10-40\%,
40-80\%, 0-80\% central Au+Au collisions at $\sqrt{s}=200$ GeV:
comparison of STAR data with theoretical predictions from the PHSD
('PHSD' - dashed lines) and the expanding fireball model ('Rapp' -
solid lines). The figures are taken from Ref.
\protect\cite{HuckQM14}.} \label{fig:dilSTAR200cd}
\end{figure}

In the low mass region $M=0-1.2$~GeV, the dilepton yield in the PHSD
is dominated by hadronic sources and roughly coincides with the
earlier HSD result in Ref.~\cite{Bratkovskaya:2008bf}. Note that the
collisional broadening scenario for the modification of the
$\rho$-meson was used in the calculations presented in
Fig.~\ref{MRHIC} that underestimates the PHENIX data from the run
2004 in the mass range from 0.2 to 0.7 GeV substantially. In
contrast, the partonic radiation as well as the yield  from
correlated $D$-meson decays dominate and saturate the mass region
$M=1-3$~GeV as seen in Fig.~\ref{MRHIC} (left panel), i.e. between
the $\phi$ and $J/\Psi$ peaks. The dileptons generated by the
quark-antiquark annihilation in the sQGP from PHSD constitute about
half of the observed yield in this intermediate{-}mass range. For
$M>2.5$~GeV the partonic yield is even higher than the D-meson
contribution. Thus, {the inclusion of the} partonic radiation in
{the} PHSD fills up the gap between the hadronic model
results~\cite{Bratkovskaya:2008bf} and the data of the PHENIX
Collaboration for $M>1$~GeV. However, the early expectation of a
strong partonic signal in the low mass dilepton spectrum is not
substantiated by the microscopic PHSD calculations.

In order to investigate the ``low-mass dilepton problem", the PHENIX
Collaboration has performed a new measurement in 2010 with a
different magnetic field setting, addition of the Hadron-Blind
Detector, and modified analysis. The results of this experimental
effort  (very recently presented in Ref.~\cite{Adare:2015ila}) are
shown in the right hand side of Fig.~\ref{MRHIC}. The new
measurements suggest that the dilepton yield in the low-mass region
from 0.2 to 0.7 GeV does no longer show such a strong enhancement
over the cocktail of hadronic decay sources as assumed based on the
earlier PHENIX analysis in
Ref.~\cite{PHENIXlast,Toia:2005vr,Toia:2006zh,Afanasiev:2007xw}. In
fact, the new PHENIX data are in agreement with the theoretical
expectations from the PHSD calculations.

In order to shed some further light on the ``PHENIX puzzle", we
compare the PHSD predictions with the data independently measured
for Au+Au collisions at $\sqrt{s{_{NN}}}$ = 200 GeV by the STAR
Collaboration. The calculations are performed for the same model
assumptions and parameters as those used for the comparison to the
PHENIX data, only the different acceptance cuts on single electron
transverse momenta $p_{eT}$, single electron pseudorapidities
$\eta_e$ and the dilepton pair rapidity $y$, i.e. $ 0.2<p_{eT}<5
\mbox{ GeV}, \ |\eta_e|<1, \ |y|<1$. The PHSD predictions for the
dilepton yield within these cuts are shown in Fig.~\ref{MSTAR} for
0-80\%. One can observe generally a good agreement with the data
from the STAR Collaboration~\cite{Zhao:2011wa} in the whole mass
regime. Notably, our calculations are also roughly in line with the
low mass dilepton spectrum from STAR in case of the most central
collisions, whereas the PHSD results severely underestimated the
PHENIX data from the Run 2004 analysis for central collisions. The
observed dilepton yield from STAR can be accounted for by the known
hadronic sources, i.e. the decays of the $\pi_0$, $\eta$, $\eta'$,
$\omega$, $\rho$, $\phi$ and $a_1$ mesons, of the $\Delta$ particle
and the semi-leptonic decays of the $D$ and $\bar D$ mesons, where
the collisional broadening of the $\rho$-meson is taken into
account.

More recently, the STAR Collaboration has released information on
the explicit centrality dependence of the dilepton spectra.   Fig.
\ref{fig:dilSTAR200cd} shows the comparison of the STAR data of
midrapidity dilepton yields (l.h.s.) and its ratios (r.h.s.) to the
'cocktail' for  0-10\%, 10-40\%, 40-80\%, 0-80\% central Au+Au
collisions at $\sqrt{s_{NN}}=200$ GeV in comparison to the
 predictions from the PHSD approach and the
expanding fireball model of Rapp and collaborators. As seen from
Fig. \ref{fig:dilSTAR200cd} the excess of the dilepton yield over
the expected cocktail is larger for very central collisions and
consistent with the model predictions including the collisional
broadening of the $\rho$-meson spectral function at low invariant
mass and QGP dominated radiations at intermediate masses.
Accordingly, the tension between the PHENIX and STAR dilepton data
(as well as PHSD predictions) no longer persists.

Moreover, the recent STAR dilepton data for Au+Au collisions from
the Beam Energy Scan (BES) program  for $\sqrt{s_{NN}}=19.6, 27, 39$
and 62.4 GeV \cite{Ruan:2014kia,HuckQM14} are also in line with the
PHSD  (as well as the expanding fireball model) predictions with a
$\rho$-meson collisional broadening. According to the PHSD
calculations the excess is increasing with decreasing energy due to
a longer $\rho$-meson propagation in the high-baryon density phase
(see Fig. 3 in Ref. \cite{Ruan:2014kia}).

\subsection{LHC energies}

On the other hand, the upcoming ALICE data \cite{dilALICE} for
heavy-ion dileptons for Pb+Pb at $\sqrt{s}$ = 2.76 TeV might give a
further access to the dileptons emitted from the QGP
\cite{Rapp13,Linnyk:2012pu}. In Fig. \ref{fig55} (l.h.s.)  we
present the PHSD predictions for central Pb+Pb collisions
\cite{Linnyk:2012pu} in the low mass sector for a lepton $p_T$ cut
of 1 GeV/c. It is clearly seen that the QGP sources and the
contribution from correlated $D{\bar D}$ pairs are non-leading in
the low mass regime where we find the conventional hadronic sources.
For a lepton $p_T$ cut of 1 GeV/c (l.h.s.) one practically cannot
identify an effect of the $\rho$-meson collisional broadening in the
dilepton spectra in the PHSD calculations. Only when applying a low
$p_T$ cut of 0.15 GeV/c a small enhancement of the dilepton yield
from 0.3 to 0.7 GeV becomes visible (r.h.s. of Fig. \ref{fig55}).
This low sensitivity to hadronic in-medium effects at LHC energies
from the PHSD is due to the fact that the hadrons appear late (after
hadronization) in central Pb+Pb collisions and are boosted to high
velocities due to the high pressure in the early partonic phase.

\begin{figure}
\hspace{0.1cm}
\includegraphics[width=0.46\textwidth]{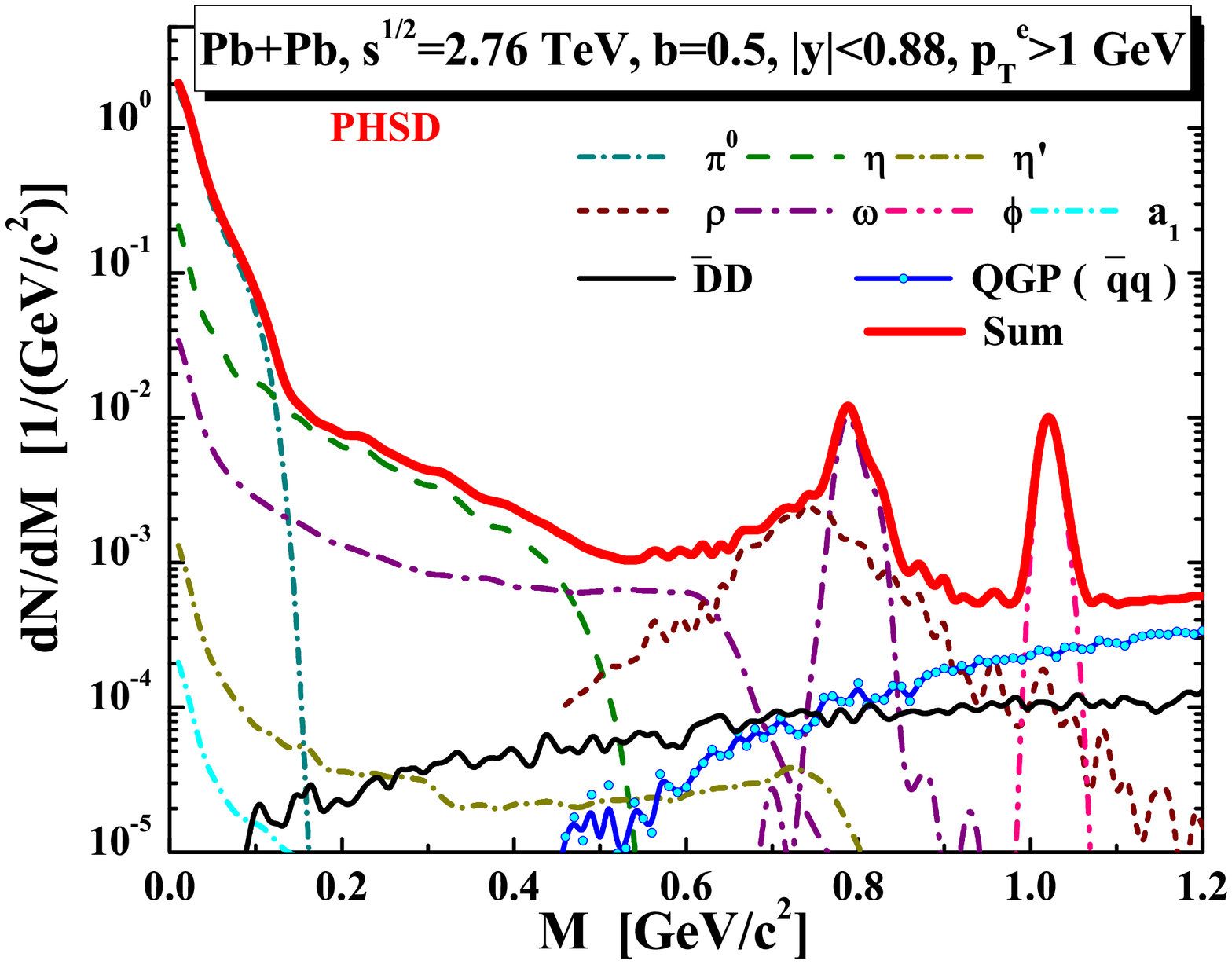} \hspace{0.1cm}
\includegraphics[width=0.48\textwidth]{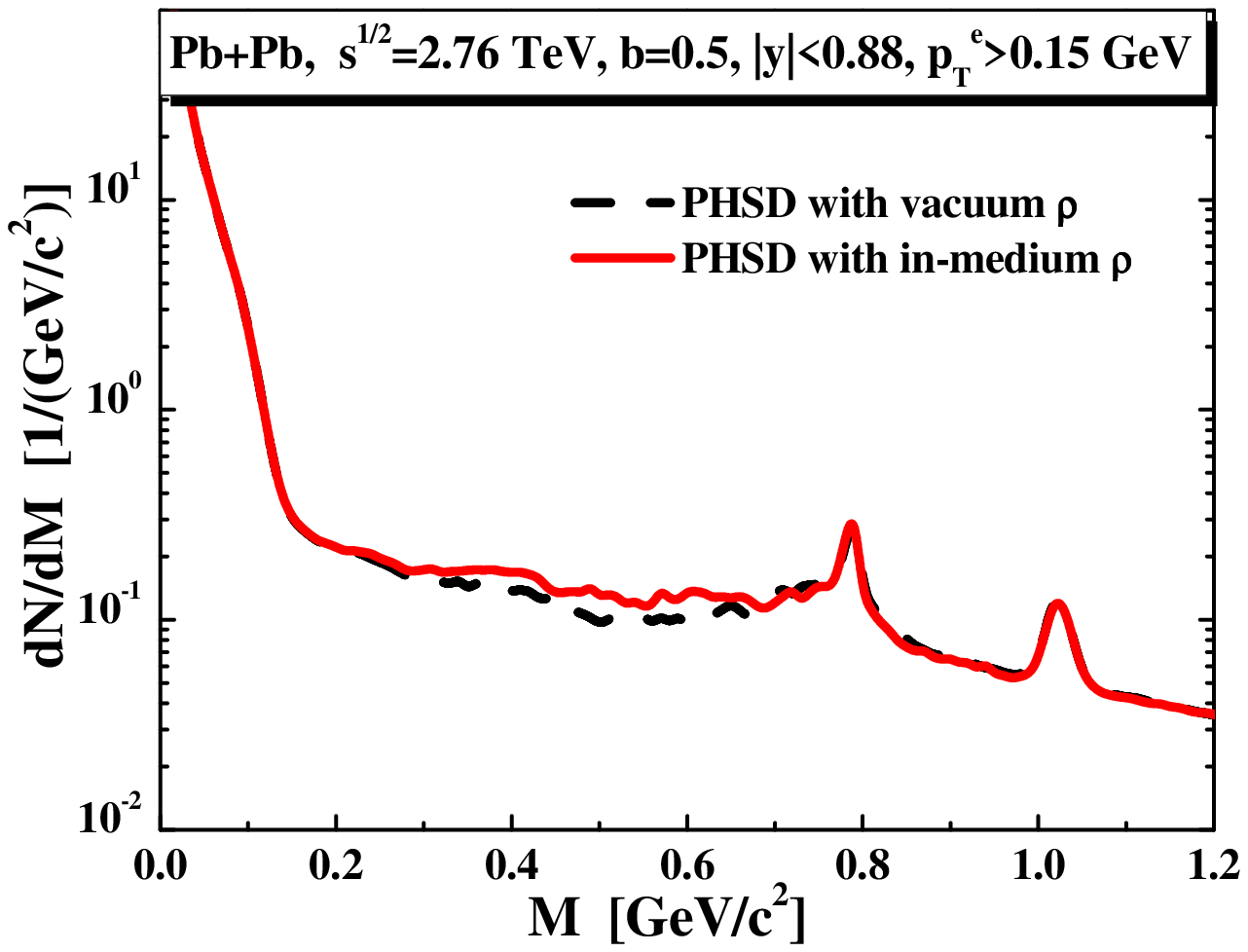}
\caption{(l.h.s.) Midrapidity dilepton yields for Pb+Pb at
$\sqrt{s_{NN}}$ = 2.76 TeV  for a lepton $p_T$ cut of 1 GeV/c. The
channel decomposition is explained in the legend. (r.h.s.) Same as
for the l.h.s. but for a lepton $p_T$ cut of 0.15 GeV/c for a 'free'
$\rho$ spectral function (dashed line) and the collisional
broadening scenario (solid line). The figures are taken from Ref.
\protect\cite{Linnyk:2012pu}.} \label{fig55}
\end{figure}

% --- v2, v3 -----
In the end, we mention that promising perspectives with dileptons
have been suggested in Ref. \cite{v3dil} to measure the flow
anisotropy coefficients  $v_n$ ($n=2,3$) similar to photons. The
calculations with the viscous (3+1)d MUSIC hydro for central Au+Au
collisions at RHIC energies show that the flow coefficients $v_2,
v_3$ are sensitive to the dilepton sources and to the EoS and
$\eta/s$ ratio. The main advantage of measuring flow coefficients
$v_n$ with dileptons compared to photons is the fact that the extra
degree-of-freedom $M$ might allow to disentangle the sources
additionally.

\section{Summary}

In this report we have addressed the dynamics of relativistic
heavy-ion reactions and in particular the information obtained from
electromagnetic probes that stem from the partonic and hadronic
phases. While the out-of-equilibrium description of strongly
interacting relativistic fields has been based on the theory of
Kadanoff and Baym (Section 2), the description of  QCD in
equilibrium has been performed within an effective dynamical
quasiparticle model (DQPM) (Section 3). The width of the dynamical
quasiparticles is controlled by transport coefficients in
equilibrium that can be compared to the same quantities from lattice
QCD (Section 4). The resulting off-shell transport approach is
denoted by Parton-Hadron-String Dynamics (PHSD) and reproduces the
equation of state, the sound velocity squared $c_s^2(T)$ as well as
the relevant transport coefficients such as the shear viscosity
$\eta$, the bulk viscosity $\zeta$ and the electrical conductivity
$\sigma_0$ in the partonic phase from lattice QCD. Furthermore, it
includes dynamical transition rates for hadronization, i.e. for the
change of colored partonic to color-neutral hadronic
degrees-of-freedom, that satisfy all conservation laws and do not
violate the second law of thermodynamics. It has been shown that the
PHSD captures the bulk dynamics of heavy-ion collisions from lower
SPS to LHC energies and thus provides a solid ground for the
evaluation of the electromagnetic emissivity on the basis of the
same dynamical propagators in the partonic phase that are employed
for the dynamical evolution of the partonic system (Section 5). The
PHSD 'tests'  indicate that the 'soft' physics at LHC in central A-A
reactions is very similar to the top RHIC energy regime although the
invariant energy is higher by more than an order of magnitude.
Furthermore, the PHSD approach is seen to work from lower SPS
energies up to LHC energies for p-p, p-A as well as A-A collisions,
i.e. over a range of more than two orders in $\sqrt{s_{NN}}$. Note
that for even lower bombarding energies the PHSD approach merges to
the HSD model, which has been successfully tested for p-A and A-A
reactions from the SIS to the SPS energy regime in the past
\cite{Cass99}.

The main messages from the {\em photon} studies in Section 6 can be
summarize in short as:
\begin{itemize}
\item %(i)
the photons  provide a critical test for the theoretical models: the
standard dynamical models - constructed to reproduce the 'hadronic
world' - fail to explain the photon experimental data;
\item
the details of the hydro models (fluctuating initial conditions,
viscosity, pre-equilibrium flow) have a small impact on the photon
observables;
\item
% (iii)
as suggested by the PHSD transport model calculations, the role of
such background
%trivial
sources as $mm$ and $mB$ bremsstrahlung has been underestimated in
the past and was found to be dominant at low photon $p_T$;
\item
% (iv)
the dynamics of the initial phases of the reaction might turn out to
be important (pre-equilibrium /'initial' flow, Glasma effect etc.).
\end{itemize}

Finally, one must conclude that the photons are one of the most
sensitive probes for the dynamics of HIC and for the role of the
partonic phase. We also mention that in an initial 'glasma' phase
the photon/dilepton production is suppressed by about an order of
magnitude since the gluon fields do not carry electric charge. In
this case the {\em direct} photons would practically stem for the
hadronic stages and carry the full
hadronic elliptic flow $v_2$. \\

The main messages from the dilepton studies in Section 7 are:
\begin{itemize}
\item
%(i)
at low masses ($M=0.2-0.6$ GeV/$c^2$) the dilepton spectra show
sizable changes due to hadronic in-medium effects, i.e. multiple
hadronic resonance formation or a modification of the properties of
vector mesons  (such as collisional broadening) in the hot and dense
hadronic medium (partially related to chiral symmetry restoration);
these effects can be observed at all energies from SIS
to LHC but are most pronounced in the FAIR/NICA energy regime; \\
\item
%(ii)
at intermediate masses the QGP ($q\bar q$ thermal radiation)
dominates for $M>1.2$ GeV/$c^2$. The fraction of QGP sources grows
with increasing energy and becomes dominant at the LHC energies. \\
\item
The tension between the PHENIX and STAR dilepton data at the top
RHIC energy (as well as PHSD predictions) no longer persists.
\end{itemize}

Finally,
the dilepton measurements within the future experimental energy and
system scan ($pp, pA, AA$) from low to top RHIC energies as well as
new ALICE data at LHC energies will extend our knowledge on the
properties of hadronic and partonic matter via its electromagnetic
radiation and show if the very initial degrees-of-freedom in relativistic
heavy-ion collisions are electrically charged (quarks and antiquarks) or not (gluons).

\section*{Acknowlegements}

The authors are grateful to J. Aichelin, H. Berrehrah, M. Bleicher,
C. Gale, G. David, A. Drees, H. v. Hees, U. Heinz, B. Jacak, B. K{\"
a}mpfer, C. M. Ko, V. Konchakovski, J. Manninen, V. Ozvenchuk, A.
Palmese, J.-F. Paquet, K. Reygers, E. Seifert, C. Shen, T. Song, J.
Stachel, T. Steinert, V. Toneev and K. Werner for valuable
discussions and their contributions to this review. The
computational resources have been provided by the LOEWE-CSC as well
as by the SKYLLA cluster at the Univ. of Giessen. This work in part
has been supported by DFG as well as by the LOEWE center HIC for
FAIR.

%----- REFERENCES --------------------

\section*{References}

%\bibliographystyle{elsarticle-num}

%\bibliographystyle{model1a-num-names}
%\bibliography{PHSDdileptEU}

\end{document}